\documentclass[showpacs,preprintnumbers,amsmath,amssymb,nofootinbib]{revtex4-1}
\usepackage{graphicx}% Include figure files
\usepackage{bm}% bold math

\usepackage{amssymb}
\usepackage{amsfonts}
\usepackage{amsmath}

\usepackage{color}
\usepackage[mathscr]{eucal}

\definecolor{nv}{rgb}{0.1,0.1,0.6}
\definecolor{pr}{rgb}{0.2,0.1,0.5}
\definecolor{mg}{rgb}{0.4,0.0,0.4}

\newcommand{\nn}{\nonumber}

\newcommand{\beq}{\begin{equation}}
\newcommand{\eeq}{\end{equation}}
\newcommand{\beqy}{\begin{eqnarray}}
\newcommand{\eeqy}{\end{eqnarray}}
\newcommand{\beqyn}{\begin{eqnarray*}}
\newcommand{\eeqyn}{\end{eqnarray*}}
\newcommand{\nl}{\newline}

\newcommand{\bs}{\begin{slide}}
\newcommand{\es}{\end{slide}}
\newcommand{\bc}{\begin{center}}
\newcommand{\ec}{\end{center}}
\newcommand{\bmin}{\begin{minipage}}
\newcommand{\emin}{\end{minipage}}

\newcommand{\bi}{\begin{itemize}}
\newcommand{\ei}{\end{itemize}}

\newcommand{\la}{\langle}
\newcommand{\ra}{\rangle}

\usepackage{latexsym}

\usepackage{dsfont}
\usepackage{multirow}

\newcommand{\bea}{\begin{eqnarray}}
\newcommand{\eea}{\end{eqnarray}}
\newcommand{\be}{\begin{equation}}
\newcommand{\ee}{\end{equation}}

\newcommand{\msc}{\mathscr}

\newcommand{\ud}{\mathrm{d}}
\newcommand{\uL}{\mathcal{L}}

\newcommand{\uTr}{\mathrm{Tr}}

\newcommand{\uslash}{/\!\!\!}

\newcommand{\barpsi}{\overline{\psi}}

\newlength\savedwidth

\newcommand\whline{\noalign{\global\savedwidth\arrayrulewidth
\global\arrayrulewidth 1pt}%
\hline
\noalign{\global\arrayrulewidth\savedwidth}}
\newcommand{\uvec}[1]{\boldsymbol{#1}}

\newcommand{\uind}[2]{^{#1}_{\phantom{#1}#2}}
\newcommand{\lind}[2]{^{\phantom{#1}#2}_{#1}}
\newcommand{\pure}{\text{pure}}
\newcommand{\phys}{\text{phys}}

\newcommand{\LRpartial}{\overset{\leftrightarrow}{\partial}\!\!\!\!\phantom{\partial}}
\newcommand{\LRD}{\overset{\leftrightarrow}{D}\!\!\!\!\!\phantom{D}}

\newcommand{\Keywords}[1]{\par\noindent
{\small{\em Keywords\/}: #1}}

\begin{document}

%\preprint{APS/123-QED}

\title{The angular momentum controversy:\\ What's it all about and does it matter?}

\author{Elliot Leader}
\email{e.leader@imperial.ac.uk}
\affiliation{Imperial College London\\ Prince Consort Road, London SW7 2AZ }

\author{C\'edric Lorc\'e}
\email{lorce@ipno.in2p3.fr}
\email{C.Lorce@ulg.ac.be}
\affiliation{IPNO, Universit\'e Paris-Sud, CNRS/IN2P3, 91406 Orsay, France}
\affiliation{IFPA,  AGO Department, Universit\'e de Li\` ege, Sart-Tilman, 4000 Li\`ege, Belgium}

\date{\today}

\begin{abstract}
The general question, crucial to an understanding of the internal structure of the nucleon, of how to split the total angular momentum of a photon or gluon into spin and orbital contributions is one of the most important and interesting challenges faced by gauge theories like Quantum Electrodynamics and Quantum Chromodynamics. This is particularly challenging since all QED textbooks state that such an splitting cannot be done for a photon (and \emph{a fortiori}  for a gluon) in a gauge-invariant way, yet experimentalists around the world are engaged in measuring what they believe is the gluon spin! This question has been a subject of intense debate and controversy, ever since, in 2008, it was claimed that such a gauge-invariant split was, in fact, possible. We explain in what sense this claim is true and how it turns out that one of the main problems is that  such a decomposition is not unique and therefore raises the question of what is the most natural or physical choice. The essential requirement of measurability does not solve the ambiguities and leads us to the conclusion that the choice of a particular decomposition is essentially a matter of taste and convenience. In this review, we provide a pedagogical introduction to the question of angular momentum decomposition in a gauge theory, present the main relevant decompositions and discuss in detail several aspects of the controversies regarding the question of gauge invariance, frame dependence, uniqueness and measurability. We stress the physical implications of the recent developments and collect into a separate section all the sum rules and relations which we think experimentally relevant . We hope that such a review will make the matter amenable to a broader community and will help to clarify the present situation.
\end{abstract}


\maketitle

\tableofcontents

\section{Introduction\label{secI}}

It is well known that the momentum density in a classical electromagnetic field is given by the Poynting vector $\uvec E\times \uvec B$, and it is therefore eminently reasonable that the angular momentum density should be given by $\uvec x\times(\uvec E\times\uvec B)$. Although this expression has the structure of an orbital angular momentum, \emph{i.e.} $\uvec r\times\uvec p$, it is, in fact, the \emph{total} photon angular momentum density. Moreover, in Quantum Chromodynamics (QCD), aside from a sum over colors, a completely analogous expression holds for the gluon angular momentum density, and this has been a cause of confusion for the following reason. Over the past four decades, several major experimental groups, the European Muon Collaboration, the New Muon Collaboration, the Spin Muon Collaboration, HERMES, COMPASS, STAR, and PHENIX have been straining themselves in an effort to measure the quantity $\Delta G(x)$, which plays a role in the perturbative QCD treatment of deep inelastic inclusive reactions like $e + p \rightarrow e' + X $ and semi-inclusive ones like $e + p \rightarrow e' + \text{hadron} + X$  and $p + p \rightarrow \text{hadron} + X $, and which is usually referred to as the \emph{polarization} or \emph{spin} of the gluon in a nucleon. Unfortunately, there is nothing like a spin term in the above expressions for total angular momentum.

However, if one starts with \emph{e.g.} the Lagrangian of Quantum Electrodynamics (QED), which is invariant under rotations, and applies Noether's theorem, one obtains a completely different expression for the angular momentum, known as the \emph{canonical} form
\beq
\uvec J_\text{photon} = \uvec S_\text{photon} + \uvec L_\text{photon},
\eeq
in which the total angular momentum is split into a spin part and an orbital part. However, as will be seen in detail later, in contradistinction to the earlier expressions, both the spin part and the orbital part depend on the vector potential $\uvec A$, and since $\uvec A\mapsto \uvec A - \uvec \nabla \alpha$ under a gauge transformation, it means that both $\uvec S$ and $\uvec L$ change under a gauge transformation. Indeed, serious textbooks on QED have, for the past 60 years, stressed that the photon total angular momentum cannot be separated into a spin part and an orbital part \emph{in a gauge-invariant} way, which is a matter of concern and intense discussions in QED \cite{Wentzel:1949,Gottfried:1966,Merzbacher:1970,Lenstra:1982,Allen:1992zz,vanEnk:1992,Nienhuis:1993,Barnett:1994,vanEnk:1994,Barnett:2002,Jauregui:2005,Calvo:2006,Hacyan:2006,Hacyan:2006,Chen:2008ag,Nieminen:2008,Li:2009,Berry:2009,Mazilu:2009,Aiello:2009a,Aiello:2009b,Barnett:2010,Stewart:2010ft,Bialynicki:2011}.

Now it is quite clear that something that is experimentally measurable cannot change under a gauge transformation. So, how is it possible that in measuring $\Delta G(x)$ we are claiming to measure the spin of the gluon? We believe the answer is absolutely straightforward. The quantity $\Delta G(x)$ that we measure is certainly gauge invariant, but it is not in general, indeed cannot be, the same as the gluon spin. What actually happens is that $\Delta G(x)$ coincides with the gauge non-invariant gluon spin, when the latter is evaluated in the particular gauge (called the light-front gauge) $A^+\equiv \tfrac{1}{\sqrt{2}}\,(A_0 + A_z)=0$. That it is the spin in a particular gauge that is measured should not be considered in a negative light, because gauge theories are very subtle and ``look different'' in different gauges. In fact, what we call the parton model, which predates QCD, is best considered as a picture of QCD in the light-front gauge. Bashinsky and Jaffe have stated this very forcefully: ``one should make clear what a quark or a gluon parton is in an interacting theory. The subtlety here is the issue of gauge invariance: a pure quark field in one gauge is a superposition of quarks and gluons in another. Different ways of gluon-field gauge fixing predetermine different decompositions of the coupled quark-gluon fields into quark and gluon degrees of freedom.''

We feel perfectly comfortable with this interpretation, but others do not, and in 2008 Chen, Lu, Sun, Wang and Goldman (later referred to as Chen \emph{et al.}) set the cat amongst the pigeons when they claimed, effectively, that all the QED textbooks were wrong, and that it was possible to split the photon angular momentum, \emph{in a gauge-invariant way}, into a spin part and an orbital part. Their publication aroused an aggressive response, with published letters flying back and forth, loaded with criticisms and rebuttals. What Chen \emph{et al.} did was to split the vector potential into two terms, which they called ``pure'' and ``physical''
\beq
\uvec A = \uvec A_\pure + \uvec A_\phys
\eeq
satisfying the constraints
\beq
\uvec\nabla\cdot \uvec A_\phys =0, \qquad  \qquad \uvec\nabla\times \uvec A_\pure =\uvec 0,
\eeq
and where $\uvec A_\phys$ is invariant under gauge transformations, whereas $\uvec A_\pure\mapsto  \uvec A_\pure - \uvec\nabla \alpha$. By adding a spatial divergence term to the classical form, $\uvec x\times (\uvec E\times \uvec B)$, which in Quantum Field Theory is referred to as the Belinfante form  $\uvec J_\text{Bel}(x)$, they were able to split $\uvec J_\text{Bel}(x)$ into a spin part and an orbital part, involving only $\uvec A_\text{phys}$, and therefore gauge invariant. Since the actual angular momentum is a space integral of the angular momentum density, one has by Gauss's theorem
\beq
\uvec J_\text{Bel} =  \int \ud^3x  \, \uvec J_\text{Bel}(x)   =  \uvec J_\text{Chen} + \text{surface integral at spatial infinity}.
\eeq
Provided the fields vanish at infinity, the surface term may be disregarded, and one has for the total angular momentum
\beq
\uvec J_\text{Bel}=\uvec J_\text{Chen}.
\eeq
So have Chen \emph{et al.} really shown that the textbooks are wrong? In fact no, as can be seen by asking, for example, for an explicit expression for $\uvec A_\phys$. It is easy to see that one can express $\uvec A_\phys$ in terms of $\uvec A$ in the following way
\beq
\uvec A_\phys = \uvec A -\uvec\nabla \frac{1}{\uvec\nabla ^2}\uvec\nabla\cdot \uvec A. \eeq
This looks innocuous, but it should be recalled that $\tfrac{1}{\uvec\nabla^2}$ is \emph{not}  a differential, but an integral operator
\beq
\frac{1}{\uvec\nabla ^2} f(\uvec x) \equiv -\frac{1}{4\pi}\int \ud^3x' \, \frac{f(\uvec x')}{| \uvec x - \uvec x'|},
\eeq
so that $\uvec A_\phys$ involves an integral over all space of a function of $\uvec A$. It is thus \emph{not} a local field and hence outside the category of fields discussed in the textbooks.

Nonetheless the Chen \emph{et al.} paper catalyzed a vast outpouring of theoretical papers \cite{Tiwari:2008nz,Ji:2009fu,Ji:2010zza,Ji:2012gc,Chen:2008gv,Chen:2008ja,Chen:2009dg,Chen:2011zzh, Wong:2010rs,Wang:2010ao,Chen:2011gn,Chen:2012vg,Goldman:2011vs,Stoilov:2010pv,Wakamatsu:2010qj,Wakamatsu:2010cb,Wakamatsu:2011mb,Wakamatsu:2012ve,Wakamatsu:2013voa,Cho:2010cw,Cho:2011ee,Hatta:2011zs,Hatta:2011ku,Hatta:2012cs,Hatta:2012jm,Zhang:2011rn,Leader:2011za,Lorce:2012ce,Lorce:2012rr,Lorce:2013gxa,Lorce:2013bja}, generalizing their original approach, which was three-dimensional, to  a four-dimensional covariant treatment, discovering several other, different ways to perform the split of $A^\mu$, and finally demonstrating that there are an infinite number of different ways to do this! The negative side to this is that there are, in principle, an infinite number of ways to define which operator should represent the momentum and angular momentum of quarks and gluons, and it seems there is no unique, compelling  argument for making any  one particular choice. It could be argued that the canonical choice is ``best'', because, as will be discussed later, the canonical angular momentum operator of, say, a quark at least generates rotations of that quark field, albeit in a slightly qualified form. But in the end, it seems to us that any choice is acceptable so long as it is made clear which definition one is using. However, we have asked ourselves whether there is really any point in going beyond the canonical and Belinfante versions and, in particular, whether there is any new physical \emph{content} in the other versions, and regrettably have come to the conclusion that there are only two fundamental versions of the angular momentum (the same is true for the linear momentum): the Belinfante and the canonical ones. We do not think that the other variants provide any further physical insight.
\newline

We wish to draw the reader's attention to the fact that these two distinct fundamental versions exist already at the level of ordinary Classical Mechanics, where the Belinfante momentum is called the ``kinetic'' momentum. It is therefore instructive to see what the different versions correspond to. Thus, the \emph{kinetic} momentum is defined as mass times velocity $\uvec p_\text{kin}=m \uvec v=m\dot{\uvec x}$. It corresponds to our classical intuition, where particles follow well-defined trajectories. It is also the momentum appearing in the non-relativistic expression for the particle kinetic energy $E_\text{kin}=\uvec p_\text{kin}^2/2m$. The other is the \emph{canonical} momentum $\uvec p$, which is used in the Hamiltonian form of Classical Mechanics and, crucially, in Quantum Mechanics. Thus, the Heisenberg uncertainty relations between position and momentum involve this form of momentum
 \beq
[x^i, p^j]= i\hbar\,\delta^{ij}.
\eeq
It is defined as $\uvec p=\partial L/\partial \dot{\uvec x}$, where $L$ is the Lagrangian of the system. Like the particle position $\uvec x$, it is a dynamical variable in the Hamiltonian formalism, which deals with coordinates and their canonically conjugate momenta. It is also the generator of translations.

For a particle moving in a potential $V(\uvec x)$
\beq
 L= E_\text{kin}- V = \tfrac{1}{2}\,m\dot{\uvec x}^2-V(\uvec x),
\eeq
so that
\beq
\uvec p= m \dot{\uvec x} = \uvec p_\text{kin},
\eeq
and there is no distinction between kinetic and canonical momentum. However, in the presence of electromagnetic fields, matters are different. To illustrate this, consider the classical problem of a charged particle, say an electron with charge $e$, moving in a fixed homogeneous \emph{external} magnetic field $\uvec B=(0,0,B)$. We know  that the particle follows a helical trajectory, so that at each instant, the particle kinetic momentum $\uvec p_\text{kin}$ points toward a different direction. The Lagrangian is given by (disregarding the electron spin)
\beq
L= \tfrac{1}{2}\,m \dot{\uvec x}^2 - e \dot{\uvec x}\cdot\uvec A,
\eeq
where $\uvec A$ is the vector potential responsible for the magnetic field $\uvec B=\uvec\nabla\times\uvec A$. It leads to
\beq
\uvec p = \uvec p_\text{kin}[\uvec x(t)] - e \uvec A[\uvec x(t)].
\eeq
A suitable vector potential is $\uvec A=\tfrac{1}{2}\,(-yB, xB, 0)$, from which  one sees, \emph{via} the Euler-Lagrange equations, that $\uvec p$ is a constant of motion.

However, exactly the same magnetic field is obtained from the vector potential $ \tilde{\uvec A} = \uvec A - \uvec\nabla\alpha$, where $\alpha(\uvec x)$ is any smooth function. This change in $\uvec A$ is, as mentioned above, a gauge transformation and does not affect the physical motion of the particle. However, it clearly changes $\uvec p$. It is said that $\uvec p$ is a gauge \emph{non-invariant} quantity, and we shall see later that one of the key issues in the controversy is whether such quantities can be measurable. It will turn out that sometimes the expectation value of a gauge non-invariant operator is \emph{gauge independent}. And sometimes it turns out that a gauge non-invariant quantity, when evaluated in a particular choice of gauge, is of fundamental interest and can be measured. An important example of the latter is precisely the gluon polarization $\Delta G(x)$, which can be measured, and which, as mentioned, coincides with the gluon helicity evaluated in the light-front gauge $A^+=0$.

In a classical picture, it is more natural to consider that the kinetic linear and angular momenta are the physical ones. The reason is that they have a direct connection with the particle motion in an external field. Moreover, one can always formulate the problems of Classical Electrodynamics in the Newtonian formalism, and therefore avoid the use of canonical quantities, as well as the problem of gauge invariance. In a quantum-mechanical picture, the canonical linear and angular momenta appear more natural. One reason is because they are the quantities which appear in the uncertainty relations. The second is that, in absence of well-defined trajectories, the only natural definition of linear and angular momenta is as the generators of translations and rotations. Thirdly, the canonical quantization rules are formulated in the Hamiltonian formalism, and so one can hardly avoid the use of canonical quantities. Nonetheless, especially in Field Theory, opinions  differ as to whether the canonical or kinetic or any other version is the more ``physical'' one.
\newline

Returning after this digression to QCD, the Belinfante and the canonical decompositions provide different and complementary information about the internal structure of the nucleon, and it is therefore important to try to measure experimentally the various terms in the decompositions given later. To this end, we shall discuss at length various sum rules and relations connecting these terms with experimentally measurable quantities.

A detailed outline of our study follows, suitable  for the reader interested in all the theoretical developments. However, note that at the end we  suggest a shortened way to read our paper, aimed at the reader principally interested in the physical implications.

In section \ref{secII}, we give a pedagogical introduction to the whole subject, reminding the reader of the concept of energy-momentum and angular momentum densities and their role in forming the momentum and angular momentum of a system in a field theory. We also remind the reader that, already in Classical Mechanics, there exist two versions of momentum, the kinetic and the canonical, and that it is the latter type that occurs in Quantum Mechanics. We then explain how these appear in Quantum Field Theory under the guise of Belinfante and canonical versions of momentum and angular momentum, and how they are related to each other. Here, and throughout the paper, we use QED, rather than QCD,  to illustrate issues, so as to minimize the technical complications. In section \ref{secIII}, we give the detailed structure of the energy-momentum and angular momentum densities for QED and QCD, in both the canonical and Belinfante versions. Section \ref{secIV} introduces the idea of Chen \emph{et al.}, which provoked the whole controversy, and explains various developments and extensions of the treatment in their original paper. It turns out their approach is just one of an infinite family of ways to achieve their aim, and the family members are related by a new, so-called Stueckelberg, symmetry. In section \ref{secV}, there is a full-scale discussion of the situation in QCD. It is shown that all the various published versions for decomposing the nucleon spin can be summarized in a ``master'' decomposition. The various versions then correspond simply to different rearrangements of the master terms. Section \ref{secVI} discusses the tricky question of the relation between the angular momentum of the system or of its constituents and the matrix elements of the energy-momentum tensor. It is explained how and why this has been the source of errors in several papers. Based on this, we are able to discuss the sum rules, involving experimentally measurable quantities like GPDs, which follow from the conservation of angular momentum, and several relations which allow the evaluation of the contributions of quarks and gluons to the nucleon spin \emph{via} measurable quantities. The developments in this section suggest the importance of orbital angular momentum. Section \ref{secVII} is therefore devoted to a discussion of orbital angular momentum and how it can be measured. Several possibilities emerge. The most practical at present is from lattice calculations, where quite beautiful results have been achieved. The orbital angular momentum can also be approached using quark models with light-front  wave-functions, and in, principle, \emph{via} twist-3 GPDs. Finally, we try in section \ref{secVIII} to gather together in one place all the relevant sum rules and relations that have practical experimental implications. Our conclusions follow.

Because many of the developments are highly technical, we would like to suggest that the reader principally interested in the physical implications should read the \textbf{Pedagogical Introduction} in section \ref{secII}, then section \ref{secIVA} to understand what the controversy is about, followed by section \ref{secVI} on \textbf{Angular Momentum Sum Rules and Relations}, and section \ref{secVII} on \textbf{ Orbital Angular Momentum and the Spin Crisis in the Parton Model}. Finally, in the \textbf{Qualitative Summary and Experimental Implications} in section \ref{secVIII} can be found a resum\'{e} of all the sum rules and relations which have direct practical implications.

\section{Pedagogical introduction\label{secII}}

\subsection{A reminder about Lorentz and Translational Invariance in Field Theory\label{IIA}}
We shall make the standard assumption that we are dealing with theories, such as QED and QCD, which are invariant under Lorentz transformations and space-time translations. The combined group of such transformations is known as the Poincar\'{e} group.

Under a space-time translation, any local field $\phi(x)$ obeys the rule
\beq \label{STTr}
\phi(x+a) = e^{iP\cdot a} \phi (x) e^{-iP\cdot a}.
\eeq
For a space translation, this becomes
\beq \label{spaceTR}
\phi (t, \uvec x+ \uvec a) = e^{-i\uvec P\cdot \uvec a} \phi (t,\uvec x) e^{i\uvec P\cdot \uvec a},
\eeq
where $\uvec P$ is the total three-momentum operator of the theory. Under a time translation, one has
\beq \label{timeTR}
\phi(t + a^0, \uvec x) =  e^{iP^0 a^0} \phi (t,\uvec x) e^{-iP^0 a^0},
\eeq
where $P^0$ is the total energy operator of the theory.

By considering  infinitesimal transformations, one finds the commutation relations
\beq  \label{Eq:Pcomm}
i \left[P^\mu,\phi\right] =  \frac{\partial \phi}{\partial x_\mu} \equiv  \partial^\mu\phi,
\eeq
the time component of which, namely
\beq
i\left[P^0, \phi\right]= \frac{\ud\phi}{\ud t},
\eeq
is simply the Heisenberg equation of motion for $\phi$. The four operators $P^\mu= (P^0, \uvec P)$ are, by the above, the \emph{generators} of space-time translations.
%\footnote{Strictly speaking, the generators are the $P_\mu$.}
Moreover, since they represent the total energy and momentum of the system, they are conserved quantities and are independent of time.

Under a homogeneous Lorentz transformation, the coordinates behave as
\beq \label{FinieLT}
x^\mu \mapsto x'^\mu = \Lambda\uind{\mu}{\nu} x^\nu.
\eeq
For an infinitesimal transformation, specified by the  infinitesimal parameters $\omega_{\mu\nu}=-\omega_{\nu\mu}$, this becomes
\beq \label{Ltrans}
x^\mu \mapsto x'^\mu = \left(\delta^\mu_\nu+\omega\uind{\mu}{\nu}\right)x^\nu,
\eeq
and the generic fields $\phi_r(x)$, where $r$ denotes the field component, transform as
\beq
\phi_r(x)\mapsto \phi'_r(x')=\phi_r(x) -\tfrac{i}{2}\,\omega_{\mu\nu}(\Sigma^{\mu\nu})\lind{r}{s}\phi_s(x),
\eeq
where $(\Sigma^{\mu\nu})\lind{r}{s}= -(\Sigma^{\nu\mu})\lind{r}{s}$ is an operator related to the spin of the particle. For example, for particles with the most common spins, one has
\begin{align} \hspace{-1cm}
\text{spin-$0$ particle}&  &&\phi(x) &(\Sigma^{\mu\nu})\lind{r}{s}&=0,\\
\text{spin-$1/2$ Dirac particle}& && \psi_r(x) & (\Sigma^{\mu\nu})\lind{r}{s}&= \tfrac{1}{2} \left(\sigma ^{\mu\nu}\right)\lind{r}{s},\label{spinop}\\
\text{spin-$1$ particle} & && A_\alpha(x)  &(\Sigma^{\mu\nu})\lind{\alpha}{\beta}  &= i \left(\delta^\mu_\alpha \,g^{\nu\beta} -\delta^\nu_\alpha \, g^{\mu\beta}\right).\label{spin1op}
\end{align}

Analogously to the case of momentum, one can introduce the six canonical generators of Lorentz transformations $M^{\mu\nu}$, antisymmetric under $\mu\leftrightarrow\nu$, whose commutation relations with any field $\phi_r(x)$ are
\beq \label{Mmunucan}
i\left[M^{\mu\nu}  ,  \phi_r\right] = \left(x^\mu\partial^\nu - x^\nu \partial^\mu\right)\phi_r - i(\Sigma^{\mu\nu})\lind{r}{s}\phi_s.
\eeq
Of the six independent operators $M^{\mu\nu}$, three, corresponding to the spatial components $M^{jk}$, are related to the conserved total angular momentum operators $\uvec J$, which generate rotations about the $x$, $y$ and $z$ axes, namely
\beq \label{J}
J^i = \tfrac{1}{2}\, \epsilon^{ijk}M^{jk},
\eeq
so that, for example, $J_z=J^3=M^{12}$, \emph{etc}. For the case of a Dirac particle, Eqs.~\eqref{spinop}, \eqref{Mmunucan} and \eqref{J} yield the well-known result
\beq \label{JDirac}
[\uvec J, \psi] = -\left(\uvec x\times\tfrac{1}{i}\uvec\nabla\right) \psi - \tfrac{1}{2}\uvec\Sigma\,\psi,
\eeq
where
\beq \label{Eq:sigma}
\Sigma^i=\tfrac{1}{2}\,\epsilon^{ijk}\sigma^{jk},  \qquad \qquad
\uvec\Sigma = \left( \begin{array}{cc}
\uvec\sigma & 0\\
0 & \uvec\sigma \\
\end{array}\right),
\eeq
with $\uvec\sigma$ the three Pauli matrices, illustrating how the total angular momentum is split into an orbital part and spin part.

The other three independent operators $M^{0i}$ are related to the so-called ``boost'' operators $\uvec K$, which generate pure Lorentz transformations along the $x$, $y$ and $z$ axes, namely
\beq
K^i= M^{0i}.
\eeq
The operator $M^{\mu\nu}$ will be somewhat loosely referred to as the  \emph{generalized angular momentum tensor} (and are sometimes written as $J^{\mu\nu}$ in the literature), but it should be remembered that it is only their spatial components $M^{jk}$ that are related to actual physical rotations.

Now the genuine angular momentum operators of the system, as already mentioned, are conserved time-independent operators, \emph{i.e.} commute with the Hamiltonian and do not contain any explicit factors of $t$. It turns out that the boost operators for the system $M^{0i}$ do not commute with the Hamiltonian, but they contain an explicit factor of $t$ in such a way as to make them, too, time-independent.

The rotation  operators for the system can be shown to satisfy the expected angular momentum commutation relations:
\beq \label{Comrel}
\left[J^i  , J^j\right]= i \epsilon^{ijk}J^k,
\eeq
while the boost operators satisfy
\beq
\left[K^i  , K^j \right] = -i \epsilon^{ijk} J^k.
\eeq
Of particular importance for the later discussions are the commutation relations between rotation and boost operators
\beq \label{JKcom}
\left[J^i  , K^j \right]= i \epsilon^{ijk} K^k,
\eeq
since they indicate immediately that the angular momentum transverse to a boost is altered by the boost. In other words, a Lorentz transformation along some direction will modify the components of the angular momentum transverse to that direction, but not the components along that direction.

The actual expressions for $P^\mu$ and $M^{\mu\nu}$ depend upon the
 theory under discussion, and follow from the structure of the Lagrangian. It is usual to consider the Lagrangian density as a scalar function of the generic fields $\phi_r$ and their derivatives $\partial_\mu\phi_r$
\beq
\uL=\uL[\phi_r,\partial_\mu\phi_r].
\eeq
The equations of motion (EOM), also known as Euler-Lagrange equations, are then given by
\beq
\partial_\mu\frac{\partial\uL}{\partial(\partial_\mu\phi_r)}-\frac{\partial\uL}{\partial\phi_r}=0.
\eeq

\subsection{The canonical energy-momentum and angular momentum densities\label{secIIB}}

\subsubsection{The canonical energy-momentum density\label{secIIB1}}

As  explained in all introductory texts on Quantum Field Theory, starting with an action
\beq
S \equiv \int \ud^4x \, \uL
\eeq
which is invariant under space-time translations and Lorentz transformations,
Noether's theorem \cite{Noether:1918zz} leads to expressions for the conserved \emph{canonical energy-momentum density}
\beq
T^{\mu\nu}(x)\equiv\frac{\partial\uL}{\partial(\partial_\mu\phi_r)}\,\partial^\nu\phi_r(x)-g^{\mu\nu}\uL,     \qquad \qquad  \partial_\mu T^{\mu\nu} (x)= 0,
\eeq
which, in general, is not symmetric under $\mu \leftrightarrow \nu$. Its detailed structure will be given for QED and QCD in section \ref{secIIIA}.

\subsubsection{The canonical angular momentum density\label{secIIB2}}

Noether's theorem also leads to expressions for the conserved \emph{canonical  generalized angular momentum density}, consisting of an orbital angular momentum (OAM) density and a spin density term
\beq
M^{\mu\nu\rho}(x)= M_\text{OAM}^{\mu\nu\rho}(x) + M_\text{spin}^{\mu\nu\rho}(x),  \qquad  \qquad \partial_\mu M^{\mu\nu\rho}(x) =0,
\eeq
where the orbital term is
\beq \label{QEDoam}
M_\text{OAM}^{\mu\nu\rho}(x)= x^\nu T^{\mu\rho} (x) - x^\rho T^{\mu\nu} (x)
\eeq
and the spin term is given in terms of the Lagrangian density $\uL(x)$  by
\beq \label{mspin}
M_\text{spin}^{\mu\nu\rho}(x) = -i \frac{\partial \uL}{\partial(\partial_\mu \phi_r)}\, (\Sigma^{\nu\rho})\lind{r}{s}\phi_s(x).
\eeq

A one-component field is necessarily spinless, \emph{i.e.} $\Sigma^{\mu\nu}=0$. In this case, the angular momentum is purely orbital $M^{\mu\nu\rho}=M^{\mu\nu\rho}_\text{OAM}$, and it follows from its conservation, \emph{via} some simple algebra, that the canonical energy-momentum density is symmetric
\beq
\partial_\mu(x^\nu T^{\mu\rho}-x^\rho T^{\mu\nu})=0\qquad\Rightarrow\qquad T^{\nu\rho}=T^{\rho\nu}.
\eeq
On the contrary, for a multi-component field with spin, \emph{i.e.} $\Sigma^{\mu\nu}\neq 0$, the associated canonical energy-momentum density is not symmetric. The antisymmetric part can then easily be related to the spin density
\beq\label{genform}
\partial_\mu M^{\mu\nu\rho}=0\qquad\Rightarrow\qquad T^{[\nu\rho]}\equiv T^{\nu\rho}-T^{\rho\nu}=-\partial_\mu M^{\mu\nu\rho}_\text{spin},
\eeq
and one has consequently
\beq
\partial_\nu T^{\rho\nu}=\partial_\mu\partial_\nu M^{\mu\nu\rho}_\text{spin}.
\eeq
The fact that the canonical energy-momentum density is generally not symmetric is often considered as a deficiency, but we feel that this issue has been over emphasized\footnote{The reason for the demand for a symmetric tensor comes from General Relativity and, more particularly, from the Einstein equations $R_{\mu\nu}-\tfrac{1}{2}\,g_{\mu\nu}R=8\pi G\,T^\text{GR}_{\mu\nu}$, where $T^\text{GR}_{\mu\nu}$ is interpreted as the matter energy-momentum tensor. To distinguish this tensor from the canonical one, we have added the label  ``GR''. In General Relativity, one \emph{assumes} that the metric is symmetric $g_{\mu\nu}=g_{\nu\mu}$ and covariantly constant $\nabla_\lambda g_{\mu\nu}=0$, where $\nabla_\lambda$ is the torsion-free covariant derivative. It follows from these assumptions that the LHS of the Einstein equations is symmetric, implying the symmetry of the RHS. It is then usually claimed that the energy-momentum \emph{has} to be symmetric. It is however important to remember that General Relativity is essentially a classical theory, where the absence of an antisymmetric part to the energy-momentum tensor signals the absence of spin-spin interactions. In natural extensions of General Relativity, like \emph{e.g.} Einstein-Cartan theory, where the assumptions mentioned above are relaxed, the energy-momentum tensor has generally an antisymmetric part that is coupled to the torsion. In conclusion, the symmetry of the energy-momentum tensor in General Relativity follows essentially from convenient assumptions, and not from strict physical requirements.}.

\subsubsection{The canonical momentum and angular momentum\label{secIIB3}}

Of great importance is the connection between the total canonical momentum $P^\mu$ and total angular momentum tensor $M^{\mu\nu}$ and the above densities. In classical dynamics, it can be shown directly that they are space integrals of the densities:
\begin{align}
P^\mu& = \int \ud^3x \, T^{0 \mu}(x),\label{Pcan} \\
M^{\mu\nu}& = \int \ud^3x \, M^{0\mu\nu}(x). \label{Mcan}
\end{align}
Usually, in Classical Mechanics, one studies the evolution of a system with time. In modern terminology this is referred to as \emph{instant form} dynamics. In Field Theory, it is sometimes more convenient to use \emph{light-front} (LF) dynamics, where the role of time is played by the light-front time $x^+=\tfrac{1}{\sqrt{2}}\,(t + z)$ and systems evolve with light-front time. In this case the analogues of Eqs.~\eqref{Pcan} and \eqref{Mcan} are
\begin{align}
P^\mu_\text{LF}& = \int dx^- \ud^2x_\perp \, T^{+\mu}(x),  \label{PLF}\\
M^{\mu\nu}_\text{LF}& = \int dx^- \ud^2x_\perp \, M^{+\mu\nu}(x), \label{MLF}
\end{align}
where $ x^\mu = [x^+, x^-, \uvec x_\perp]$ with $x^-=\tfrac{1}{\sqrt{2}}\,(t - z)$ and $\uvec x_\perp = (x^1, x^2)$. Unless explicitly stated, we shall henceforth consider instant-form expressions.

In a quantum theory, where we are dealing with operators, it is necessary to check that the above expressions are compatible with the commutation relations given in Eqs.~\eqref{Eq:Pcomm} and \eqref{Mmunucan}. It should be noted that $P^\mu$ and $M^{\mu\nu}$ refer to the \emph{total} momentum and angular momentum of the system, and that they generate the relevant transformations on all the different fields in the system, \emph{e.g.} both electron and photon fields in QED, both quark and gluon fields in QCD. Of crucial interest later will be the problem of defining the momenta and angular momenta of the individual quanta in the system, \emph{e.g.} $P^\mu(\text{electron})$ and $P^\mu(\text{photon})$, \emph{etc}. For example, can we define $P^\mu(\text{electron})$ so that it generates the relevant space-time translations on the \emph{electron field}, and something analogous for the other fields in the system?

It will turn out later that it is important to distinguish these canonical operators from others which share some, but not all of their properties. When referring to such operators, we shall always add a subscript label to distinguish them from the canonical versions. Henceforth, then, \emph{all operators referring to momentum, angular momentum, \emph{etc.}, which do not carry such an additional subscript label should be read as the canonical versions of these operators}. In particular the crucial fact, that it is the canonical operators which are the generators of the relevant transformations, should be remembered.

\subsection{Angular momentum in a relativistic theory\label{secIIC}}

In non-relativistic Quantum Mechanics, the spin of a particle is introduced as an additional, independent degree of freedom. The pioneering work of Dirac \cite{Dirac:1928hu} showed that spin emerges automatically in a relativistic theory and cannot be treated as an independent degree of freedom -- for a pedagogical treatment, see section $\bm{1.2}$ of \cite{Leader:2001gr}.

The intertwining of angular momentum and linear momentum can be seen immediately from the commutation relations between the boost operators $\uvec K$ and the angular momentum operators $\uvec J$ given in Eq.~\eqref{JKcom}, which follow from the commutation relations of the $M^{\mu\nu}$, which in turn can be derived by considering the effect of a sequence of two Poincar\'{e} transformations. As already stressed, this shows that the angular momentum will change under a boost, the only exception being the component of $\uvec J$ along the direction of the boost. Put another way, this indicates that the components of $\uvec J$ transverse to the boost direction are momentum dependent, or in the jargon of recent papers on this subject, are not \emph{frame-independent}. This can be seen intuitively from a classical picture of orbital angular momentum given in \cite{Landau:1951}. In Fig.~1, the vectors $\uvec r$ and $\uvec p$ which form $L_z$ via $\uvec L=\uvec r \times \uvec p$ are perpendicular to the boost direction $OZ$ and so are unaffected by the boost. The vectors $\uvec r$ and $\uvec p$ which form $L_x$ have components along the boost direction $OZ$, and so are changed as a result of the boost.
\begin{figure}[ h ]
\begin{center}
\includegraphics[height=6cm  ]{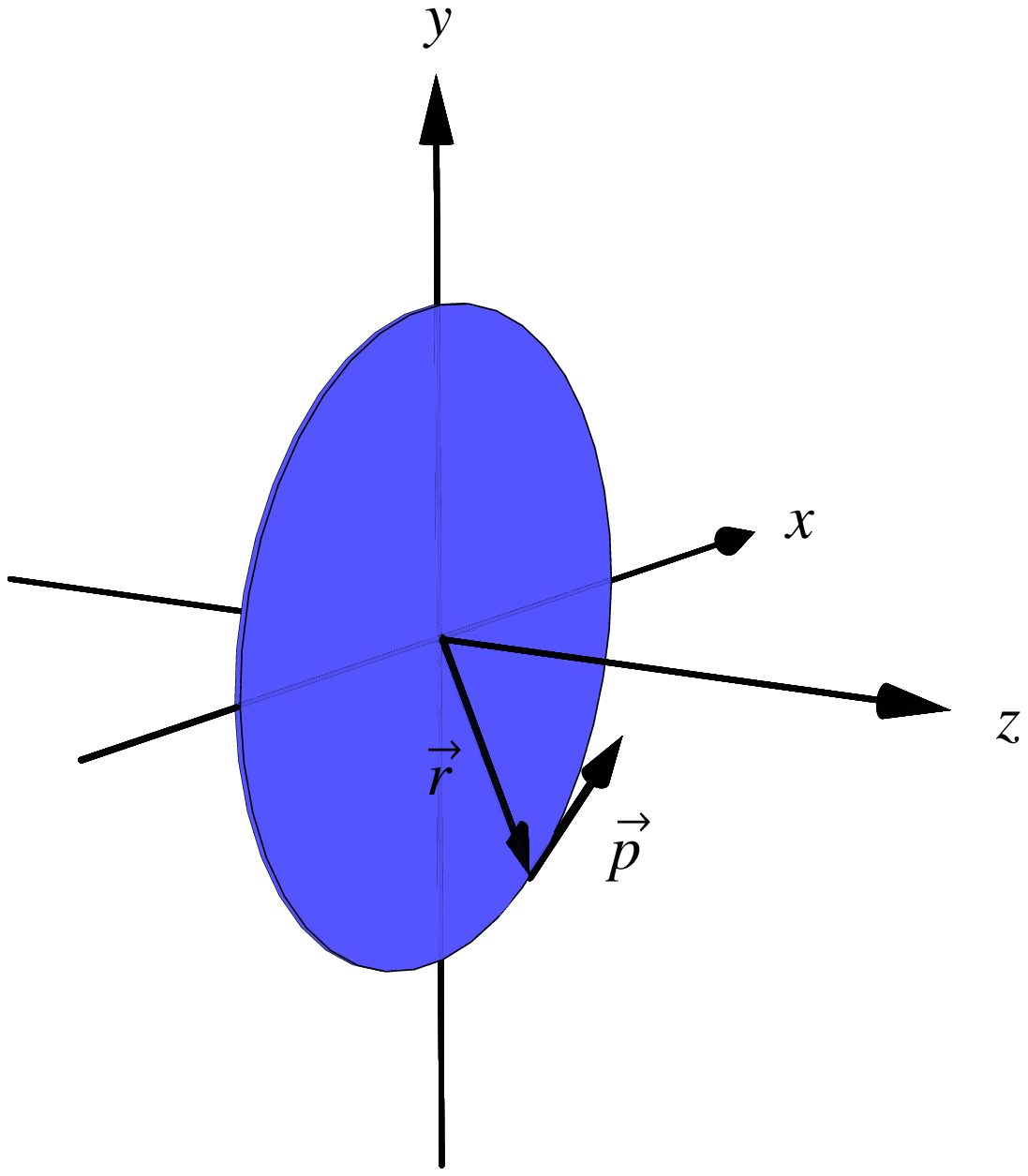}\hspace{2cm}
\includegraphics[ height=6cm ]{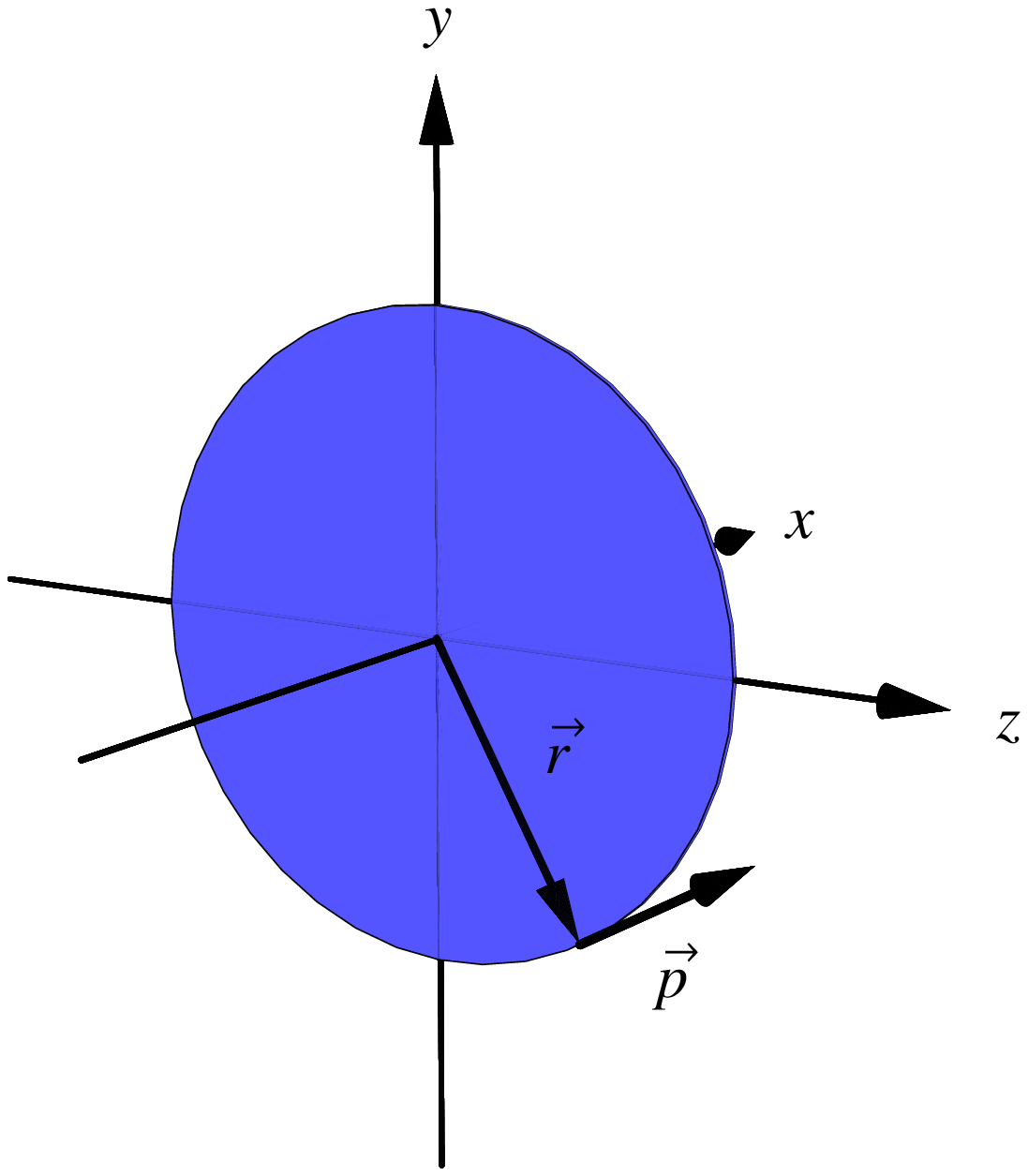}
\caption{The longitudinal $L_z$ (left) and transverse $L_x$ (right) components of the orbital angular momentum.}
%%% filename in {  };  in [  ]  can have:
%%width=  cm, height=  cm, angle = %%
\end{center}
\end{figure}

A consequence of this, as will be seen later, is that any sum rule relating the transverse angular momentum of a nucleon to the transverse angular momentum of its constituents, or any division of the total transverse angular momentum into spin and orbital parts, must necessarily involve energy dependent factors, \emph{i.e.} cannot be frame-independent.

\subsection{Lorentz transformation properties of the gauge field $A^\mu(x)$ and its consequences\label{secIID}}

In Classical Electrodynamics, the fields $\uvec E$ and $\uvec B$ are physical, and the photon vector potential $A^\mu$ largely plays the role of a mathematical aid, \emph{i.e.} it is often simpler to calculate $A^\mu$ than $\uvec E$ and $\uvec B$ from the currents and sources, and the physical fields can then be extracted \emph{via}
\beq \label{EB}
\uvec E = - \uvec\nabla A^0 -\frac{\partial \uvec A}{\partial t},  \qquad \qquad \uvec B = \uvec\nabla\times\uvec A.
\eeq
But $A^\mu$ is not determined \emph{a priori} by the currents and sources, since making a classical, local gauge transformation on it\footnote{Note that we shall consistently indicate gauge transformed quantities by a tilde sign, whereas Lorentz transformed quantities will be indicated by a prime.}
\beq \label{CGT}
A^\mu(x) \mapsto \tilde A^\mu(x) =  A^\mu(x) + \partial^\mu \alpha (x),
\eeq
where $\alpha(x)$ is any reasonably behaved scalar function, does not alter the physical fields. Thus, in order to carry out any kind of calculation, one has to put some extra condition on $A^\mu$, and this is described as choosing a gauge. Now, as its appearance suggests, $A^\mu$ is usually treated as a four-vector under Lorentz transformations, but  choosing the gauge may contradict this. In Classical Field Theory, one can impose the beautiful Lorenz condition
\beq \label{LorCon}
\partial_\mu A^\mu(x) =0,
\eeq
which is manifestly covariant and thus does not affect the Lorentz four-vector transformation properties of $A^\mu$, and which, moreover, yields uncoupled equations for each component of $A^\mu$ in terms of the currents and sources. But sometimes it is more convenient to fix the gauge in some other way. For example, one might use the Coulomb gauge, in which one imposes
\beq \label{CoulG}
\uvec\nabla\cdot\uvec A(x)=0.
\eeq
Clearly, this is not a manifestly covariant condition and if it is to hold in all frames, then $A^\mu$ cannot transform as a Lorentz four-vector. Indeed, one finds that under a boost of reference frame, $A^\mu$ undergoes a combined Lorentz four-vector transformation and a gauge transformation\footnote{Lorc\'{e} argues that these should be regarded as  ``generalized Lorentz transformations'' \cite{Lorce:2012rr}.}, as explained in section $\bm{14.3}$ of \cite{Bjorken:1965zz} and in section $\bm{5.9}$ of \cite{Weinberg:1995mt}. Note that this does not affect the undisputed Lorentz transformation properties of $\uvec E$ and $\uvec B$, which follow from Eq.~\eqref{EB} if $A^\mu$ transforms as a four-vector, since the extra term, being a gauge transformation, does not affect the electric and magnetic fields.

To construct a Quantum Field Theory, one starts with a classical Lagrangian density. For example, in QED one might begin with the classical Lagrangian density ($e$ is the charge of the electron)
 \beq \label{Eq:LClas}
\uL_\text{clas} = -\tfrac{1}{4}\,F_{\mu\nu}F^{\mu\nu}+\tfrac{1}{2}\left[\barpsi (i\,\uslash\! D - m)\psi +  \text{h.c.}\right],
\eeq
where the \emph{covariant} derivative is defined as
\beq \label{QEDCovD}
D_\mu= \partial_\mu + ie A_\mu.
\eeq
The Lagrangian in Eq.~\eqref{Eq:LClas} is invariant under the combined classical, local gauge transformation
 \begin{align} \label{Eq:GT}
\psi(x)&\mapsto\tilde\psi(x)= e^{-ie\alpha(x)} \,\psi(x),\\
A_\mu(x)&\mapsto\tilde A_\mu(x)= A_\mu(x)+\partial_\mu\alpha(x),
\end{align}
but in order to quantize the theory, one has to first choose a gauge. The canonical quantization process amounts to fixing the form of the \emph{equal-time} commutators between the fields and their conjugates. It is crucial to check that these commutator conditions are compatible with the conditions that fixed the gauge. Unfortunately, it turns out that the canonical commutation relations are incompatible with the \emph{operator} $A^\mu$ satisfying the Lorenz condition. There are several possibilities. One can decide to work in a non-manifestly covariant gauge like the Coulomb gauge, but then one should check that physical quantities thus calculated have the correct Lorentz transformation properties. Or one can work in a manifestly covariant gauge at the expense of introducing additional so-called gauge-fixing fields in QED \cite{Lautrup,Nakanishi:66} and also ghost fields in QCD \cite{Kugo:78}. For example, in the Lautrup-Nakanishi covariantly quantized version of QED \cite{Lautrup,Nakanishi:66}, one adds a \emph{gauge-fixing} part to the classical Lagrangian density in Eq.~\eqref{Eq:LClas}
 \beq \label{Eq:Gf}
\uL_\text{gf} = B(\partial_{\mu}A^{\mu})+ \tfrac{\textsf{a}}{2}\,B^2,
\eeq
where $B(x)$ is the gauge-fixing field and the parameter $\textsf{a}$ determines the structure of the photon propagator and is irrelevant for the present discussion. $B(x)$ is taken to be unaffected by gauge transformations.

The reason we are emphasizing this is that we will need to write down the most general structure, allowed by symmetry principles, for matrix elements of certain operators like $T^{\mu\nu}$, and this structure will be governed by the transformation properties of the operators. For example, does $T^{\mu\nu}$ transform as a rank-2 Lorentz tensor? Since it is a function of the gauge potential $A^\mu$, this will apparently depend on the transformation properties of $A^\mu$ imposed by the gauge choice. If $A^\mu$ transforms as a Lorentz four-vector, then manifestly $T^{\mu\nu}$ will transform as a second rank tensor, but perhaps surprisingly, as will be discussed in section \ref{secIVD}, in some cases  $T^{\mu\nu}$ will transform as a second rank tensor even though $A^\mu$ does not transform as a Lorentz four-vector. It turns out that in most papers this issue has not been recognized, though it is clearly explained in \cite{Lorce:2012rr}.

\subsection{Quantization of a gauge theory\label{secIIE}}

One of the major issues in the controversy about defining quark and gluon angular momentum is the question as to whether a quantity which can be measured experimentally must necessarily be represented in the theory by a gauge-invariant operator. Leader \cite{Leader:2011za} argued that what one actually measures is the expectation value of an operator, and it can certainly happen that the expectation value of a gauge non-invariant operator, taken between physical states yields a gauge-independent result. The problem is that it seems to be extremely difficult to actually prove such a result for some particular operator. In this section we wish to explain briefly why this is so.

In ordinary Quantum Mechanics in one dimension, the fundamental quantum condition is the canonical commutation relation between the position operator $x$ and its canonical conjugate momentum $p$, namely
\beq \label{CR}
[x,p] = i\hbar,
\eeq
where, if $L$ is the Lagrangian,
\beq
p = \frac{\ud L}{\ud{\dot{x}}}.
\eeq

In a Quantum Field Theory, one starts with a Classical Field Theory in which the role of the position operator is played by the fundamental fields $\phi(x)$ and the role of the canonical momentum is played by the conjugate fields $\pi(x)$, defined by
\beq
\pi= \frac{\partial \uL}{\partial(\partial_0\phi)},
\eeq
where $\uL$ is the Lagrangian density (we shall henceforth simply refer to it as the ``Lagrangian''). To quantize the theory, one tries to mimic Eq. \eqref{CR} as closely as possible, by imposing the form of the \emph{equal-time commutator}
\beq \label{ETC}
[\phi(t , \uvec x),\pi(t , \uvec y) ] = i \hbar\,\delta^{(3)}(\uvec x - \uvec y).
\eeq

\subsubsection{Problems in the quantization procedure in Gauge Theories\label{secIIE1}}

Once $\uL$ is chosen, it can happen that $\uL$ does not depend on $\partial_0 \phi$ for some particular field $\phi$, so that its conjugate field is zero, contradicting Eq.~\eqref{ETC} for that field. An example is the classical QED Lagrangian given in Eq.~\eqref{Eq:LClas} which does not contain the variable $\partial_0 A^0$, \emph{i.e.} there is no momentum conjugate to $A^0$.

Just as in Classical Electrodynamics, to actually work with the vector potential $A^\mu$, one has to choose a gauge, \emph{i.e.} impose some condition on $A^\mu$, and it may turn out that that condition contradicts the equal-time commutation relations. As an example, consider once again the classical QED Lagrangian given in Eq. \eqref{Eq:LClas}. Suppose we start with this fully gauge-invariant Lagrangian. Fixing the gauge is tantamount to working with a modified Lagrangian, say $\uL'$. Such a Lagrangian is known as a gauge-fixed Lagrangian. In principle, there are an infinite number of ways to fix the gauge, so that there are an infinite number of possible gauge-fixed Lagrangians $\uL' , \uL'', \uL''',\cdots$ and the key issue is to show that physical results are the same when working with the various gauge-fixed Lagrangians. For example, to show that $\uL'$ and $\uL''$ yield the same result for some physical experimentally measurable quantity $\la\la O\ra\ra_\phys$, one has to show that the expression for $\la\la O\ra\ra$ does not change under the gauge transformation that changes $\uL'$ into $\uL''$, \emph{i.e.} that $\la\la O'\ra\ra_\phys = \la\la O''\ra\ra_\phys$.

Now it may happen that the gauge-fixed Lagrangian $\uL'$ is still invariant under a \emph{residual} class of gauge transformations. An example is the Lagrangian in the covariantly quantized version of QED due to Lautrup and Nakanishi mentioned earlier,  which is a combination of the classical Lagrangian $\uL_\text{clas}$ in Eq.~\eqref{Eq:LClas} and the gauge-fixing part $\uL_\text{gf}$ given in Eq.~\eqref{Eq:Gf}
\beq \label{Eq:LN}
\uL' = \uL_\text{clas}  +\uL_\text{gf}.
\eeq
The theory is invariant under the residual infinitesimal $c$-number gauge transformation
\beq \label{Eq:GT}
A_\mu \mapsto A_\mu + \partial_\mu\alpha_\text{res}, \qquad \psi \mapsto \psi-ie\alpha_\text{res}\psi, \qquad B \mapsto B,
\eeq
where, here, $\alpha_\text{res}(x)$ is not an \emph{arbitrary} smooth function, but a function satisfying $\Box\alpha_\text{res}\equiv \partial_\mu \partial^\mu \alpha_\text{res}= 0$ and vanishing at infinity.

It is of course necessary that $\la\la O\ra\ra_\phys$ does not change under these residual gauge transformations, but that is in no way equivalent to demanding that $\la\la O\ra\ra_\phys$ does not change under the gauge transformation that changes $\uL'$ into some other $\uL''$. For example, it was shown in Ref. \cite{Leader:2011za} that, surprisingly, the \emph{total} canonical momentum in covariantly quantized QED is not gauge invariant under this residual class of gauge transformations, and it was then proven that its expectation value for physical states is unchanged by such gauge transformations. If the expectation value of the total momentum is a physical quantity, then the latter is certainly necessary, but that is not equivalent to the more demanding requirement of proving that its expectation value is gauge independent.

\subsubsection{Gauge invariance \emph{vs.} gauge independence\label{secIIE2}}

One of the least clearly explained concepts in Quantum Field Theory is that of a gauge transformation on a quantum operator. The point is that there is no fully developed theory of operator gauge changes. All the discussion of gauge invariance is based on making \emph{classical} gauge changes to the fields, \emph{i.e.} the functions $\alpha_\text{res}(x)$ mentioned above, or the more general functions $\alpha(x)$ mentioned in section \ref{secI}, are all ordinary numerical functions, called, in Field Theory, $c$-number functions. Whenever it is stated that a certain operator is \emph{gauge invariant}, what is meant is that it is invariant under a $c$-number gauge transformation.

Now in a quantum theory, classical dynamical variables are represented by operators, and when we say that we measure a particular dynamical variable, we mean that we measure the expectation value or a physical matrix element of that operator. If then, in our theory, we calculate the value of such a matrix element, its value must not depend upon the choice of gauge we have made in order to carry out the calculation. In other words the matrix elements involved should be \emph{gauge independent}. Collins (see section $\bm{2.12}$ of \cite{Collins:1984xc}) has stressed that it is important to distinguish the concepts of gauge invariance and gauge independence. Gauge invariance is the property of a quantity which does not change under a $c$-number gauge transformation. Gauge independence is a property of a quantum variable whose value does not depend on the method used for fixing the gauge.

It should be clear that demanding gauge independence is not the same as demanding the gauge invariance of the operator, because, as we have emphasized above, gauge invariance of an operator only refers to $c$-number transformations. If, for example, one wanted to compare a calculation in the Coulomb gauge $\uvec\nabla\cdot \uvec A =0$ with one in the light-front gauge $A^+=0$, one could not make the comparison using gauge invariance, because the corresponding operators $\hat O_\text{Coulomb}$ and $\hat O_\text{light-front}$ do not simply differ by a $c$-number function. The easiest way to see this is to note that if you add a $c$-number to an operator, you do not change its commutation relations with other operators, whereas the canonical commutation relations are different in different gauge choices.

So we would need to be able to handle \emph{operator-valued} gauge transformations, about which almost nothing is known. Several papers in the 1990s (see for example \cite{Chen:1998iu}) claimed to prove that the physical matrix elements of the canonical angular momentum operators were gauge independent, using methods very similar to those in \cite{Leader:2011za}. But unlike the approach in \cite{Leader:2011za}, it appears that these proofs are valid for a wide class of \emph{operator} gauge transformations, namely those which are not functions of $\partial_0 A^\mu$ and which therefore commute with $A^\mu$. Unfortunately, these papers were considered controversial and never appeared in a journal.

Supposedly, the only safe solution is to express the relevant expectation values in terms of Feynman path integrals, since these involve strictly classical fields. It can be shown \emph{e.g.} that the physical matrix elements of gauge-invariant operators are gauge independent\footnote{It is not clear to us whether this holds for the most general transformations imaginable.}. However, even this approach is far from clear and has been the subject of much debate, and claims that it can lead to unreliable results \cite{Hoodbhoy:1998bt,Chen:1999kz, Sun:2000gc}. Regrettably we are unable to offer any clarification.

It can happen that the physical matrix elements of a gauge non-invariant operator are gauge independent. Crucially, this means that one cannot automatically demand that every measurable dynamical variable should be represented by a gauge-invariant operator. Indeed, Leader showed in Ref. \cite{Leader:2011za} that even the \emph{total} momentum operator in covariantly quantized QED is not invariant under a certain class of $c$-number gauge transformations, yet it ought surely to be a measurable quantity.

\subsection{The Belinfante-improved energy-momentum and angular momentum densities\label{secIIF}}

\subsubsection{The Belinfante energy-momentum density\label{secIIF1}}

As already mentioned, the canonical energy-momentum density $T^{\mu\nu}(x)$ is generally not symmetric under interchange of $\mu$ and $\nu$. It is also not gauge invariant. It is possible to construct from $T^{\mu\nu}(x)$ a so-called Belinfante-improved density $T^{\mu\nu}_\text{Bel}(x)$, which is symmetric and which is usually gauge invariant \cite{Belinfante:1939,Rosenfeld:1940}. It differs from $T^{\mu \nu}(x)$ by a divergence term of the following form:
\beq\label{TBel}
T^{\mu\nu}_\text{Bel}\equiv T^{\mu\nu}+\partial_\lambda G^{\lambda\mu\nu},
\eeq
where the so-called \emph{superpotential} reads
\beq \label{superpot}
G^{\lambda\mu\nu}=\tfrac{1}{2}\left(M^{\lambda\mu\nu}_\text{spin}+M^{\mu\nu\lambda}_\text{spin}+M^{\nu\mu\lambda}_\text{spin}\right),
\eeq
and, crucially, is antisymmetric w.r.t. its first two indices $G^{\lambda\mu\nu}=-G^{\mu\lambda\nu}$. Alternatively, one can write
\beq
T^{\mu\nu}_\text{Bel}\equiv \tfrac{1}{2}\left(T^{\mu\nu}+T^{\nu\mu}\right)+\tfrac{1}{2}\,\partial_\lambda\left(M^{\mu\nu\lambda}_\text{spin}+M^{\nu\mu\lambda}_\text{spin}\right)
\eeq
which shows how the Belinfante-improved density generally differs from the symmetric part of the canonical density. The Belinfante-improved density is conserved $\partial_\mu T^{\mu\nu}_\text{Bel}=0$ and  symmetric $T^{\mu\nu}_\text{Bel}=T^{\nu\mu}_\text{Bel}$. Detailed expressions for $T^{\nu\mu}_\text{Bel}$ for QED and QCD will be given in section \ref{secIIIA}.

\subsubsection{The Belinfante  angular momentum density\label{secIIF2}}

In a similar way, one can define a Belinfante-improved generalized angular momentum density
\beq \label{MBdef}
M^{\mu\nu\rho}_\text{Bel} = M^{\mu\nu\rho} + \partial_\lambda(x^\nu G^{\lambda\mu\rho}-x^\rho G^{\lambda\mu\nu}).
\eeq
Now from Eq.~\eqref{superpot}, one sees that
\beq \label{Mpsuper}
M^{\mu\nu\rho}_\text{spin}=G^{\rho\mu\nu}-G^{\nu\mu\rho}
\eeq
so that, for the added term in Eq.~\eqref{MBdef},
\begin{align} \label{MBor}
\partial_\lambda(x^\nu G^{\lambda\mu\rho}-x^\rho G^{\lambda\mu\nu}) &= (x^\nu\partial_\lambda G^{\lambda\mu\rho}-x^\rho\partial_\lambda G^{\lambda\mu\nu})-( G^{\rho\mu\nu}-G^{\nu\mu\rho}) \nn \\
&=  (x^\nu\partial_\lambda G^{\lambda\mu\rho}-x^\rho\partial_\lambda G^{\lambda\mu\nu}) -  M^{\mu\nu\rho}_\text{spin}.
\end{align}
It follows from Eq.~\eqref{MBdef} that
\begin{align} \label{MBorb}
M^{\mu\nu\rho}_\text{Bel} &=  M_\text{OAM}^{\mu\nu\rho}+ (x^\nu\partial_\lambda G^{\lambda\mu\rho}-x^\rho\partial_\lambda G^{\lambda\mu\nu}) \nn \\
&=  ( x^\nu T^{\mu\rho}-x^\rho T^{\mu\nu}) + (x^\nu\partial_\lambda G^{\lambda\mu\rho}-x^\rho\partial_\lambda G^{\lambda\mu\nu}) \nn \\\
 &=     x^\nu (T^{\mu\rho}+ \partial_\lambda G^{\lambda\mu\rho}])-  x^\rho( T^{\mu\nu} +  \partial_\lambda G^{\lambda\mu\nu}) \nn \\
 &=   x^\nu T^{\mu\rho}_\text{Bel}-x^\rho T^{\mu\nu}_\text{Bel}.
\end{align}
 Hence the surprising result that the Belinfante-improved generalized angular momentum density  has the structure of a purely orbital angular momentum.
It is conserved, $\partial_\mu M^{\mu\nu\rho}_\text{Bel}=0$, as a consequence of the symmetry of $T^{\mu\nu}_\text{Bel}$.

\subsubsection{The Belinfante momentum and angular momentum\label{secIIF3}}

It follows from Eq.~\eqref{TBel} that
\begin{align}
P^\mu_\text{Bel}& \equiv \int\ud^3x\,T^{0\mu}_\text{Bel}  \nn \\
&= \int\ud^3x\,T^{0\mu}+\int\ud^3x\,\partial_\lambda G^{\lambda 0\mu} \nn \\
&= \int\ud^3x\,T^{0\mu}+\int\ud^3x\,\partial_iG^{i 0\mu},
\end{align}
where, in the last line, we used the fact that $G^{0 0\mu}=0$. Thus, provided that one is allowed to drop the surface term $\int_V\ud^3x\,\partial_iG^{i0\mu}=\int_S\ud^2\sigma_i\,G^{i0\mu}$, which is tantamount to assuming that the fields vanish at spatial infinity, one has, apparently,
\beq \label{Eq:PBC}
P^\mu_\text{Bel} = P^\mu .
\eeq

Now for a classical $c$-number field, it is meaningful to argue that the field vanishes at infinity and that Eq.~\eqref{Eq:PBC} holds as a numerical equality\footnote{Note that non-trivial topological effects could prevent the fields from vanishing at infinity.}. It is much less obvious what this means for a quantum operator. The correct way to tell whether a divergence term can be neglected is to check what its role is in the relevant physical \emph{matrix elements} involving the operator. In the case of Eq.~\eqref{Eq:PBC}, one can readily check that the matrix elements between \emph{any} normalizable physical states $| \Phi \rangle $ and $| \Phi' \rangle $ are the same\footnote{This is not true for all operators which differ by a divergence term. Singularities can affect the result.}, \emph{i.e.}
\beq \label{Eq:ExpVal}
\langle  \Phi' | P^\mu_\text{Bel}  |  \Phi \rangle = \langle  \Phi' | P^\mu  |  \Phi \rangle .
\eeq
However, the operators cannot be identical, because one, for example, may be gauge invariant and the other not, so that the equality would be contradicted upon performing a gauge transformation. On the other hand, the operators are essentially equivalent, and they generate the same transformations on the fields\footnote{This is true only for the total momentum of the system. It does not necessarily hold for the momenta of the individual constituents.}. We feel therefore that Eq.~\eqref{Eq:PBC} is somewhat misleading and prefer to indicate the equivalence between $P_\text{Bel}$ and $P$ as
\beq \label{Eq:defEq}
P^\mu_\text{Bel} \sim   P^\mu .
\eeq

It should be noted that it would be impossible to construct a consistent theory if it were not permissible, in certain cases, to ignore the spatial integral of the divergence of a local operator. For example we could not even establish the obvious requirement that the momentum operator commutes with itself! For one has, (no sum over $j$)
\beq \label{Eq:consist}
i\left[P^j,P^j\right] = \int \ud^3x \,i\left[P^j, T^{ 0  j}(x)\right]= \int \ud^3x \, \partial^j T^{ 0  j}(x),
\eeq
where we have used Eq.~\eqref{Eq:Pcomm}, and this vanishes only if $T^{0j}(x)$ vanishes sufficiently fast at spatial infinity.

The Belinfante angular momentum density leads to the same generalized  angular momentum tensor as the canonical one
\begin{align}
M^{\mu\nu}_\text{Bel}&=\int\ud^3x\,M^{0\mu\nu}_\text{Bel} \nn \\
&= \int\ud^3x\,M^{0\mu\nu}+\int\ud^3x\,\partial_i(x^\mu G^{i0\nu}-x^\nu G^{i0\mu}) \nn \\
&= M^{\mu\nu},
\end{align}
provided once more that one is allowed to drop the surface term $\int_V\ud^3x\,\partial_i[x^\mu G^{i0\nu}-x^\nu G^{i0\mu}]=\int_S\ud^2\sigma_i\,[x^\mu G^{i0\nu}-x^\nu G^{i0\mu}]$. However, as for the momentum, we prefer to indicate the equivalence of $ M^{\mu\nu}$ and $ M^{\mu\nu}_\text{Bel}$ by
\beq \label{eqBC}
M^{\mu\nu}_\text{Bel} \sim  M^{\mu\nu}.
\eeq
Operators like  $M^{\mu\nu}$, which involve the product of $x$ with a local operator, do not transform like local operators under space-time translations, see Eq.~\eqref{STTr}, and have been called \emph{compound} operators in Ref. \cite{Bakker:2004ib}\footnote{A simple exercise shows that treating $L(x)=xO(x)$ as a local operator leads to the absurd conclusion that $L(x)=0$ for all $x$.}. For compound operators like the  angular momentum, it is a much more difficult task to show the equivalence of the \emph{total} angular momentum generators $M^{ij}$ and $M^{ij}_\text{Bel}$, and care has to be exercised to always use normalizable states. This has been done by Shore and White \cite{Shore:1999be}.

Interestingly, it seems that as early as 1921, Bessel-Hagen \cite{Bessel:1921} found a way to obtain what is now called the Belinfante decomposition, using Noether's theorem. The trick is to make a \emph{combined} infinitesimal Lorentz and gauge transformation in the Lagrangian. Such generalized Lorentz transformations are discussed in section \ref{secIVD}. Recently this trick was rediscovered by Guo and Schmidt \cite{Guo:2013jia}, who were able to show that all the new decompositions to be discussed in the next few sections could be obtained using Noether's theorem.

\subsubsection{Example of the difference between canonical and Belinfante angular momentum in Classical Electrodynamics\label{secIIF4}}

Here is an amusing example, in Classical Electrodynamics, to show that $M^{ij}$ and $M^{ij}_\text{Bel}$ or, equivalently, that $\uvec J$ and $\uvec J_\text{Bel}$ do not always agree with each other. The general expressions for $\uvec J$ and $\uvec J_\text{Bel}$ in terms of fields will be given in section \ref{secIVA}. For a free classical electromagnetic field, one has
\beq
\uvec J =  \underbrace{\int \ud^3x \,(\uvec E \times \uvec A)}_{\text{spin term}}  + \underbrace{\int\ud^3x \,E^i (\uvec x \times \uvec \nabla) A^i}_{\text{orbital term }}
\eeq
and
\beq
\uvec J_\text{Bel}= \int\ud^3x \left[\uvec x \times (\uvec E \times \uvec B)\right].
 \eeq

Consider a left-circularly polarized (= positive helicity) beam, with angular frequency $\omega$, and amplitude proportional to $E_0$,  propagating along $OZ$, \emph{i.e.} along the unit vector $\uvec e_{z}$. Then
\beq
A^\mu =\left(0,\frac{E_0}{\omega}\, \cos(kz-\omega t),\frac{E_0}{\omega}\,\sin(kz-\omega t), 0 \right)
\eeq
gives the correct electric and magnetic fields.  $\uvec E$, $\uvec B$ and $\uvec A$ all rotate in the $XY$ plane. Now consider the component of $\uvec J$ along $OZ$. Note that
\beq
\uvec\nabla A_{x,y} \propto\uvec e_{z}\qquad \text{so that} \qquad (\uvec x \times \uvec\nabla)_z A_{x,y} = 0,
\eeq
so only the spin term contributes to $J_z$. One finds
\beq
J_z \text{ per unit volume}= \frac{E_0^2}{\omega}.
\eeq
For one photon per unit volume, one requires $E_0^2 = \hbar \omega $ so that
\beq
J_z \text{ per photon}= \hbar
\eeq
as expected.

For the Belinfante case $ \uvec E \times \uvec B \propto \uvec e_z  $, so that one obtains the incorrect result
\beq
J_{\text{Bel}, z} \text{ per unit volume} = \frac{1}{V}\int \ud^3x \left[ \uvec x \times (\uvec E \times \uvec B)\right]_z = 0 \,\,!
\eeq
Of course, the failure of the two versions to agree with each other is simply due to the fact that the light beam is here described by a plane wave, so that the fields do not vanish at spatial infinity. Computing explicitly the surface integral over the boundary of a cylinder with symmetry axis $\uvec e_z$, one finds for (infinitely) large length $L$ and radius $R$
\beq
J_{\text{surf},z}=-\pi R^2L\,\frac{E^2_0}{\omega}=-V\,\frac{E^2_0}{\omega},
\eeq
which explains why
\beq
J_{\text{Bel}, z}=J_z+J_{\text{surf},z}=0.
\eeq
This example may seem a bit academic, but it is a warning that some care must be utilized when discarding integrals of spatial divergences.

%It is important to keep in mind that this procedure is completely \emph{ad hoc} and solely motivated by the wish of dealing with a symmetric and conserved energy-momentum tensor. Moreover, if one is interested in quantities localized in a finite volume, canonical and Belinfante-improved tensors clearly give %different physical answers.

\section{Detailed structure of the canonical and Belinfante energy-momentum and angular momentum densities\label{secIII}}

\subsection{Structure and Lorentz transformation properties of the energy-momentum density\label{secIIIA}}

In section \ref{secVI}, we will need to relate the matrix elements of the angular momentum to the matrix elements of the energy-momentum density. To do this, we will need to write down the most general structure for the matrix elements of $T^{\mu\nu}$, and this will  depend upon its behaviour under Lorentz transformations. Major simplifications occur if, as is usually assumed, $T^{\mu\nu}$ transforms as a second-rank Lorentz tensor, but since it is a function of the vector potential $A^\mu$, this will only be \emph{manifest} if $A^\mu$ transforms as a Lorentz four-vector. Thus, for simplicity, we are forced to consider covariantly quantized versions of QED and QCD. The problem then is that, besides the electron and photon fields and the quark and gluon fields, one has to introduce gauge-fixing and ghost fields, and these appear in the expressions for $T^{\mu\nu}$. However, in all the recent papers on the angular momentum controversy, with the exception of \cite{Leader:2011za}, this issue has been completely ignored and the $T^{\mu\nu}$ have been expressed entirely in terms of the electron, photon, quark and gluon fields. In this section we explain why this is actually correct and give explicit expressions for the various versions of the energy-momentum density in terms of the electron, photon, quark and gluon fields.

\subsubsection{Structure of the canonical and Belinfante energy-momentum densities in covariantly quantized QED\label{secIIIA1}}

For covariantly quantized QED, using the Lagrangian given in Eqs.~\eqref{Eq:LClas} and \eqref{Eq:Gf}, one finds \cite{Leader:2011za} for the conserved canonical energy-momentum  density
\beq \label{Eq:canQED}
T^{\mu\nu}= T^{\mu\nu}_\text{clas}+ T^{\mu\nu}_\text{gf},
\eeq
where
\begin{align}
T^{\mu\nu}_\text{clas} &= \tfrac{1}{2}\, \barpsi\gamma^\mu i\LRpartial^\nu \psi - F^{\mu\alpha}\partial^\nu A_\alpha  - g^{\mu\nu}{\cal{L}}_\text{clas},\label{Eq:thetacQED}\\
T^{\mu\nu}_\text{gf} &= B \, \partial^\nu A^\mu - g^{\mu\nu}\uL_\text{gf}\label{Eq:tcanGf}
\end{align}
with  $\LRpartial^\nu \equiv \overset{\rightarrow}{\partial}\!\!\!\!\phantom{\partial}^\nu - \overset{\leftarrow}{\partial}\!\!\!\!\phantom{\partial}^\nu $.

For the conserved Belinfante density, one finds
\beq \label{Eq:tbQED}
T^{\mu\nu}_\text{Bel}  = T^{\mu\nu}_\text{Bel,clas} + T^{\mu\nu}_\text{Bel,gf},
\eeq
where
\begin{align}
T^{\mu\nu}_\text{Bel,clas} &=  \tfrac{1}{4}\, \barpsi(\gamma^\mu i\LRD^\nu + \gamma^\nu i\LRD^\mu )\psi - F^{\mu\alpha}F\uind{\nu}{\alpha} - g^{\mu\nu}\uL_\text{clas},\label{Eq:tClas}\\
T^{\mu\nu}_\text{Bel,gf}& = - (\partial^\mu B)A^\nu -(\partial^\nu B)A^\mu - g^{\mu\nu}{\cal{L}}_\text{gf},\label{Eq:tGf}
\end{align}
where $\LRD^\nu = \LRpartial^\nu+2ieA^\nu $.

The conservation of an energy-momentum density depends on the equations of motion, which are a consequence of the Lagrangian. Thus, for example, $T^{\mu\nu}_\text{Bel}$ is conserved, but $T^{\mu\nu}_\text{Bel,clas}$ is not, when the Lagrangian is $\uL_\text{clas}  +  \uL_\text{gf}$. On the other hand, $T^{\mu\nu}_\text{Bel,clas}$ would be conserved \emph{if} the Lagrangian were $\uL_\text{clas}$. Now often in the literature, the Belinfante energy-momentum density is simply taken to be $T^{\mu\nu}_\text{Bel,clas}$ and is treated as if it were conserved, \emph{i.e.} the momentum operator based on it is taken to be independent of time (equivalently: to remain unrenormalized), which would imply that the Lagrangian is just $\uL_\text{clas}$. But it is well known that one cannot quantize QED \emph{covariantly} using just $\uL_\text{clas}$. It turns out, however, that this is innocuous, since it can be shown \cite{Leader:2011za} that for physical matrix elements, for both the canonical and Belinfante versions,
\beq \label{Eq:NonCont}
\langle  \Phi' |T^{\mu\nu}_\text{gf} | \Phi \rangle = \langle  \Phi' |T^{\mu\nu}_\text{Bel,gf} | \Phi \rangle = 0.
\eeq
Hence
\beq
\begin{split}
\langle  \Phi' |\,\partial_{\mu} T^{\mu\nu}_\text{clas}| \Phi \rangle &= \langle  \Phi' |\,\partial_{\mu} T^{\mu\nu}| \Phi \rangle =0,\\
\langle  \Phi' |\partial_\mu T^{\mu\nu}_\text{Bel,clas}| \Phi \rangle &= \langle  \Phi' |\,\partial_\mu T^{\mu\nu}_\text{Bel}| \Phi \rangle =0.
\end{split}
\eeq

In summary, covariant quantization of QED complicates some aspects and there is no compelling reason to insist on it. Indeed, the non-manifestly covariant Coulomb gauge leads to a perfectly good Lorentz-invariant theory. However, for our purposes it is helpful to work with a covariantly quantized theory, and since we will only consider physical matrix elements, $T^{\mu\nu}_\text{clas}$ and $T^{\mu\nu}_\text{Bel,clas}$ may be treated as conserved tensor operators. Consequently, in the following, we may take
\begin{align}
T^{\mu\nu} &\sim T^{\mu\nu}_\text{clas},\label{QEDTcan}\\
T^{\mu\nu}_\text{Bel}& \sim T^{\mu\nu}_\text{Bel,clas}.\label{QEDTBel}
\end{align}

\subsubsection{Structure of the canonical and Belinfante energy-momentum densities in covariantly quantized QCD\label{secIIIA2}}

The situation in QCD is somewhat more complicated. The infinitesimal gauge transformations on the gluon vector potential and on the quark fields, under which the pure quark-gluon Lagrangian $\uL_{\textrm{QCD}}$ (the QCD analogue of the QED $\uL_\text{clas})$
\beq \label{Eq:LqG}
\uL_{\textrm{QCD}}= \uL_{\textrm{D}}+\uL_{\textrm{YM}}+\uL_{\textrm{int}}, \eeq
where the Dirac (D), Yang-Mills (YM) and interaction terms (int) are given by
\begin{align}
\uL_\text{D}&=\barpsi_r(\tfrac{i}{2}\,\uslash\!\LRpartial-m)\delta_{rs} \psi_s,\\
\uL_\text{YM}&=  -\tfrac{1}{4}\,G^a_{\mu\nu}G^{\mu\nu}_a,  \\
\uL_\text{int}& = g \,\barpsi_r \,\uslash\! A^a t^a_{rs} \psi_s,
\end{align}
is invariant, are determined by eight scalar $c$-number fields $\theta^a(x)$,
\begin{align}
\delta A^a_\mu (x)&=\left[\partial_\mu \,\delta^{ab}- gf^{abc}A^c_\mu(x)\right] \theta^b(x)\equiv D^{ab}_\mu\theta^b(x), \label{Eq:GluonGT}\\
\delta \psi_r(x)&=ig\theta^a(x)t^a_{rs}\psi_s(x).\label{Eq:quarkGT}
\end{align}
Here $a,b,c =1,2...8$  and $r,s=1,2,3$ are color labels, the matrices $t^a$ satisfy $[t^a, t^b]=if^{abc}t^c$, and the gluon field-strength tensor is given by
\beq \label{Eq:GQCD}
G^a_{\mu\nu} =\partial_\mu A^a_\nu-\partial_\nu A^a_\mu+gf^{abc}A^b_\mu A^c_\nu.
\eeq
A sum over quark flavors is to be understood here and in what follows.

However, in order to quantize the theory covariantly, one has to introduce both a gauge-fixing field $B(x)$ and Faddeev-Popov anti-commuting fermionic ghost fields $c(x)$ and $\overline{c}(x)$. The Kugo-Ojima Lagrangian \cite{Kugo:78} for the covariantly quantized theory is then
\beq \label{Eq:KO}
\uL = \uL_{\textrm{QCD}} + \uL_{\text{gf}+\text{gh}},
\eeq
where
\beq\label{Eq:Gf+Gh}
\uL_{\text{gf}+\text{gh}}=-i(\partial^\mu\overline{c}^a)D^{ab}_\mu c^b  - (\partial^\mu B^a) A^a_\mu + \tfrac{\textsf{a}}{2}\,B^aB^a
\eeq
with $\textsf{a}$ a parameter which fixes the structure of the gluon propagator, and which is irrelevant for the present discussion.  The extra term $\uL_{\text{gf}+\text{gh}}$ is not invariant under the original infinitesimal gauge transformations given by Eqs.~\eqref{Eq:GluonGT} and \eqref{Eq:quarkGT}. Instead the theory is invariant under the BRST transformations \cite{Becchi:1975nq,*Tyutin}
\beq \label{Eq:BRST}
\begin{split}
\delta A^a_\mu(x) &=\theta D^{ab}_\mu c^b(x), \\
\delta \psi_r(x)&= ig\theta \,t^a_{rs}c^a(x)\psi_s(x), \\
\delta c^a(x) &=-\tfrac{g}{2}\, \theta f^{abc}c^b(x)c^c(x),  \\
\delta \overline{c}^a(x)&=  i\theta B^a(x),  \\
\delta B(x)& = 0,
\end{split}
\eeq
where $\theta$ is a constant operator which commutes with bosonic fields and anti-commutes with fermionic fields. The BRST transformation is generated by $\theta Q_B$, \emph{i.e.} for any of the above fields $\phi$
\beq \label{Eq:QBtr}
i[\theta Q_B, \phi] = \delta \phi,
\eeq
where the conserved, hermitian charge $Q_B$ is given by
\beq \label{Eq:QBdef}
Q_B= \int \ud^3x \left[ B^a \LRpartial_0c^a+gB^af^{abc}A^b_0c^c+i\tfrac{g}{2}\,(\partial_0\overline{c}^a) f^{abc}c^bc^c \right].
\eeq
There is also a conserved charge
\beq \label{Eq:Qc}
Q_c=\int \ud^3x \left[\overline{c}^a\LRpartial_0 c^a+g \overline{c}^af^{abc}A^b_0c^c \right]
\eeq
which ``measures'' the \emph{ghost number}
\beq \label{Eq:GNo}
i[Q_c, \phi] = N \phi,
\eeq
where $N=1$  for  $\phi=c$, $-1$ for $\phi=\overline{c}$ and $0$ for all other fields. The physical states $|\Phi \rangle $ are defined by the subsidiary conditions
\beq \label{Eq:Phys}
Q_B |\Phi \rangle =  0,  \qquad \qquad Q_c |\Phi \rangle  = 0.
\eeq
%\beq  \label{Eq:Phys'} Q_c |\Psi \rangle  = 0. \eeq

One finds for the canonical energy-momentum tensor density
\beq \label{Eq:tcQCD}
T^{\mu\nu}= T^{\mu\nu}_{\textrm{QCD}} + T^{\mu\nu}_{\text{gf}+\text{gh}},
\eeq
where
\begin{align}
T^{\mu\nu}_{\textrm{QCD}} &= \tfrac{1}{2}\,\barpsi_r \gamma^\mu  i\LRpartial^\nu  \psi_r - G^{\mu\alpha}_a \partial^\nu A^a_\alpha - g^{\mu\nu}\uL_{\textrm{QCD}},\label{Eq:tcanqG}\\
T^{\mu\nu}_{\textrm{gf+gh}}&= -A^{\mu}_a\partial^\nu B^a  -i(\partial^\nu \overline{c}^a)D^{\mu}_{ab}c^b  -i (\partial^\mu\overline{c}^a)(\partial^\nu c^a) - g^{\mu\nu} \uL_{\text{gf}+\text{gh}}. \label{Eq:tcG}
\end{align}
The Belinfante version is
\beq \label{Eq:tbQCD}
T^{\mu\nu}_\text{Bel}=  T^{\mu\nu}_{\text{Bel},\textrm{QCD}} + T^{\mu\nu}_{\text{Bel,gf}+\text{gh}},
\eeq
where
\beq \label{Eq:tbelqG}
T^{\mu\nu}_{\text{Bel},\textrm{QCD}}= \tfrac{1}{4}\,\barpsi_r (\gamma^\mu i\LRD^\nu_{rs}+\gamma^\nu i\LRD^\mu_{rs})\psi_s -G^{\mu\alpha}_a G\uind{a\nu}{\alpha}- g^{\mu\nu}\uL_{\textrm{QCD}}
\eeq
is BRST invariant, \emph{i.e.} commutes with $Q_B$. Here $\LRD^\nu_{rs}$ is a matrix in color space
\beq \label{DQCD}
\LRD^\nu_{rs} = \LRpartial^\nu\delta_{rs}-2ig A^{\nu}_a t^a_{rs}.
\eeq
The gauge-fixing and ghost terms are given by
\beq \label{Eq:tbelGf}
T^{\mu\nu}_{\text{Bel,gf}+\text{gh}}= - (A^{\mu}_a \partial^\nu B^a + A^{\nu}_a\partial^\mu B^a) -i[(\partial^\mu\overline{c}^a)D^{\nu}_{ab} c^b +(\partial^\nu\overline{c}^a)D^{\mu}_{ab} c^b ] - g^{\mu\nu}\uL_{\text{gf}+\text{gh}}.
\eeq
This can be rewritten as an anti-commutator with $Q_B$ \cite{Kugo:1979gm}
\beq \label{tbelGfnew}
T^{\mu\nu}_{\text{Bel,gf}+\text{gh}}= - \left\{ Q_B, \left[ (\partial^{\mu} \overline{c}^a) A^{\nu}_a+ (\partial^\nu \overline{c}^a) A^{\mu}_a+ g^{\mu\nu}\left(\tfrac{\textsf{a}}{2}\,\overline{c}^aB^a - (\partial^\rho\overline{c}^a) A^a_\rho \right) \right]\right\}.
\eeq
It follows that $T^{\mu\nu}_{\text{Bel,gf}+\text{gh}}$ is BRST invariant (because $Q_B$ is nilpotent, \emph{i.e.} $Q_B^2=0$) and does not contribute to physical matrix elements
\beq \label{tbelQCD}
\langle  \Phi'   | T^{\mu\nu}_\text{Bel} |  \Phi \rangle = \langle  \Phi'   | T^{\mu\nu}_{\text{Bel},\textrm{QCD}}  |  \Phi \rangle.
\eeq
The situation with the canonical energy-momentum tensor $T^{\mu\nu}$ is somewhat more complicated, but it can be shown \cite{Leader:2011za} that contrary to the statement in Ref. \cite{Shore:1999be}, $T^{\mu\nu}_{\text{gf}+\text{gh}}$ does \emph{not} contribute to physical matrix elements.

Consequently, analogous to the QED case, in the following we may use
\begin{align}
T^{\mu\nu} &\sim T^{\mu\nu}_{\textrm{QCD}},\label{QCDTcan}\\
T^{\mu\nu}_\text{Bel}& \sim T^{\mu\nu}_{\text{Bel},\textrm{QCD}},\label{QCDTBel}
\end{align}
where, we remind the reader, a sum over quark flavors is implied.

We can now define separate quark  and gluon  parts of $T^{\mu\nu}$:
\begin{align} \label{Tq}
T^{\mu\nu}_q &\sim  \tfrac{1}{2}\,\barpsi_r \gamma^\mu  i\LRpartial^\nu  \psi_r - g^{\mu\nu}\uL_\text{D },\\
\label{TG} T^{\mu\nu}_G &\sim - G^{\mu\alpha}_a \partial^\nu A^a_\alpha - g^{\mu\nu}(\uL_\text{YM}+\uL_\text{int}),
\end{align}
and similarly, for $T^{\mu\nu}_\text{Bel}$:
\begin{align}
T^{\mu\nu}_{\text{Bel},q} &\sim \tfrac{1}{4}\,\barpsi_r (\gamma^\mu i\LRD^\nu_{rs}+\gamma^\nu i\LRD^\mu_{rs})\psi_s- g^{\mu\nu}(\uL_\text{D}+\uL_\text{int}), \label{TqBel}\\
T^{\mu\nu}_{\text{Bel},G}& \sim  - G^{\mu\alpha}_a G\uind{a\nu}{\alpha}- g^{\mu\nu}\uL_\text{YM}.  \label{TGBel}
\end{align}
Note that for considerations of the genuine angular momentum, one always has $\mu \neq \nu$, so the terms proportional to $g^{\mu\nu}$ are irrelevant\footnote{In order to agree with the later discussion  in section \ref{secV}, the ``int'' term  has been incorporated into the gluon term in the canonical expression and into the quark term in the Belifnate expression. This is largely a matter of convenience.}.

\subsection{Structure of the angular momentum operators in QED\label{secIIIB}}

In this section we give a pedagogical demonstration of how the principal angular momentum operators are derived from the corresponding expressions for the energy-momentum tensors, given above. So as not to drown the essential ideas in a mass of algebra, we shall limit ourselves to QED. The derivation for QCD is similar but more tedious.

\subsubsection{The Belinfante angular momentum in QED and the form used by Ji\label{secIIIB1}}

According to Eq.~\eqref{MBorb}, the Belinfante angular momentum is given by
\beq
M^{jk}_\text{Bel}=\int \ud^3 x\, M^{0jk}_\text{Bel}(x),
\eeq
where
\beq \label{BOrbagain}
M^{\mu\nu\rho}_\text{Bel}=x^\nu T^{\mu\rho}_\text{Bel}-x^\rho T^{\mu\nu}_\text{Bel}.\eeq
For QED, the electron part of this is, from Eq.~\eqref{Eq:tClas} and keeping only the relevant terms for the actual angular momentum,
\beq
M^{\mu\nu\rho}_{\text{Bel},e} =  x^\nu \,\tfrac{1}{4}\, \barpsi(\gamma^\mu i\LRD^\rho + \gamma^\rho i\LRD^\mu )\psi - (\nu \leftrightarrow \rho),
\eeq
and is not split into a spin part and orbital part. Such a split can be achieved as follows. We can write
\beq \label{Bel1}
\tfrac{1}{4}\, \barpsi(\gamma^\mu i\LRD^\nu + \gamma^\nu i\LRD^\mu )\psi= \tfrac{1}{2}\,\barpsi\gamma^\mu i\LRD^\nu \psi + \tfrac{1}{4}\, \barpsi(\gamma^\nu i\LRD^\mu - \gamma^\mu i\LRD^\nu )\psi.
\eeq
Then, using the following gamma matrix identities ($\epsilon_{0123}=1$)
\beq
\begin{split}
\sigma^{\mu\nu}\gamma^\rho&= ig^{\nu\rho}\gamma^\mu -ig^{\mu\rho}\gamma^\nu + \epsilon^{\mu\nu\rho\sigma}\gamma_\sigma\gamma_5,  \\
\gamma^\rho \sigma^{\mu\nu}&= ig^{\mu\rho}\gamma^\nu -ig^{\nu\rho}\gamma^\mu + \epsilon^{\mu\nu\rho\sigma}\gamma_\sigma\gamma_5,
\end{split}
\eeq
and multiplying the first by $i\overset{\rightarrow}{D}\!\!\!\!\!\phantom{D}_\rho= i\overset{\rightarrow}{\partial}\!\!\!\!\phantom{\partial}_\rho - eA_\rho$ and the second by $i\overset{\leftarrow}{D}\!\!\!\!\!\phantom{D}_\rho= i\overset{\leftarrow}{\partial}\!\!\!\!\phantom{\partial}_\rho + eA_\rho$, one obtains
\beq
\begin{split}
\sigma^{\mu\nu}i\,\uslash\!\overset{\rightarrow}{D}\!\!\!\!\!\phantom{D}&=    \gamma^\nu \overset{\rightarrow}{D}\!\!\!\!\!\phantom{D}^\mu -    \gamma^\mu \overset{\rightarrow}{D}\!\!\!\!\!\phantom{D}^\nu  +i \, \epsilon^{\mu\nu\rho\sigma}\gamma_\sigma\gamma_5  \overset{\rightarrow}{D}\!\!\!\!\!\phantom{D}_\rho,\\
 i\,\uslash\!\overset{\leftarrow}{D}\!\!\!\!\!\phantom{D}\,\sigma^{\mu\nu} &= \overset{\leftarrow}{D}\!\!\!\!\!\phantom{D}^\nu \gamma^\mu-\overset{\leftarrow}{D}\!\!\!\!\!\phantom{D}^\mu \gamma^\nu  + i\, \epsilon^{\mu\nu\rho\sigma}\gamma_\sigma\gamma_5 \overset{\leftarrow}{D}\!\!\!\!\!\phantom{D}_\rho.
\end{split}
\eeq
Sandwiching these between $\barpsi$ and $\psi$ and using the equations of motion for the electron field $i\,\uslash\!\overset{\rightarrow}{D}\!\!\!\!\!\phantom{D}\psi = m\psi$ and $ i\barpsi\,\uslash\!\overset{\leftarrow}{D}\!\!\!\!\!\phantom{D}= -m \barpsi$, one obtains the useful identity
\beq \label{Bel2}
\barpsi(\gamma^\nu i\LRD^\mu - \gamma^\mu i\LRD^\nu )\psi = \epsilon^{\mu\nu\rho\sigma}\partial_\rho\!\left(\barpsi\gamma_\sigma \gamma_5 \psi\right),
\eeq
which is consistent with the generic form given in Eq. \eqref{genform}. The contribution of the antisymmetric term in Eq.~\eqref{Bel1} to $M^{jk}_{\text{Bel},e}$ is then
\begin{align} \label{Bel3}
\int \ud^3 x\,x^j \,\tfrac{1}{4}\, \barpsi(\gamma^k i\LRD^0 - \gamma^0 i\LRD^k )\psi -(j\leftrightarrow k)&=\tfrac{1}{4}\,\epsilon^{0k\rho\sigma}  \int \ud^3 x\,x^j\partial_\rho\!\left(\barpsi\gamma_\sigma \, \gamma_5 \psi\right) -(j\leftrightarrow k)  \nn \\
&=\tfrac{1}{4}\,\epsilon^{0k\rho\sigma} \int \ud^3 x\left[ \partial_\rho\!\left(x^j \,\barpsi\gamma_\sigma \gamma_5 \psi\right)-\delta^j_\rho\,\barpsi\gamma_\sigma \gamma_5 \psi\right] - (j\leftrightarrow k) \nn \\
&= \tfrac{1}{2}\, \epsilon^{jkl}\int \ud^3 x\,\barpsi\gamma^l \gamma_5 \psi \nn \\
&=\tfrac{1}{2}  \int \ud^3 x \, \psi^\dag \sigma^{jk}  \psi,
\end{align}
where we have discarded the integral of a spatial divergence. The contribution to $M^{jk}_{\text{Bel},e}$ from the first term on the RHS of Eq.~\eqref{Bel1} is
\beq \label{Bel4}
\tfrac{1}{2}\int \ud^3 x\,x^j\, \barpsi\gamma^0 i\LRD^k \psi -(j\leftrightarrow k)= \int \ud^3 x\,\psi^\dag x^j iD^k\psi  - (j\leftrightarrow k),
\eeq
where, again, we have discarded a surface term. Finally then, from Eqs.~\eqref{Bel3} and \eqref{Bel4}, $J^i_{\text{Bel},e} = \tfrac{1}{2} \,\epsilon^{ijk} M^{jk}_{\text{Bel},e}$ can be written in the form of a spin term plus an orbital term in the form used by Ji
\beq \label{Jiel}
\uvec J^e_\text{Ji}=\int \ud^3 x \,\psi^\dag \tfrac{1}{2}\uvec\Sigma\psi + \int \ud^3 x\,\psi^\dag(\uvec x\times i \uvec D)\psi,
\eeq
where we have used Eq.~\eqref{Eq:sigma}. We shall refer to the above terms as $\uvec S^e_\text{Ji}$ and $\uvec L^e_\text{Ji}$, respectively. The photon part of the Belinfante angular momentum follows directly upon substituting in terms of the electric and magnetic fields, \emph{i.e.} using $F^{i0}=E^i$ and $F^{ij}=-\epsilon^{ijk}B^k$, and one arrives at the Ji decomposition:
\beq \label{QEDJi}
\uvec J_\text{QED}=\underbrace {\int\ud^3x\,\psi^\dag\tfrac{1}{2}\uvec\Sigma\psi}_{\uvec S^e_\text{Ji}}+\underbrace{\int\ud^3x\,\psi^\dag(\uvec x\times i\uvec D)\psi}_{\uvec L^e_\text{Ji}}+\underbrace{\int\ud^3x\,\uvec x\times(\uvec E\times\uvec B)}_{\uvec J^\gamma_\text{Ji}},
\eeq
which will be discussed in section \ref{secIVA}. Here and everywhere in the following, $e$ means electron plus positron.

\subsubsection{The canonical angular momentum in QED used by Jaffe and Manohar\label{secIIIB2}}

As explained in section \ref{secIIB2}, the canonical angular momentum density is automatically split into orbital and spin parts for both electrons and photons. From Eqs.~\eqref{QEDoam} and \eqref{Eq:thetacQED}, the electron orbital angular momentum density is given by
\beq \label{eloam}
M^{0jk}_{e,\text{OAM}}=   \tfrac{1}{2}\,x^j(\barpsi\gamma^0 i\LRpartial^k \psi) -(j\leftrightarrow k)=  \psi^\dag x^j i\partial^k\psi  - (j\leftrightarrow k)\, + \,\text{divergence term}.
\eeq
The photon orbital angular momentum density is given by
\begin{align} \label{gammaoam}
M^{0jk}_{\gamma,\text{OAM}}&= -x^j  F^{0l}\partial^k A_l  - (j\leftrightarrow k)\nn\\
&=x^j E^l \partial^k A_l  - (j\leftrightarrow k)  \nn \\
&= E^l x^j\nabla^k A^l  - (j\leftrightarrow k).
\end{align}
The electron spin density term, from Eqs.~\eqref{mspin} and \eqref{spinop} reads
\begin{align} \label{elspin}
M^{0jk}_{e, \text{spin}}&= -i \,\frac{\partial \uL_{\text{clas}}}{\partial(\partial_0 \psi_r)}\, \tfrac{1}{2}(\sigma ^{jk})\lind{r}{s}\psi_s \nn \\
&= \psi^\dag \tfrac{1}{2}\sigma ^{jk}\psi,
\end{align}
where $r$ and $s$ are here Dirac spinor indices, and where we have used $\frac{\partial \uL_{\text{clas}}}{\partial(\partial_0 \psi_r)}=i \psi^{\dag r} $. Finally, the photon spin density term is, from Eqs.~\eqref{spin1op} and \eqref{mspin}
\begin{align} \label{gammaspin}
M^{0jk}_{\gamma, \text{spin}}&=   -i \,\frac{\partial \uL_{\text{clas}}}{\partial(\partial_0 A_\alpha)}\,i(\delta^j_\alpha \,g^{k\beta} -\delta^k_\alpha \, g^{j\beta})A_\beta\nn\\
& =  \frac{\partial \uL_{\text{clas}}}{\partial(\partial_0 A_j)}\,A^k -  (j\leftrightarrow k)  \nn \\
&= E^jA^k-E^kA^j,
\end{align}
where we have used $\frac{\partial \uL_{\text{clas}}}{\partial(\partial_0 A_j)}=E^j$.  Putting these together to form $J^i_\text{QED} = \tfrac{1}{2}\, \epsilon^{ijk} M^{jk}$, one obtains the canonical form of the QED angular momentum used by Jaffe and Manohar:
\beq
\uvec J_\text{QED}=\underbrace {\int\ud^3x\,\psi^\dag\tfrac{1}{2}\uvec\Sigma\psi}_{\uvec S^e_\text{JM}}+\underbrace{\int\ud^3x\,\psi^\dag(\uvec x\times\tfrac{1}{i}\uvec\nabla)\psi}_{\uvec L^e_\text{JM}}+\underbrace{\int\ud^3x\,\uvec E\times\uvec A}_{\uvec S^\gamma_\text{JM}}+\underbrace{\int\ud^3x\,E^i(\uvec x\times\uvec\nabla) A^i}_{\uvec L^\gamma_\text{JM}},
\eeq
which will be discussed in section \ref{secIVA}.

\section{The controversy in detail\label{secIV}}

To keep the presentation as simple as possible, we consider in this section the case of QED. The expressions for more general gauge theories, including QCD, will be given in the next section. We confine also our discussions mainly to \emph{Classical Field Theory}, so that we can spare the complications originating from the quantization procedure. Whenever we refer to quantum operators, these correspond to the \emph{naive} operators obtained by replacing the classical fields in the classical expressions by their quantum counterpart. Note that, to the best of our knowledge, a proper and complete treatment at the quantum level has unfortunately never been achieved in the literature.

\subsection{The main decompositions of the angular momentum in a nutshell\label{secIVA}}

Here, we present and compare the main different decompositions of the angular momentum proposed in the literature and comment on their advantages and disadvantages. Similar decompositions exist for the linear momentum and most of our discussions can easily be transposed.

\subsubsection{The Belinfante decomposition\label{secIVA1}}

As shown in Eq.~\eqref{MBorb}, the Belinfante angular momentum density has a purely orbital appearance, and following the procedure utilized by Belinfante and Rosenfeld \cite{Belinfante:1939,Rosenfeld:1940}, one explicitly obtains the following decomposition
\beq
\uvec J_\text{QED}=\underbrace{\int\ud^3x\,\barpsi\left[\uvec x\times\tfrac{1}{2}\,(\gamma^0 \,i\uvec D+\uvec\gamma\,iD^0)\right]\psi}_{\uvec J^e_\text{Bel}}+\underbrace{\int\ud^3x\,\uvec x\times(\uvec E\times\uvec B)}_{\uvec J^\gamma_\text{Bel}},
\eeq
where the covariant derivative is given by $\uvec D=\uvec\partial+ie\uvec A \equiv -\uvec\nabla+ie\uvec A$ and $D^0=\partial_t+ieA^0$ in accordance with Eq. \eqref{QEDCovD}. The quantities $\uvec J^e_\text{Bel}$ and $\uvec J^\gamma_\text{Bel}$ are interpreted as the electron and photon total angular momentum, respectively.
\newline

\emph{Advantages}
\bi
\item Each separate term is a gauge-invariant quantity and therefore measurable in principle;
\ei

\emph{Disadvantages}
\bi
\item There is no decomposition of the total angular momentum into spin and OAM contributions;
\item The individual contributions $\uvec J^e_\text{Bel}$ and $\uvec J^\gamma_\text{Bel}$, seen as operators, do not satisfy the generic equal-time commutation relations $[J^i,J^j]=i\epsilon^{ijk}J^k$ defining angular momentum operators in a quantum theory;
\item $\uvec J^e_\text{Bel}$ and $\uvec J^\gamma_\text{Bel}$ are \emph{not} generators of rotations.
\ei

\subsubsection{The Ji decomposition\label{secIVA2}}

In section \ref{secIIIB1}, we showed how the Belinfante angular momentum could be rewritten in such a way that the electron angular momentum was split into a sum of a spin and orbital term. This is the form used by Ji in his seminal paper relating the quark and gluon angular momenta to GPDs and, following the tradition in the literature, we shall therefore refer to it as the ``Ji decomposition''. One has thus
\beq \label{QEDJi}
\uvec J_\text{QED}=\underbrace {\int\ud^3x\,\psi^\dag\tfrac{1}{2}\uvec\Sigma\psi}_{\uvec S^e_\text{Ji}}+\underbrace{\int\ud^3x\,\psi^\dag(\uvec x\times i\uvec D)\psi}_{\uvec L^e_\text{Ji}}+\underbrace{\int\ud^3x\,\uvec x\times(\uvec E\times\uvec B)}_{\uvec J^\gamma_\text{Ji}}.
\eeq
Interestingly, this decomposition can be obtained by adding the surface term $-\int\ud^3x\,\nabla^i[E^i(\uvec x\times\uvec A)]$ to the Jaffe-Manohar decomposition given in the next subsection, and by using the equation of motion $\uvec\nabla\cdot\uvec E=e\psi^\dag\psi$. The quantities $\uvec S^e_\text{Ji}$,  $\uvec L^e_\text{Ji}$, and $\uvec J^\gamma_\text{Ji}$ are interpreted as the electron spin, electron OAM, and photon total angular momentum, respectively.
\newline

\emph{Advantages}
\bi
\item Each separate term is a gauge-invariant quantity and therefore measurable in principle;
\item The presence of the covariant derivative $\uvec D=-\uvec\nabla+ie\uvec A$ suggests that the electron OAM $\uvec L^e_\text{Ji}$ is kinetic, \emph{i.e.} corresponds to the classical definition $\uvec x\times\uvec p_\text{kin}$, where $\uvec p_\text{kin}=m\uvec v$ with $\uvec v$ the velocity of the particle, according to the common understanding of Classical Electrodynamics;
\item The photon total angular momentum coincides with the corresponding Belinfante expression $\uvec J^\gamma_\text{Ji}=\uvec J^\gamma_\text{Bel}$. This holds also for the electron total angular momentum $\uvec S^e_\text{Ji}+\uvec L^e_\text{Ji}=\uvec J^e_\text{Bel}$ up to a surface term. The Ji decomposition generalizes therefore the Belinfante decomposition by providing an explicit gauge-invariant decomposition of the electron total angular momentum into spin and OAM contributions up to a surface term.
\ei

\emph{Disadvantages}
\bi
\item There is no decomposition of the photon total angular momentum into spin and OAM contributions;
\item The individual contributions $\uvec L^e_\text{Ji}$ and $\uvec J^\gamma_\text{Ji}$, seen as operators, do not satisfy the generic equal-time commutation relations $[J^i,J^j]=i\epsilon^{ijk}J^k$ defining angular momentum operators in a quantum theory. Only $\uvec S^e_\text{Ji}$ and the combination $\uvec L^e_\text{Ji}+\uvec J^\gamma_\text{Ji}$ can be considered as quantum angular momentum operators;
\item Contrary to $\uvec S^e_\text{Ji}$, the operators $\uvec L^e_\text{Ji}$ and $\uvec J^\gamma_\text{Ji}$ are \emph{not} generators of rotations.
\ei

\subsubsection{The Jaffe-Manohar decomposition\label{secIVA3}}

The Jaffe-Manohar (JM) decomposition of angular momentum \cite{Jaffe:1989jz} simply corresponds to the well-known canonical angular momentum decomposition which follows from Noether's theorem as shown in section \ref{secIIIB2}:
\beq
\uvec J_\text{QED}=\underbrace {\int\ud^3x\,\psi^\dag\tfrac{1}{2}\uvec\Sigma\psi}_{\uvec S^e_\text{JM}}+\underbrace{\int\ud^3x\,\psi^\dag(\uvec x\times\tfrac{1}{i}\uvec\nabla)\psi}_{\uvec L^e_\text{JM}}+\underbrace{\int\ud^3x\,\uvec E\times\uvec A}_{\uvec S^\gamma_\text{JM}}+\underbrace{\int\ud^3x\,E^i(\uvec x\times\uvec\nabla) A^i}_{\uvec L^\gamma_\text{JM}}.
\eeq
The quantities $\uvec S^e_\text{JM}$,  $\uvec L^e_\text{JM}$, $\uvec S^\gamma_\text{JM}$, and $\uvec L^\gamma_\text{JM}$ are interpreted as the electron spin, electron OAM, photon spin, and photon OAM, respectively.
\newline

\emph{Advantages}
\bi
\item The decomposition into electron/photon and spin/OAM contributions is complete;
\item Each of the individual Jaffe-Manohar terms, seen as operators, satisfies the generic equal-time commutation relations $[J^i,J^j]=i\epsilon^{ijk}J^k$ defining angular momentum operators in a quantum theory;
\item As follows from Noether's theorem, the Jaffe-Manohar operators (being simply the canonical operators) are the generators of rotations for the electron and photon fields\footnote{The global minus sign on the RHS comes from the fact that rotating the point in some direction is equivalent to rotating the field in the opposite direction. The field conjugate to $\psi$ is $\pi_\psi=i\psi^\dag$, and the field conjugate to $A^i$ is $\pi_{A^i}=-E^i$.}
\begin{align}
[S^{e,i}_\text{JM},\psi]&=-\tfrac{1}{2}\Sigma^i\psi,\\
[L^{e,i}_\text{JM},\psi]&=-(\uvec x\times\tfrac{1}{i}\uvec\nabla)^i\psi,\\
[S^{\gamma,i}_\text{JM},A^j]&=-(-i\epsilon^{ijk})A^k,\\
[L^{\gamma,i}_\text{JM},A^j]&=-(\uvec x\times\tfrac{1}{i}\uvec\nabla)^iA^j,
\end{align}
where all the fields are considered at equal time. We recall that for an infinitesimal transformation of a quantum field $\phi$, we have $e^{iJ\varepsilon}\phi e^{-iJ\varepsilon}=\phi+\delta\phi$ with $\delta\phi=i[J,\phi]\varepsilon$ to first order in $\varepsilon$.
\ei

\emph{Disadvantages}
\bi
\item The individual contributions $\uvec L^e_\text{JM}$, $\uvec S^\gamma_\text{JM}$, and $\uvec L^\gamma_\text{JM}$ are gauge non-invariant quantities and therefore not obviously measurable\footnote{This will be discussed in section \ref{secIVC4}.}. Only $\uvec S^e_\text{JM}$ and the combination $\uvec L^e_\text{JM}+\uvec S^\gamma_\text{JM}+\uvec L^\gamma_\text{JM}$ are gauge invariant, and therefore measurable in principle.
\ei

\subsubsection{The Chen \emph{et al.} decomposition\label{secIVA4}}

More recently, Chen \emph{et al.} emphasized that the gauge potential plays a dual role. On the one hand, it allows one to define a covariant derivative. On the other hand, it provides the coupling between the charged particles and the electromagnetic field. The first role is just related to the issue of gauge symmetry and so has to do only with the unphysical gauge degrees of freedom. The second role clearly involves the physical degrees of freedom, namely the two polarizations of the photon. Chen \emph{et al.} then proposed to split the gauge potential into so-called \emph{pure-gauge} and \emph{physical} terms playing, respectively, the first and second role
\be\label{decomposition}
\uvec A=\uvec A_\pure+\uvec A_\phys.
\ee
The two terms are defined by the constraints
\beq\label{Apurephysdef}
\uvec\nabla\times\uvec A_\pure=\uvec 0, \qquad
\uvec\nabla\cdot\uvec A_\phys=0,
\eeq
implying in particular that $\uvec A_\pure=-\uvec\nabla\alpha_\pure$, where $\alpha_\pure$ is some scalar function. Note that one has, on account of Eq.~\eqref{Apurephysdef},
\beq \label{ChenMag}
\uvec B = \uvec \nabla\times\uvec A = \uvec\nabla\times\uvec A_\phys.
\eeq
Under a gauge transformation, one has $\alpha_\pure\mapsto \tilde\alpha_\pure=\alpha_\pure+\alpha$ leading to the transformation laws
\beq\label{Gtransformation}
\begin{split}
\uvec A_\pure&\mapsto\tilde{\uvec A}_\pure=\uvec A_\pure-\uvec\nabla\alpha,\\
\uvec A_\phys&\mapsto\tilde{\uvec A}_\phys=\uvec A_\phys.
\end{split}
\eeq

The Chen \emph{et al.} decomposition of angular momentum \cite{Chen:2008ag,Chen:2009mr} reads
\beq \label{CHEN}
\uvec J_\text{QED}=\underbrace {\int\ud^3x\,\psi^\dag\tfrac{1}{2}\uvec\Sigma\psi}_{\uvec S^e_\text{Chen}}+\underbrace{\int\ud^3x\,\psi^\dag(\uvec x\times i\uvec D_\pure)\psi}_{\uvec L^e_\text{Chen}}+\underbrace{\int\ud^3x\,\uvec E\times\uvec A_\phys}_{\uvec S^\gamma_\text{Chen}}+\underbrace{\int\ud^3x\,E^i(\uvec x\times\uvec\nabla) A^i_\phys}_{\uvec L^\gamma_\text{Chen}},
\eeq
where the pure-gauge covariant derivative is defined as $\uvec D_\pure=-\uvec\nabla+ie\uvec A_\pure$. This decomposition can be obtained by adding the surface term $-\int\ud^3x\,\nabla^i[E^i(\uvec x\times\uvec A_\pure)]$ to the Jaffe-Manohar decomposition, and by using the equation of motion $\uvec\nabla\cdot\uvec E=e\psi^\dag\psi$. The quantities $\uvec S^e_\text{Chen}$,  $\uvec L^e_\text{Chen}$, $\uvec S^\gamma_\text{Chen}$, and $\uvec L^\gamma_\text{Chen}$ are interpreted as the electron spin, electron OAM, photon spin, and photon OAM, respectively.
\newline

\emph{Advantages}
\bi
\item The decomposition into electron/photon and spin/OAM contributions is complete;
\item Each of the individual Chen \emph{et al.} terms, seen as operators, satisfy the generic equal-time commutation relations $[J^i,J^j]=i\epsilon^{ijk}J^k$ defining angular momentum operators in a quantum theory;
\item The Chen \emph{et al.} operators are the generators of rotations of the ``physical'' photon field and a ``physical'', gauge-invariant version of the electron field \cite{Chen:2012vg,Lorce:2012rr,Lorce:2013gxa}
\beq \label{ElPhys}
\hat\psi\equiv e^{ie\alpha_\pure}\psi,
\eeq
\emph{i.e.} satisfy the standard canonical commutation relations
\begin{align}
[S^{e,i}_\text{Chen},\hat\psi]&=-\tfrac{1}{2}\Sigma^i\hat\psi,\\
[L^{e,i}_\text{Chen},\hat\psi]&=-(\uvec x\times \tfrac{1}{i}\uvec\nabla)\hat\psi,\label{LeChen}\\
[S^{\gamma,i}_\text{Chen},A^j_\phys]&=-(-i\epsilon^{ijk})A^k_\phys,\\
[L^{\gamma,i}_\text{Chen},A^j_\phys]&=-(\uvec x\times\tfrac{1}{i}\uvec\nabla)^iA^j_\phys,
\end{align}
where all the fields are considered at equal time;
\item Thanks to the pure-gauge covariant derivative $\uvec D_\pure=-\uvec\nabla+ie\uvec A_\pure$ and the gauge transformation laws given in Eq. \eqref{Gtransformation}, one can easily check that each separate term is a gauge-invariant quantity, and therefore measurable in principle. Note in particular that, just like the partial derivatives $\uvec\nabla$, the pure-gauge covariant derivatives commute with each other $[D^i_\pure,D^j_\pure]=0$. Moreover, since we may choose the gauge, we can take $\uvec A_\pure=0$ \emph{i.e.}  $\uvec A_\phys = \uvec A$, in which case we are in the Coulomb gauge $\uvec\nabla\cdot\uvec A=0$. It then appears that the Chen \emph{et al.} decomposition gives the same physical answer as the Jaffe-Manohar decomposition in the Coulomb gauge. In other words, the Chen \emph{et al.} decomposition is \emph{physically equivalent} to the Jaffe-Manohar decomposition in the Coulomb gauge.
\ei

\emph{Disadvantages}
\bi
\item Although gauge-invariant, the Chen \emph{et al.} decomposition makes the Coulomb gauge special, which seems to contradict the spirit of gauge invariance;
\item The fields involved are non-local in terms of $\uvec A$. For example, $\uvec A_\phys $ is given by
\beq \label{ChenNonLoc}
\uvec A_\phys = \uvec A -  \uvec\nabla\frac{1}{\uvec\nabla ^2}\uvec\nabla\cdot \uvec A,   \eeq
where it should be remembered that $\frac{1}{\uvec\nabla ^2}$ is an integral operator
\beq \label{inverseLaplace}
\frac{1}{\uvec\nabla ^2}f(\uvec x)  \equiv -\frac{1}{4\pi}\int \ud^3x' \,\frac{f(\uvec x')}{| \uvec x - \uvec x'|}.
\eeq
\ei

\subsubsection{The Wakamatsu decomposition\label{secIVA5}}

The Wakamatsu decomposition of angular momentum \cite{Wakamatsu:2010qj} starts from the Chen \emph{et al.} decomposition (\ref{CHEN}) and subtracts the so-called \emph{potential} angular momentum\footnote{In the literature, this term is also known as the \emph{bound} angular momentum.}
\beq \label{POT1}
\uvec L_\text{pot}=\int\ud^3x\,e\psi^\dag\psi\,(\uvec x\times\uvec A_\phys)
\eeq
from the electron OAM and compensates for this by adding the term
\beq \label{POT2} \uvec L_\text{pot}= \int\ud^3x\,(\uvec\nabla\cdot\uvec E)\,\uvec x\times\uvec A_\phys \eeq
to the photon OAM, where use has been made of the fact that
the two expressions for $\uvec L_\text{pot}$ coincide as a consequence of the equation of motion $\uvec\nabla\cdot\uvec E=e\psi^\dag\psi$. The result is
\beq
\uvec J_\text{QED}=\underbrace {\int\ud^3x\,\psi^\dag\tfrac{1}{2}\uvec\Sigma\psi}_{\uvec S^e_\text{Wak}}+\underbrace{\int\ud^3x\,\psi^\dag(\uvec x\times i\uvec D)\psi}_{\uvec L^e_\text{Wak}}+\underbrace{\int\ud^3x\,\uvec E\times\uvec A_\phys}_{\uvec S^\gamma_\text{Wak}}+\underbrace{\int\ud^3x\left[E^i(\uvec x\times\uvec\nabla) A^i_\phys+(\uvec\nabla\cdot\uvec E)\,\uvec x\times\uvec A_\phys\right]}_{\uvec L^\gamma_\text{Wak}}.
\eeq
Thus the Wakamatsu and Chen \emph{et al.} decompositions simply differ in the way $\uvec L_\text{pot}$ is attributed to either the electron or the photon
\begin{align}
\uvec L^e_\text{Wak}&=\uvec L^e_\text{Chen}-\uvec L_\text{pot},\\
\uvec L^\gamma_\text{Wak}&=\uvec L^\gamma_\text{Chen}+\uvec L_\text{pot}.
\end{align}
Lorc\'e \cite{Lorce:2013fpa} recently noticed that $\uvec L^\gamma_\text{Wak}$ can be more compactly written as
\beq\label{LWak}
\uvec L^\gamma_\text{Wak}=-\int\ud^3x\,\uvec x\times\left[(\uvec A_\phys\times\uvec\nabla)\times\uvec E\right]
\eeq
after dropping a surface term $\int\ud^3x\,\uvec x\times\uvec \nabla(\uvec E\cdot\uvec A_\phys)=\int\ud^3x\,\nabla^i(\epsilon^{ijk}\uvec e_jx^k \uvec E\cdot\uvec A_\phys)$. This has the form of an OAM term since the photon kinetic momentum, usually given in terms of the integral of the Poynting vector $\uvec E \times \uvec B$, can be rewritten as
\beq\label{PWak}
\uvec P^\gamma_\text{kin}=\int\ud^3x\,\uvec E\times\uvec B=-\int\ud^3x\,(\uvec A_\phys\times\uvec\nabla)\times\uvec E,
\eeq
where, according to Eq.~\eqref{ChenMag}, $\uvec B=\uvec\nabla\times\uvec A_\phys$ has been used and a surface term $-\int\ud^3x\,(\uvec\nabla\times\uvec A_\phys)\times\uvec E=\int\ud^3x\,\nabla^i(\uvec e_i\,\uvec E\cdot\uvec A_\phys-E^i\uvec A_\phys)$ has been dropped\footnote{Here $\uvec\nabla$ acts on $\uvec E$ as well.} in the last expression. The quantities $\uvec S^e_\text{Wak}$,  $\uvec L^e_\text{Wak}$, $\uvec S^\gamma_\text{Wak}$, and $\uvec L^\gamma_\text{Wak}$ are interpreted as the electron spin, electron OAM, photon spin, and photon OAM, respectively.
\newline

\emph{Advantages}
\bi
\item The decomposition into electron/photon and spin/OAM contributions is complete;
\item Each separate term is a gauge-invariant quantity, and therefore measurable in principle. Note that the potential angular momentum $\uvec L_\text{pot}$ is also gauge invariant;
\item The presence of the covariant derivative $\uvec D=-\uvec\nabla+ie\uvec A$ suggests that the electron OAM $\uvec L^e_\text{Wak}$ is kinetic, \emph{i.e.} corresponds to the classical definition $\uvec x\times\uvec p_\text{kin}$, where $\uvec p_\text{kin}=m\uvec v$ with $\uvec v$ the velocity of the particle, according to the common understanding of Classical Electrodynamics. Similarly, Eqs. \eqref{LWak} and \eqref{PWak} also indicate that $\uvec L^\gamma_\text{Wak}$ is kinetic;
\item Since the electron spin and orbital terms coincide exactly with the Ji expressions, the photon total angular momentum must coincide with the corresponding Ji and Belinfante expressions $\uvec S^{\gamma}_\text{Wak}+\uvec L^{\gamma}_\text{Wak}=\uvec J^{\gamma}_\text{Ji}=\uvec J^{\gamma}_\text{Bel}$ up to a surface term. The Wakamatsu decomposition generalizes therefore the Ji decomposition by providing an explicit gauge-invariant decomposition of the photon total angular momentum into spin and OAM contributions up to a surface term.
\ei

\emph{Disadvantages}
\bi
\item Although gauge-invariant, the Wakamatsu decomposition (just like the Chen \emph{et al.} decomposition) makes the Coulomb gauge special, which seems to contradict the spirit of gauge invariance;
\item The individual contributions $\uvec L^e_\text{Wak}$ and $\uvec L^\gamma_\text{Wak}$, seen as operators, do not satisfy the generic equal-time commutation relations $[J^i,J^j]=i\epsilon^{ijk}J^k$ defining angular momentum operators in a quantum theory. Only the spin operators $\uvec S^e_\text{Wak}$ and $\uvec S^\gamma_\text{Wak}$, and the total OAM operator $\uvec L^e_\text{Wak}+\uvec L^\gamma_\text{Wak}$ can be considered as quantum angular momentum operators;
\item Contrary to the spin operators $\uvec S^e_\text{Wak}$ and $\uvec S^\gamma_\text{Wak}$, the OAM operators $\uvec L^e_\text{Wak}$ and $\uvec L^\gamma_\text{Wak}$ are not generators of rotations;
\item As in the Chen \emph{et al.} decomposition, the ``physical'' photon field is a non-local expression in terms of $\uvec A$.
\ei

\subsubsection{A classification of the different decompositions\label{secIVA6}}

Apart from the Belinfante decompostion, all the other decompositions presented above share a common piece, namely the electron spin contribution $\uvec S^e_\text{JM}=\uvec S^e_\text{Ji}=\uvec S^e_\text{Chen}=\uvec S^e_\text{Wak}$. They then just differ in the way the rest of the total angular momentum is shared between the electron OAM and the photon angular momentum.

%\subsubsection{Two families of decompositions}

\begin{figure}[t!]
	\centering
		\includegraphics[width=.6\textwidth]{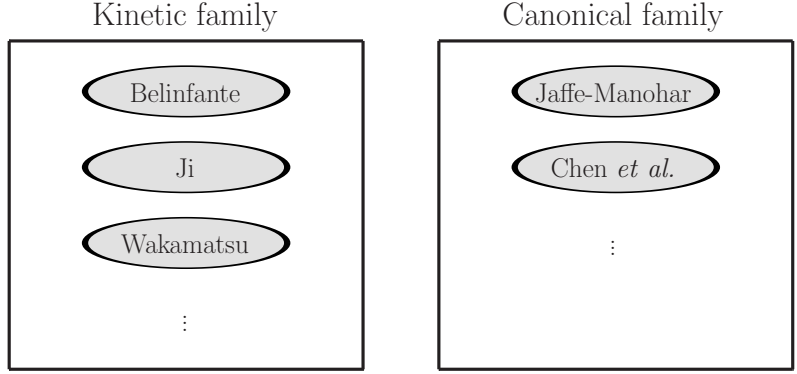}
\caption{\footnotesize{The Wakamatsu classification of proton spin decompositions into two families. See text for more details.}}
		\label{classification_short}
\end{figure}

As summarized by Wakamatsu \cite{Wakamatsu:2010cb}, all these decompositions can be sorted into two families\footnote{Wakamatsu did not consider the Belinfante decomposition in his classification. We have added it for completeness.}, see Fig. \ref{classification_short}:
\begin{itemize}
\item The \emph{kinetic} family (Wakamatsu's family I), where the potential angular momentum is attributed to the photon. The Belinfante, Ji and Wakamatsu decompositions are members of the kinetic family.
\item The \emph{canonical} family (Wakamatsu's family II), where the potential angular momentum is attributed to the electron. The Jaffe-Manohar and Chen \emph{et al.} decompositions are members of the canonical family.
\end{itemize}
Since the potential angular momentum contribution is likely non-vanishing, decompositions belonging to different families are expected to be physically inequivalent. While the difference is small in non-relativistic systems like the atom \cite{Burkardt:2008ua,Wakamatsu:2012ve,Ji:2012gc}, it becomes significant for relativistic systems like the proton \cite{Chen:2009mr,Cho:2011ee}.

The potential angular momentum is itself a gauge-invariant quantity. Therefore, the splitting of the gauge potential into pure-gauge and physical terms allows one to decompose the proton spin into \emph{five} gauge-invariant contributions, instead of the expected four. Based on this observation, Leader \cite{Leader:2011za} criticized Wakamatsu's classification arguing that one could in fact consider an infinite number of families by attributing a fraction $\alpha$ of the potential term to the electrons and the remaining fraction $(1-\alpha)$ to the photons. Note however that only the values $\alpha=0,1$ are natural as they simply correspond to the kinetic and canonical OAM, respectively. Leader favors the canonical version  because the operators, at least at equal time, generate the expected rotations of the relevant fields, and this seems a reasonable property to demand for an angular momentum operator.

There is no totally convincing answer as to which family should be preferred. Deciding which family is the ``physical'' one appears to be essentially a matter of taste. This is somehow analogous to the scheme dependence in parton distribution functions, where it has been  understood for some time that it is meaningless to claim that one has measured a quark distribution $q(x)$, but must specify which scheme \emph{e.g.} $\overline{MS}$ or $DIS$ has been used. As long as one indicates clearly which version of the angular momentum one is using, it is irrelevant which one chooses.

\subsection{The covariant form of the decompositions\label{secIVB}}

The Chen \emph{et al.} split raised some concerns about the Lorentz symmetry as the definition of the pure-gauge and physical terms given in Eq. \eqref{Apurephysdef} does not look Lorentz covariant. More precisely, the question is: does the split of the Lorentz-transformed gauge potential coincide with the Lorentz transform of the split?
\beq\label{question}
\begin{split}
(\uvec A')_\pure&\stackrel{?}{=}(\uvec A_\pure)',\\
(\uvec A')_\phys&\stackrel{?}{=}(\uvec A_\phys)',
\end{split}
\eeq
where the prime indicates that the field is Lorentz transformed. To address this question, Wakamatsu developed a covariant version of the Chen \emph{et al.} decomposition \cite{Wakamatsu:2010cb}.

The starting point is the split of the four-component gluon potential into pure-gauge and physical terms
\begin{equation}\label{Adecomp}
A_\mu(x)=A^\pure_\mu(x)+A^\phys_\mu(x).
\end{equation}
This is simply the covariant version of the split of the gauge potential proposed by Chen \emph{et al.} \cite{Chen:2008ag}. As noticed in Ref. \cite{Lorce:2013gxa}, this approach is not new, since it had already been adopted in 1962 by Schwinger \cite{Schwinger:1962zz,Schwinger:1962fg} and followed by Arnowitt and Fickler \cite{Arnowitt:1962cv}. Moreover, the same idea reappeared in the works of \emph{e.g.} Goto \cite{Goto:1966}, Treat \cite{Treat:1973yc,Treat:1975dz}, Duan \cite{Duan:1979,Duan:1984cb,Duan:1998um,Duan:2002vh}, Fulp \cite{Fulp:1983bt}, and Kashiwa and Tanimura \cite{Kashiwa:1996rs,Kashiwa:1996hp}. However, since Chen \emph{et al.} revived this idea in the context of the controversy about the angular momentum decomposition, we shall refer to the generic split \eqref{Adecomp} as the ``Chen \emph{et al.} approach''.

By definition, the pure-gauge field is unphysical and therefore cannot contribute to the field-strength tensor
\begin{equation}\label{Apuredef}
F^\pure_{\mu\nu}=\partial_\mu A^\pure_\nu-\partial_\nu A^\pure_\mu=0.
\end{equation}
Moreover, it is \emph{assumed} to have the same gauge transformation law as the original gauge potential\footnote{Note that, in the non-covariant case, this followed from the conditions \eqref{Apurephysdef} on $\uvec A_\pure$ and $\uvec A_\phys$.}
\beq\label{gaugeTpure}
A_\mu^\pure(x)\mapsto\tilde A_\mu^\pure(x)=A_\mu^\pure(x)+\partial_\mu\alpha(x).
\eeq
Note that the precise definition of the physical field is postponed until a later stage. Nonetheless the conditions \eqref{Adecomp}-\eqref{gaugeTpure} are actually sufficient for achieving a gauge-invariant decomposition of the angular momentum. The defining constraint \eqref{Apuredef} implies that the pure-gauge field can be put it the form
\begin{equation}\label{puregauge}
A_\mu^\pure(x)=\partial_\mu\alpha_\pure(x),
\end{equation}
where $\alpha_\pure(x)$ is some scalar function of spacetime. From Eq. \eqref{gaugeTpure}, one easily derives the gauge transformation laws of the scalar and the physical fields
\beq
\begin{split}\label{gaugeT}
\alpha_\pure(x)&\mapsto\tilde\alpha_\pure(x)=\alpha_\pure(x)+\alpha(x),\\
A_\mu^\phys(x)&\mapsto\tilde A_\mu^\phys(x)=A_\mu^\phys(x).
\end{split}
\eeq
Note, in particular, that the physical field in QED is gauge invariant, just like the field-strength tensor. Because of Eq.~\eqref{Apuredef}, the latter can simply be expressed as
\beq \label{Fphys}
F_{\mu\nu}=\partial_\mu A^\phys_\nu-\partial_\nu A^\phys_\mu.
\eeq

\subsubsection{The gauge-invariant canonical decomposition\label{secIVB1}}

The Wakamatsu covariant generalization of the Chen \emph{et al.} decomposition will be referred to as the \emph{gauge-invariant canonical} (gic) decomposition. It reads at the density level \cite{Wakamatsu:2010cb}
\beq
M^{\mu\nu\rho}=S^{\mu\nu\rho}_{\text{gic},e}+L^{\mu\nu\rho}_{\text{gic},e}+S^{\mu\nu\rho}_{\text{gic},\gamma}+L^{\mu\nu\rho}_{\text{gic},\gamma}+M^{\mu\nu\rho}_{\text{gic},\text{boost}}+ \text{four-divergence},
\eeq
where the electron spin, electron OAM, photon spin, and photon OAM densities are given by
\beq\label{Chenetaldec}
\begin{split}
S^{\mu\nu\rho}_{\text{gic},e}&=\tfrac{1}{2}\,\epsilon^{\mu\nu\rho\sigma}\barpsi\gamma_\sigma\gamma_5\psi,\\
L^{\mu\nu\rho}_{\text{gic},e}&=\barpsi \gamma^\mu x^{[\nu}iD^{\rho]}_\pure\psi,\\
S^{\mu\nu\rho}_{\text{gic},\gamma}&=-F^{\mu[\nu} A^{\rho]}_\phys,\\
L^{\mu\nu\rho}_{\text{gic},\gamma}&=-F^{\mu\alpha}x^{[\nu} \partial^{\rho]} A^\phys_\alpha,
\end{split}
\eeq
with  $D^\pure_\mu=\partial_\mu+ie A^\pure_\mu$ the pure-gauge covariant derivative, and $\epsilon^{\mu\nu\rho\sigma}$ the totally antisymmetric Levi-Civita tensor satisfying $\epsilon_{0123}=+1$. We do not need to write down explicitly the boost term, because it does not contribute to the angular momentum expressions $M^{0ij}_{\text{gic},\text{boost}}=0$. We also do not need to write down explicitly the four-divergence term as it corresponds, once integrated over spatial coordinates, to a surface term assumed to vanish. The complete expressions for a generic gauge theory will however be given in section \ref{secV}.

This decomposition is clearly gauge invariant and has a strong resemblance with the covariant form of the Jaffe-Manohar decomposition. Indeed, starting from the gauge-invariant canonical decomposition, one can choose to work in the gauge where $A_\mu^\pure=0$, obtained with the gauge transformation function $\alpha(x)=-\alpha_\pure(x)$. In that gauge, one has $A_\mu=A^\phys_\mu$ and $D^\pure_\mu=\partial_\mu$, so that the gauge-invariant canonical decomposition takes the same mathematical form as the Jaffe-Manohar decomposition
\beq\label{JaffeManohardec}
\begin{split}
S^{\mu\nu\rho}_{\text{gic},e}\big|_{A^\pure_\mu=0}&=S^{\mu\nu\rho}_{\text{JM},e},\\
L^{\mu\nu\rho}_{\text{gic},e}\big|_{A^\pure_\mu=0}&=L^{\mu\nu\rho}_{\text{JM},e},\\
S^{\mu\nu\rho}_{\text{gic},\gamma}\big|_{A^\pure_\mu=0}&=S^{\mu\nu\rho}_{\text{JM},\gamma},\\
L^{\mu\nu\rho}_{\text{gic},\gamma}\big|_{A^\pure_\mu=0}&=L^{\mu\nu\rho}_{\text{JM},\gamma}.
\end{split}
\eeq
The gauge-invariant canonical decomposition can then be thought of as a \emph{gauge-invariant extension} (GIE) of the Jaffe-Manohar decomposition  \cite{Hoodbhoy:1998bt,Ji:2012gc}. However, as stressed \emph{e.g.} by Hatta, in this scheme there is no actual definition given for the field $A^\mu_\phys$. The issue of the freedom in choosing  $A^\mu_\phys$ will be discussed in the next section. The concept of GIE can be applied to any gauge non-invariant quantity like \emph{e.g.} the Chern-Simons current \cite{Guo:2012wv}. It consists in finding a gauge-invariant quantity that gives the same \emph{physical results} as a gauge non-invariant quantity evaluated in a specific gauge \cite{Hoodbhoy:1998bt,Ji:2012gc}.

\subsubsection{The gauge-invariant kinetic decomposition\label{secIVB2}}

Wakamatsu also proposed a second type of gauge-invariant covariant decomposition, which will be referred to as the \emph{gauge-invariant kinetic} (gik) decomposition, and is simply related to the gauge-invariant canonical decomposition as follows \cite{Wakamatsu:2010cb}
\beq\label{Wakamatsudec}
\begin{split}
S^{\mu\nu\rho}_{\text{gik},e}&=S^{\mu\nu\rho}_{\text{gic},e},\\
L^{\mu\nu\rho}_{\text{gik},e}&=L^{\mu\nu\rho}_{\text{gic},e}-L^{\mu\nu\rho}_\text{pot},\\
S^{\mu\nu\rho}_{\text{gik},\gamma}&=S^{\mu\nu\rho}_{\text{gic},\gamma},\\
L^{\mu\nu\rho}_{\text{gik},\gamma}&=L^{\mu\nu\rho}_{\text{gic},\gamma}+L^{\mu\nu\rho}_\text{pot},
\end{split}
\eeq
where, following Konopinski's terminology \cite{Konopinski}, the covariant potential angular momentum is given by
\beq
\begin{split}
L^{\mu\nu\rho}_\text{pot}&=e\barpsi\gamma^\mu \psi\,x^{[\nu}A_\text{phys}^{\rho]}\\
&=(\partial_\alpha F^{\alpha\mu})\,x^{[\nu}A^{\rho]}_\text{phys}.
\end{split}
\eeq
Note that the QED equation of motion $\partial_\alpha F^{\alpha\mu}=e\barpsi\gamma^\mu\psi$ has been used in order to be able to write the potential angular momentum as either an electron or a photon contribution.  The sum of the photon spin and OAM appearing in the gauge-invariant kinetic decomposition coincides (up to a four-divergence term) with the photon total angular momentum appearing in the covariant form of the Ji decomposition
\beq\label{Wakamatsudec}
\begin{split}
S^{\mu\nu\rho}_{\text{Ji},e}&=S^{\mu\nu\rho}_{\text{gik},e},\\
L^{\mu\nu\rho}_{\text{Ji},e}&=L^{\mu\nu\rho}_{\text{gik},e},\\
J^{\mu\nu\rho}_{\text{Ji},\gamma}&=S^{\mu\nu\rho}_{\text{gik},\gamma}+L^{\mu\nu\rho}_{\text{gik},\gamma}+\text{four-divergence}.
\end{split}
\eeq
Moreover, following Lorc\'e's observation \cite{Lorce:2013fpa}, we note that the gauge-invariant kinetic photon OAM can alternatively be written as
%\trd{\beq
%L^{\mu\nu\rho}_{\gamma,\text{gik}}=x^{[\nu}g^{\rho][\alpha}\delta^{\beta]}_\lambda\,A^\phys_\alpha\partial_\beta F^{\lambda\mu}+\text{four-divergence},
%\eeq}
\beq
L^{\mu\nu\rho}_{\text{gik},\gamma}=x^\nu \left(A^\rho_\phys \partial_\lambda -A^\phys_\lambda\partial^\rho\right) F^{\lambda\mu} - (\nu \leftrightarrow \rho)+\text{four-divergence},
\eeq
where one recognizes the gauge-invariant kinetic photon momentum
\beq
T^{\mu\rho}_{\text{gik},\gamma}=F^{\mu\lambda}F_\lambda^{\phantom{\alpha}\rho}=\left(A^\rho_\phys \partial_\lambda -A^\phys_\lambda\partial^\rho\right) F^{\lambda\mu} +\text{four-divergence},
\eeq
where we have used Eq.~\eqref{Fphys} and ignored the term $g^{\mu\rho}\,\tfrac{1}{4} F^{\alpha\beta}F_{\alpha\beta}$, because it does not contribute to the momentum expressions obtained with $\mu=0$ and $\rho=i$. Complete expressions for a general gauge theory will be given in the next section.

\subsection{The ambiguity in defining $A^\phys_\mu$\label{secIVC}}

Let us pause for a moment and discuss a new issue raised by the Wakamatsu covariant form. While the conditions \eqref{Apuredef} and \eqref{gaugeTpure} on the pure-gauge term are sufficient to construct complete gauge-invariant decompositions of the angular momentum in a seemingly covariant form, it is actually not sufficient to determine the precise form of $A^\pure_\mu$ and $A^\phys_\mu$. As observed by Stoilov \cite{Stoilov:2010pv} and discussed in more detail by Lorc\'e \cite{Lorce:2012rr,Lorce:2012ce}, the split of the gauge potential into pure-gauge and physical terms introduces a \emph{new symmetry}.

\subsubsection{The Stueckelberg symmetry\label{secIVC1}}

For a given split into pure-gauge and physical terms $A_\mu=A^\pure_\mu+A^\phys_\mu$, it is always possible to define a new split $A_\mu=\bar A^\pure_\mu+\bar A^\phys_\mu$ where
\beq
\begin{split}\label{StueckabT}
\bar A^\pure_\mu(x)&=A^\pure_\mu(x)+\partial_\mu C(x),\\
\bar A^\phys_\mu(x)&=A^\phys_\mu(x)-\partial_\mu C(x)
\end{split}
\eeq
with $C(x)$ an arbitrary scalar function of spacetime. Notice that this is \emph{not} a gauge transformation since $\bar A_\mu = A_\mu$. The new pure-gauge term $\bar A^\pure_\mu$ automatically satisfies the condition \eqref{Apuredef}
\beq
F^\pure_{\mu\nu}=0\qquad\Rightarrow\qquad \bar F^\pure_{\mu\nu}=0,
\eeq
and also the gauge transformation law \eqref{gaugeTpure}, provided that $C(x)$ is gauge invariant.

\begin{figure}[t!]
	\centering
		\includegraphics[width=.5\textwidth]{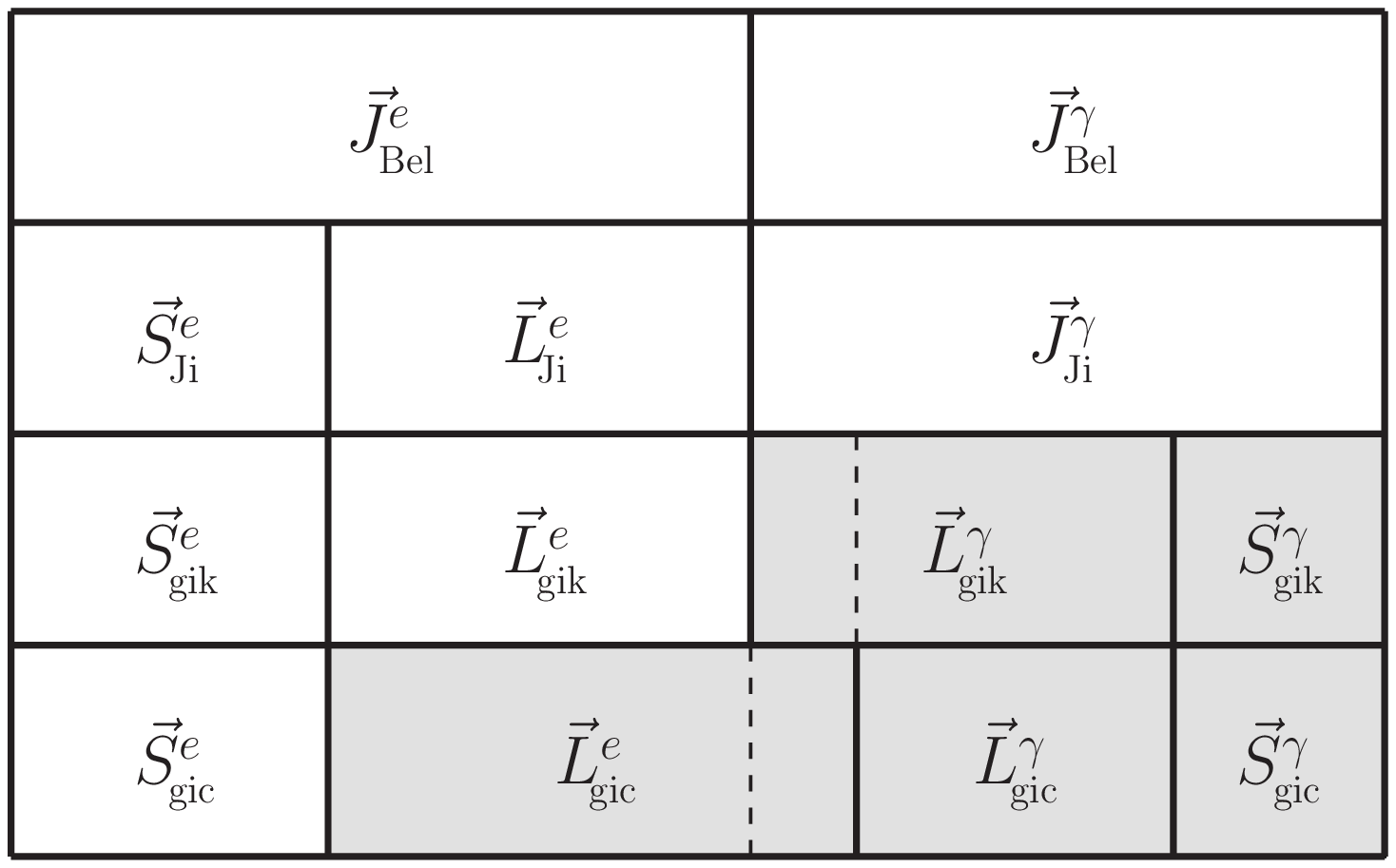}
\caption{\footnotesize{A schematic view of the Belinfante, Ji, gauge-invariant kinetic and gauge-invariant canonical decompositions. In white are depicted the pieces that are Stueckelberg invariant, and in gray are depicted the pieces that are Stueckelberg non-invariant. The potential OAM corresponds to the piece delimited by a dashed line common to both $\uvec L^e_\text{gic}$ and $\uvec L^\gamma_\text{gik}$.}}
		\label{decompositions}
\end{figure}

So, if one starts with a gauge-invariant Lagrangian involving only the full gauge potential $A_\mu$, this Lagrangian will automatically be invariant under the new transformation \eqref{StueckabT}. Because of the similarity with the famous Stueckelberg mechanism\footnote{It is well known that the introduction of a mass term for the photon breaks explicitly the $U(1)$ gauge symmetry of the massless theory. However, Stueckelberg found a mechanism where such a term can be introduced without breaking the gauge invariance \cite{Stueckelberg:1900zz,Stueckelberg:1938zz,Stueckelberg:1957zz}. The idea consists in increasing the number of fields without increasing the number of degrees of freedom thanks to an additional symmetry \cite{Pauli:1941zz,Ruegg:2003ps}. Writing the abelian pure-gauge field as $A^\pure_\mu(x)=\partial_\mu\alpha_\pure(x)$, one sees that the scalar function $\alpha_\pure(x)$ plays a role similar to the Stueckelberg field $B(x)/m$.}, this symmetry is referred to as the Stueckelberg symmetry. The Stueckelberg symmetry group turns out to be a simple copy of the gauge symmetry group, though acting in a different manner on the fields. Just like the QED Lagrangian, the Belinfante and Ji decompositions involve only the full gauge potential, and so are automatically invariant under Stueckelberg transformations. On the contrary, the gauge-invariant canonical and kinetic decompositions rely on our ability to separate the gauge potential into pure-gauge and physical terms. While gauge invariant by construction, these decompositions are obviously not Stueckelberg invariant. This means that for different explicit definitions of $A^\phys_\mu$, different physical values will be attributed to the electron and photon contributions. For a schematic view, see Fig. \ref{decompositions}.

\subsubsection{Towards a more refined classification\label{secIVC2}}

Since the gauge-invariant canonical and kinetic decompositions are Stueckelberg non-invariant, they should rather be seen as representing two different \emph{classes} of decompositions. There are in principle infinitely many possible gauge-invariant canonical and kinetic decompositions, differing in the way one explicitly splits the gauge potential into pure-gauge and physical contributions. In order to single out a particular element in a class, one has to impose an additional constraint on the physical field that fixes the Stueckelberg symmetry. Note that it is necessary to impose a constraint that does not look Lorentz covariant to get completely rid of the Stueckelberg symmetry. Indeed, any explicitly covariant condition, like \emph{e.g.} the Lorenz constraint $\partial_\mu A^\mu_\phys=0$ \cite{Cho:2010cw,Cho:2011ee,Guo:2012wv,Guo:2013jia} or the Fock-Schwinger constraint $x_\mu A_\phys^\mu=0$, leaves some residual Stueckelberg symmetry \cite{Zhang:2011rn}. In order to specify a unique decomposition, this residual symmetry must also be fixed. The fixing of the Stueckelberg symmetry simply mirrors the fixing of the gauge symmetry, and so all the difficulties encountered with the latter have a counterpart with the former.

\begin{figure}[t!]
	\centering
		\includegraphics[width=.6\textwidth]{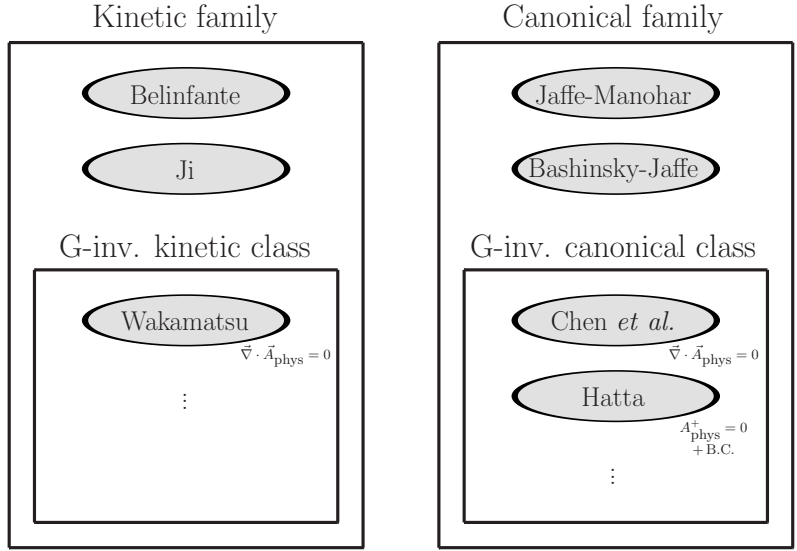}
\caption{\footnotesize{Refinement of the Wakamatsu classification depicted in Fig. \ref{classification_short}. See text for more details.}}
		\label{classification}
\end{figure}

In the Chen \emph{et al.} and Wakamatsu decompositions, the physical field was defined by the Coulomb constraint $\uvec\nabla\cdot\uvec A_\phys=0$. These decompositions therefore simply correspond to the Coulomb gauge-invariant canonical and gauge-invariant kinetic decompositions, respectively. This choice was motivated by the desire to recover the well-known Helmholtz decomposition in QED of the vector potential into longitudinal $\uvec A_\pure=\uvec A_\parallel$ and transverse $\uvec A_\phys=\uvec A_\perp$ parts, which is unique once the Lorentz frame is fixed \cite{Jauch:1955jr,Berestetskii,Cohen}. However, it is perfectly legitimate to impose a different constraint. For example, with the motivation of making contact with the parton model of QCD, Hatta imposed the light-front constraint $A^+_\phys=0$ with additional boundary conditions at infinity \cite{Hatta:2011ku,Lorce:2012ce}. The decomposition he obtained corresponds to the light-front gauge-invariant canonical decomposition. A schematic picture of the refined classification is illustrated in Fig. \ref{classification}.

\begin{figure}[t!]
	\centering
		\includegraphics[width=.4\textwidth]{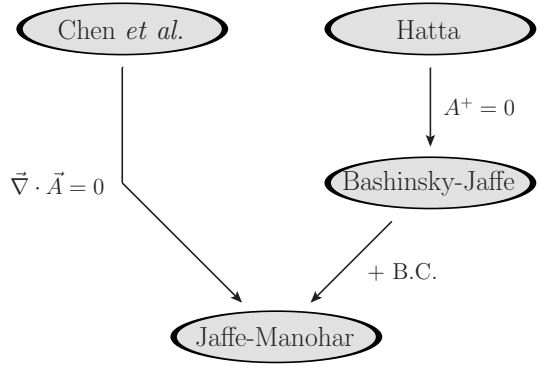}
\caption{\footnotesize{Relations between the Chen \emph{et al.}, Hatta, Bashinsky-Jaffe and Jaffe-Manohar decompositions. The Chen \emph{et al.} decomposition reduces to the Jaffe-Manohar decomposition in the Coulomb gauge. The Hatta decomposition reduces to the Bashinsky-Jaffe decomposition in the light-front gauge and reduces further to the Jaffe-Manohar decomposition once appropriate boundary conditions are imposed on the gauge potential.}}
		\label{GIEs}
\end{figure}

\begin{figure}[t!]
	\centering
		\includegraphics[width=.4\textwidth]{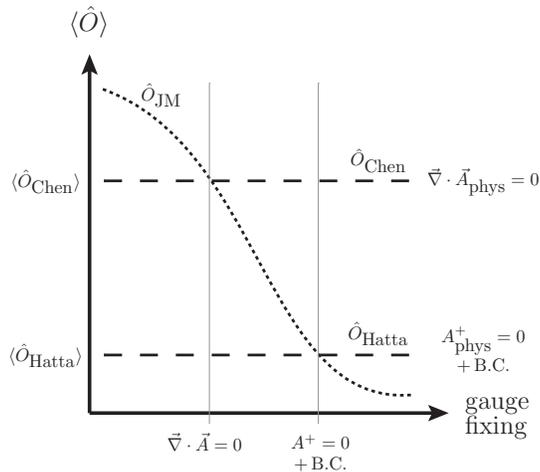}
\caption{\footnotesize{Schematic visualization of the effect of gauge fixing (vertical lines) and Stueckelberg fixing (horizontal lines). While the gauge non-invariant operators $\hat O_\text{JM}$ belonging to the Jaffe-Manohar decomposition  lead to different physical results in different gauges (dotted line), the operators $\hat O_\text{Chen}$ and $\hat O_\text{Hatta}$ belonging, respectively, to the Coulomb and light-front GIEs are gauge invariant (dashed lines). Since one has $\hat O_\text{Chen}=\hat O_\text{JM}$ in the Coulomb gauge and $\hat O_\text{Hatta}=\hat O_\text{JM}$ in the light-front gauge with appropriate boundary conditions, different GIEs of the same gauge non-invariant quantity are physically inequivalent $\la\hat O_\text{Chen}\ra\neq\la\hat O_\text{Hatta}\ra$.}}
		\label{GIE}
\end{figure}

Except for Wakamatsu, there is a general agreement about the fact that different elements in a same class will generally lead to different physical results\footnote{We certainly do not mean that the physical measurements depend on how one defines $A^\phys_\mu$. What we mean is that different GIEs are associated with different measurable quantities.}, \emph{i.e.} give different numerical answers for the expectation values of the operators involved. One should be careful not to confuse a gauge fixing with a Stueckelberg fixing. For example, in the covariant form of the Chen \emph{et al.} decomposition, we can generalise their three-dimenisonal constraints by demanding that the pure-gauge field vanishes only in the Coulomb gauge [for the explicit form of $A^\pure_\mu$ and $A^\phys_\mu$, see Eq.~\eqref{explicitChen}]
\beq
A^{\mu,\text{Chen}}_\pure\big|_{\uvec\nabla\cdot\uvec A=0}=0,\qquad A^{\mu,\text{Chen}}_\pure\big|_{A^+=0\,+\, \text{B.C.}}\neq 0,
\eeq
while in the Hatta decomposition, the pure-gauge field vanishes only in the light-front gauge with appropriate boundary conditions
\beq
A^{\mu,\text{Hatta}}_\pure\big|_{\uvec\nabla\cdot\uvec A=0}\neq 0,\qquad A^{\mu,\text{Hatta}}_\pure\big|_{A^+=0\,+\, \text{B.C.}}=0.
\eeq
Explicit expressions for $A_\mu^\pure$ and $A_\mu^\phys$ in both Chen \emph{et al.} and Hatta decompositions are given in the next subsection. Even though the Chen \emph{et al.} and Hatta decompositions share the same abstract mathematical structure in terms of $A^\pure_\mu$ and $A^\phys_\mu$, they generally lead to different physical results. Indeed, the Chen \emph{et al.} decomposition coincides with the Jaffe-Manohar decomposition in the Coulomb gauge. In the light-front gauge, the Hatta decomposition reduces to the Bashinsky-Jaffe decomposition \cite{Bashinsky:1998if}, which is a residual\footnote{The light-front gauge does not exhaust the gauge symmetry as $x^-$-independent gauge transformations leave the condition $A^+=0$ unchanged. The set of these transformations consitute the residual gauge symmetry group.} GIE of the Jaffe-Manohar decomposition \cite{Wakamatsu:2010cb,Lorce:2012ce}. Only when imposing further boundary conditions does the Hatta decomposition reduce to the Jaffe-Manohar decomposition itself. These relations are schematically represented in Fig. \ref{GIEs}. Since the Jaffe-Manohar decomposition is not gauge invariant, it is expected to give different results in the Coulomb gauge and in the light-front gauge with some boundary conditions, see Fig. \ref{GIE} for an illustration. It then follows that different gauge-invariant canonical decompositions, \emph{i.e.} GIEs of the Jaffe-Manohar decomposition based on different constraints, are generally not physically equivalent. Wakamatsu's discussions about the relation between the gauge-invariant canonical decomposition, the Chen \emph{et al.} decomposition and the Jaffe-Manohar decomposition in the light-front gauge is rather confusing, as there is no clear distinction between the gauge-fixing procedure and the Stueckelberg-fixing procedure \cite{Wakamatsu:2010cb,Wakamatsu:2012ve}. It is because of this that he wrongly concluded that all the GIEs are physically equivalent.

We would like to stress here the similarities with the more familiar procedure of Lorentz-invariant extension (LIE). In relativity, the mass $m$ of a particle is not frame independent. It is given by $m=\gamma m_0$, where $m_0$ is the inertial mass in the rest frame and $\gamma=1/\sqrt{1-v^2}$ is the boost factor with $v$ the velocity of the particle in units of the speed of light. One can consider the Lorentz-invariant quantity $p^2=E^2-\bm{p}^2$ as a LIE of the quantity $m^2$, which agrees with the value of $m^2$ in the rest frame, \emph{i.e.} in the frame where $m=m_0$.  Similarly, the Mandelstam variable $s$ is the LIE of the square of the center-of-mass energy $E^2_\text{CM}$.

Strictly speaking, one does not measure the rest mass of particles produced in high-energy experiments, because these particles are usually far from being at rest. What is actually measured is the LIE $p^2$ which can be interpreted in the rest frame as the square of the particle inertial mass. It is in this sense that one has to understand the ``measurement'' of a particle rest mass. Similarly, a gauge non-invariant quantity is strictly speaking not measurable. However, it is possible to measure its GIE, \emph{i.e.} a gauge-invariant quantity which coincides in a certain fixed gauge with the gauge non-invariant quantity \cite{Hoodbhoy:1998bt,Ji:2012gc}. For example, even if the Jaffe-Manohar decomposition is gauge non-invariant, it is considered to be ``measurable'' in the light-front gauge \cite{Jaffe:1989jz}. This is justified by the fact that what is actually measurable is the Hatta decomposition, which is its light-front GIE.

\subsubsection{Origin and geometrical interpretation of the Stueckelberg symmetry\label{secIVC3}}

Even though \emph{formally} gauge invariant, the Chen \emph{et al.} approach gives a special or privileged role to some gauge. It is the gauge where the full gauge potential satisfies the same constraint as the physical gauge potential, \emph{i.e.} the gauge where $A^\pure_\mu=0$. For example, in the Chen \emph{et al.} and Wakamatsu three-dimensional decompositions, this special gauge is simply the Coulomb gauge $\uvec\nabla\cdot\uvec A=0$, since the physical field was defined by the Coulomb constraint $\uvec\nabla\cdot\uvec A_\phys=0$. In the Hatta decomposition, it is the light-front gauge $A^+=0$ with advanced, retarded or antisymmetric boundary conditions. The fact that there exists a preferred or natural gauge in the Chen \emph{et al.} approach seems at odds with the spirit of gauge symmetry, in the sense that all the gauges should in principle be physically equivalent. Moreover, since writing a GIE consists in finding the gauge-invariant operator which leads to the same physical results as a gauge non-invariant operator in some fixed gauge, determining a specific GIE is somehow equivalent to fixing a gauge for the gauge non-invariant quantity. It appears therefore that gauge symmetry and Stueckelberg symmetry are tightly connected. Our purpose here is to reveal the geometrical meaning of the Stueckelberg symmetry and its relation with  gauge symmetry.

In order to better understand the origin and the physical meaning of the Stueckelberg symmetry, it is useful to consider first some explicit example. As shown explicitly by Hatta with the light-front GIE of the Jaffe-Manohar decomposition \cite{Hatta:2011ku} and generalized later by Lorc\'e \cite{Lorce:2012ce}, the Chen \emph{et al.} approach amounts in many cases to adding Wilson lines along some path to preserve the gauge symmetry. The presence of a Wilson line clearly indicates that explicit expressions for $A^\pure_\mu$ and $A^\phys_\mu$ in terms of $A_\mu$ are usually non-local and path dependent. The freedom in the choice of the path appears to be at the origin of the Stueckelberg symmetry. Namely, Stueckelberg-invariant quantities are path independent, whereas path-dependent quantities are Stueckelberg non-invariant. A choice of path automatically renders one gauge special, namely the gauge where the Wilson line factor reduces to unity. For a generic path, this particular gauge is known as the \emph{contour gauge} \cite{Ivanov:1985np}.  For example, the Wilson lines in the light-front GIE run along the light-front direction $n^\mu=\tfrac{1}{\sqrt{2}}\,(1,0,0,-1)$
\beq
\mathcal W_\text{LF}(x,\pm\infty)=e^{-ie\int^0_{\pm\infty}A^+(x+\lambda n)\,\ud\lambda},
\eeq
and reduce to unity in the light-front gauge $A^+=n\cdot A=0$. The pure-gauge and physical fields in the Hatta decomposition can then be written in an explicitly non-local form
\beq\label{explicitHatta}
\begin{split}
A^{\mu,\text{Hatta}}_\pure(x)&=-\frac{i}{e}\,\mathcal W_\text{LF}(x,\pm\infty)\partial^\mu\mathcal W_\text{LF}(\pm\infty,x)=\int^0_{\pm\infty}\partial^\mu A^+(x+\lambda n)\,\ud\lambda,\\
A^{\mu,\text{Hatta}}_\phys(x)&=A^\mu(\pm\infty)+\int^0_{\pm\infty}F^{+\mu}(x+\lambda n)\,\ud\lambda,
\end{split}
\eeq
where $A^+(\pm\infty)=0$. In particular, one has
\begin{align}
\text{in any gauge}&&A^{+,\text{Hatta}}_\pure(x)&=A^+(x),&&A^{+,\text{Hatta}}_\phys(x)=0,\\
\text{in the light-front gauge}&&A^{\mu,\text{Hatta}}_\pure(x)&=0,&&A^{\mu,\text{Hatta}}_\phys(x)=A^\mu(x).
\end{align}
To keep the example simple, we have not included the transverse gauge links at spatial light-front infinity $x^-=\pm\infty$. A more detailed discussion including these contributions can be found in Ref. \cite{Lorce:2012ce}.

The Stueckelberg dependence is actually more general than the path dependence, since some GIEs, like \emph{e.g.} the Coulomb GIE or the Lorenz GIE, cannot be written simply in terms of Wilson lines, and are in this sense path independent \cite{Belinfante:1962zz,Mandelstam:1962mi,Rohrlich:1965,Yang:1985,Kashiwa:1994jn,Kashiwa:1997we}.  The explicit expressions for $A^\pure_\mu$ and $A^\phys_\mu$ in terms of $A_\mu$ in these path-independent GIEs are nevertheless again non-local \cite{Zhang:2011rn}. For example, in the Chen \emph{et al.} decomposition (and therefore also in the Wakamatsu decomposition), the pure-gauge and physical fields explicitly read
\beq\label{explicitChen}
\begin{split}
A^{\mu,\text{Chen}}_\pure(x)&=-\partial^\mu\frac{1}{\uvec\nabla^2}\uvec\nabla\cdot\uvec A(x),\\
A^{\mu,\text{Chen}}_\phys(x)&=-\frac{1}{\uvec\nabla^2}\partial_iF^{i\mu}(x),
\end{split}
\eeq
where the non-local expressions arise owing to the inverse Laplace operator, see Eq. \eqref{inverseLaplace}. In particular, one has
\begin{align}
\text{in any gauge}&&\uvec\nabla\cdot\uvec A^\text{Chen}_\pure(x)&=\uvec\nabla\cdot\uvec A(x),&&\uvec\nabla\cdot\uvec A^\text{Chen}_\phys(x)=0,\\
\text{in the Coulomb gauge}&&A^{\mu,\text{Chen}}_\pure(x)&=0,&&A^{\mu,\text{Chen}}_\phys(x)=A^\mu(x).
\end{align}
Gauge-invariant, Stueckelberg non-invariant quantities are therefore intrinsically non-local, \emph{i.e.} cannot be written as gauge-invariant local expressions in terms of the full gauge potential $A_\mu$. On the contrary, quantities that are both gauge and Stueckelberg invariant can in principle be written as gauge-invariant local expressions in terms of $A_\mu$. Non-local expressions are often considered to be dangerous, because they may lead to violations of causality. Note however that unphysical pure-gauge degrees of freedom are allowed to propagate faster than light since they are decoupled from the physical observables. In the Chen \emph{et al.} approach, one can always work in the gauge where $A^\pure_\mu(x)=0$ so that $A^\phys_\mu(x)=A_\mu(x)$. This means that there always exists a gauge where the Stueckelberg non-invariant expressions reduce to a local form in terms of $A_\mu$. No genuine violation of causality is therefore expected in the Chen \emph{et al.} approach.

In some sense, the Chen \emph{et al.} approach can be seen as a convenient rewriting of complicated non-local expressions (RHS of Eqs. \eqref{explicitHatta} and \eqref{explicitChen}) in terms of simpler seemingly local expressions (LHS of Eqs. \eqref{explicitHatta} and \eqref{explicitChen}). This means that, contrary to what they claim, Chen \emph{et al.} \cite{Chen:2008ag} do not really contradict the textbook statement that there exist no gauge-invariant local operators for the photon spin and OAM \cite{Cohen,Simmons}. Alternatively, one could say that the Chen \emph{et al.} approach allows one to get around the textbook statement by considering two four-vector fields at one point, $A^\pure_\mu(x)$ and $A^\phys_\mu(x)$, instead of a single one, $A_\mu(x)$.
\newline

We have seen that Stueckelberg symmetry is related to the ability to write local expressions in terms of the full gauge potential. Lorc\'e showed that Stueckelberg symmetry can also be understood from a different point of view, closer to the discussions in General Relativity and String Theory \cite{Lorce:2013bja}. He also observed that the Chen \emph{et al.} approach is pretty similar to the background field method introduced by DeWitt \cite{DeWitt:1967ub,DeWitt:1967uc,DeWitt:1980jv,'tHooft:1975vy,Grisaru:1975ei,Boulware:1980av,Abbott:1980hw}, which has been extensively used in gravity and supergravity \cite{'tHooft:1973us,'tHooft:1974bx,Deser:1977nt,Abbott:1981ff,Petrov:2007xva,Petrov:2012qn},  as well as in both continuum and lattice gauge theories \cite{Dashen:1980vm,Abbott:1982jh,Luscher:1995vs,Freire:2000bq,Binosi:2009qm,Binosi:2012st}. A nice introduction to the background field method has been provided by Abbott in Ref. \cite{Abbott:1981ke}.

When discussing gauge transformations, one usually does not specify whether these are considered as \emph{active} transformations (\emph{i.e.} the system is modified) or \emph{passive} transformations (\emph{i.e.} the reference axes are modified) \cite{Guay:2004zz}. The reason for this is that there is usually no way to distinguish active gauge transformations from passive gauge transformations\footnote{Similarly, in General Relativity there is \emph{a priori} no way to distinguish diffeomorphisms (active point transformations) from coordinate transformations (passive point transformations).}. As a matter of fact, physicists tend to think of gauge transformations as passive transformations, and therefore consider that gauge symmetry is just a mere redundancy of the mathematical description. On the contrary, mathematicians think of gauge transformations as active transformations, and so consider that gauge symmetry is a physical property of the system.

The situation changes however once one introduces some \emph{background} field in the theory \cite{DeWitt:1967ub,Abbott:1981ke}, as the latter transforms differently under active and passive transformations \cite{Smolin:2005mq,Rickles:2008,Rozali:2008ex,Barenz:2012av}. In the Chen \emph{et al.} approach, it is the pure-gauge field $A^\pure_\mu(x)$ that plays the role of a non-dynamical, background field \cite{Zhang:2011rn,Lorce:2013gxa,Lorce:2013bja}. Stueckelberg dependence is therefore simply equivalent to background dependence. Notice that, from the gauge transformation law of the electron field, it is not possible to derive the gauge transformation law for the pure-gauge and physical fields, separately. This is the reason why, in the covariant version of  the Chen \emph{et al.} approach, it is \emph{assumed} that the pure-gauge field transforms in the same way as the full gauge potential. This assumption corresponds actually to \emph{passive} gauge transformations, as the latter consist in a mere change of coordinates, which affects in the same way dynamical and background fields. On the contrary, under active gauge transformations, the dynamical and background fields are transformed in a different way \cite{Rickles:2008,Barenz:2012av,Lorce:2013bja}.
\begin{table}[t!]
\begin{center}
\caption{\footnotesize{The expressions for the different fields after passive gauge transformations, active gauge transformations and Stueckelberg transformations. The interesting case is $C(x)=-\beta(x)=\alpha(x)$.}}\label{GTFs}
\begin{tabular}{@{\quad}c@{\quad}|@{\quad}c@{\quad}|@{\quad}c@{\quad}|@{\quad}c@{\quad}}\whline
Fields&Passive gauge&Active gauge&Stueckelberg\\
\hline
$\psi(x)$&$e^{-ie\alpha(x)}\psi(x)$&$e^{-ie\beta(x)}\psi(x)$&$\psi(x)$\\
$A_\mu(x)$&$A_\mu(x)+\partial_\mu\alpha(x)$&$A_\mu(x)+\partial_\mu\beta(x)$&$A_\mu(x)$\\
$A^\pure_\mu(x)$&$A^\pure_\mu(x)+\partial_\mu\alpha(x)$&$A^\pure_\mu(x)$&$A^\pure_\mu(x)+\partial_\mu C(x)$\\
$A^\phys_\mu(x)$&$A^\phys_\mu(x)$&$A^\phys_\mu(x)+\partial_\mu\beta(x)$&$A^\phys_\mu(x)-\partial_\mu C(x)$\\
\whline
\end{tabular}
\end{center}
\end{table}
In Table \ref{GTFs}, we compare the expressions for the transformed fields under passive gauge transformations, active gauge transformations and Stueckelberg transformations. Using $C(x)=-\beta(x)=\alpha(x)$, it appears that the Stueckelberg transformations simply correspond to passive gauge transformations followed by the corresponding inverse active gauge transformations \cite{Lorce:2013bja}. In particular, one can now easily understand why the Stueckelberg symmetry group is a simple copy of the gauge symmetry group. The set of passive and active gauge transformations (second and third columns of Table \ref{GTFs}) is therefore equivalent to the set of (passive) gauge and Stueckelberg transformations (second and fourth columns of Table \ref{GTFs}). Since most of the recent papers on the proton spin decomposition deal with the second set, we adopt it also in our subsequent discussions. Therefore, unless explicitly stated, the gauge transformations in the following always refer to the passive ones (second column of Table \ref{GTFs}).

\subsubsection{Measurability and the controversy about Stueckelberg symmetry\label{secIVC4}}

Ji criticized the Chen \emph{et al.} approach arguing that their notion of gauge invariance does not coincide with the ``usual textbook type'' \cite{Ji:2009fu,Ji:2010zza}, where physical observables are usually constructed from gauge-invariant local operators. Moreover, having stressed that the Chen \emph{et al.} decomposition simply corresponds to the Coulomb GIE of the Jaffe-Manohar decomposition, Ji, Xu and Zhao wrote in Ref. \cite{Ji:2012gc} that ``the GIE of an intrinsically gauge-noninvariant quantity is not naturally gauge invariant'' and that ``GIE operators are in general unmeasurable. So far, the only example is offered in high-energy scattering in which certain partonic GIE operators may be measured.'' This kind of statement is pretty confusing as it seems at first sight self-contradictory. Indeed, how can a measurable quantity be ``not naturally gauge invariant'' or, using Wakamatsu's words \cite{Wakamatsu:2013voa}, ``not a gauge-invariant quantity in a \emph{true} or \emph{traditional} sense''?

This aspect of the controversy can easily be clarified by using a more precise terminology. Indeed, the discussions in the previous subsection suggest the importance of distinguishing two forms of gauge invariance, weak and strong \cite{Lorce:2013bja}:
\bi
\item \emph{Weak gauge symmetry} refers to the invariance under gauge transformations only. Intrinsically non-local (\emph{i.e.} Stueckelberg non-invariant) expressions are allowed as long as they are gauge invariant;
\item \emph{Strong gauge symmetry} refers to the invariance under both gauge and Stueckelberg transformations. In other words, only the gauge-invariant expressions that can be written locally in terms of $A_\mu$ are allowed.
\ei
Clearly, when Ji invokes the usual textbook gauge invariance, he refers to the strong form of gauge symmetry, and the claim that ``the GIE of an intrinsically gauge-noninvariant quantity is not naturally gauge invariant'' should be understood as ``a Stueckelberg non-invariant GIE is gauge invariant in a weak sense only''. Wakamatsu appears to be less restrictive. Indeed, he writes in Ref. \cite{Wakamatsu:2013voa} that ``if a quantity in question is seemingly gauge-invariant but path-dependent, it is not a gauge-invariant quantity in a \emph{true} or \emph{traditional} sense, which in turn indicates that it may not correspond to genuine observable.'' This is motivated by the claim in Refs. \cite{Belinfante:1962zz,Mandelstam:1962mi,Rohrlich:1965,Yang:1985} that path dependence is a reflection of gauge dependence. He therefore considers that physical quantities may be intrinsically non-local, provided that they are path independent. Wakamatsu therefore requires something that is in between the weak and strong forms of gauge symmetry. Note however that, from the discussions in the previous subsection, it should be clear that path dependence \emph{is not} gauge dependence, but rather some instance of Stueckelberg dependence. Otherwise, this would mean that all the parton high-energy physics, based on Wilson lines running along the light-front direction, is also not gauge invariant. This shows once more the importance of distinguishing gauge symmetry from Stueckelberg symmetry.

Now comes the essential question of \emph{measurability}, which we think has been grossly misconstrued in the literature. Measurable quantities are quantities that can in principle be extracted from experimental data. It turns out that a \emph{sufficient}  condition for measurability is that an operator satisfies just the condition of the weak form of gauge invariance. For example, the parton distribution functions are defined as particular matrix elements of gauge-invariant non-local partonic operators. Such quantities can be extracted from experiments thanks to the so-called factorization theorems that allow one to approximate a cross section by a convolution of these (process-independent) parton distribution functions with a perturbatively calculable (process-dependent) partonic cross section \cite{Collins}.  We would also like to stress that many physical phenomena, like \emph{e.g.} the Aharonov-Bohm effect, originate from a topological problem which cannot be addressed from a purely local point of view. This shows that the measurable quantities are not necessarily intrinsically local. Requiring strong form of gauge invariance to decide whether a quantity is measurable or not is therefore excessive. Put in different words, it is sufficient that a measurable operator is gauge invariant in the weak sense but, as we have seen, not necessarily ``naturally gauge invariant'' (\emph{i.e.} gauge invariant in a strong sense). However this is far from being \emph{necessary}. Collins \cite{Collins:1984xc} has given explicit examples of gauge non-invariant operators whose expectation values are gauge independent. And Leader \cite{Leader:2011za} has shown that, in covariantly quantized QED, even the \emph{total} momentum and angular momentum, \emph{i.e.} operators which have none of the problems linked to the separation into constituent contributions, are not gauge invariant, yet, crucially, their expectation values are gauge independent. The key point is that in a gauge theory, care has to be exercised in considering matrix elements only between physical states.

Following Lorc\'e's suggestion \cite{Lorce:2013bja}, measurable quantities that can be written locally in terms of $A_\mu$ (\emph{i.e.} that are Stueckelberg invariant or gauge invariant in a strong sense) are called \emph{observables}, while those that are intrinsically non-local (\emph{i.e.} Stueckelberg non-invariant or gauge invariant in a weak sense) are called \emph{quasi-observables}. While the Belinfante and Ji decompositions involve only observables, quasi-observables appear in the gauge-invariant canonical and gauge-invariant kinetic decompositions, see the gray contributions in Fig. \ref{decompositions}. All these decompositions are in principle measurable, since they fulfill the requirement of weak gauge symmetry. This makes some people feel uncomfortable because, in some sense, there should be only one ``truly physical'' picture.
 %One can avoid this conceptual difficulty by requiring that the gauge symmetry should be understood in a strong sense. From this point of view, quasi-observables are not considered as truly physical quantities, and so only the Belifante and Ji decompositions obtain a physical status. Notice it is pretty hard to literally stick to this point of view since, otherwise, one would be forced to admit that electrons and photons are also not physical.
However, these arguments are somewhat metaphysical.
%\trd{This brings us back to the problem of the \emph{physical interpretation} in a gauge theory. The essential requirement }
What counts in Physics should be measurability.
%\trd{We feel that arguing whether a measurable quantity is really ``physical'' or not is meaningless, and goes somewhat beyond physical considerations.}
From this point of view, there is no reason to disregard quasi-observables and not to consider the gauge-invariant canonical and gauge-invariant kinetic decompositions. In addition, this sort of argument about the properties of the operators completely ignores the point of view stressed by Collins \cite{Collins:1984xc} and Leader \cite{Leader:2011za} that what is finally relevant is whether the expectation value of an operator is or is not independent of the gauge choice. If it is gauge independent, then that is all that matters and it is in principle measurable. Unfortunately, it is difficult to see, apart from explicit calculations, whether or not a gauge non-invariant operator leads to gauge-independent matrix elements.

As mentioned previously, the Chen \emph{et al.} approach is a typical example of a GIE. Considering any gauge non-invariant object in a specific gauge, one can easily construct a gauge-invariant object leading to the same physical result in any gauge. Formally, it is sufficient to replace the full gauge potential $A_\mu$ by the physical field $A^\phys_\mu$, and the ordinary derivative $\partial_\mu$ by the appropriate pure-gauge covariant derivative\footnote{Since the electron field transforms as an internal vector, one has to perform the substitution $\partial_\mu\psi\mapsto D^\pure_\mu\psi$. On the contrary, $A^\phys_\mu$ is gauge invariant in the abelian theory, and so the substitution reads simply $\partial_\mu A_\nu\mapsto\partial_\mu A^\phys_\nu$.}. Starting from a single gauge non-invariant local quantity, one can construct infinitely many GIEs. There exist therefore formally infinitely many gauge-invariant canonical and gauge-invariant kinetic decompositions that are in principle measurable. According to Chen \emph{et al.} \cite{Chen:2008ag,Chen:2009mr} and Wakamatsu \cite{Wakamatsu:2010cb,Wakamatsu:2012ve,Wakamatsu:2013voa}, the Stueckelberg symmetry does not exist from the very beginning. They indeed claim that the Coulomb gauge $\uvec\nabla\cdot\uvec A=0$ is the ``physical'' gauge, since it is the one compatible with the Helmholtz decomposition. Wakamatsu further argues that the Coulomb constraint is the ``physical'' one because its explicit non-local expression in terms of the full gauge potential is path independent, contrary to other constraints like \emph{e.g.} the light-front constraint \cite{Wakamatsu:2013voa}. Accordingly, the GIE defined by the Coulomb constraint $\uvec\nabla\cdot\uvec A_\phys=0$ is considered by them as the only ``physical'' one\footnote{Incidentally, Chen \emph{et al.} emphasized contrary to Wakamatsu that different GIEs are not physically equivalent.}. We strongly disagree with such a statement. It is certainly true that, in electrodynamics, the Helmholtz decomposition appears particularly appealing. One should however keep in mind that there cannot be any experimental way to \emph{prove} that the Helmholtz decomposition is the ``physical'' one, otherwise it would mean that the gauge symmetry is explicitly broken in Nature. As emphasized in Ref. \cite{Ji:2013fga}, there is also no charge that responds separately to the transverse part $\uvec E_\perp$ and the longitudinal part $\uvec E_\parallel$ of the electric field. Moreover, the Helmholtz decomposition cannot easily be generalized to non-abelian gauge theories, because the Coulomb gauge is not sufficient to fix the gauge freedom completely in these theories. There is a residual gauge freedom usually associated with homotopically non-trivial gauge transformations, a problem known as the Gribov copies issue \cite{Gribov:1977wm}. The Coulomb gauge is not more ``physical'' than the other gauges. In conclusion, we do not see any good reason for giving the Helmholtz decomposition a special physical status.

As stressed \emph{e.g.} by Ji, Xu and Zhao \cite{Ji:2012gc} and Lorc\'e \cite{Lorce:2012rr,Lorce:2012ce,Lorce:2013gxa,Lorce:2013bja}, the breaking of the Stueckelberg symmetry is actually dictated by the framework used to describe the actual experimental process. In theory, any GIE is as good as any other one. In practice, however, a particular GIE can be singled out. In order to describe experimental data, one often relies on some controlled expansion scheme suited to the experimental conditions. Because of this expansion scheme, one particular GIE turns out to be much more \emph{convenient}, simply because the corresponding operators contribute only at certain order in the expansion. The choice of a Stueckelberg breaking is therefore purely a matter of convenience. Even if one can, in theory, define infinitely many quasi-observables associated with a single gauge non-invariant local quantity, it is in practice rarely known how to access them experimentally. For example, non-relativistic systems, like atoms, and free radiation are most conveniently described in the instant form of dynamics and the Coulomb gauge, where it preserves the natural power counting \cite{Ji:2012gc}. For such systems, it is therefore more natural to work with the Coulomb GIE. On the contrary, the proton is a relativistic system and its internal structure is essentially probed in high-energy experiments involving large momentum transfer, where a parton model picture is very convenient \cite{Feynman:1969ej} and theoretically justified by the factorization theorems\footnote{It is often considered that gauge symmetry is simply a mathematical redundancy of the theory. Gauge non-invariant quantities are therefore not considered as physical. However, as nicely stressed by Rovelli in Ref. \cite{Rovelli:2013fga}, by coupling gauge non-invariant quantities from different systems, one can form new gauge-invariant quantities. One can therefore in some sense ``measure'' gauge non-invariant quantities of a system as long as it is \emph{relative} to another system. The QCD factorization theorems allow one to separate the leading contribution of a scattering amplitude into hard and soft parts. The Wilson lines entering the definition of the parton distributions represent in some sense the relative phase between these hard and soft parts.}. For this reason, it appears more natural to describe the proton in the framework of light-front dynamics \cite{Lepage:1980fj,Brodsky:1997de}. The contact with the parton model picture can then be achieved in the light-front gauge \cite{Jaffe:1989jz,Bashinsky:1998if}. In this context, it is clearly more convenient to work with the light-front GIE.

In summary, even though the QED Lagrangian is both gauge and Stueckelberg invariant, for a quantity to be measurable it is sufficient that the operator is only gauge invariant, or more strictly, that its expectation value is gauge independent. Many different decompositions of the angular momentum may therefore coexist, as long as they are coherent and have a clear connection with measurable quantities. One has just to specify with which decomposition one is working. There is no point in trying to argue that a particular decomposition is more physical than the other ones, simply because measurability is the only fundamental requirement. In the context of the proton spin decomposition, the relevant decompositions are therefore
\bi
\item The Belinfante decomposition;
\item The Ji decomposition;
\item The light-front gauge-invariant kinetic decomposition defined by the light-front constraint $A^+_\phys=0$ supplemented by boundary conditions;
\item The light-front gauge-invariant canonical decomposition defined by the light-front constraint $A^+_\phys=0$ supplemented by boundary conditions,  \emph{i.e.} the Hatta decomposition which is physically equivalent to the Jaffe-Manohar decomposition considered in the light-front gauge with appropriate boundary conditions.
\ei
We also stress that, despite the fact that Stueckelberg non-invariant quantities are physically equivalent to (\emph{i.e.} give the same answer for the measurable quantities as) gauge non-invariant quantities considered in a particular gauge, Stueckelberg fixing is \emph{different} from gauge fixing. Claiming that Stueckelberg and gauge symmetry are the same thing results from either superficial understanding of gauge invariance or the unnecessary requirement of strong gauge invariance.

\subsection{The Lorentz transformation properties\label{secIVD}}

Since Wakamatsu succeeded in writing gauge-invariant decompositions in a covariant form \cite{Wakamatsu:2010cb}, one might be tempted to conclude that the Chen  \emph{et al.} approach is frame independent. While the conclusion appears to be correct, it is actually a bit premature. To conclude that a formalism is covariant, it is not sufficient to put it in a tensorial form. One has also to care about the Lorentz tranformation properties of the fields involved.

\subsubsection{The standard approach\label{secIVD1}}

To stress the Lorentz covariance of a theory and to deal with expressions transforming in a simple way from one Lorentz frame to another, one usually tries to reformulate the theory in terms of Lorentz tensors. The reason is that an equation between tensors is automatically Lorentz covariant when the uncontracted indices on both sides of the equation match with each other. This makes the Lorentz covariance of the theory \emph{manifest} or \emph{explicit}.

In Classical Electrodynamics, the electric $\uvec E$ and magnetic $\uvec B$ fields have rather complicated Lorentz transformation laws, and it is quite  tedious to check that the Maxwell equations are Lorentz covariant. However, when one combines these electric and magnetic fields into an antisymmetric matrix $F^{\mu\nu}$ such that $F^{i0}=E^i$ and $F^{ij}=-\epsilon^{ijk}B^k$, the Maxwell's equations and the Lorentz force can be written in a much more compact form
\begin{align}
\partial_\mu F^{\mu\nu}&=j^\nu,\label{inhomogeneous}\\
\partial_\mu \tilde F^{\mu\nu}&=0,\qquad\qquad \tilde F^{\mu\nu}=\tfrac{1}{2}\,\epsilon^{\mu\nu\alpha\beta} F_{\alpha\beta},\label{homogeneous}\\
\frac{\ud\pi^\mu}{\ud\tau}&=\frac{e}{m}\,F^{\mu\nu}\pi_\nu,\label{Lorentzforce}
\end{align}
where the current $j^\mu=(\rho,\uvec j)$ is a Lorentz four-vector owing to the fact that the electric charge $e$ is a Lorentz scalar, $\pi^\mu=(m\gamma,m\gamma\uvec\beta)$ is the \emph{kinetic} four-momentum proportional to the rest mass $m$ of the particle, and $\tau$ is the proper time. Clearly, these equations will be Lorentz covariant if $F^{\mu\nu}$ transforms as a Lorentz tensor
\begin{equation}\label{EMtensor}
F^{\mu\nu}(x)\mapsto F'^{\mu\nu}(x')=\Lambda\uind{\mu}{\alpha}\Lambda\uind{\nu}{\beta}F^{\alpha\beta}(x).
\end{equation}
Most textbooks simply \emph{assume} that $F^{\mu\nu}$ transforms as a Lorentz tensor, and then derive the Lorentz transformation laws of the electric and magnetic fields. However, in order to \emph{prove} that the classical laws of electromagnetism are Lorentz covariant, one has to do it the other way around, namely first establish experimentally the Lorentz transformation laws of the electric and magnetic fields, and then check whether $F^{\mu\nu}$ transforms as Eq. \eqref{EMtensor}.

Instead of dealing with the electric and magnetic fields, it appears more economical to deal with the electromagnetic potentials
\beq
\begin{aligned}
\uvec E&=-\uvec\nabla\Phi-\partial_t\uvec A,\\
\uvec B&=\uvec\nabla\times\uvec A.
\end{aligned}
\eeq
The scalar potential $\Phi$ and the vector potential $\uvec A$ are conveniently combined into a single four-component object $A^\mu=(\Phi,\uvec A)$. One can then express the electromagnetic tensor $F^{\mu\nu}$ as
\begin{equation}\label{AtoF}
F^{\mu\nu}=\partial^\mu A^\nu-\partial^\nu A^\mu,
\end{equation}
which is consistent with the homogeneous Maxwell's equation~\eqref{homogeneous}. In terms of the four-component gauge potential $A^\mu$, the inhomogeneous Maxwell's equation~\eqref{inhomogeneous} reads
\begin{equation}\label{inhomogeneous2}
\partial_\mu\partial^\mu A^\nu-\partial^\nu \partial_\mu A^\mu=j^\nu.
\end{equation}
Because of the presence of the index $\mu$, it seems obvious that $A^\mu$ transforms as a Lorentz four-vector, and indeed treating the gauge potential as a Lorentz four-vector is almost universal and is perfectly consistent and particularly \emph{convenient}. The main difficulty comes only with the quantization procedure where a special treatment of the unphysical modes is required.

Some standard textbooks on Classical Electrodynamics, like \emph{e.g.} \cite{Jackson,Feynman}, actually argue that the gauge potential \emph{must} be a Lorentz four-vector. The argument is the following: restricting ourselves to the class of gauge potentials satisfying the Lorenz condition
\begin{equation}\label{Lorenz}
\partial_\mu A^\mu=0,
\end{equation}
the inhomogeneous Maxwell's equation reduces to
\begin{equation}\label{Lorenzreduced}
\partial_\mu\partial^\mu A^\nu=j^\nu.
\end{equation}
Since $\partial_\mu\partial^\mu$ is a Lorentz scalar operator and $j^\nu$ is a Lorentz four-vector, it follows that $A^\mu$ has to transform as a Lorentz four-vector.

\subsubsection{Critique of the standard approach\label{secIVD2}}

All the fields involved in Eqs. \eqref{inhomogeneous}-\eqref{Lorentzforce} are gauge invariant, and so the issues of Lorentz covariance  and gauge invariance are completely disentangled. But things are not so simple anymore in Eq. \eqref{inhomogeneous2}, because $A^\mu$ changes also under gauge transformations. In order to ``prove'' that $A^\mu$ is a Lorentz four-vector, the standard argument relies on a partial fixing of the gauge freedom by means of the Lorenz condition. Note however that a gauge transformation
\beq
A^\mu(x)\mapsto \tilde A^\mu(x)=A^\mu(x)+\partial^\mu \alpha(x)
\eeq
satisfying $\partial_\mu\partial^\mu\alpha=0$ leaves both \eqref{Lorenz} and \eqref{Lorenzreduced} invariant. This means that one cannot conclude that the only possible Lorentz transformation law for the gauge potential is the four-vector one, unless one removes the residual gauge freedom with \emph{e.g.} some additional boundary conditions. Another more subtle but crucial point in the standard argument is that one \emph{implicitly} assumes that the Lorenz condition is Lorentz covariant \cite{Lorce:2012rr}. Indeed, if one imposes a non-covariant condition on a covariant equation, one ends up with a non-covariant equation, and Lorentz covariance cannot be invoked anymore to draw a conclusion about the Lorentz transformation properties of the fields involved. Assuming that the Lorenz condition is Lorentz invariant amounts to assuming that the gauge potential transforms as a Lorentz four-vector (\emph{modulo} residual gauge transformations). The standard argument is therefore \emph{not} a proof of the Lorentz four-vector nature of the gauge potential, since this is implicitly assumed in the intermediate steps. In other words, the standard argument is just an \emph{argumentum in circulo}.

To the best of our knowledge, there exists no consistent proof that the gauge potential has to transform as a Lorentz four-vector. We even strongly doubt that such a proof could exist. The most general Lorentz transformation law for $A^\mu$ that is consistent with Eqs.~\eqref{EMtensor} and \eqref{AtoF} is actually
\begin{equation}\label{AgenLT}
A^\mu(x)\mapsto A'^\mu(x')=\Lambda\uind{\mu}{\nu}\left[A^\nu(x)+\partial^\nu\omega_\Lambda(x)\right],
\end{equation}
where $\omega_\Lambda(x)$ is a scalar function of space and time associated with the Lorentz transformation $\Lambda$. So, in general, $A^\mu$ transforms as a Lorentz four-vector only \emph{up to} a gauge transformation \cite{Bjorken:1965zz,Weinberg:1995mt,Manoukian:1987hy,Moriyasu:1984mh}. This does not contradict the Coleman-Mandula no-go theorem \cite{Coleman:1967ad}, since the latter does not forbid the Lorentz transformations to act also in the internal space as long as the full symmetry group consists locally in the direct product of the Poincar\'e group with the internal symmetry group
\beq
G_\text{full}\stackrel{\text{locally}}{=}G_\text{Poincar\'e}\times G_\text{internal}.
\eeq
Now, because of gauge symmetry, there seems to be no possible unique determination of the actual function $\omega_\Lambda(x)$. In gauge theories, there is \emph{a priori} an infinite number of physically equivalent Lorentz transformation laws.

Bjorken and Drell discussed QED in the Coulomb gauge $\uvec\nabla\cdot\uvec A=0$ \cite{Bjorken:1965zz}. Using the Lorentz transformation operator derived from Noether's theorem, they concluded that the gauge potential does not transform as a four-vector, but according to Eq. \eqref{AgenLT}. At first sight, it seems that the canonical formalism \emph{implies} that the gauge potential cannot transform as a four-vector. Actually, what the canonical formalism does, is to provide not the Lorentz transformation law of fields, but the conserved operator that generates a \emph{given} Lorentz transformation of the fields. What happens in the canonical formalism, is that one imposes that the gauge-fixing condition is preserved by the Lorentz transformation law, analogously to what was done by Weinberg in his book \cite{Weinberg:1995mt}. It is therefore not a surprise that the gauge-fixing condition is preserved by the canonical Lorentz transformations, since it is just what \emph{defines} the latter. So arguing that the canonical formalism provides the Lorentz transformation law of fields is another example of \emph{argumentum in circulo}. Thanks to gauge symmetry, one has in fact the \emph{freedom} to choose which gauge-fixing condition is preserved by Lorentz transformations, and this determines a particular function $\omega_\Lambda(x)$. Using the gauge-covariant canonical formalism developed in Ref. \cite{Lorce:2013gxa} based on the Chen \emph{et al.} approach, we expect a similar conclusion with the Stueckelberg-fixing condition.

Similar things naturally happen with the electron field. The Lorentz transformation law of the Dirac spinor is usually derived in the free theory, and reads
\beq\label{DiracLT}
\psi(x)\mapsto\psi'(x')=S[\Lambda]\,\psi(x),
\eeq
where $S[\Lambda]$ is the transformation matrix in Dirac space associated with the Lorentz transformation $\Lambda$. In QED, it is often implicitly assumed that the Lorentz transformation of the electron field is the same as in the free Dirac theory. Again, because of gauge symmetry, one can in fact only say that the electron field transforms in general as a Dirac spinor \eqref{DiracLT} up to a gauge transformation
\beq\label{DiracgenLT}
\psi(x)\mapsto\psi'(x')=S[\Lambda]\,e^{-ie\omega_\Lambda(x)}\,\psi(x).
\eeq
Imposing that the covariant derivative of the electron field transforms in a natural way
\beq
D_\mu\psi(x)\mapsto D'_\mu\psi'(x')=\Lambda\lind{\mu}{\nu}\,S[\Lambda]\,e^{-ie\omega_\Lambda(x)}\,D_\nu\psi(x)
\eeq
then implies that it is the same function $\omega_\Lambda(x)$ that enters the Lorentz transformation law of the photon \eqref{AgenLT} and the electron \eqref{DiracgenLT}.

One may feel uncomfortable with the existence of infinitely many possible Lorentz transformations laws. Fortunately, because of gauge symmetry, we are free to choose the most convenient one for our purpose. Since it is obviously much simpler to deal with Lorentz tensors and since the latter allow one to maintain the Lorentz covariance explicitly, one usually \emph{decides} to work with $\omega_\Lambda=0$. This rather natural choice makes the Lorenz gauge condition ``special'', in the sense that it appears to be Lorentz invariant. As stressed earlier, the existence of a ``special'' gauge condition seems at odds with the spirit of gauge symmetry. There is of course no problem since one could have chosen to work with another Lorentz transformation law associated with $\omega_\Lambda\neq 0$, which would have then rendered another gauge condition special. As an example, consider the Coulomb constraint $\uvec\nabla\cdot\uvec A=0$. Since one is allowed to impose such a condition in any Lorentz frame, there exists a function $\omega_\Lambda(x)$ such that the associated Lorentz transformation \eqref{AgenLT} leaves the Coulomb gauge condition invariant. Unfortunately, the explicit expression for the function $\omega_\Lambda(x)$ is usually pretty complicated. In numerous textbooks and papers, one finds the claim that the Coulomb gauge condition is not Lorentz invariant. The reason is that one has (implicitly) decided to work from the beginning with a gauge potential transforming as a Lorentz four-vector. But such a statement is somewhat misleading as one could have chosen to work precisely with the Lorentz transformation law that preserves the Coulomb gauge condition, and would have then concluded that the Coulomb gauge condition \emph{is} Lorentz invariant. A more careful formulation of the claim would then be the following: the Coulomb gauge condition is not invariant under Lorentz four-vector transformations. Alternatively, one could also say that the Coulomb gauge condition is not \emph{explicitly} or \emph{manifestly} Lorentz invariant.

\subsubsection{Lorentz transformation law of the pure-gauge and physical fields\label{secIVD3}}

In his critique of the Chen \emph{et al.} approach, Ji claimed that the latter is not consistent with Lorentz symmetry \cite{Ji:2009fu,Ji:2010zza}. He stresses, in particular, that if a gauge potential $A^\mu_\phys$ satisfies a non manifestly Lorentz-covariant condition (like \emph{e.g.} the Coulomb constraint) in one frame, the transformed potential $A'^\mu_\phys$ no longer satisfies that condition in a different frame. In their reply \cite{Chen:2008ja,Chen:2009dg}, Chen \emph{et al.} explained that this criticism is based on the wrong implicit assumption that the physical field transforms as a Lorentz four-vector. Ji, Xu and Zhao then state in Ref. \cite{Ji:2012gc} that ``According to Einstein's theory of special relativity, all physical quantities should be Lorentz tensors, \emph{i.e.}, when coordinates transform, they behave either as scalars, four-vectors, or high-order tensors.'' Accordingly, since $A^\mu_\phys$ does not transform as a Lorentz four-vector, it does not correspond to a physical quantity. This is clearly a misunderstanding of what Lorentz symmetry is. Indeed, Lorentz invariance requires only that the physical laws take the same form in all Lorentz frames. It does not force the physical quantities to transform as Lorentz tensors. Lorentz tensors are very convenient and ubiquitous, but are not the only physical quantities. It is for example well known in General Relativity that the connection or Christoffel symbol (which is the analogue of the gauge potential) does not transform as a tensor. One might be tempted to argue that the Christoffel symbol is not really ``physical'', but this would immediately imply that the metric itself is also not really ``physical'', because the former can be obtained uniquely from the latter.  Once again, we feel that arguing whether a quantity is really ``physical'' or not is meaningless, and goes somewhat beyond physical considerations. The only essential point is whether a quantity is measurable or not.

Similarly to the full gauge potential, the pure-gauge and physical fields are expected to transform in general according to \cite{Lorce:2012rr}
\begin{align}
A^\mu_\pure(x)\mapsto  (A_\pure)'^\mu(x')&=\Lambda\uind{\mu}{\nu}\left[A^\nu_\pure(x)+\partial^\nu\omega^\pure_\Lambda(x)\right],\label{pureabL}\\
A^\mu_\phys(x)\mapsto (A_\phys)'^\mu(x')&=\Lambda\uind{\mu}{\nu}\left[A^\nu_\phys(x)+\partial^\nu\omega^\phys_\Lambda(x)\right].\label{physabL}
\end{align}
As emphasized in some standard textbooks on Quantum Field Theory like \emph{e.g.} Weinberg \cite{Weinberg:1995mt} and Bjorken and Drell \cite{Bjorken:1965zz}, the set of physical degrees of freedom of the photon field cannot form a Lorentz four-vector\footnote{According to the theory of massless representations of the Lorentz group, the only physical massless four-vector field is the gradient of a scalar field $\partial^\mu\phi$ which has consequently spin 0.}, but necessarily transforms in a way such that the new physical field $(A_\phys)'^\mu$ satisfies the same condition in the new frame as the original physical field $A^\mu_\phys$ in the original frame. In order to fix \emph{completely} the Stueckelberg symmetry, one is forced to use a non manifestly covariant condition, just like with gauge symmetry.
%\trd{The new physical field $(A_\phys)'^\mu$ therefore coincides with the physical part of the new gauge potential $(A'^\mu)_\phys$.}
This implies that the function $\omega^\phys_\Lambda$ is in general different from zero.

For the Chen \emph{et al.} approach to be completely consistent with Lorentz symmetry, what remains is to check that the sum of the new pure-gauge and physical fields coincides with the new gauge potential
\beq
A'^\mu(x')=(A_\pure)'^\mu(x')+(A_\phys)'^\mu(x').
\eeq
This is obviously achieved by imposing the condition \cite{Lorce:2012rr}
\beq
\omega_\Lambda=\omega^\pure_\Lambda+\omega^\phys_\Lambda.
\eeq
Thanks to the gauge symmetry, one has the freedom to choose the representation of $A^\mu$, \emph{i.e.} the function $\omega_\Lambda$. Two options appear particularly interesting:
\begin{enumerate}
\item To maintain Lorentz covariance explicitly, one may choose to work with $\omega_\Lambda=0$, \emph{i.e.} with $A^\mu$ transforming as a Lorentz four-vector. In this case, the pure-gauge field does not transform as a Lorentz four-vector, since it has to compensate for the non-tensorial nature of the physical field $\omega^\pure_\Lambda=-\omega^\phys_\Lambda$. This means in particular that these Lorentz transformations will generally mix physical and gauge degrees of freedom. Indeed, if one starts with $A^\mu_\pure=0$ in one frame, one ends up in general with $(A_\pure)'^\mu\neq 0$ in another Lorentz frame. In other words, every time one changes the Lorentz frame, one needs to perform an additional gauge transformation in order to recover the physical polarizations.
\item To avoid a mixing between physical and gauge degrees of freedom under Lorentz transformations, one may choose to work with $\omega_\Lambda=\omega^\phys_\Lambda$. Indeed, in this case one has $\omega^\pure_\Lambda=0$, which means that $A^\mu_\pure$ transforms as a Lorentz four-vector or, equivalently, that $\alpha_\pure$ (the gauge degree of freedom) transforms as a Lorentz scalar owing to Eq. \eqref{puregauge}. If $A^\mu_\pure=0$ in one frame, it remains zero in any other Lorentz frame. In other words, physical polarizations remain physical after these Lorentz transformations, but the cost is that $A^\mu$ necessarily transforms in a complicated way.
\end{enumerate}
In conclusion, we can give a positive answer to the question posed at the beginning of section \ref{secIVB}:
\beq
\begin{split}
(A_\pure)'^\mu(x')&=(A'^\mu)_\pure(x'),\\
(A_\phys)'^\mu(x')&=(A'^\mu)_\phys(x').
\end{split}
\eeq
The Chen \emph{et al.} approach is consistent with Lorentz symmetry, and one can write without ambiguity $A_\pure'^\mu(x')$ and $A_\phys'^\mu(x')$.

\section{The proton spin decomposition\label{secV}}

In the previous section, we have discussed the angular momentum decomposition in the case of an abelian gauge theory like QED. The abelian case gave us the opportunity to discuss and illustrate the main aspects of the controversy in the simplest case. The same discussions naturally apply to non-abelian gauge theories as well, though with additional complications due to the non-abelian nature of the gauge group. Note that the non-abelian expressions are generalizations of the abelian ones. The latter can therefore easily be recovered from the former.

We are going to discuss in this section the proton spin decomposition in QCD. We shall not repeat in detail the discussions made in the previous section, but simply give the corresponding non-abelian expressions, and occasionally develop some aspects not addressed so far.

\subsection{The QCD energy-momentum and covariant angular momentum tensors\label{secVA}}

As mentioned in section \ref{secIIIA2}, the QCD Lagrangian can be seen as made of three terms
\beq\label{LQCD}
\uL_\text{QCD}=\uL_\text{D}+\uL_\text{YM}+\uL_\text{int},
\eeq
where the so-called Dirac, Yang-Mills and interaction terms are given in the fundamental representation of the color group by
\beq
\begin{split}
\uL_\text{D}&=\barpsi(\tfrac{i}{2}\,\uslash\!\LRpartial-m)\psi,\\
\uL_\text{YM}&=-\tfrac{1}{2}\,\uTr[G^{\mu\nu}G_{\mu\nu}],\\
\uL_\text{int}&=g\barpsi\,\uslash\!A\psi.
\end{split}
\eeq
For convenience, we have omitted the sum over quark flavors, which is irrelevant for our discussion, and we are using a matrix notation in color space, as explained below. Note that $\overset{\leftrightarrow}{\partial}=\overset{\rightarrow}{\partial}-\overset{\leftarrow}{\partial}$ where the arrow indicates whether the derivative operator acts on the left or on the right. In color space, the quark fields are vectors whereas the gauge potential $A_\mu$ and the gluon field-strength tensor
\beq
G_{\mu\nu}=\partial_\mu A_\nu-\partial_\nu A_\mu-ig[A_\mu,A_\nu]
\eeq
are $N_c\times N_c$ matrices with $N_c$ the number of colors. Using the generator matrices $t^a$ of color rotations satisfying the relations
\beq
\uTr[t^at^b]=\tfrac{1}{2}\,\delta^{ab},\qquad [t^a,t^b]=if^{abc}t^c,
\eeq
where $f^{abc}$ are the totally antisymmetric structure constants of the color gauge group, one can write
\beq
A_\mu=A^a_\mu\,t^a,\qquad G_{\mu\nu}=G^a_{\mu\nu}\,t^a
\eeq
with
\beq
\qquad G^a_{\mu\nu}=\partial_\mu A^a_\nu-\partial_\nu A^a_\nu+gf^{abc}A^b_\mu A^c_\nu.
\eeq
The QED expressions can formally be recovered by first rewriting everything in terms of $A^a_\mu$ and $G^a_{\mu\nu}$ (\emph{i.e.} in the adjoint representation), and then by making the substitutions $f^{abc}\mapsto 0$ and $g\mapsto -e$.

One can associate to the QCD Lagrangian \eqref{LQCD} the following three kinds of (hermitian) energy-momentum and generalized angular momentum densities:
\begin{itemize}
\item The \emph{canonical} densities obtained directly from Noether's theorem\footnote{The extra anti-hermitian term $\tfrac{i}{2}\,g^{\mu[\nu}\barpsi\gamma^{\rho]}\psi$ appearing in $M^{\mu\nu\rho}$ of Ref. \cite{Lorce:2012rr} is a remainder of a non-hermitian treatment, and should of course have been dropped in the final hermitian expressions. Note however that it is harmless as long as one considers angular momentum operators only.}
\begin{align}
T^{\mu\nu}&=\tfrac{1}{2}\,\barpsi \gamma^\mu i\LRpartial^\nu\psi-2\,\uTr[G^{\mu\alpha}\partial^\nu A_\alpha]-g^{\mu\nu}\uL_\text{QCD},\\
M^{\mu\nu\rho}&=\tfrac{1}{2}\,\epsilon^{\mu\nu\rho\sigma}\,\barpsi\gamma_\sigma\gamma_5\psi+\tfrac{1}{2}\,\barpsi \gamma^\mu x^{[\nu}i\LRpartial^{\rho]}\psi-2\,\uTr[G^{\mu[\nu} A^{\rho]}]-2\,\uTr[G^{\mu\alpha}x^{[\nu} \partial^{\rho]}A_\alpha]-x^{[\nu}g^{\rho]\mu}\uL_\text{QCD};
\end{align}
\item The \emph{gauge-invariant canonical} densities obtained from Noether's theorem in conjunction with the Chen \emph{et al.} approach \cite{Lorce:2013gxa}
\begin{align}
T^{\mu\nu}_\text{gic}&=\tfrac{1}{2}\,\barpsi \gamma^\mu i\LRD^\nu_\pure\psi-2\,\uTr[G^{\mu\alpha}\mathcal D^\nu_\pure A^\phys_\alpha]-g^{\mu\nu}\uL_\text{QCD},\\
M^{\mu\nu\rho}_\text{gic}&=\tfrac{1}{2}\,\epsilon^{\mu\nu\rho\sigma}\,\barpsi\gamma_\sigma\gamma_5\psi+\tfrac{1}{2}\,\barpsi \gamma^\mu x^{[\nu}i\LRD^{\rho]}_\pure\psi-2\,\uTr[G^{\mu[\nu} A^{\rho]}_\phys]-2\,\uTr[G^{\mu\alpha}x^{[\nu} \mathcal D^{\rho]}_\pure A^\phys_\alpha]-x^{[\nu}g^{\rho]\mu}\uL_\text{QCD};
\end{align}
\item The \emph{Belinfante}-improved densities obtained from the canonical ones using the Belinfante-Rosenfeld prescription
\begin{align}
T^{\mu\nu}_\text{Bel}&=\tfrac{1}{4}\,\barpsi (\gamma^\mu i\LRD^\nu+\gamma^\nu i\LRD^\mu)\psi-2\,\uTr[G^{\mu\alpha}G^\nu_{\phantom{\nu}\alpha}]-g^{\mu\nu}\uL_\text{QCD},\\
M^{\mu\nu\rho}_\text{Bel}&=\tfrac{1}{4}\,\barpsi\gamma^\mu x^{[\nu}i\LRD^{\rho]}\psi+\tfrac{1}{4}\,x^{[\nu}(\barpsi\gamma^{\rho]}i\LRD^\mu\psi)-2\,\uTr[G^{\mu\alpha}x^{[\nu}G^{\rho]}_{\phantom{\rho}\alpha}]-x^{[\nu}g^{\rho]\mu}\uL_\text{QCD}.
\end{align}
\end{itemize}
For convenience, we have used the notation $a^{[\mu}b^{\nu]}=a^\mu b^\nu-a^\nu b^\mu$. The covariant derivatives acting on the quark fields are defined as $D_\mu=\partial_\mu-igA_\mu$ and $D^\pure_\mu=\partial_\mu-igA^\pure_\mu$, and those acting on the gluon fields are defined as $\mathcal D_\mu=\partial_\mu-ig[A_\mu,\quad]$ and $\mathcal D^\pure_\mu=\partial_\mu-ig[A^\pure_\mu,\quad]$. The non-abelian physical field is related to the field-strength tensor as \cite{Lorce:2012rr}
\beq
G_{\mu\nu}=\mathcal D^\pure_\mu A^\phys_\nu-\mathcal D^\pure_\nu A^\phys_\mu-ig[A^\phys_\mu,A^\phys_\nu].
\eeq

All these densities simply differ by superpotential terms \cite{Jaffe:1989jz}, \emph{i.e.} terms of the generic form $\partial_\alpha B^{[\alpha\mu]\cdots}$ where the dots stand for additional indices,
\beq
\begin{split}
T^{\mu\nu}&=T^{\mu\nu}_\text{gic}-2\partial_\alpha\uTr[G^{\mu\alpha}A^\nu_\pure]=T^{\mu\nu}_\text{Bel}+\tfrac{1}{4}\,\partial_\alpha(\epsilon^{\nu\mu\alpha\beta}\,\barpsi\gamma_\beta\gamma_5\psi)-2\partial_\alpha\uTr[G^{\mu\alpha}A^\nu],\\
M^{\mu\nu\rho}&=M^{\mu\nu\rho}_\text{gic}-2\partial_\alpha\uTr[G^{\mu\alpha}x^{[\nu}A^{\rho]}_\pure]=M^{\mu\nu\rho}_\text{Bel}+\tfrac{1}{4}\,\partial_\alpha(x^{[\nu}\epsilon^{\rho]\mu\alpha\beta}\,\barpsi\gamma_\beta\gamma_5\psi)-2\partial_\alpha\uTr[G^{\mu\alpha}x^{[\nu}A^{\rho]}],
\end{split}
\eeq
and are conserved thanks to the QCD equations of motion
\beq
\begin{split}
\partial_\mu T^{\mu\nu}=\partial_\mu T^{\mu\nu}_\text{gic}=\partial_\mu T^{\mu\nu}_\text{Bel}&=0,\\
\partial_\mu M^{\mu\nu\rho}=\partial_\mu M^{\mu\nu\rho}_\text{gic}=\partial_\mu M^{\mu\nu\rho}_\text{Bel}&=0.
\end{split}
\eeq
Moreover, under gauge transformations the fields change as follows:
\beq\label{nonabGT}
\begin{split}
\psi(x)&\mapsto\tilde\psi(x)=U(x)\psi(x),\\
A_\mu(x)&\mapsto\tilde A_\mu(x)=U(x)\left[A_\mu(x)+\tfrac{i}{g}\,\partial_\mu\right]U^{-1}(x),\\
A^\pure_\mu(x)&\mapsto\tilde A^\pure_\mu(x)=U(x)\left[A^\pure_\mu(x)+\tfrac{i}{g}\,\partial_\mu\right]U^{-1}(x),\\
A^\phys_\mu(x)&\mapsto\tilde A^\phys_\mu(x)=U(x)A^\phys_\mu(x)U^{-1}(x),\\
G_{\mu\nu}(x)&\mapsto\tilde G_{\mu\nu}(x)=U(x)G_{\mu\nu}(x)U^{-1}(x),
\end{split}
\eeq
where $U(x)$ is an $N_c\times N_c$ matrix in color space\footnote{In the abelian case, it reduces to a simple complex phase factor $U(x)=e^{-ie\alpha(x)}$.}. The Belinfante-improved and gauge-invariant canonical densities are gauge invariant, whereas the canonical densities change by a superpotential term
\beq
\begin{split}
T^{\mu\nu}&\mapsto\tilde T^{\mu\nu}=T^{\mu\nu}-\tfrac{2i}{g} \,\partial_\alpha\uTr[G^{\mu\alpha}(\partial^\nu U^{-1})U],\\
M^{\mu\nu\rho}&\mapsto\tilde M^{\mu\nu\rho}=M^{\mu\nu\rho}-\tfrac{2i}{g} \,\partial_\alpha\uTr[G^{\mu\alpha}x^{[\nu}(\partial^{\rho]}U^{-1})U].
\end{split}
\eeq
Upon integration, the superpotential terms turn into surface terms which are usually assumed to vanish\footnote{Note that this might not be justified in QCD because of non-perturbative ambiguities, like \emph{e.g.} Gribov copies, associated with gluon field configurations with non-trivial topology.}. One then concludes that the three different sets of conserved densities lead to the \emph{same} set of time-independent gauge-invariant tensors
\beq\label{charges}
\begin{split}
P^\mu&=\int\ud^3x\,T^{0\mu}(x)\sim\int\ud^3x\,T^{0\mu}_\text{gic}(x)\sim\int\ud^3x\, T^{0\mu}_\text{Bel}(x),\\
M^{\mu\nu}&=\int\ud^3x\,M^{0\mu\nu}(x)\sim\int\ud^3x\, M^{0\mu\nu}_\text{gic}(x)\sim\int\ud^3x\, M^{0\mu\nu}_\text{Bel}(x).
\end{split}
\eeq
According to Noether's theorem, since these charges are obtained from the canonical densities, they are \emph{total} generators of space-time translations and Lorentz transformations. They are consequently identified with the total energy-momentum $P^\nu$ and generalized angular momentum $M^{\nu\rho}$ operators.

Playing around with superpotential terms, many more densities leading to the same tensors could be defined. Clearly, while the total energy-momentum and generalized angular momentum are uniquely defined\footnote{Provided one can drop the contribution of the superpotential terms, \emph{i.e.} identify the operators via the $\sim$ relation discussed in section \ref{secIIF3}.}, their associated densities are not. Because of these ambiguities, it is sometimes thought that the densities are not really ``physical'' quantities. If one adopts the strong notion of gauge invariance discussed in the previous section, one would conclude that among the three sets of densities, only  the Belinfante-improved ones are ``physical'', since they are both local and gauge-invariant expressions. However, we remind the reader that a sufficient condition for measurability is weak gauge invariance, and that non-local expressions in terms of $A_\mu$ are perfectly acceptable as long as the non-locality is confined in the unphysical gauge sector, \emph{i.e.} can be removed by a suitable gauge transformation.

\subsection{Decompositions of the proton momentum and the proton spin\label{secVB}}

We quickly discuss here the main types of decompositions of the linear and angular momentum in QCD. First we present the decompositions belonging to the canonical family, and then those belonging to the kinetic family. We keep track of the superpotential terms since they contribute at the density level. As stressed by Leader \cite{Leader:2011za}, it is important to remember that, contrary to the total densities, the individual contributions are not conserved, and therefore lead to time-dependent tensors. The corresponding matrix elements are however time independent as long as one considers initial and final states with the same energy.

In the following we give the complete expressions for the densities, \emph{i.e.} not just for the actual angular momentum operators which would correspond to the values for the indices $\mu=0$, $\nu=i$ and $\rho=j$ with $i\neq j$. Note also that throughout the following $q$ means quark plus antiquark.

\subsubsection{The canonical decompositions\label{secVB1}}

The Jaffe-Manohar decomposition corresponds to the natural decomposition of the canonical densities
\beq
\begin{split}
T^{\mu\nu}&=T^{\mu\nu}_{\text{JM},q,\text{M}}+T^{\mu\nu}_{\text{JM},q,\text{E}}+T^{\mu\nu}_{\text{JM},G,\text{M}}+T^{\mu\nu}_{\text{JM},G,\text{E}},\\
M^{\mu\nu\rho}&=M^{\mu\nu\rho}_{\text{JM},q,\text{spin}}+M^{\mu\nu\rho}_{\text{JM},q,\text{OAM}}+M^{\mu\nu\rho}_{\text{JM},q,\text{boost}}+M^{\mu\nu\rho}_{\text{JM},G,\text{spin}}+M^{\mu\nu\rho}_{\text{JM},G,\text{OAM}}+M^{\mu\nu\rho}_{\text{JM},G,\text{boost}},
\end{split}
\eeq
where the quark and gluon contributions are given by\footnote{We used the identity $\tfrac{1}{2}\,\barpsi\left\{\gamma^\mu,\Sigma^{\nu\rho}\right\}\psi=\tfrac{1}{2}\,\epsilon^{\mu\nu\rho\sigma}\,\barpsi\gamma_\sigma\gamma_5\psi$.}
\beq\label{JMdec}
\begin{aligned}
T^{\mu\nu}_{\text{JM},q,\text{M}}&=\tfrac{1}{2}\,\barpsi\gamma^\mu i\LRpartial^\nu\psi,&T^{\mu\nu}_{\text{JM},G,\text{M}}&=-2\,\uTr[G^{\mu\alpha}\partial^\nu A_\alpha],\\
T^{\mu\nu}_{\text{JM},q,\text{E}}&=-g^{\mu\nu}\uL_\text{D},&T^{\mu\nu}_{\text{JM},G,\text{E}}&=-g^{\mu\nu}(\uL_\text{YM}+\uL_\text{int}),\\
M^{\mu\nu\rho}_{\text{JM},q,\text{spin}}&=\tfrac{1}{2}\,\epsilon^{\mu\nu\rho\sigma}\,\barpsi\gamma_\sigma\gamma_5\psi,&M^{\mu\nu\rho}_{\text{JM},G,\text{spin}}&=-2\,\uTr[G^{\mu[\nu} A^{\rho]}],\\
M^{\mu\nu\rho}_{\text{JM},q,\text{OAM}}&=\tfrac{1}{2}\,\barpsi\gamma^\mu x^{[\nu}i\LRpartial^{\rho]}\psi,&M^{\mu\nu\rho}_{\text{JM},G,\text{OAM}}&=-2\,\uTr[G^{\mu\alpha}x^{[\nu} \partial^{\rho]}A_\alpha],\\
M^{\mu\nu\rho}_{\text{JM},q,\text{boost}}&=-x^{[\nu}g^{\rho]\mu}\uL_\text{D},&M^{\mu\nu\rho}_{\text{JM},G,\text{boost}}&=-x^{[\nu}g^{\rho]\mu}(\uL_\text{YM}+\uL_\text{int}),
\end{aligned}
\eeq
with the convention $\epsilon_{0123}=+1$. The terms labeled ``E'' contribute only to energy operators, but energy operators receive also a contribution from the momentum (M) terms. Similarly, the terms labeled ``boost'' contribute only to boost operators, but boost operators receive also a contribution from the angular momentum (AM) terms.

The gauge-invariant canonical decomposition corresponds to the natural decomposition of the gauge-invariant canonical densities
\beq
\begin{split}
T^{\mu\nu}_\text{gic}&=T^{\mu\nu}_{\text{gic},q,\text{M}}+T^{\mu\nu}_{\text{gic},q,\text{E}}+T^{\mu\nu}_{\text{gic},G,\text{M}}+T^{\mu\nu}_{\text{gic},G,\text{E}},\\
M^{\mu\nu\rho}_\text{gic}&=M^{\mu\nu\rho}_{\text{gic},q,\text{spin}}+M^{\mu\nu\rho}_{\text{gic},q,\text{OAM}}+M^{\mu\nu\rho}_{\text{gic},q,\text{boost}}+M^{\mu\nu\rho}_{\text{gic},G,\text{spin}}+M^{\mu\nu\rho}_{\text{gic},G,\text{OAM}}+M^{\mu\nu\rho}_{\text{gic},G,\text{boost}},
\end{split}
\eeq
where the quark and gluon contributions are given by
\beq\label{gicdec}
\begin{aligned}
T^{\mu\nu}_{\text{gic},q,\text{M}}&=\tfrac{1}{2}\,\barpsi\gamma^\mu i\LRD^\nu_\pure\psi,&T^{\mu\nu}_{\text{gic},G,\text{M}}&=-2\,\uTr[G^{\mu\alpha}\mathcal D^\nu_\pure A^\phys_\alpha],\\
T^{\mu\nu}_{\text{gic},q,\text{E}}&=-g^{\mu\nu}(\uL_\text{D}+\uL^\pure_\text{int}),&T^{\mu\nu}_{\text{gic},G,\text{E}}&=-g^{\mu\nu}(\uL_\text{YM}+\uL^\phys_\text{int}),\\
M^{\mu\nu\rho}_{\text{gic},q,\text{spin}}&=\tfrac{1}{2}\,\epsilon^{\mu\nu\rho\sigma}\,\barpsi\gamma_\sigma\gamma_5\psi,&M^{\mu\nu\rho}_{\text{gic},G,\text{spin}}&=-2\,\uTr[G^{\mu[\nu} A^{\rho]}_\phys],\\
M^{\mu\nu\rho}_{\text{gic},q,\text{OAM}}&=\tfrac{1}{2}\,\barpsi\gamma^\mu x^{[\nu}i\LRD^{\rho]}_\pure\psi,&M^{\mu\nu\rho}_{\text{gic},G,\text{OAM}}&=-2\,\uTr[G^{\mu\alpha}x^{[\nu} \mathcal D^{\rho]}_\pure A^\phys_\alpha],\\
M^{\mu\nu\rho}_{\text{gic},q,\text{boost}}&=-x^{[\nu}g^{\rho]\mu}(\uL_\text{D}+\uL^\pure_\text{int}),&M^{\mu\nu\rho}_{\text{gic},G,\text{boost}}&=-x^{[\nu}g^{\rho]\mu}(\uL_\text{YM}+\uL^\phys_\text{int}).
\end{aligned}
\eeq
The interaction Lagrangian has been decomposed into pure-gauge and physical contributions $\uL_\text{int}=\uL^\pure_\text{int}+\uL^\phys_\text{int}$ with $\uL^\pure_\text{int}=g\barpsi\,\uslash\!A_\pure\psi$ and $\uL^\phys_\text{int}=g\barpsi\,\uslash\! A_\phys\psi$. The gauge-invariant canonical decomposition reduces formally to the Jaffe-Manohar decomposition in the natural gauge $A^\pure_\mu=0$, and can therefore be considered as the natural gauge-invariant extension of the latter.

\subsubsection{The kinetic decompositions\label{secVB2}}

The Belifante decomposition corresponds to the natural decomposition of the Belinfante-improved densities
\beq
\begin{split}
T^{\mu\nu}_\text{Bel}&=T^{\mu\nu}_{\text{Bel},q,\text{M}}+T^{\mu\nu}_{\text{Bel},q,\text{E}}+T^{\mu\nu}_{\text{Bel},G,\text{M}}+T^{\mu\nu}_{\text{Bel},G,\text{E}},\\
M^{\mu\nu\rho}_\text{Bel}&=M^{\mu\nu\rho}_{\text{Bel},q,\text{AM}}+M^{\mu\nu\rho}_{\text{Bel},q,\text{boost}}+M^{\mu\nu\rho}_{\text{Bel},G,\text{AM}}+M^{\mu\nu\rho}_{\text{Bel},G,\text{boost}},
\end{split}
\eeq
where the quark and gluon contributions are given by
\beq
\begin{aligned}
T^{\mu\nu}_{\text{Bel},q,\text{M}}&=\tfrac{1}{4}\,\barpsi (\gamma^\mu i\LRD^\nu+\gamma^\nu i\LRD^\mu)\psi,&
T^{\mu\nu}_{\text{Bel},G,\text{M}}&=-2\,\uTr[G^{\mu\alpha}G^\nu_{\phantom{\nu}\alpha}],\\
T^{\mu\nu}_{\text{Bel},q,\text{E}}&=-g^{\mu\nu}(\uL_\text{D}+\uL_\text{int}),&
T^{\mu\nu}_{\text{Bel},G,\text{E}}&=-g^{\mu\nu}\uL_\text{YM},\\
M^{\mu\nu\rho}_{\text{Bel},q,\text{AM}}&=\tfrac{1}{4}\,\barpsi\gamma^\mu x^{[\nu}i\LRD^{\rho]}\psi+\tfrac{1}{4}\,x^{[\nu}(\barpsi\gamma^{\rho]}i\LRD^\mu\psi),&M^{\mu\nu\rho}_{\text{Bel},G,\text{AM}}&=-2\,\uTr[G^{\mu\alpha}x^{[\nu}G^{\rho]}_{\phantom{\rho}\alpha}],\\
M^{\mu\nu\rho}_{\text{Bel},q,\text{boost}}&=-x^{[\nu}g^{\rho]\mu}(\uL_\text{D}+\uL_\text{int}),&
M^{\mu\nu\rho}_{\text{Bel},G,\text{boost}}&=-x^{[\nu}g^{\rho]\mu}\uL_\text{YM}.
\end{aligned}
\eeq
Note that by the QCD equation of motion $(i\gamma^\mu D_\mu-m)\psi=0$, one has $\uL_\text{D}+\uL_\text{int}=0$.

The Ji decomposition\footnote{Ji actually discussed only the decomposition of the angular momentum. For completeness, we added the expressions for the boost terms and the energy-momentum.} uses the modified Belinfante decomposition explained in section \ref{secIIIB1} by providing a separation of the quark angular momentum into spin and orbital angular momentum contributions
\beq
\begin{split}
T^{\mu\nu}_\text{Ji}&=T^{\mu\nu}_{\text{Ji},q,\text{M}}+T^{\mu\nu}_{\text{Ji},q,\text{E}}+T^{\mu\nu}_{\text{Ji},G,\text{M}}+T^{\mu\nu}_{\text{Ji},G,\text{E}},\\
M^{\mu\nu\rho}_\text{Ji}&=M^{\mu\nu\rho}_{\text{Ji},q,\text{spin}}+M^{\mu\nu\rho}_{\text{Ji},q,\text{OAM}}+M^{\mu\nu\rho}_{\text{Ji},q,\text{boost}}+M^{\mu\nu\rho}_{\text{Ji},G,\text{AM}}+M^{\mu\nu\rho}_{\text{Ji},G,\text{boost}},
\end{split}
\eeq
where the quark and gluon contributions are given by\footnote{Note that in Ji's paper \cite{Ji:1996ek} the quark orbital angular momentum is written with the derivative acting only to the right, $\,\barpsi\gamma^\mu x^{[\nu}i\overrightarrow{D}^{\rho]}\psi$, which differs from the expression below by a surface term. We prefer the manifestly hermitian form.}
\beq
\begin{aligned}
T^{\mu\nu}_{\text{Ji},q,\text{M}}&=\tfrac{1}{2}\,\barpsi\gamma^\mu i\LRD^\nu\psi,&T^{\mu\nu}_{\text{Ji},G,\text{M}}&=-2\,\uTr[G^{\mu\alpha}G^\nu_{\phantom{\nu}\alpha}],\\
T^{\mu\nu}_{\text{Ji},q,\text{E}}&=-g^{\mu\nu}(\uL_\text{D}+\uL_\text{int}),&T^{\mu\nu}_{\text{Ji},G,\text{E}}&=-g^{\mu\nu}\uL_\text{YM},\\
\begin{split}
M^{\mu\nu\rho}_{\text{Ji},q,\text{spin}}\\
M^{\mu\nu\rho}_{\text{Ji},q,\text{OAM}}
\end{split}
&\!\begin{split}&=\tfrac{1}{2}\,\epsilon^{\mu\nu\rho\sigma}\,\barpsi\gamma_\sigma\gamma_5\psi,\\
&=\tfrac{1}{2}\,\barpsi\gamma^\mu x^{[\nu}i\LRD^{\rho]}\psi,
\end{split}
&M^{\mu\nu\rho}_{\text{Ji},G,\text{AM}}&=-2\,\uTr[G^{\mu\alpha}x^{[\nu}G^{\rho]}_{\phantom{\rho}\alpha}],\\
M^{\mu\nu\rho}_{\text{Ji},q,\text{boost}}&=-x^{[\nu}g^{\rho]\mu}(\uL_\text{D}+\uL_\text{int}),&
M^{\mu\nu\rho}_{\text{Ji},G,\text{boost}}&=-x^{[\nu}g^{\rho]\mu}\uL_\text{YM}.
\end{aligned}
\eeq

The gauge-invariant kinetic decomposition improves the Ji decomposition by providing a separation of the gluon angular momentum into spin and orbital angular momentum contributions based on the Chen \emph{et al.} approach
\beq
\begin{split}
T^{\mu\nu}_\text{gik}&=T^{\mu\nu}_{\text{gik},q,\text{M}}+T^{\mu\nu}_{\text{gik},q,\text{E}}+T^{\mu\nu}_{\text{gik},G,\text{M}}+T^{\mu\nu}_{\text{gik},G,\text{E}},\\
M^{\mu\nu\rho}_\text{gik}&=M^{\mu\nu\rho}_{\text{gik},q,\text{spin}}+M^{\mu\nu\rho}_{\text{gik},q,\text{OAM}}+M^{\mu\nu\rho}_{\text{gik},q,\text{boost}}+M^{\mu\nu\rho}_{\text{gik},G,\text{spin}}+M^{\mu\nu\rho}_{\text{gik},G,\text{OAM}}+M^{\mu\nu\rho}_{\text{gik},G,\text{boost}},
\end{split}
\eeq
where the quark and gluon contributions are given by
\beq
\begin{aligned}
T^{\mu\nu}_{\text{gik},q,\text{M}}&=\tfrac{1}{2}\,\barpsi\gamma^\mu i\LRD^\nu\psi,&T^{\mu\nu}_{\text{gik},G,\text{M}}&=-2\,\uTr[G^{\mu\alpha}\mathcal D^\nu_\pure A^\phys_\alpha-(\mathcal D_\alpha G^{\alpha\mu})A^\nu_\phys],\\
T^{\mu\nu}_{\text{gik},q,\text{E}}&=-g^{\mu\nu}(\uL_\text{D}+\uL_\text{int}),&T^{\mu\nu}_{\text{gik},G,\text{E}}&=-g^{\mu\nu}\uL_\text{YM},\\
M^{\mu\nu\rho}_{\text{gik},q,\text{spin}}&=\tfrac{1}{2}\,\epsilon^{\mu\nu\rho\sigma}\,\barpsi\gamma_\sigma\gamma_5\psi,&M^{\mu\nu\rho}_{\text{gik},G,\text{spin}}&=-2\,\uTr[G^{\mu[\nu} A^{\rho]}_\phys],\\
M^{\mu\nu\rho}_{\text{gik},q,\text{OAM}}&=\tfrac{1}{2}\,\barpsi\gamma^\mu x^{[\nu}i\LRD^{\rho]}\psi,&M^{\mu\nu\rho}_{\text{gik},G,\text{OAM}}&=-2\,\uTr[G^{\mu\alpha}x^{[\nu} \mathcal D^{\rho]}_\pure A^\phys_\alpha-(\mathcal D_\alpha G^{\alpha\mu})x^{[\nu}A^{\rho]}_\phys],\\
M^{\mu\nu\rho}_{\text{gik},q,\text{boost}}&=-x^{[\nu}g^{\rho]\mu}(\uL_\text{D}+\uL_\text{int}),&M^{\mu\nu\rho}_{\text{gik},G,\text{boost}}&=-x^{[\nu}g^{\rho]\mu}\uL_\text{YM}.
\end{aligned}
\eeq

These three decompositions are related by superpotential terms and the following identities (see section \ref{secIIIB1})
\beq
\begin{split}
\barpsi\gamma^\nu i\LRD^\mu\psi&=\barpsi\gamma^\mu i\LRD^\nu\psi-\partial_\alpha(\epsilon^{\nu\mu\alpha\beta}\,\barpsi\gamma_\beta\gamma_5\psi),\\
x^{[\nu}(\barpsi \gamma^{\rho]} i\LRD^\mu\psi)&=\barpsi\gamma^\mu x^{[\nu}i\LRD^{\rho]}\psi+2\,\epsilon^{\mu\nu\rho\sigma}\,\barpsi\gamma_\sigma\gamma_5\psi-\partial_\alpha(x^{[\nu}\epsilon^{\rho]\mu\alpha\beta}\,\barpsi\gamma_\beta\gamma_5\psi),
\end{split}
\eeq
based on the QCD equation of motion $(i\,\uslash\!D-m)\psi=0$. More precisely, one has
\beq
\begin{split}
T^{\mu\nu}_{\text{Bel},q,\text{M}}+\tfrac{1}{4}\,\partial_\alpha(\epsilon^{\nu\mu\alpha\beta}\,\barpsi\gamma_\beta\gamma_5\psi)&=T^{\mu\nu}_{\text{Ji},q,\text{M}}=T^{\mu\nu}_{\text{gik},q,\text{M}},\\
T^{\mu\nu}_{\text{Bel},G,\text{M}}&=T^{\mu\nu}_{\text{Ji},G,\text{M}}=T^{\mu\nu}_{\text{gik},G,\text{M}}+2\partial_\alpha\uTr[G^{\mu\alpha} A^\nu_\phys],\\
T^{\mu\nu}_{\text{Bel},q,\text{E}}&=T^{\mu\nu}_{\text{Ji},q,\text{E}}=T^{\mu\nu}_{\text{gik},q,\text{E}},\\
T^{\mu\nu}_{\text{Bel},G,\text{E}}&=T^{\mu\nu}_{\text{Ji},G,\text{E}}=T^{\mu\nu}_{\text{gik},G,\text{E}},\\
M^{\mu\nu\rho}_{\text{Bel},q,\text{AM}}+\tfrac{1}{4}\,\partial_\alpha(x^{[\nu}\epsilon^{\rho]\mu\alpha\beta}\,\barpsi\gamma_\beta\gamma_5\psi)&=M^{\mu\nu\rho}_{\text{Ji},q,\text{spin}}+M^{\mu\nu\rho}_{\text{Ji},q,\text{OAM}}=M^{\mu\nu\rho}_{\text{gik},q,\text{spin}}+M^{\mu\nu\rho}_{\text{gik},q,\text{OAM}},\\
M^{\mu\nu\rho}_{\text{Bel},G,\text{AM}}&=M^{\mu\nu\rho}_{\text{Ji},G,\text{AM}}=M^{\mu\nu\rho}_{\text{gik},G,\text{spin}}+M^{\mu\nu\rho}_{\text{gik},G,\text{OAM}}+2\partial_\alpha\uTr[G^{\mu\alpha} x^{[\nu}A^{\rho]}_\phys],\\
M^{\mu\nu\rho}_{\text{Bel},q,\text{boost}}&=M^{\mu\nu\rho}_{\text{Ji},q,\text{boost}}=M^{\mu\nu\rho}_{\text{gik},q,\text{boost}},\\
M^{\mu\nu\rho}_{\text{Bel},G,\text{boost}}&=M^{\mu\nu\rho}_{\text{Ji},G,\text{boost}}=M^{\mu\nu\rho}_{\text{gik},G,\text{boost}}.
\end{split}
\eeq

Following Lorc\'e's observation \cite{Lorce:2013fpa}, one could consider alternatively the following gluon kinetic linear and orbital angular momentum contributions
\beq
\begin{split}
T^{\mu\nu}_{\text{Lor},G}&=T^{\mu\nu}_{\text{gik},G}+2\partial_\lambda\uTr[g^{\nu[\lambda}G^{\mu]\alpha}A^\phys_\alpha]\\
&=\underbrace{g^{\nu[\alpha}\delta^{\beta]}_\lambda\, 2\,\uTr[A^\phys_\alpha\mathfrak D^\pure_\beta G^{\lambda\mu}]}_{T^{\mu\nu}_{\text{Lor},G,\text{M}}}\underbrace{-g^{\mu\nu}(\uL_\text{YM}+2\partial_\alpha\uTr[G^{\alpha\beta}A^\phys_\beta])}_{T^{\mu\nu}_{\text{Lor},G,\text{E}}},\\
M^{\mu\nu\rho}_{\text{Lor},G,\text{OAM}+\text{boost}}&=M^{\mu\nu\rho}_{\text{gik},G,\text{OAM}+\text{boost}}+2\partial_\lambda\uTr[x^{[\nu}g^{\rho][\lambda}G^{\mu]\alpha}A^\phys_\alpha]\\
&=\underbrace{x^{[\nu}g^{\rho][\alpha}\delta^{\beta]}_\lambda\, 2\,\uTr[A^\phys_\alpha\mathfrak D^\pure_\beta G^{\lambda\mu}]}_{M^{\mu\nu\rho}_{\text{Lor},G,\text{OAM}}}\underbrace{-x^{[\nu}g^{\rho]\mu}(\uL_\text{YM}+2\partial_\alpha\uTr[G^{\alpha\beta}A^\phys_\beta])-2\,\uTr[g^{\mu[\rho}G^{\nu]\alpha}A^\phys_\alpha]}_{M^{\mu\nu\rho}_{\text{Lor},G,\text{boost}}},
\end{split}
\eeq
where we have introduced for convenience the hybrid covariant derivative $\mathfrak D^\pure_\mu=\tfrac{1}{2}\left(\mathcal D_\mu+\mathcal D^\pure_\mu\right)$. These expressions have the advantage of exhibiting the standard orbital structure
\beq
M^{\mu\nu\rho}_{\text{Lor},G,\text{OAM}}=x^\nu T^{\mu\rho}_{\text{Lor},G,\text{M}}-x^\rho T^{\mu\nu}_{\text{Lor},G,\text{M}}.
\eeq
Moreover, they simply differ from the gauge-invariant kinetic expressions by superpotential terms, implying that the integrated quantities are the same.

\subsubsection{The master decomposition\label{secVB3}}

Wakamatsu observed \cite{Wakamatsu:2010qj} that the so-called \emph{potential momentum} and \emph{potential angular momentum} terms, following Konopinski's terminology \cite{Konopinski}, can be written either as a quark or a gluon contribution
\beq
\begin{split}
T^{\mu\nu}_\text{pot,M}&=-g\barpsi\gamma^\mu A^\nu_\phys\psi\\
&=2\,\uTr[(\mathcal D_\alpha G^{\alpha\mu})A^\nu_\phys],\\
T^{\mu\nu}_\text{pot,E}&=g^{\mu\nu}\uL^\phys_\text{int},\\
M^{\mu\nu\rho}_\text{pot,OAM}&=-g\barpsi\gamma^\mu x^{[\nu}A^{\rho]}_\phys\psi\\
&=2\,\uTr[(\mathcal D_\alpha G^{\alpha\mu}) x^{[\nu}A^{\rho]}_\phys],\\
M^{\mu\nu\rho}_\text{pot,boost}&=x^{[\nu}g^{\rho]\mu}\uL^\phys_\text{int},
\end{split}
\eeq
with
\beq
\begin{split}
\uL^\phys_\text{int}&=g\barpsi\,\uslash\! A_\phys\psi\\
&=-2\uTr[(\mathcal D_\alpha G^{\alpha\mu}) A^\phys_\mu],
\end{split}
\eeq
using the QCD equation of motion $2(\mathcal D_\alpha G^{\alpha\mu})^a_{\phantom{a}b}=-g\barpsi_b\gamma^\mu\psi^a$ where $a,b$ are color indices in the fundamental representation. In the canonical decompositions, these potential terms are attributed to the quarks, whereas in the kinetic decompositions, they are attributed to the gluons
\beq
\begin{aligned}
T^{\mu\nu}_{\text{gik},q,\text{M}}&=T^{\mu\nu}_{\text{gic},q,\text{M}}-T^{\mu\nu}_\text{pot,M},&T^{\mu\nu}_{\text{gik},G,\text{M}}&=T^{\mu\nu}_{\text{gic},G,\text{M}}+T^{\mu\nu}_\text{pot,M},\\
T^{\mu\nu}_{\text{gik},q,\text{E}}&=T^{\mu\nu}_{\text{gic},q,\text{E}}-T^{\mu\nu}_\text{pot,E},&T^{\mu\nu}_{\text{gik},G,\text{E}}&=T^{\mu\nu}_{\text{gic},G,\text{E}}+T^{\mu\nu}_\text{pot,E},\\
M^{\mu\nu\rho}_{\text{gik},q,\text{OAM}}&=M^{\mu\nu\rho}_{\text{gic},q,\text{OAM}}-M^{\mu\nu\rho}_\text{pot,OAM},\qquad&M^{\mu\nu\rho}_{\text{gik},G,\text{OAM}}&=M^{\mu\nu\rho}_{\text{gic},G,\text{OAM}}+M^{\mu\nu\rho}_\text{pot,OAM},\\
M^{\mu\nu\rho}_{\text{gik},q,\text{boost}}&=M^{\mu\nu\rho}_{\text{gic},q,\text{boost}}-M^{\mu\nu\rho}_\text{pot,boost},\qquad&M^{\mu\nu\rho}_{\text{gik},G,\text{boost}}&=M^{\mu\nu\rho}_{\text{gic},G,\text{boost}}+M^{\mu\nu\rho}_\text{pot,boost}.
\end{aligned}
\eeq
In Ref. \cite{Wakamatsu:2012ve}, Wakamatsu commented that the attribution of the potential terms to the quarks is closer to the concept of ``action at a distance'', while its attribution to the gluons is closer to the concept of ``action through a medium''.

The present situation regarding the proton spin decomposition appears to be quite confusing, particularly because of the number of possible decompositions. All the gauge-invariant decompositions can in fact be considered as different groupings of terms belonging to a general \emph{gauge-invariant master} (gim) decomposition. This master decomposition reads
\beq
\begin{split}
T^{\mu\nu}_\text{gim}&=\sum_{i=1}^8T^{\mu\nu}_i,\\
M^{\mu\nu\rho}_\text{gim}&=\sum_{i=1}^{10} M^{\mu\nu\rho}_i,
\end{split}
\eeq
where the gauge-invariant terms are given by
\beq
\begin{aligned}
T^{\mu\nu}_1&=\tfrac{1}{2}\,\barpsi\gamma^\mu i\LRD^\nu\psi,&T^{\mu\nu}_2&=-g^{\mu\nu}(\uL_\text{D}+\uL_\text{int}),\\
T^{\mu\nu}_3&=-2\,\uTr[G^{\mu\alpha}\mathcal D^\nu_\pure A^\phys_\alpha],&T^{\mu\nu}_4&=-g^{\mu\nu}(\uL_\text{YM}+\uL^\phys_\text{int}),\\
T^{\mu\nu}_5&=-g\barpsi\gamma^\mu A^\nu_\phys\psi,&T^{\mu\nu}_6&=g^{\mu\nu}\uL^\phys_\text{int},\\
T^{\mu\nu}_7&=-\tfrac{1}{4}\,\partial_\alpha(\epsilon^{\nu\mu\alpha\beta}\,\barpsi\gamma_\beta\gamma_5\psi),&T^{\mu\nu}_8&=2\partial_\alpha\uTr[G^{\mu\alpha} A^\nu_\phys],\\
M^{\mu\nu\rho}_1&=\tfrac{1}{2}\,\epsilon^{\mu\nu\rho\sigma}\,\barpsi\gamma_\sigma\gamma_5\psi,&M^{\mu\nu\rho}_2&=-2\,\uTr[G^{\mu[\nu} A^{\rho]}_\phys],\\
M^{\mu\nu\rho}_3&=\tfrac{1}{2}\,\barpsi\gamma^\mu x^{[\nu}i\LRD^{\rho]}\psi,&M^{\mu\nu\rho}_4&=-x^{[\nu}g^{\rho]\mu}(\uL_\text{D}+\uL_\text{int}),\\
M^{\mu\nu\rho}_5&=-2\,\uTr[G^{\mu\alpha}x^{[\nu} \mathcal D^{\rho]}_\pure A^\phys_\alpha],&M^{\mu\nu\rho}_6&=-x^{[\nu}g^{\rho]\mu}(\uL_\text{YM}+\uL^\phys_\text{int}),\\
M^{\mu\nu\rho}_7&=-g\barpsi\gamma^\mu x^{[\nu}A^{\rho]}_\phys\psi,&M^{\mu\nu\rho}_8&=x^{[\nu}g^{\rho]\mu}\uL^\phys_\text{int}.\\
M^{\mu\nu\rho}_9&=-\tfrac{1}{4}\,\partial_\alpha(x^{[\nu}\epsilon^{\rho]\mu\alpha\beta}\,\barpsi\gamma_\beta\gamma_5\psi),&M^{\mu\nu\rho}_{10}&=2\partial_\alpha\uTr[G^{\mu\alpha} x^{[\nu}A^{\rho]}_\phys].
\end{aligned}
\eeq
Note that the terms $T^{\mu\nu}_7$, $T^{\mu\nu}_8$, $M^{\mu\nu\rho}_9$ and $M^{\mu\nu\rho}_{10}$ are superpotentials, and therefore do not contribute to the tensors. The relations between the gauge-invariant master, gauge-invariant canonical, gauge-invariant kinetic, Ji and Belinfante decompositions are given explicitly in Tables \ref{Mdec} and \ref{AMdec}.

\begin{table}[t!]
\begin{center}
\caption{\footnotesize{The relations between the gauge-invariant master, gauge-invariant canonical, gauge-invariant kinetic, Ji and Belinfante linear momentum decompositions, for the various quark and gluon contributions. Note that the sums of the terms in each column differ from each other only by superpotentials.}}\label{Mdec}
\begin{tabular}{@{\quad}c@{\quad}|@{\quad}c@{\quad}|@{\quad}c@{\quad}|@{\quad}c@{\quad}|@{\quad}c@{\quad}}\whline
&gic&gik&Ji&Belinfante\\
\hline
$T^{\mu\nu}_{q,\text{M}}$&$T^{\mu\nu}_1+T^{\mu\nu}_5$&$T^{\mu\nu}_1$&$T^{\mu\nu}_1$&$T^{\mu\nu}_1+T^{\mu\nu}_7$\\
$T^{\mu\nu}_{q,\text{E}}$&$T^{\mu\nu}_2+T^{\mu\nu}_6$&$T^{\mu\nu}_2$&$T^{\mu\nu}_2$&$T^{\mu\nu}_2$\\
$T^{\mu\nu}_{G,\text{M}}$&$T^{\mu\nu}_3$&$T^{\mu\nu}_3+T^{\mu\nu}_5$&$T^{\mu\nu}_3+T^{\mu\nu}_5+T^{\mu\nu}_8$&$T^{\mu\nu}_3+T^{\mu\nu}_5+T^{\mu\nu}_8$\\
$T^{\mu\nu}_{G,\text{E}}$&$T^{\mu\nu}_4$&$T^{\mu\nu}_4+T^{\mu\nu}_6$&$T^{\mu\nu}_4+T^{\mu\nu}_6$&$T^{\mu\nu}_4+T^{\mu\nu}_6$\\
\whline
\end{tabular}
\end{center}
\end{table}

\begin{table}[t!]
\begin{center}
\caption{\footnotesize{The relations between the gauge-invariant master, gauge-invariant canonical, gauge-invariant kinetic, Ji and Belinfante angular momentum decompositions, for the various quark and gluon contributions. Note that the sums of the terms in each column differ from each other only by superpotentials.}}\label{AMdec}
\begin{tabular}{@{\quad}c@{\quad}|@{\quad}c@{\quad}|@{\quad}c@{\quad}|@{\quad}c@{\quad}|@{\quad}c@{\quad}}\whline
&gic&gik&Ji&Belinfante\\
\hline
$M^{\mu\nu\rho}_{q,\text{spin}}$&$M^{\mu\nu\rho}_1$&$M^{\mu\nu\rho}_1$&$M^{\mu\nu\rho}_1$&\multirow{2}{*}{\Bigg\}\,$M^{\mu\nu\rho}_1+M^{\mu\nu\rho}_3+M^{\mu\nu\rho}_9$}\\
$M^{\mu\nu\rho}_{q,\text{OAM}}$&$M^{\mu\nu\rho}_3+M^{\mu\nu\rho}_7$&$M^{\mu\nu\rho}_3$&$M^{\mu\nu\rho}_3$&\\
$M^{\mu\nu\rho}_{q,\text{boost}}$&$M^{\mu\nu\rho}_4+M^{\mu\nu\rho}_8$&$M^{\mu\nu\rho}_4$&$M^{\mu\nu\rho}_4$&$M^{\mu\nu\rho}_4$\\
$M^{\mu\nu\rho}_{G,\text{spin}}$&$M^{\mu\nu\rho}_2$&$M^{\mu\nu\rho}_2$&\multirow{2}{*}{\Bigg\}\,$M^{\mu\nu\rho}_2+M^{\mu\nu\rho}_5+M^{\mu\nu\rho}_7+M^{\mu\nu\rho}_{10}$}&\multirow{2}{*}{\Bigg\}\,$M^{\mu\nu\rho}_2+M^{\mu\nu\rho}_5+M^{\mu\nu\rho}_7+M^{\mu\nu\rho}_{10}$}\\
$M^{\mu\nu\rho}_{G,\text{OAM}}$&$M^{\mu\nu\rho}_5$&$M^{\mu\nu\rho}_5+M^{\mu\nu\rho}_7$&&\\
$M^{\mu\nu\rho}_{G,\text{boost}}$&$M^{\mu\nu\rho}_6$&$M^{\mu\nu\rho}_6+M^{\mu\nu\rho}_8$&$M^{\mu\nu\rho}_6+M^{\mu\nu\rho}_8$&$M^{\mu\nu\rho}_6+M^{\mu\nu\rho}_8$\\
\whline
\end{tabular}
\end{center}
\end{table}

\subsection{Non-abelian Stueckelberg and Lorentz transformations\label{secVC}}

Like in the abelian case, the non-abelian $A^\pure_\mu$ field does not contribute by definition to the field-strength tensor
\beq
G^\pure_{\mu\nu}=\partial_\mu A^\pure_\nu-\partial_\nu A^\pure_\mu-ig[A^\pure_\mu,A^\pure_\nu]=0,
\eeq
and can consequently be written as a pure-gauge term
\beq\label{nonabpuregauge}
A_\mu^\text{pure}(x)=\tfrac{i}{g}\,U_\text{pure}(x)\partial_\mu U^{-1}_\text{pure}(x),
\eeq
where $U_\pure(x)$ is an $N_c\times N_c$ matrix in color space. In the abelian case with $g=-e$, it consists in a simple complex phase factor $U(x)=e^{-ie\alpha_\pure(x)}$, so that $A^\pure_\mu(x)=\partial_\mu\alpha_\pure(x)$. For the pure-gauge field $A^\pure_\mu$ to transform like the full gauge potential $A_\mu$ under gauge transformation \eqref{nonabGT}, the matrix $U_\pure$ is required to transform like the quark field
\beq\label{nonabUpureGT}
U_\pure(x)\mapsto\tilde U_\pure(x)=U(x)U_\pure(x).
\eeq

In the following, we briefly present the generic expressions for the non-abelian Stueckelberg and Lorentz transformations. They are the natural generalizations of the abelian expressions discussed in section \ref{secIV}.

\subsubsection{Generic Stueckelberg transformations\label{secVC1}}

In the non-abelian case, the Stueckelberg transformations read \cite{Lorce:2012rr}
\beq\label{nonabST}
\begin{split}
\psi(x)&\mapsto\bar\psi(x)=\psi(x),\\
U_\pure(x)&\mapsto\bar U_\pure(x)=U_\pure(x)U^{-1}_S(x),\\
A_\mu(x)&\mapsto\bar A_\mu(x)=A_\mu(x),\\
A^\pure_\mu(x)&\mapsto\bar A^\pure_\mu(x)=A^\pure_\mu(x)+C_\mu(x),\\
A^\phys_\mu(x)&\mapsto\bar A^\phys_\mu(x)=A^\phys_\mu(x)-C_\mu(x),\\
G_{\mu\nu}(x)&\mapsto\bar G_{\mu\nu}(x)=G_{\mu\nu}(x),
\end{split}
\eeq
with $U_S(x)$ a gauge-invariant $N_c\times N_c$ matrix in color space. The function $C_\mu(x)$ represents a transfer between the pure-gauge and physical fields. It is given by
\beq
C_\mu(x)=\tfrac{i}{g}\,U_\text{pure}(x)U_S^{-1}(x)\left[\partial_\mu U_S(x)\right]U^{-1}_\text{pure}(x)
\eeq
and satisfies
\beq
\mathcal D^\pure_\mu C_\nu-\mathcal D^\pure_\nu C_\mu-ig[C_\mu,C_\nu]=0.
\eeq
This is easily understood since the part $C_\mu$ which is transfered from the physical field $A^\phys_\mu$ to the pure-gauge field $A^\pure_\mu$ should not contribute to the field-strength tensor. Contrary to the canonical and Belinfante densities, the gauge-invariant canonical densities are not invariant under Stueckelberg transformations
\beq
\begin{split}
T^{\mu\nu}_\text{gic}&\mapsto\bar T^{\mu\nu}_\text{gic}=T^{\mu\nu}_\text{gic}+2\partial_\alpha\uTr[G^{\mu\alpha}C^\nu],\\
M^{\mu\nu\rho}_\text{gic}&\mapsto\bar M^{\mu\nu\rho}_\text{gic}=M^{\mu\nu\rho}_\text{gic}+2\partial_\alpha\uTr[G^{\mu\alpha}x^{[\nu}C^{\rho]}].
\end{split}
\eeq
These variations, having the form of superpotential terms, it follows that the corresponding charges are nonetheless Stueckelberg invariant, in agreement with Eq. \eqref{charges}.\newline

As already stressed many times, it is important not to confuse gauge transformations with Stueckelberg transformations. This is particularly clear with the matrix $U_\pure$ since gauge transformations act on its left \eqref{nonabUpureGT}, whereas Stueckelberg transformations act on its right \eqref{nonabST}. This aspect is not reflected in QED because the fields $U$, $U_\pure$ and $U_S$ commute (at the classical level) with each other in the abelian case. For a detailed discussion of the gauge  and Stueckelberg transformations from a geometrical point of view, see Ref. \cite{Lorce:2012rr}. The gauge and Stueckelberg symmetries have complementary roles. On the one hand, the gauge symmetry allows us to \emph{set to zero} a component of the gauge potential $A_\mu$ by means of a gauge transformation. On the other hand, the Stueckelberg symmetry gives us the freedom to \emph{choose} which component is not physical. Since most of the time one determines the physical components by setting directly some component to zero \emph{via} a gauge transformation, the two roles are usually not distinguished and attributed to the sole gauge symmetry. This explains further why Stueckelberg symmetry (and therefore path dependence) is sometimes claimed to be just gauge symmetry.

\begin{figure}[t!]
	\centering
		\includegraphics[width=.4\textwidth]{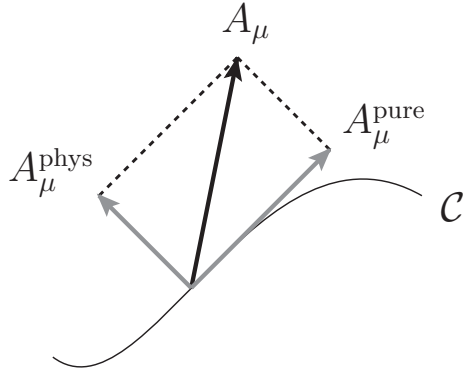}
\caption{\footnotesize{The split of the gauge potential $A_\mu$ defined by a contour $\mathcal C$. The tangential component is identified with the pure-gauge contribution $A^\pure_\mu$, and the orthogonal component is identified with the physical contribution $A^\phys_\mu$.}}
		\label{contour}
\end{figure}

The process of Stueckelberg fixing consists in choosing which component of the gauge potential is not physical. This is achieved by imposing a linear constraint $C$ on the physical field
\beq
C[A^\phys_\mu]=0.
\eeq
For example, the constraint could be the (generalized) Coulomb constraint $C[A^\phys_\mu]=\uvec{\mathcal D}_\pure\cdot\uvec A_\phys$ or the light-front constraint $C[A^\phys_\mu]=A^+_\phys$. By linearity of the constraint, one has in particular
\beq
C[A_\mu]=C[A^\pure_\mu].
\eeq
Choosing a particular constraint $C$ therefore simply amounts to choosing which component of the gauge potential $A_\mu$ is treated as a pure-gauge contribution. Consider, for example, the following Stueckelberg-fixing constraint \cite{Lorce:2012ce}
\beq\label{contourGIE}
C[A^\phys_\mu]=\frac{\partial s^\mu}{\partial\lambda}\,A^\phys_\mu=0,
\eeq
where $s^\mu(\lambda)$ is a path parametrizing some contour $\mathcal C$ connecting the point $x$ to some fixed reference point $x_0$. By linearity, one has
\beq
\frac{\partial s^\mu}{\partial\lambda}\,A_\mu=\frac{\partial s^\mu}{\partial\lambda}\,A^\pure_\mu.
\eeq
In this example, it is the particular choice of the contour $\mathcal C$ that determines the split of the gauge potential into pure-gauge and physical terms. More precisely, the component of the gauge potential \emph{along} $\mathcal C$ is identified with the pure-gauge field, whereas the component \emph{orthogonal} to $\mathcal C$ is identified with the physical field, see Fig. \ref{contour}.
Changing the contour modifies the split of the gauge potential into pure-gauge and physical contributions. In other words, path dependence is an aspect of Stueckelberg dependence.

From Eq. \eqref{nonabpuregauge}, it is straightforward to see that the matrix $U_\pure$ is covariantly conserved
\beq
D^\pure_\mu U_\pure=0.
\eeq
This shows, in particular, that the matrix $U_\pure$ has no dynamics, and therefore plays the role of a background field\footnote{Note that in General Relativity the metric $g_{\mu\nu}$ is by definition covariantly conserved $\nabla_\lambda g_{\mu\nu}=0$. But contrary to $U_\pure$, it is a dynamical object. The reason is that, unlike $D^\pure_\mu$, the covariant derivatives $\nabla_\mu$ do not commute with each other and are therefore not canonical. Their commutator is related to the Riemann curvature tensor $R^\lambda_{\phantom{\lambda}\mu\nu\rho}$ which gives dynamics to the metric \emph{via} the Einstein equation $R_{\mu\nu}-\tfrac{1}{2}\,g_{\mu\nu}R=8\pi G\,T^\text{GR}_{\mu\nu}$.}. Moreover, for the component along the contour $\mathcal C$, one has
\beq
\frac{\partial s^\mu}{\partial \lambda}\,D_\mu U_\pure=0
\eeq
which is the familiar parallel transport equation expressing the fact that $U_\pure$ does not change along the contour $\mathcal C$. The solution of this equation is well known\footnote{One can use invariance under global transformations to set $U_\pure(x_0)=\mathds 1$.}
\beq\label{solUpure}
U_\text{pure}(x)=\mathcal W_{\mathcal C}(x,x_0)\,U_\text{pure}(x_0)
\eeq
and involves the path-ordered exponential or Wilson line
\beq\label{Wilsonline}
\mathcal W_{\mathcal C}(b,a)=\mathcal P\left[e^{ig\int_a^bA_\mu(s)\,\ud s^\mu}\right]\equiv\mathds 1+ig\int_a^bA_\mu(s)\,\ud s^\mu+(ig)^2\int_a^b\int_a^{s_1}A_\mu(s_1)A_\nu(s_2)\,\ud s^\mu_1\,\ud s^\nu_2+\cdots.
\eeq
Using the generic formula for the derivative of the Wilson line
\beq
\frac{\partial}{\partial z^\mu}\mathcal W_{\mathcal C}(x,y)=ig\,\mathcal W_{\mathcal C}(x,s)\,A_\alpha(s)\,\frac{\partial s^\alpha}{\partial z^\mu}\,\mathcal W_{\mathcal C}(s,y)\Big|_{s=y}^{s=x}+ig\int_y ^x\mathcal W_{\mathcal C}(x,s)\,G_{\alpha\beta}(s)\,\mathcal W_{\mathcal C}(s,y)\,\frac{\partial s^\alpha}{\partial z^\mu}\,\ud s^\beta,\label{derivativeW}
\eeq
and the fact that the inverse of the Wilson line is simply given by $\mathcal W^{-1}_{\mathcal C}(x,y)=\mathcal W_{\mathcal C}(y,x)$, the pure-gauge and physical fields can be expressed as \cite{Lorce:2012ce}
\beq
\begin{split}\label{Apurephysx}
A^\text{pure}_\mu(x)&=\frac{i}{g}\,\mathcal W_{\mathcal C}(x,x_0)\,\frac{\partial}{\partial x^\mu}\mathcal W_{\mathcal C}(x_0,x),\\
A^\text{phys}_\mu(x)&=-\int_{x_0}^x \mathcal W_{\mathcal C}(x,s)\,G_{\alpha\beta}(s)\,\mathcal W_{\mathcal C}(s,x)\,\frac{\partial s^\alpha}{\partial x^\mu}\,\ud s^\beta.
\end{split}
\eeq
From the gauge transformation law of the Wilson line
\beq\label{Wilsong}
\mathcal W_{\mathcal C}(x,y)\mapsto\tilde{\mathcal W}_{\mathcal C}(x,y)=U(x)\mathcal W_{\mathcal C}(x,y)U^{-1}(y),
\eeq
it is straightforward to check that the pure-gauge and physical fields given in Eq. \eqref{Apurephysx} transform in accordance with Eq. \eqref{nonabGT}. Moreover, different choices of contour $\mathcal C$ and reference point $x_0$ are easily related to each other. Indeed, denoting the fields obtained with a different contour $\bar{\mathcal C}$ and possibly different reference point $\bar x_0$ with a bar, one can write
\beq\label{stueck3}
\bar U_\pure(x)=U_\pure(x)U^{-1}_S(x),
\eeq
where $U_S(x)=\bar U^{-1}_\pure(\bar x_0)\mathcal W_{\bar{\mathcal C}}(\bar x_0,x)\mathcal W_{\mathcal C}(x,x_0)U_\pure(x_0)$ is obviously unitary and gauge invariant. Clearly, Eq. \eqref{stueck3} has the generic form of a Stueckelberg transformation \eqref{nonabST}.

This explicit example illustrates once more that the Chen \emph{et al.} approach leads to non-local expressions in terms of the full gauge potential $A_\mu$. It is  because of this non-locality that one can define infinitely many gauge-invariant expressions, as nicely expressed by Bashinsky and Jaffe in Ref. \cite{Bashinsky:1998if}: ``Different ways of gluon-field gauge fixing predetermine different decompositions of the coupled quark-gluon fields into quark and gluon degrees of freedom. Similarly, one can generalize a gauge-variant non-local operator [\ldots] to more than one gauge-invariant expressions, raising the problem of deciding which is the \emph{true} one.''

\subsubsection{Generic Lorentz transformations\label{secVC2}}

Let us now discuss the Lorentz transformation properties in non-abelian gauge theories. It is usually implicitly assumed that the quark field has the same Lorentz transformation law in both free and interacting cases
\beq
\psi(x)\mapsto\psi'(x')=S[\Lambda]\psi(x),
\eeq
where $S[\Lambda]$ is the standard matrix representing the Lorentz transformation in Dirac space. Similarly, the ordinary covariant derivative of the quark field is assumed to transform according to
\beq
D_\mu\psi(x)\mapsto D'_\mu\psi'(x')=\Lambda_\mu^{\phantom{\mu}\nu}S[\Lambda]D_\nu\psi(x).
\eeq
One deduces immediately that the gauge potential transforms as a Lorentz four-vector
\beq
A_\mu(x)\mapsto A'_\mu(x')=\Lambda\lind{\mu}{\nu}A_\nu(x).
\eeq

As explained in section \ref{secIVD2}, the quark field in a gauge theory transforms as a Dirac spinor only up to a gauge transformations. The generic Lorentz transformation law of the quark and gluon fields are therefore
\beq
\begin{split}
\psi(x)&\mapsto\psi'(x')=U_\Lambda(x)S[\Lambda]\psi(x),\\
A_\mu(x)&\mapsto A'_\mu(x')=\Lambda\lind{\mu}{\nu}U_\Lambda(x)\left[A_\nu(x)+\tfrac{i}{g}\,\partial_\nu\right]U_\Lambda^{-1}(x),
\end{split}
\eeq
where $U_\Lambda$ is some $N_c\times N_c$ unitary matrix in color space associated with the Lorentz transformation $\Lambda$. Clearly, the gauge potential $A_\mu$ generally transforms as a \emph{connection}. Indeed, writing explicitly the internal indices
\beq
-igA^a_{\mu b}\mapsto -igA'^a_{\mu b}=(\Lambda^{-1})\uind{\nu}{\mu}(U^{-1}_\Lambda)\uind{d}{b}(U_\Lambda)\uind{a}{c}(-igA^c_{\nu d})+(U_\Lambda)\uind{a}{e}\left[\partial'_\mu(U^{-1}_\Lambda)\uind{e}{b}\right]
\eeq
reveals exactly the same structure as the Lorentz transformation law of the Christoffel symbols in General Relativity
\beq
\Gamma^\lambda_{\mu\nu}\mapsto \Gamma'^\lambda_{\mu\nu}=\frac{\partial x^\alpha}{\partial x'^\mu}\,\frac{\partial x^\beta}{\partial x'^\nu}\,\frac{\partial x'^\lambda}{\partial x^\gamma}\,\Gamma^\gamma_{\alpha\beta}+\frac{\partial x'^\lambda}{\partial x^\rho}\,\frac{\partial^2x^\rho}{\partial x'^\mu\partial x'^\nu},
\eeq
see \emph{e.g.} Ref. \cite{Soper}. The corresponding field-strength tensor generally transforms as
\beq
G_{\mu\nu}(x)\mapsto G'_{\mu\nu}(x')=\Lambda\lind{\mu}{\alpha}\Lambda\lind{\nu}{\beta}U_\Lambda(x)G_{\alpha\beta}(x)U^{-1}_\Lambda,
\eeq
or more explicitly
\beq
G^a_{\mu\nu b}\mapsto G'^a_{\mu\nu b}=(\Lambda^{-1})\uind{\alpha}{\mu}(\Lambda^{-1})\uind{\beta}{\nu}(U^{-1}_\Lambda)\uind{d}{b}(U_\Lambda)\uind{a}{c}G^c_{\alpha\beta d},
\eeq
which has the same structure as the Lorentz transformation law of the Riemann curvature tensor
\beq
R^\lambda_{\phantom{\lambda}\mu\nu\rho}\mapsto R'^\lambda_{\phantom{\lambda}\mu\nu\rho}=\frac{\partial x^\alpha}{\partial x'^\mu}\,\frac{\partial x^\beta}{\partial x'^\nu}\,\frac{\partial x^\delta}{\partial x'^\rho}\,\frac{\partial x'^\lambda}{\partial x^\gamma}\,R^\gamma_{\phantom{\gamma}\alpha\beta\delta}.
\eeq

The generic Lorentz transformation laws for the non-abelian pure-gauge and physical fields compatible with the Chen \emph{et al.} approach are given by \cite{Lorce:2012rr}
\beq
\begin{split}
U_\pure(x)&\mapsto U'_\pure(x')=U_\Lambda(x)U_\pure(x)U^{\phys,-1}_\Lambda(x),\\
A^\mu_\pure(x)&\mapsto A'^\mu_\pure(x')=\Lambda\uind{\mu}{\nu}U_\Lambda(x)\left[A^\nu_\pure(x)+\tfrac{i}{g}\,\partial^\nu+C^\nu_\Lambda(x)\right]U^{-1}_\Lambda(x),\\
A^\mu_\phys(x)&\mapsto A'^\mu_\phys(x')=\Lambda\uind{\mu}{\nu}U_\Lambda(x)\left[A^\nu_\phys(x)-C^\nu_\Lambda(x)\right]U^{-1}_\Lambda(x),
\end{split}
\eeq
where $C^\nu_\Lambda(x)=\frac{i}{g}\,U_\pure(x)U^{\phys,-1}_\Lambda(x)\left[\partial^\nu U^\phys_\Lambda(x)\right]U^{-1}_\pure(x)$. Similarly to the function $\omega^\phys_\Lambda(x)$ in Eq. \eqref{physabL}, the unitary matrix $U^\phys_\Lambda(x)$ allows one to preserve the condition defining the physical field under Lorentz transformations. In the abelian case, it simply reads $U^\phys_\Lambda(x)=e^{ig\omega^\phys_\Lambda(x)}$.

As in the abelian case, one has the freedom to choose the unitary function $U_\Lambda$ thanks to the gauge symmetry. Once again, two options are particularly interesting:
\begin{enumerate}
\item To maintain Lorentz covariance explicitly, one may choose to work with $U_\Lambda=\mathds{1}$, \emph{i.e.} with $A^\mu$ transforming as a Lorentz four-vector. In this case, the pure-gauge field does not transform as a Lorentz four-vector, since it has to compensate for the non-tensorial nature of the physical field $C^\nu_\Lambda(x)$.   Again, with this choice, Lorentz transformations will generally mix physical and gauge degrees of freedom. For example, suppose that in a given Lorentz frame one has chosen to work in the natural gauge, \emph{i.e.} with $U_\pure=\mathds 1$ and consequently $A^\mu_\pure=0$. After a Lorentz transformation, the pure-gauge part becomes $A'^\mu_\pure=\tfrac{i}{g}\,\Lambda\uind{\mu}{\nu}U^{\phys,-1}_\Lambda\partial^\nu U^\phys_\Lambda$. One therefore needs to perform an additional gauge transformation with $U=U^{\phys,-1}_\Lambda$ in order to recover a vanishing pure-gauge part $\tilde A'^\mu_\pure=0$ in the new Lorentz frame.
\item To avoid a mixing between physical and gauge degrees of freedom under Lorentz transformations, one may choose to work with $U_\Lambda$ satisfying the condition $(\partial_\mu U_\Lambda)U_\pure U^{\phys,-1}_\Lambda+U_\Lambda U_\pure(\partial_\mu U^{\phys,-1}_\Lambda)=0$. In this case, the gauge potential transforms in the same way as its physical part, while the pure-gauge part undergoes a simple rotation in the internal space on top of a four-vector transformation in the physical space $A'^\mu_\pure=\Lambda\uind{\mu}{\nu} U_\Lambda A^\nu_\pure U^{-1}_\Lambda$. Consequently, the physical polarizations remain physical under Lorentz transformations, but are generally rotated in the internal space. This internal rotation comes from the fact that different observers may not agree on the color of a quark. Note that when the observers manage to agree on what is ``red'', ``green'' and ``blue'', the matrix $U^\phys_\Lambda$ is expected to reduce to a simple phase factor like in the abelian case.
\end{enumerate}

The generic conclusion in non-abelian gauge theories is therefore the same as in the abelian ones:
\beq
\begin{split}
(A_\pure)'^\mu(x')&=(A'^\mu)_\pure(x'),\\
(A_\phys)'^\mu(x')&=(A'^\mu)_\phys(x').
\end{split}
\eeq
The Chen \emph{et al.} approach is consistent with Lorentz symmetry, and one can write without ambiguity $A_\pure'^\mu(x')$ and $A_\phys'^\mu(x')$. Note that the matrix $U^\phys_\Lambda$ can be determined by the Noether's theorem in the gauge-covariant canonical formalism \cite{Lorce:2013gxa}. It is such that the Stueckelberg-fixing constraint transforms covariantly under the canonical Lorentz transformations. To the best of our knowledge, this has never been checked explicitly in the non-abelian case so far, and would be an interesting non-trivial result.

\subsection{Equivalence with the Dirac formalism based on gauge-invariant variables\label{secVD}}

We discuss in this section the relation between the Chen \emph{et al.} approach and a popular approach based on gauge-invariant variables. While they appear rather different at a first sight, they turn out to be related by a unitary transformation, and are in this sense mathematically equivalent.

\subsubsection{Generalized Dirac variables\label{secVD1}}

Dirac soon realized that one of the main obstacles in the quantization of a gauge theory is the gauge dependence of the fields. He therefore introduced new variables constructed by adjoining phase factors to the dynamical fields and referred to as \emph{Dirac variables} \cite{Dirac:1955uv}. These new variables are non-local functionals of $\psi$ and $A_\mu$, but have the advantage of being gauge invariant by construction. They have been rediscovered and generalized several times under different names and in different contexts \cite{DeWitt:1962mg,Mandelstam:1962mi,Mandelstam:1968hz,Mandelstam:1968ud,BialynickiBirula:1963,Steinmann:1983ar,Steinmann:1985id,Skachkov:1985cz,Haagensen:1997pi,Horan:1998im,Masson:2010vx,Fournel:2012cr,Chen:2012vg}. We refer to \cite{Pervushin:2001kq} for a review of the subject. For a comparison between the quantization based on Dirac variables and the more standard Faddeev-Popov approach, see Ref. \cite{Lantsman:2006ry}.

As already mentioned, the unitary matrix $U_\pure$ plays the role of a non-dynamical background field. It can be seen as a non-local functional of the gauge potential which has, in many cases, the form of a Wilson line \eqref{solUpure}. More importantly, it has the same gauge transformation law as the $\psi$ field. In the spirit of the Dirac approach, one can use $U_\pure$ as a phase factor and construct the following new variables \cite{Chen:2012vg,Lorce:2012rr,Lorce:2013gxa}
\beq\label{generalizedDiracvariable}
\begin{split}
\hat\psi(x)&\equiv U^{-1}_\pure(x)\psi(x),\\
\hat A_\mu(x)&\equiv U^{-1}_\pure(x)\left[A_\mu(x)+\tfrac{i}{g}\,\partial_\mu\right]U_\pure(x).
\end{split}
\eeq
Clearly, following Eq.~\eqref{nonabUpureGT}, these variables are \emph{by construction} gauge invariant and represent the generic form of any kind of Dirac variable. In this sense, these new variables are referred to as \emph{generalized Dirac variables}. Sometimes, the gauge-invariant fields $\hat\psi$ and $\hat A_\mu$ are interpreted as \emph{dressed} quark and gluon fields. Accordingly, the matrix $U_\pure$ is called the dressing field. From a geometrical point of view, $U_\pure$ simply specifies a reference configuration in the internal space \cite{Lorce:2012rr}. The gauge-invariant fields $\hat\psi$ and $\hat A_\mu$ then represent ``physical'' deviations from this reference configuration. This is supported by the following observations
\beq\label{Diracvar}
\begin{split}
\hat A^\pure_\mu(x)&\equiv U^{-1}_\pure(x)\left[A^\pure_\mu(x)+\tfrac{i}{g}\,\partial_\mu\right]U_\pure(x)\\
&=0,\\
\hat A^\phys_\mu(x)&\equiv U^{-1}_\pure(x)A^\phys_\mu(x)U_\pure(x)\\
&=\hat A_\mu(x),\\
\hat G_{\mu\nu}(x)&\equiv U^{-1}_\pure(x)G_{\mu\nu}(x)U_\pure(x)\\
&=\partial_\mu\hat A_\nu(x)-\partial_\nu\hat A_\mu(x)-ig[\hat A_\mu(x),\hat  A_\nu(x)].
\end{split}
\eeq

We stress that, despite appearances, Eq.~\eqref{generalizedDiracvariable} does not represent a gauge transformation, but is a \emph{definition} of new variables in any gauge. This should not be confused with the fact that after a gauge transformation with $U(x)=U^{-1}_\pure(x)$, one has
\beq
\begin{split}
\tilde\psi(x)&=\hat\psi(x),\\
\tilde A_\mu(x)&=\hat A_\mu(x).
\end{split}
\eeq
The generalized Dirac variables $\hat\psi$ and $\hat A_\mu$, being defined in any gauge, can therefore be interpreted as the natural gauge-invariant extensions of the original gauge non-invariant fields $\psi$ and $A_\mu$.

\subsubsection{The Dirac gauge-invariant approach\label{secVD2}}

In the Dirac gauge-invariant approach, one first rewrites the standard QCD Lagrangian
\beq
\uL_\text{QCD}(x)=f[\psi(x),\partial_\mu\psi(x),A_\nu(x),\partial_\mu A_\nu(x)]
\eeq
in terms of the new gauge-invariant variables $\hat\psi$ and $\hat A_\mu$
\beq
\uL_\text{QCD}(x)=f[\hat\psi(x),\partial_\mu\hat\psi(x),\hat A_\nu(x),\partial_\mu \hat A_\nu(x)].
\eeq
This is of course always possible thanks to the gauge symmetry and the fact that the generalized Dirac variables are the gauge-invariant extensions of the original quark and gluon fields.

Since the Lagrangian now involves only gauge-invariant variables, one can simply apply the standard canonical formalism to derive the QCD equations of motion
\begin{align}
(i\,\uslash\!\hat D-m)\hat\psi&=0,\\
2(\hat{\mathcal D}_\alpha \hat G^{\alpha\mu})\uind{a}{b}&=-g\hat\barpsi_b\gamma^\mu\hat\psi^a,
\end{align}
where the gauge-invariant derivatives are given by $\hat D_\mu=\partial_\mu-ig\hat A_\mu$ and $\hat{\mathcal D}_\mu=\partial_\mu-ig[\hat A_\mu,\quad]$. From Noether's theorem, one obtains the following gauge-invariant expressions for the generators of Poincar\'e transformations
\begin{align}
\hat T^{\mu\nu}&=\tfrac{1}{2}\,\hat\barpsi \gamma^\mu i\LRpartial^\nu\hat\psi-2\,\uTr[\hat G^{\mu\alpha}\partial^\nu \hat A_\alpha]-g^{\mu\nu}\uL_\text{QCD},\\
\hat M^{\mu\nu\rho}&=\tfrac{1}{2}\,\epsilon^{\mu\nu\rho\sigma}\,\hat\barpsi\gamma_\sigma\gamma_5\hat\psi+\tfrac{1}{2}\,\hat\barpsi \gamma^\mu x^{[\nu}i\LRpartial^{\rho]}\hat\psi-2\,\uTr[\hat G^{\mu[\nu} \hat A^{\rho]}]-2\,\uTr[\hat G^{\mu\alpha} x^{[\nu} \partial^{\rho]}\hat A_\alpha]-x^{[\nu}g^{\rho]\mu}\uL_\text{QCD}.
\end{align}
They are naturally split into quark and gluon contributions as follows
\beq\label{physdec}
\begin{aligned}
\hat T^{\mu\nu}_{q,\text{M}}&=\tfrac{1}{2}\,\hat\barpsi\gamma^\mu i\LRpartial^\nu\hat\psi,&\hat T^{\mu\nu}_{G,\text{M}}&=-2\,\uTr[\hat G^{\mu\alpha}\partial^\nu \hat A_\alpha],\\
\hat T^{\mu\nu}_{q,\text{E}}&=-g^{\mu\nu}\hat\uL_\text{D},&\hat T^{\mu\nu}_{G,\text{E}}&=-g^{\mu\nu}(\hat\uL_\text{YM}+\hat\uL_\text{int}),\\
\hat M^{\mu\nu\rho}_{q,\text{spin}}&=\tfrac{1}{2}\,\epsilon^{\mu\nu\rho\sigma}\,\hat\barpsi\gamma_\sigma\gamma_5\hat\psi,&\hat M^{\mu\nu\rho}_{G,\text{spin}}&=-2\,\uTr[\hat G^{\mu[\nu} \hat A^{\rho]}],\\
\hat M^{\mu\nu\rho}_{q,\text{OAM}}&=\tfrac{1}{2}\,\hat\barpsi \gamma^\mu x^{[\nu}i\LRpartial^{\rho]}\hat\psi,&\hat M^{\mu\nu\rho}_{G,\text{OAM}}&=-2\,\uTr[\hat G^{\mu\alpha} x^{[\nu} \partial^{\rho]}\hat A_\alpha],\\
\hat M^{\mu\nu\rho}_{q,\text{boost}}&=-x^{[\nu}g^{\rho]\mu}\hat\uL_\text{D},&\hat M^{\mu\nu\rho}_{G,\text{boost}}&=-x^{[\nu}g^{\rho]\mu}(\hat\uL_\text{YM}+\hat\uL_\text{int}),
\end{aligned}
\eeq
where
\beq
\begin{split}
\hat\uL_\text{D}&=\hat\barpsi(\tfrac{i}{2}\,\uslash\!\LRpartial-m)\hat\psi,\\
\hat\uL_\text{YM}&=-\tfrac{1}{2}\,\uTr[\hat G^{\mu\nu}\hat G_{\mu\nu}],\\
\hat\uL_\text{int}&=g\hat\barpsi\,\uslash\!\hat A\hat\psi,
\end{split}
\eeq
which have exactly the structure of the canonical expressions, with $\psi$ and $A^\mu$ replaced by the gauge-invariant fields $\hat\psi$ and $\hat A_\mu$, respectively. In particular, the corresponding gauge-invariant linear and angular momentum operators are, respectively, generators of translations and rotations for the gauge-invariant quark and gluon fields, \emph{i.e.} satisfy the following equal-time commutation relations
\beq\label{CRphys}
\begin{aligned}
\, [\hat P^i_q,\hat\psi]&=-\tfrac{1}{i}\nabla^i\hat\psi,&[\hat P^i_G,\hat A^j]&=-\tfrac{1}{i}\nabla^i\hat A^j,\\
[\hat S^i_q,\hat\psi]&=-\tfrac{1}{2}\Sigma^i\hat\psi,&[\hat S^i_G,\hat A^j]&=-(-i\epsilon^{ijk})\hat A^k,\\
[\hat L^i_q,\hat\psi]&=-(\uvec x\times\tfrac{1}{i}\uvec\nabla)^i\hat\psi,&[\hat L^i_G,\hat A^j]&=-(\uvec x\times\tfrac{1}{i}\uvec\nabla)^i\hat A^j,
\end{aligned}
\eeq
where
\beq \label{CRphys1}
\begin{aligned}
\hat P^i_q&=\int\ud^3x\,\hat T^{0i}_{q,\text{M}},&\hat P^i_G&=\int\ud^3x\,\hat T^{0 i}_{G,\text{M}},\\
\hat S^i_q&=\tfrac{1}{2}\,\epsilon^{ijk}\int\ud^3x\,\hat M^{0jk}_{q,\text{spin}},&\hat S^i_G&=\tfrac{1}{2}\,\epsilon^{ijk}\int\ud^3x\,\hat M^{0jk}_{G,\text{spin}},\\
\hat L^i_q&=\tfrac{1}{2}\,\epsilon^{ijk}\int\ud^3x\,\hat M^{0 jk}_{q,\text{OAM}},&\hat L^i_G&=\tfrac{1}{2}\,\epsilon^{ijk}\int\ud^3x\,\hat M^{0jk}_{G,\text{OAM}}.
\end{aligned}
\eeq
One sees that choosing gauge-invariant fields $\hat\psi$ and $\hat A_\mu$ as the canonical variables leads naturally to gauge-invariant canonical linear and angular momentum operators. The rule of thumb is particularly simple: it suffices to replace in every standard canonical expression the gauge non-invariant fields $\psi$ and $A_\mu$ by the corresponding generalized Dirac variable $\hat\psi$ and $\hat A_\mu$, in order to obtain the gauge-invariant expressions. Interestingly, Chen \cite{Chen:2012vg} rediscovered Dirac variables precisely while trying to derive gauge-invariant equations of the type  \eqref{CRphys} and \eqref{CRphys1}.

\subsubsection{Equivalence with the Chen \emph{et al.} approach\label{secVD3}}

The decomposition \eqref{physdec} mimics perfectly the Jaffe-Manohar decomposition \eqref{JMdec}, except that the fields involved are now gauge invariant. As observed by Chen \cite{Chen:2012vg} and further explored by Lorc\'e \cite{Lorce:2012rr,Lorce:2013gxa}, the decomposition \eqref{physdec} is just the gauge-invariant canonical decomposition \eqref{gicdec} written in terms of the gauge-invariant variables $\hat \psi$ and $\hat A_\mu$
\beq
\begin{aligned}
\hat T^{\mu\nu}&=T^{\mu\nu}_\text{gic}\Big|_{\begin{subarray}{l} \quad\,\,\,\psi\mapsto U^{-1}_\pure\hat\psi\\
A^\pure_\mu\mapsto\tfrac{i}{g}\,U_\pure\partial_\mu U^{-1}_\pure\\
A^\phys_\mu\mapsto U_\pure\hat A_\mu U^{-1}_\pure\end{subarray}},&\qquad
\hat M^{\mu\nu\rho}&=M^{\mu\nu\rho}_\text{gic}\Big|_{\begin{subarray}{l} \quad\,\,\,\psi\mapsto U^{-1}_\pure\hat\psi\\
A^\pure_\mu\mapsto\tfrac{i}{g}\,U_\pure\partial_\mu U^{-1}_\pure\\
A^\phys_\mu\mapsto U_\pure\hat A_\mu U^{-1}_\pure\end{subarray}}.
\end{aligned}
\eeq
Indeed, one can easily see that
\beq
\begin{split}
\partial_\mu\hat\psi&=U^{-1}_\text{pure}D^\text{pure}_\mu\psi,\\
\partial_\mu\hat A_\nu&=U^{-1}_\pure[\mathcal D^\pure_\mu A^\phys_\nu]U_\pure.
\end{split}
\eeq
The commutation relations \eqref{CRphys} can therefore be rewritten as
\beq
\begin{aligned}
\,[P^{q,i}_\text{gic},\psi]&=-iD^i_\pure\psi,&[P^{G,i}_\text{gic},A^j_\phys]&=-i\mathcal D^i_\pure A^j_\phys,\\
[S^{q,i}_\text{gic},\psi]&=-\tfrac{1}{2}\Sigma^i\psi,&[S^{G,i}_\text{gic},A^j_\phys]&=-(-i\epsilon^{ijk})A^k_\phys,\\
[L^{q,i}_\text{gic},\psi]&=-(\uvec x\times i\uvec D_\pure)^i\psi,&[L^{G,i}_\text{gic},A^j_\phys]&=-(\uvec x\times i\uvec{\mathcal D}_\pure)^iA^j_\phys,
\end{aligned}
\eeq
and are consistent with the gauge-covariant canonical formalism based on the Chen  \emph{et al.} approach \cite{Lorce:2013gxa}.

Clearly, the Chen \emph{et al.} and Dirac gauge-invariant approaches are mathematically equivalent, since they are simply related by a unitary transformation of the variables. This also means that the issue of uniqueness raised by the Stueckelberg symmetry affects the Dirac approach as well. Indeed, even if the generalized Dirac variables are gauge invariant, they change under the Stueckelberg transformations\footnote{Note that one has in particular $\hat A^\pure_\mu(x)=\hat{\bar A}^\pure_\mu(x)=0$ owing to Eq. \eqref{Diracvar}.}
\beq
\begin{split}
\hat\psi(x)&\mapsto\hat{\bar\psi}(x)=U_S(x)\hat\psi(x),\\
\hat A_\mu(x)&\mapsto\hat{\bar A}_\mu(x)=U_S(x)\left[\hat A_\mu(x)+\tfrac{i}{g}\,\partial_\mu\right]U^{-1}_S(x),\\
\hat A^\pure_\mu(x)&\mapsto\hat{\bar A}^\pure_\mu(x)=U_S(x)\hat A^\pure_\mu(x)U^{-1}_S(x),\\
\hat A^\phys_\mu(x)&\mapsto\hat{\bar A}^\phys_\mu(x)=U_S(x)\left[\hat A^\phys_\mu(x)+\tfrac{i}{g}\,\partial_\mu\right]U^{-1}_S(x).
\end{split}
\eeq
There is exactly the same freedom in defining precisely the generalized Dirac variables $\hat\psi$ and $\hat A_\mu$ as in defining a precise split $A_\mu=A^\pure_\mu+A^\phys_\mu$. The existence of an entire class of composite gauge-invariant fields was already pointed out by Dirac and Steinmann \cite{Dirac:1955uv,Steinmann:1983ar,Steinmann:1985id}. Note in particular that the Dirac gauge-invariant formulation of QED \cite{Dirac:1955uv} is equivalent to the original three-dimensional Chen \emph{et al.} approach based on the Coulomb constraint $\uvec\nabla\cdot\uvec A_\phys=0$ \cite{Chen:2008ag}, since both make use of the phase factor
\beq
 U_\text{pure}(x)=e^{ie\frac{\uvec\nabla\cdot\uvec A(x)}{\uvec\nabla^2}},
\eeq
which is the dressing factor associated with the Coulomb GIE. By construction, they give the same results as the standard approach to QED restricted to the Coulomb gauge. For the contour-based gauge-invariant extensions defined by Eq. \eqref{contourGIE}, the explicit non-local expressions for the generalized Dirac variables are
\beq
\begin{split}
\hat\psi(x)&=\mathcal W_{\mathcal C}(x_0,x)\psi(x),\\
\hat A_\mu(x)&=-\int_{x_0}^x \hat G_{\alpha\beta}(s)\,\frac{\partial s^\alpha}{\partial x^\mu}\,\ud s^\beta\\
&=-\int_{x_0}^x\mathcal W_{\mathcal C}(x_0,s)G_{\alpha\beta}(s)\mathcal W_{\mathcal C}(s,x_0)\,\frac{\partial s^\alpha}{\partial x^\mu}\,\ud s^\beta,
\end{split}
\eeq
with $\mathcal W_{\mathcal C}(b,a)$ given by Eq. \eqref{Wilsonline} and where we have set $U_\pure(x_0)=\mathds 1$ for simplicity thanks to the invariance under global (\emph{i.e.} $x$-independent) rotations in internal space.

\section{Angular momentum sum rules and relations\label{secVI}}

It is of great interest to try to derive sum rules, which follow rigorously from QCD, relating experimentally measurable quantities, because checking  such sum rules then provides a searching test of the theory upon which the derivation is based. Sometimes it is not possible to measure every term in a sum rule. Strictly speaking such a relation should then not be called a sum rule, but it may still be of value in checking the validity of model calculations of some of the terms in the relation. Angular momentum sum rules, relating the angular momentum of a hadron, usually a nucleon\footnote{There are some results on deuterons, see  Ref. \cite{Taneja:2011sy}}, to the angular momentum of its constituents, are of particular interest at  present and much effort, both theoretical and experimental, is being devoted to studying them.

\subsection{General overview\label{secVIA}}

There are now five angular momentum  relations or sum rules for the nucleon in the literature: the Jaffe-Manohar (JM) relation for a longitudinally polarized nucleon \cite{Jaffe:1989jz}, and the Bakker-Leader-Trueman (BLT) result for the case of transverse polarization \cite{Bakker:2004ib}; the Ji relation for longitudinal polarization \cite{Ji:1997pf, Ji:1996ek, Ji:1996nm}, and the Leader result for transverse polarization \cite{Leader:2011cr}, both the latter involving generalized parton distributions; and a new sum rule due to Ji, Xiong and Yuan dealing with the transverse component of the Pauli-Lubanski vector \cite{Ji:2012vj}. In addition to these, Harindranath and Kundu  \cite{Harindranath:1998ve} have shown how the JM sum rule can be derived from a study of the light-front rotation operator about the $OZ$ axis, and Harindranath, Kundu and Ratabole \cite{Harindranath:2001rc, Harindranath:1999ve} have discussed an interesting  sum rule based on the transverse light-front spin operators. Since these are interaction dependent, the analysis involves a perturbative QCD treatment and we shall not discuss this paper here.

We shall discuss these various relations and examine their precise interpretation in the light of the ``angular momentum controversy''. In particular, we shall show that the claim of Ji, Xiong and Yuan (JXY) that their Pauli-Lubanski relation is frame or energy independent is incorrect, and that they have discarded an energy-dependent term in their expression. A different point of view is taken in Ref. \cite{Harindranath:2013goa} and this will be discussed later.

Generally speaking, the above sum rules fall into two classes. In the JM and BLT relations, one obtains an expression for the \emph{nucleon} matrix elements of the angular momentum operators and then substitutes a Fock expansion for the nucleon state in terms of quarks and gluons. In the Ji and Leader relations, one expresses the nucleon matrix elements of the angular momentum operators in terms of nucleon matrix elements of the energy-momentum density, and then relates these matrix elements to quark and gluon generalized parton  distributions (GPDs). The JXY case is somewhat different, as will become clear later.

\subsection{Expectation values of operators\label{secVIB}}

In order to derive sum rules, we will need to evaluate the expectation values of the momentum and angular momentum operators. Using the conventional relativistic state normalization
\beq \label{norm}
\la  P'  |  P  \ra = 2 P^0\, (2\pi)^3 \, \delta ^{(3)}(\uvec P' - \uvec P),
\eeq
the expectation value of any operator $\hat{O}$ is given in terms of its forward matrix elements as
\beq \label{expct}
\la \la   P  | \hat{O}  |  P  \ra \ra = \frac{\la   P  | \hat{O} |  P  \ra}{2 P^0\, (2\pi)^3 \, \delta^{(3)} (\uvec 0 )}.
\eeq
For the total momentum and angular momentum of the system, in this case the nucleon, the operators are time-independent, so that their expectation values are simply numerical functions of the momentum and spin of the nucleon. The situation appears, at first sight, to be completely different when we separate, say, the momentum $P^\mu$ of the nucleon into contributions from quarks and gluons. Thus the oft written equation
\beq \label{Eq:split}
P^\mu=\sum_qP^\mu_q +P^\mu_G
\eeq
as it stands, is somewhat misleading, and should be written
\beq \label{Eq:Spli}
P^\mu=\sum_qP^\mu_q(t) +P^\mu_G(t)
\eeq
to reflect the fact that the  quark and gluon momentum operators are not separately conserved, since the quarks and gluons exchange momentum as a result of their interaction.

Nonetheless, as we shall now show, the  matrix elements between \emph{states of the same energy}, and thus the expectation values, of $P^\mu_q(t)$ and $P^\mu_G(t)$ are time-independent, so that it \emph{does} make sense to talk about \emph{e.g.} the quark contribution to the nucleon momentum. To see this for the momentum $P_q^\mu(t)$, which is expressed as a spatial integral of the quark part of the momentum density $T^{0\mu}_q(x)$, let
\beq
|\psi \rangle = \int \ud^3 p' \, \psi (\uvec p') \, |\uvec p'\rangle \quad \textrm{and} \quad |\phi\rangle = \int \ud^3 p \, \phi (\uvec p) \, |\uvec p\rangle
\eeq
be arbitrary states of a free particle of mass $m$, with the same energy, so that
\beq
(p^0)^2= \uvec p^2 + m^2 = (p'^0)^2= \uvec p'^2 + m^2.
\eeq
Then
\beqy
\langle\psi|P_q^\mu(t)  |\phi\rangle &= &\int \ud^3p'\,\ud^3p \,\ud^3x\,\psi^*(\uvec p')\,\phi(\uvec p) \,\langle\uvec p'|T_q^{0\mu}(x)  |\uvec p\rangle  \nn \\
&=& \int \ud^3p'\,\ud^3p \,\ud^3x\,\psi^*(\uvec p')\,\phi(\uvec p)\,e^{-i(\uvec p'-\uvec p)\cdot\uvec x} \, e^{i(p'^0-p^0)t}\, \langle\uvec p'| T_q^{0\mu}(0) |\uvec p\rangle \nn \\
&=& (2\pi)^3 \int \ud^3p'\,\ud^3p \,\psi^*(\uvec p')\,\phi(\uvec p) \,\delta^{(3)}(\uvec p' - \uvec p)\, e^{i(p'^0-p^0)t} \,\langle\uvec p'| T_q^{0\mu}(0)  |\uvec p\rangle \nn \\
&=& (2\pi)^3 \int \ud^3p \,\psi^*(\uvec p)\,\phi(\uvec p) \, \langle\uvec p|T_q^{0\mu}(0)  |\uvec p\rangle
\eeqy
which is independent of time because $p'^0=p^0$.

A similar, though more complicated, argument shows that the matrix elements of the angular momentum operator $\uvec J_{q,G}(t)$, between states of equal energy, are also time-independent.

\subsection{Relation between the matrix elements of angular momentum and energy-momentum density\label{secVIC}}

One of the key steps in deriving angular momentum sum rules involves relating the matrix elements of the  angular momentum operator to those of the energy-momentum density, and this turns out to be a very subtle issue, with incorrect results appearing in several papers in the literature. Whether one is considering the orbital part of the canonical version $M^{\nu\rho}$ or the Belinfante version $M^{\nu\rho}_\text{Bel}$ of the angular momentum operator, one is faced with the problem of evaluating the expectation value of a compound operator of the form $x^\nu T^{\mu\rho}-x^\rho T^{\mu\nu}$. For example consider Eq.~\eqref{MBorb}
\beqy
\la P, \mathscr{S} | M^{\nu\rho}_\text{Bel}  |  P, \mathscr{S}\ra
&=&\int \ud^3x \,\la  P,\mathscr{S} |M^{0\nu\rho}_\text{Bel}(x)  | P,\mathscr{S}\ra  \nn \\
&=&\int \ud^3x \, \la P, \mathscr{S} |x^\nu T^{0\rho}_\text{Bel}(x) -x^\rho T^{0\nu}_\text{Bel}(x)  | P, \mathscr{S} \ra .
\eeqy
Using the fact that $T^{\mu\nu}_\text{Bel}(x)$ is a local operator, this becomes
\beqy
  \la P, \mathscr{S} | M^{\nu\rho}_\text{Bel}  |  P, \mathscr{S}\ra  & =& \int \ud^3x \, x^\nu \,\langle P ,\mathscr{S}  | e^{i P \cdot x} \,T^{0\rho}_\text{Bel} (0)\, e^{-iP \cdot x} | P ,\mathscr{S} \rangle- (\nu \leftrightarrow  \rho) \nonumber \\
& =& \int \ud^3x \, x^\nu  \,\langle P, \mathscr{S} | T^{0\rho}_\text{Bel}(0) | P , \mathscr{S}\rangle - (\nu \leftrightarrow \rho).\label{ambiguous}
\eeqy
Now the matrix element in Eq.~\eqref{ambiguous} is independent of $x$, so that the integral  $\int\ud^3x \, x^\nu $ is totally ambiguous, being either infinite or, by symmetry, zero.

This kind of problem is  typical of what happens, even in non-relativistic Quantum Mechanics, when one uses plane-wave states, and the solution is well known. One must use normalized  wave packets which are  superpositions of physical plane-wave states. A possible example  is a packet with momentum localized around the value $\bar{\uvec P}$
\beq
\label{NRW} |\phi_{\bar{\uvec P}} \ra = \int \ud^3P
\, \phi_{\bar{\uvec P}}(\uvec P) \,|\uvec P \ra = \mathcal N \int \ud^3P\,
e^{-\lambda^2(\uvec P- \bar{\uvec P})^2} \,|\uvec P\ra,
\eeq
where $\mathcal N$ is a normalization constant, and where, at the end, one takes the limit $\lambda \rightarrow \infty$.

The generalization of this to particles with spin is not discussed in non-relativistic Quantum Mechanics and one has to be aware of certain subtleties arising in the relativistic case. For example, for a spin-$1/2$ particle it is often convenient, as above,  to specify its state in the form $| P,  \msc{S}\ra$ where $P^\mu=(\sqrt{\uvec P^2 + M^2}, \uvec P)$ and $\msc{S}^\mu$ is the \emph{covariant} spin vector. It is then incorrect to regard the following as an acceptable physical wave packet centered around the momentum $\bar P$:
\beq \label{badWP}
|\phi_{\bar{P}, \msc{S}} \ra =\mathcal N \,\int \ud^3P \,
\phi_{\bar{P}}(\uvec P) \,|P,\msc{S}\ra
\eeq
for the simple reason that $\msc{S}^\mu$ is a function of $\uvec P$ satisfying
\beq
\label{PS} P\cdot \msc{S}=0,
\eeq
and so cannot be kept fixed if one integrates freely over the momenta in the  superposition. The desire, in certain papers, to use the above form with fixed $\msc{S}^\mu$, stems from the fact that it simplifies calculations, which, however,  are then  incorrect. Failure to recognize this and to respect Eq.~\eqref{PS} has led to errors in the literature \cite{Jaffe:1989jz, Wakamatsu:2010cb}.

 The correct way to build a physical wave packet for a spin-$1/2$ particle was given in Bakker, Leader and Trueman (BLT) \cite{Bakker:2004ib} and utilizes the relation between $\msc{S}^\mu$, normalized to $\msc{S}^2=-M^2$, and the rest-frame spin vector $\uvec s$, namely\footnote{In a relativistic theory, there are several different ways to define the state corresponding to ``spin $\uvec s$'' for a moving particle \cite{Leader:2001gr}. The expression in Eq.~\eqref{S} corresponds actually to the \emph{canonical} covariant spin four-vector associated with the standard  definition of spin in instant form dynamics. The covariant spin four-vector associated with the light-front definition of spin is different \cite{Polyzou:2012ut}. Using the notation in terms of light-front components $a^\mu=[a^+,a^-,\uvec a_\perp]$ with $a^\pm=\tfrac{1}{\sqrt{2}}\,(a^0\pm a^3)$, one has for the light-front covariant spin four-vector
\begin{equation*}
\msc{S}^\mu_\text{LF}= \mathfrak{S}^\mu_\text{LF}(P,\uvec s) \equiv \left[s_zP^+,-s_zP^-+\tfrac{\uvec P_\perp}{P^+}\cdot(M\uvec s_\perp+\uvec P_\perp s_z),M\uvec s_\perp+\uvec P_\perp s_z\right],
\end{equation*}
where $P^-=\tfrac{M^2+\uvec P^2_\perp}{2P^+}$. It is therefore important to specify what definition of spin, for a moving particle, one is using.}
 \beq
\label{S} \msc{S}^\mu= \mathfrak{S}^\mu(P,\uvec s) \equiv\left(\uvec P \cdot \uvec s, M \uvec s +\frac{\uvec P \cdot \uvec s}{P^0 + M}\, \uvec P \right)
\eeq
 to build a wave packet as a superposition of momentum eigenstates with fixed rest-frame spin vector \emph{i.e.}
 \beq
\label{RelWP} | \phi_{\bar{P}, \uvec s} \ra =\mathcal N \,\int \ud^3P \,
\phi_{\bar{P}}(\uvec P) \,|P, \uvec s\ra .
\eeq
However, the use of wave packets to regulate the integral in Eq.~\eqref{ambiguous}, while rigorous, is extremely cumbersome, so we shall now turn to a different approach suggested by Jaffe and Manohar  \cite{Jaffe:1989jz} but, unfortunately, incorrectly used by them. They begin their analysis by considering the non-forward quantity
\beq
\mathcal M^{\mu\nu\rho} (P,k,P',\mathscr{S}) \equiv \int \ud^4 x\,e^{i k\cdot x}\,\langle P', \mathscr{S}|M^{\mu\nu\rho} (x) | P, \mathscr{S}\rangle. \label{JM}
\eeq
One may wonder why a four-dimensional Fourier transform is introduced in dealing with a three-dimensional integral. The reason is that $\mathcal M^{\mu\nu\rho} (P,k,P',\mathscr{S})$ \emph{seems} to transform as a Lorentz tensor, but that is an illusion, because the non-forward matrix element of a tensor operator is not a tensor \cite{Bakker:2004ib}. This misunderstanding is partly responsible for the error in Ref. \cite{Jaffe:1989jz} mentioned above. Note  that strictly speaking, as discussed above, the covariant four-vector $\mathscr{S}$ in the final state cannot be the same as in the initial state, since for a physical nucleon one must have $P \cdot \mathscr{S}= P' \cdot \mathscr{S}' =0 $ leading generally to $ \mathscr{S}\neq \mathscr{S}'$. The correct way to handle Eq.~\eqref{JM} is to specify the same \emph{rest-frame vector} $\uvec s$ in both initial and final states.

Since nothing is gained by using a four-dimensional Fourier transform, we modify the JM approach by rewriting Eq.~\eqref{ambiguous} in the form
\beqy
\la P, \mathscr{S}  |M^{\nu\rho}_\text{Bel} |  P, \mathscr{S} \ra  & =& \int \ud^3x \, x^\nu \,\langle  P, \mathscr{S}  |e^{i P^0 t} \,T^{0\rho}_\text{Bel} (0, \uvec x) \,e^{-iP^0 t} | P, \mathscr{S}  \rangle - (\nu \leftrightarrow  \rho) \nonumber \\
& =& \int \ud^3x \, x^\nu  \,\langle P, \mathscr{S} | T^{0\rho}_\text{Bel}(0, \uvec x) |P , \mathscr{S}  \rangle - (\nu \leftrightarrow \rho),
\eeqy
and then evaluating the RHS \emph{via}\footnote{Of course, this yields the same result as the one obtained from the correctly implemented four-dimensional Fourier transform approach.}
 \beqy \label{MB'}
 \la \, P, \mathscr{S}  | M^{\nu\rho}_\text{Bel}  | P, \mathscr{S} \ra &=& \lim_{\uvec\Delta \rightarrow \uvec 0} \int\ud^3x \, \langle  P + \tfrac{\Delta}{2}, \mathscr{S}_{f}  | x^\nu  T^{0\rho}_\text{Bel}(0, \uvec x)- x^\rho T^{0\nu}_\text{Bel}(0, \uvec x)| P -\tfrac{\Delta}{2}, \mathscr{S}_{i}  \rangle \nn \\
&=&\lim_{\uvec\Delta\rightarrow \uvec 0} \int \ud^3x \, e^{-i\uvec\Delta\cdot\uvec x} \langle  P +\tfrac{\Delta}{2}, \mathscr{S}_f  |  x^\nu  T^{0\rho}_\text{Bel}(0)- x^\rho  T^{0\nu}_\text{Bel}(0)| P - \tfrac{\Delta}{2}, \mathscr{S}_i  \rangle,
\eeqy
where, see Eq.~\eqref{S},
\beq
\mathscr{S}_i^\mu=  \mathfrak{S}^\mu(P-\tfrac{\Delta}{2}, \uvec s), \qquad \mathscr{S}_f^\mu=  \mathfrak{S}^\mu(P+\tfrac{\Delta}{2}, \uvec s).
\eeq
It is crucial to understand that the four components of $\Delta^\mu$ are not independent. This follows from the mass-shell conditions $(P\pm\tfrac{\Delta}{2})^2 = M^2$ which imply
\beq \label{delnon}
P\cdot \Delta =0  \qquad \text{or} \qquad \Delta^0 = \frac{\uvec P\cdot \uvec\Delta}{P^0},
\eeq
so that the limit $\uvec\Delta \rightarrow \uvec 0$ implies that $\Delta^\mu \rightarrow 0$, and also that $ \msc{S}_i \rightarrow \msc{S}$ and $\msc{S}_f \rightarrow \msc{S}$. Note that the physical requirements $(P-\tfrac{\Delta}{2})\cdot \msc{S}_i =0$ and $(P+\tfrac{\Delta}{2})\cdot \msc{S}_f =0$ are satisfied automatically and do not put any further constraints on the components of $\Delta^\mu$. In summary, the essential point is that the three spatial components of $\Delta^\mu$ should be taken as independent variables and so, when differentiating with respect to say $\Delta^j$ ($j=1,2,3$), it must not be forgotten, as has happened in some analyses,  that $\Delta^0$ is a function of $\Delta^j$.

Consider now the expression for the actual angular momentum operators
%\footnote{\trd{We have considered the non-forward element at fixed time $t=0$ where the nucleon position coincides with the origin of axes. For $t\neq 0$, one has to take the motion of the nucleon contained in the extra $e^{i\Delta_0t}$ factor into account. The final result is however the same: one trades $x^j$ for $i\tfrac{\partial}{\partial\Delta^j}$.}}, \emph{i.e.} Eq.~\eqref{MB'} with $\mu=j$ and $\nu=k$
\beqy  \label{AM}
\la P, \mathscr{S}  | M^{jk}_\text{Bel}  |  P, \mathscr{S} \ra &=& \lim_{\uvec\Delta\rightarrow \uvec 0} \int \ud^3x \, e^{-i\uvec\Delta\cdot\uvec x} \langle  P +\tfrac{\Delta}{2}, \mathscr{S}_f  |  x^j  T^{0k}_\text{Bel}(0)- x^k  T^{0j}_\text{Bel}(0)| P - \tfrac{\Delta}{2}, \mathscr{S}_i  \rangle \nn \\
&=&   \lim_{\uvec\Delta\rightarrow \uvec 0} \int \ud^3x \left[i\,\frac{\partial}{\partial \Delta^j}\,e^{-i\uvec\Delta\cdot\uvec x}\right] \langle P +\tfrac{\Delta}{2}, \mathscr{S}_f  |   T^{0k}_{Bel}(0)  | P - \tfrac{\Delta}{2}, \mathscr{S}_i  \rangle  - (j\leftrightarrow k).
\eeqy
We now use the Leibniz product rule to write the derivative of the exponential as the derivative of the entire expression minus the exponential times the derivative of the matrix element. The former, as shown by the wave-packet analysis of BLT, measures  the angular momentum of the wave packet about the origin of coordinates,  and the latter measures   the internal angular momentum of the nucleon. Only the latter is of interest and yields
\beq \label{Jfinal}
\la P, \mathscr{S}  | M^{jk}_\text{Bel}  |  P, \mathscr{S}  \ra  =   \lim_{\uvec\Delta \rightarrow \uvec 0} \int \ud^3x \, e^{-i\uvec\Delta\cdot\uvec x} \,(-i)\,\frac{\partial}{\partial \Delta^j}\langle P + \tfrac{\Delta}{2}, \mathscr{S}_f |  T^{0k}_\text{Bel}(0)  | P -\tfrac{\Delta}{2}, \mathscr{S}_i \rangle    - (j\leftrightarrow k),
\eeq
so that using Eq.~\eqref{expct}
\beq \label{JFF}
\la\la  P, \mathscr{S}  | M^{jk}_\text{Bel}  |  P, \mathscr{S} \ra \ra=\frac{1}{2P^0}\left[-i\, \frac{\partial}{\partial \Delta^j}\langle  P + \tfrac{\Delta}{2}, \mathscr{S}_f |  T^{0k}_\text{Bel}(0)  | P - \tfrac{\Delta}{2}, \mathscr{S}_i  \rangle    - (j\leftrightarrow k)\right]_{\uvec\Delta= \uvec 0}.
\eeq
A similar expression relates the matrix elements of the canonical orbital angular momentum $M^{jk}$ to the matrix elements of the canonical momentum tensor $T^{0k}$. The difference between the canonical and Belinfante cases lies in the different structure of the energy-momentum densities, and was discussed in section \ref{secIII}. Eq.~\eqref{JFF} is a key equation and will be used several times in what follows.

\subsection{Expressions for the total angular momentum in terms of the matrix elements of the energy-momentum density\label{secVID}}

It is clear from Eq.~\eqref{JFF} that we only require an expression for the matrix elements of $T^{\mu\nu}_\text{Bel}$ accurate to first order in $\uvec\Delta$, but this is a little tricky and was given incorrectly in \cite{Jaffe:1989jz}. We write
\beq \label{StrucT}
\langle  P+\tfrac{\Delta}{2},\mathscr{S}_f | T^{\mu\nu}_\text{Bel}(0) | P-\tfrac{\Delta}{2},  \mathscr{S}_i \rangle =\overline u(P+\tfrac{\Delta}{2},\mathscr S_f) \mathfrak{M}^{\mu\nu}(P,\Delta) u(P-\tfrac{\Delta}{2},\mathscr S_i)
\eeq
and consider the most general Dirac structure of $\mathfrak{M}^{\mu\nu}(P,\Delta)$ consistent with parity conservation, time-reversal invariance, hermiticity, and the fact that under a Lorentz transformation the operator $T^{\mu\nu}_\text{Bel}(0)$ transforms as a second rank tensor. We then expand in terms of $\uvec\Delta$. The correct result for this expansion, derived for the first time by BLT\footnote{The connection with  the notation in Ref. \cite{Leader:2011cr} is : $A=\mathbb{D}$, $B=2\mathbb{S}-\mathbb{D}$, $C=(\mathbb G-\mathbb H)/4$, $\bar C=\mathbb R/2$.} \cite{Bakker:2004ib}, is, in the convention $\epsilon_{0123}=+1$,
\begin{multline} \label{Tmunu}
\langle  P+\tfrac{\Delta}{2},\mathscr{S}_f |T^{\mu\nu}_\text{Bel}(0) | P-\tfrac{\Delta}{2},  \mathscr{S}_i \rangle = 2\left(A\, P^\mu P^\nu +  M^2\bar{C}\, g^{\mu\nu}\right)\\
+\frac{i\Delta_\rho}{M}\left[(A+ B)\,\frac{P^\mu \epsilon^{\rho\nu\alpha\beta} + P^\nu \epsilon^{\rho\mu\alpha\beta}}{2}+ \frac{A\,P^\mu P^\nu + M^2 \bar{C} \, g^{\mu\nu}}{P^0 + M}\,\epsilon^{0\rho\alpha\beta}\right] \frac{\mathscr{S}_\alpha}{M} \,P_\beta + \mathcal O(\uvec\Delta^2),
\end{multline}
where we have written $A$, $B$, and $\bar{C}$ for $A(\Delta^2=0)$, $B(\Delta^2=0)$, and $\bar{C}(\Delta^2=0)$, respectively, and have included the term $\bar{C} g^{\mu\nu}$ which is only present if $T^{\mu\nu}_\text{Bel}$ is a \emph{non-conserved} operator, \emph{e.g.} when using Eq.~\eqref{Tmunu} for the individual quark and gluon contributions. Note that the RHS of Eq.~\eqref{Tmunu}, because it refers to a non-forward matrix element, is, as was already mentioned, not a Lorentz tensor in the indices $\mu$ and $\nu$. The non-covariant last term comes from the expansion of the product of spinors\footnote{Using the light-front spinors instead of the standard Dirac spinors, the non-covariant term reads $\epsilon^{+\rho\alpha\beta}/P^+$ instead of $\epsilon^{0\rho\alpha\beta}/(P^0+M)$.} $\overline{u}(P+\tfrac{\Delta}{2}, \msc{S}_f) u(P-\tfrac{\Delta}{2}, \msc{S}_i)$ and was missed in Ref. \cite{Jaffe:1989jz}.

Consider now $T^{\mu\nu}_\text{Bel}(x)$ for the nucleon itself moving with momentum along $OZ$ so that
\beq \label{Zmom}
P^\mu=(P^0 , 0,0,P_z) .
\eeq
From Eq.~\eqref{Pcan}, using the fact that $P_{\text{Bel},z}$ is the $z$-component of the total momentum operator, one has
\begin{eqnarray}
\langle P', \mathscr{S}'|\int \ud^3x\, T^{03}_\text{Bel} (0,\uvec x) | P,\mathscr{S}\rangle & = &\langle  P',\mathscr{S}' |  P_{\text{Bel},z}  | \, P,\mathscr{S}\, \rangle  \nonumber \\
 & = & 2P^0P_z \,(2\pi)^3\, \delta^{(3)} (\uvec P' - \uvec P),\label{eq.28}
\end{eqnarray}
where we have put $t=0$, since, being  a conserved operator, $T^{\mu\nu}_\text{Bel}(x)$ is independent of time. But from Eq.~\eqref{Tmunu}, we have
\begin{equation}\label{eq.29}
\int \ud^3x \,e^{-i(\uvec P'-\uvec P)\cdot\uvec x}\,\langle  P',\mathscr{S}'|T^{03}_\text{Bel} (0) | P, \mathscr{S} \rangle=  2 A \,P^0 P_z\,(2\pi)^3\, \delta^{(3)} (\uvec P' - \uvec P),
\end{equation}
so that comparing Eqs.~\eqref{eq.28} and \eqref{eq.29} yields
\begin{equation}
A = 1.\label{eq.30}
\end{equation}

Next, putting Eq.~\eqref{Tmunu} into Eq.~\eqref{JFF} and carrying out the differentiation, not forgetting Eq.~\eqref{delnon}, we obtain, after some algebra, for the expectation value of $\uvec J_\text{Bel}$
\beq \label{Jexp}
\la\la P,\mathscr{S} | \uvec J_\text{Bel} (0) |P, \mathscr{S}\ra \ra = \frac{A}{2}\, \uvec s + \frac{B}{2M} \left( P^0 \uvec s - \frac{(\uvec P\cdot \uvec s)}{P^0 + M}\, \uvec P\right).
\eeq
Now consider a  state with momentum along $OZ$ and helicity $\lambda = 1/2$ so that
\beq \label{Zmom}
P^\mu=( P^0 ,0,0,P_z), \qquad \qquad \uvec s = ( 0,0,1).
\eeq
This is an eigenstate of $J_{\text{Bel},z}$ with eigenvalue $1/2$. Substituting Eq.~\eqref{Zmom} into Eq.~\eqref{Jexp} gives
\beq \label{JqAB}
\tfrac{1}{2}= \la \la  J_{\text{Bel}, z} \ra \ra=\tfrac{1}{2}\,(A+ B)
\eeq
which implies $B=0$ by Eq.~\eqref{eq.30}.

Hence, finally, we have the remarkably simple result for the \emph{total} angular momentum of a spin-1/2 nucleon
\beq \label{Jexpect}
\la\la P,\mathscr{S} |\uvec J_\text{Bel} (0) | P, \mathscr{S} \ra \ra  = \tfrac{1}{2}\, \uvec s.
\eeq
Note that this is completely general, \emph{i.e.} holds for both longitudinal and transverse polarization, and was first derived for a spin-1/2 nucleon by BLT \cite{Bakker:2004ib} using a wave-packet approach. The generalization of the above derivative method, or of the  wave-packet approach, to higher spin particles is forbiddingly complicated, so BLT found a totally different derivation based on the fact that we know how a state changes under a rotation and that $\uvec J$ is the generator of rotations. We shall only quote the, again, remarkably simple result. For the state of a particle with arbitrary spin $s$ quantized along the $OZ$ axis, one finds $(-s \leq m,m' \leq s)$
\beq
\la \la P,m'|\uvec J_\text{Bel}(0)  | P,m  \ra \ra = \left(\uvec S\right)_{m'm},
\eeq
where the $S^i$ are the three $(2s+1)$-dimensional spin matrices for spin $s$ satisfying
\beq
\left[S^i,S^j\right]= i\epsilon^{ijk} S^k.
\eeq

\subsection{Expressions for the quark and gluon angular momentum in terms of matrix elements of the energy-momentum density\label{secVIE}}

Based on the split of $T^{\mu\nu}$ into quark and gluon contributions given in Eqs.~\eqref{TqBel} and \eqref{TGBel}, we can split $\uvec J_\text{Bel}$ into contributions  $\uvec J^q_\text{Bel}$ and $\uvec J^G_\text{Bel}$ from quarks and antiquarks of a given flavor, and gluons, respectively, so that
\beq \label{Jsum}
\uvec J_\text{Bel} =\sum_q\uvec J^q_\text{Bel} + \uvec J^G_\text{Bel}.
\eeq
Analogously to Eq.~\eqref{Jexp}, we will have
\beq \label{Jqexp}
\la\la P,\mathscr{S}| \uvec J^q_\text{Bel} (0) | P, \mathscr{S} \rangle \ra = \frac{A_q}{2}\,\uvec s + \frac{B_q}{2M} \left( P^0 \uvec s - \frac{(\uvec P\cdot \uvec s)}{P^0 + M}\, \uvec P\right)
\eeq
and a similar relation for gluons, so that
\beq \label{ABsum}
A= \sum_qA_q + A_G = 1, \qquad  \qquad B=\sum_qB_q + B_G =0.
\eeq
In the next section, we shall explore the connections between these and the generalized parton distributions.

\subsection{Belinfante type angular momentum sum rules and relations involving generalized parton distributions (GPDs)\label{secVIF}}

Let us write the expression for the quark\footnote{Note that throughout this section ``quark'' means quark plus antiquark of a given flavor.} contribution to the matrix elements of $T^{\mu\nu}_\text{Bel}(0)$ in a more general form than Eq.~\eqref{Tmunu} \cite{Ji:1996ek}
\begin{align} \label{TMN}
\langle P',\mathscr{S}' |T^{\mu\nu}_{\text{Bel},q}(0) | P, \mathscr{S}\rangle &=\overline{u}(P',\mathscr S')\left[\frac{\bar{P}^{\{\mu}\gamma^{\nu\}}}{2} \,A_q(\Delta^2) +\frac{\bar{P}^{\{\mu}i\sigma^{\nu\}\rho}\Delta_\rho}{4M}\,B_q(\Delta^2)\right. \nn \\
&\hspace{2cm}\left.+\frac{\Delta^\mu \Delta^\nu - \Delta^2 g^{\mu\nu}}{M} \,C_q(\Delta^2)+Mg^{\mu\nu} \bar{C}_q(\Delta^2)\right]u(P,\mathscr S),
\end{align}
with the notations $a^{\{\mu}b^{\nu\}}=a^\mu b^\nu+a^\nu b^\mu$, $\bar{P} = \tfrac{P'+P}{2}$, $\Delta = P'-P$, and where the spinors are normalized to $\overline{u}u=2M$.

In the standard notation (see \emph{e.g.} the review of Diehl \cite{Diehl:2003ny}) the GPDs, for a nucleon moving in the positive $z$-direction, are defined by
\beq \label{GPD}
\frac{1}{2}\int \frac{\ud z^-}{2\pi}\,e^{i x \bar{P}^+ z^-} \langle P', \msc{S}' | \barpsi(-\tfrac{z^-}{2}) \gamma^+ \mathcal W\psi (\tfrac{z^-}{2})  |  P,  \msc{S} \rangle
= \frac{1}{2 \bar{P}^+}\,\overline{u}(P',\mathscr S')\left[\gamma^+  H_q(x, \xi, t) +\tfrac{i\sigma^{+ \rho}\Delta_\rho}{2M}\, E_q(x,\xi, t) \right]u(P,\mathscr S),
\eeq
where
\beq
t= \Delta^2, \qquad \Delta^+= -2\xi\bar{P}^+,
\eeq
and we are using the standard definition of the $\pm$ components of a four-vector, \emph{i.e.}
\beq
a^\pm = \frac{a^0 \pm a^3}{\sqrt{2}}.
\eeq
The factor $\mathcal W$ is the Wilson line operator
\beq \label{Wilson}
\mathcal W\equiv \mathcal W[-\tfrac{z^-}{2},\tfrac{z^-}{2}]_{rs}= \mathcal{P} \,exp\left[ig \int^{-z^-/2}_{z^-/2} \ud\lambda\, A^+_a( \lambda n)\, t^a_{rs}\right],
\eeq
a matrix in color space, with $n^\mu=\tfrac{1}{\sqrt{2}}\,(1,0,0,-1)$.

Multiplying by $x$ and integrating, Eq.~\eqref{GPD} yields after some manipulation
 \beq \label{intGPD}
\frac{1}{2 \bar{P}^+}  \left[(\overline{u}'\gamma^+ u)\int \ud x \,x\, H_q(x, \xi, t) +\left( \overline{u}'\tfrac{i\sigma^{+ \rho}\Delta_\rho}{2M} u \right) \int \ud x\, x\, E_q(x,\xi, t) \right]= \frac{1}{4(\bar{P}^+)^2}\,\langle  P' ,\msc{S}' | \barpsi(0)  \gamma^+ i\LRD^+ \psi (0)  |  P, \msc{S}\rangle.
\eeq
 Comparing with Eq.~\eqref{TqBel} we see that the RHS of Eq.~\eqref{intGPD} is, up to a factor, the matrix element of $T^{++}_{\text{Bel},q}$, so that
\beq \label{IntGPD}
(\overline{u}'\gamma^+ u)\int \ud x \,x\, H_q(x, \xi, t) +\left( \overline{u}'\tfrac{i\sigma^{+ \rho}\Delta_\rho}{2M} u \right) \int \ud x\, x\, E_q(x,\xi, t)  = \frac{1}{\bar{P}^+} \,\langle  P',\msc{S}'  | T^{++}_{\text{Bel},q}(0) |  P,\msc{S}\rangle.
\eeq
From Eq.~\eqref{TMN}, remembering that $g^{++}=0$ and that $\Delta^+= -2\xi \bar{P}^+$, one obtains
\beq \label{tplusplus}
\langle P',\msc{S}'  | T^{++}_{\text{Bel},q}(0) |  P,\msc{S }\rangle = (\overline{u}'\gamma^+ u)\left[A_q(\Delta^2) +4 \xi^2 C_q(\Delta^2)\right]+\left( \overline{u}'\tfrac{i\sigma^{+ \rho}\Delta_\rho}{2M} u \right) \left[ B_q(\Delta^2) - 4\xi^2  C_q(\Delta^2)\right].
\eeq
Upon taking the limit $\Delta^\mu \rightarrow 0 $, Eqs.~\eqref{intGPD} and \eqref{tplusplus}  yield
\begin{align} \label{Hsum}
\int_{-1}^{1} \ud x\, x\, H_q(x,0,0) &= A_q,\\
\label{Esum} \int_{-1}^{1} \ud x\, x\, E_q(x,0,0)&= B_q,
\end{align}
and consequently
\beq \label{HEsum}
A_q + B_q=\int_{-1}^{1} \ud x \,x \left[H_q(x,0,0) + E_q(x,0,0) \right].
\eeq

\subsubsection{Belinfante type relations and sum rules for longitudinally polarized nucleons\label{secVIF1}}

For the case of a \emph{longitudinally} polarized  nucleon, $\msc{S}=\msc{S}_L\equiv(P_z,P^0\uvec e_z)$, moving in the $z$-direction, Eq.~\eqref{Jqexp} yields for the longitudinal \emph{i.e.} $z$-component of $\uvec J^q_\text{Bel}$
\beq \label{JqzAB}
\la\la  P,\mathscr{S}_L | J^q_{\text{Bel},z}  | P, \mathscr{S}_L  \rangle \ra = \tfrac{1}{2}\left(A_q+ B_q\right).
\eeq
Hence, Eq.~\eqref{HEsum} can be written
\beq \label{Jzsum}
\la\la   P,\mathscr{S}_L | J^q_{\text{Bel},z}  | P, \mathscr{S}_L  \ra \ra=\tfrac{1}{2}\int_{-1}^{1} \ud x\, x \left[H_q(x,0,0) +E_q(x,0,0)\right]
\eeq
which is the relation first derived by Ji\footnote{Note that in Ji's original paper, it is not indicated that the nucleon is longitudinally polarized. However, in the review of Ji and Filippone \cite{Filippone:2001ux}, the relation is explicitly stated for the $z$-component of $\uvec J^q_\text{Bel}$.} \cite{Ji:1996nm}. \nl

Now from the analogue of Eq.~\eqref{eq.29}, for the longitudinal $z$-component of the quark momentum, one obtains
\beq \label{PLexp}
\la\la   P,\mathscr{S} | P^q_{\text{Bel},z}  | P, \mathscr{S}  \ra \ra =  P_z A_q,
\eeq
so that Eq.~\eqref{Hsum} becomes
\beq \label{HsumP}
\int_{-1}^{1} \ud x \,x \,H_q(x,0,0) = \frac{\la\la   P,\mathscr{S} | P^q_{\text{Bel}, z}  | P, \mathscr{S} \rangle \ra}{P_z},
\eeq
which is in accordance with the interpretation of $x \,H_q(x,0,0)=x\,q(x)$ as a measure of the fraction of the nucleon momentum carried by quarks and antiquarks of a given flavor and given value of $x$. It should not be forgotten that neither $ \uvec J^q_\text{Bel}$ nor $\uvec P^q_\text{Bel}$ are conserved operators. Consequently, they are renormalization scale dependent and, strictly speaking, both sides of Eqs.~\eqref{Jzsum} and \eqref{HsumP} should carry a label $Q^2$.

Hence, summing over flavors and adding the analogous gluon contribution\footnote{For gluons the integrals run from $0$ to $1$.}, one must have the sum rule
\beq \label{Hfrac}
\sum_q \int_{-1}^{1} \ud x\, x \,H_q(x,0,0) + \int_{0}^{1} \ud x\,x\, H_G(x,0,0) =1.
\eeq
Next, using this and summing Eq.~\eqref{Jzsum} over flavors and adding the analogous equation for gluons, one obtains
\beq\label{qG}
\tfrac{1}{2} + \tfrac{1}{2}\left[\sum_q \int_{-1}^{1} \ud x\, x \,E_q(x,0,0) + \int_{0}^{1} \ud x\, x\, E_G(x,0,0)\right]= \la\la   P,\mathscr{S}_L | J_{\text{Bel},z} | P, \mathscr{S}_L  \rangle \ra=\tfrac{1}{2},
\eeq
so that
\beq \label{EqG}
\sum_q \int_{-1}^{1} \ud x\, x\, E_q(x,0,0) + \int_{0}^{1} \ud x \,x \,E_G(x,0,0) =0.
\eeq
This is a fundamental sum rule, since in principle all terms in it can be measured. It has wide ramifications and can be shown to correspond to the vanishing of the nucleon anomalous gravitomagnetic moment \cite{Brodsky:2000ii,Teryaev:1999su}.

In summary, since the GPDs $H$ and $E$ are in principle measurable, the relation in Eq.~\eqref{Jzsum} and its gluon analogue, while not really sum rules, provide a beautiful way to measure the longitudinal components of the quark and gluon angular momentum in a longitudinally polarized nucleon. On the other hand, Eqs.~\eqref{Hfrac} and \eqref{EqG} are true sum rules, so that testing them would provide a fundamental test of QCD.\newline

In the above we have been considering the Belinfante version of the \emph{total} angular momentum carried by a quark or by a gluon. Recall that the Belinfante form of the angular momentum given by Eq.~\eqref{MBorb}, which has the structure of an orbital angular momentum with no separation of the quark total angular momentum into a spin and orbital part, can be rewritten, as explained in section \ref{secIIIB1}, in the form used by Ji, in which, after use of the equations of motion and discarding a surface term, $\uvec J^q_\text{Bel}$ for a quark of some given flavor is split in a gauge-invariant way into a spin term $\uvec S^q_\text{Ji}$ and an orbital term $\uvec L^q_\text{Ji}$, as presented in Eq.~\eqref{Jiel}. \emph{Via} a little gamma matrix algebra, the matrix elements of the spin term between states of the same energy, can be related to the matrix elements of the axial current
\beqy \label{SJi5}
\langle  P,\mathscr{S}|  \uvec S^q_\text{Ji}  |  P,\mathscr{S} \rangle  &= &\langle  P,\mathscr{S} | \int \ud^3x \,\psi^\dag (0, \uvec x) \,\tfrac{1}{2}\uvec\Sigma\, \psi(0, \uvec x) |  P,\mathscr{S} \rangle\nn\\
&=& \tfrac{1}{2}\,\langle  P,\mathscr{S} |\int \ud^3x\, \barpsi(0,\uvec x)\uvec\gamma\gamma_5 \psi(0, \uvec x) |  P,\mathscr{S}\rangle  \nn \\
 &=& (2\pi)^3\, \delta^{(3)}(\uvec 0) \,\tfrac{1}{2} \,\langle  P,\mathscr{S} |  \barpsi(0) \uvec\gamma\gamma_5 \psi(0)  |  P,\mathscr{S} \rangle,
\eeqy
where $\psi$ is a quark field of some given flavor. Now the matrix elements of the axial current are given by
\beq \label{axchg}
\langle  P,\mathscr{S} |   \barpsi(0) \gamma^\mu \gamma_5 \psi(0)  |  P,\mathscr{S}\rangle =2\,a_0^q\,\msc{S}^\mu,
\eeq
where $a_0^q$ is the contribution to $a_0$ (or $g_A^{(0)}$), the flavor-singlet axial charge of the nucleon,  from a quark plus antiquark of given flavor. Hence, for the expectation value\footnote{Note that the polarization dependence is entirely contained in the factor $\uvec{\msc{S}}$, implying that $a_0$ appears for both longitudinal and transverse polarizations.}
\beq \label{SJiexp}
\la\la \,  P,\mathscr{S} \,|\, \bm{S}^q_{Ji} \, |\, P, \mathscr{S} \, \rangle \ra = \frac{a_0^q}{ 2 P^0} \, \bm{\msc{S}}.
\eeq
For longitudinal polarization, from Eq.~\eqref{S}
\beq \label{Sz}
\msc{S}_z= P^0
\eeq
so that
\beq \label{SLrel}
\la\la P,\mathscr{S}_L | S^q_{\text{Ji}, z}  | P, \mathscr{S}_L  \ra \ra = \tfrac{1}{2}\,a_0^q .
\eeq
Combining this with Eq.~\eqref{Jzsum} yields an expression, experimentally measurable,  for the longitudinal component of the Ji version of the orbital angular momentum of a quark plus antiquark of given flavor
 \beq \label{LsumL}
\la\la P,\mathscr{S}_L | L^q_{\text{Ji}, z}  | P, \mathscr{S}_L \ra \ra = \tfrac{1}{2}   \int_{-1}^{1}  \ud x \,x \left[H_q(x,0,0) +E_q(x,0,0)\right] - \tfrac{1}{2}\,a_0^q.
\eeq
Again, one should be aware that each term in Eq.~\eqref{LsumL} depends on the renormalization scale.

The term $a_0^q$ is obtained from inclusive polarized deep inelastic lepton-hadron scattering and is related to the polarized parton densities. However this relation is factorization scheme dependent, and its connection with  $\Delta\Sigma$ and $\Delta G$ is scheme dependent (for a detailed discussion, see Refs. \cite{Bass:1992ti} and \cite{Leader:1998nh}). For example, one has in the $\overline{MS}$, $AB$ and $JET$ schemes
 \beq \label{a0q}
\begin{split}
a_0^q(Q^2) &=\int_0^1 \ud x \left[ \Delta q(x,Q^2) + \Delta \overline{q}(x,Q^2)\right]_{\overline{MS}}\nn \\
&= \int_0^1 \ud x \left[ \Delta q(x,Q^2) + \Delta \overline{q}(x,Q^2)\right]_{AB/JET}   - \frac{\alpha_s(Q^2)}{2\pi}\int_0^1 \ud x \, \Delta G(x,Q^2)_{AB/JET}.
\end{split}
\eeq
In the section \ref{secVII}, which deals in more detail with orbital angular momentum, we shall show that $\la\la  P,\mathscr{S}_L | L^q_{\text{Ji}, z}  | P, \mathscr{S}_L  \ra \ra $ can be related to the 2nd moment of a twist-3 GPD. We shall also present results for the angular momentum and orbital angular momentum obtained from Lattice studies and models.

\subsubsection{Belinfante type relations and a sum rule for transversely polarized nucleons\label{secVIF2}}

For the case of a nucleon \emph{transversely} polarized, say in the $x$-direction, $\msc{S}=\msc{S}_x= \msc{S}_T\equiv(0,M\uvec s)$ with $\uvec s=(1,0,0)$, and moving in the $z$-direction, Eq.~\eqref{Jqexp} yields for the transverse component\footnote{Clearly the same result holds for the $x$ or $y$ components.} of $\uvec J^q_\text{Bel}$
\beq \label{JqxAB}
\la\la  P,\mathscr{S}_T |J^q_{\text{Bel},T} | P, \mathscr{S}_T  \ra \ra = \tfrac{1}{2}\left(A_q+\tfrac{P^0}{M}\,B_q\right),
\eeq
so that from Eqs.~\eqref{Hsum} and \eqref{Esum}, one obtains the result first derived by Leader in Ref. \cite{Leader:2011cr}
\beq \label{JqHET}
 \la\la   P,\mathscr{S}_T | J^q_{\text{Bel},T}  | P, \mathscr{S}_T  \ra \ra =  \frac{1}{2M}\left[ P^0\int_{-1}^{1} \ud x\, x\, E_q(x,0,0) + M \int_{-1}^{1} \ud x\, x\, H_q(x,0,0) \right].
\eeq
The energy factor $P^0$ may seem unintuitive. However, if we go the rest frame, Eq.~\eqref{JqHET} reduces to the Ji result \eqref{Jzsum}, as it should, since in the rest frame there is no distinction between $x$, $y$ and $z$-directions. Moreover, as explained in section \ref{secIIC} for a classical relativistic system of particles, if one calculates the orbital angular momentum about the center of inertia for the system at rest, and then boosts the system, one finds that the transverse angular momentum grows like $P^0$ \cite{Landau:1951}. Finally, if one sums Eq.~\eqref{JqHET} over flavors and adds the analogous gluon equation, one finds, as a consequence of Eq.~\eqref{EqG}, that the term proportional to $P^0$ disappears, as it ought to, and using Eq.~\eqref{Hfrac}, one obtains the correct result for a transversely polarized nucleon
\beq \label{sumJT}
\sum_q\la\la P,\mathscr{S}_T | J^q_{\text{Bel},T}  | P, \mathscr{S}_T \ra \ra + \la\la P,\mathscr{S}_T | J^G_{\text{Bel},T}  | P, \mathscr{S}_T  \ra \ra
= \tfrac{1}{2}.
\eeq
For the Ji spin and orbital terms, from Eq.~\eqref{SJiexp} one obtains, since $\msc{S}_x=M$, one has
\beq \label{STrel}
\la\la  P,\mathscr{S}_T | S^q_{\text{Ji},T} | P, \mathscr{S}_T \ra \ra = \frac{M}{2P^0}\,a_0^q,
\eeq
and from Eq.~\eqref{JqHET}, one obtains a relation for the orbital angular momentum in terms of GPDs:
\beq \label{LsumT}
\la\la   P,\mathscr{S}_T | L^q_{\text{Ji},T}  | P, \mathscr{S}_T \ra \ra = \frac{1}{2M}\left[ P^0\int_{-1}^{1} \ud x\, x\, E_q(x,0,0) + M\int_{-1}^{1} \ud x\, x\, H_q(x,0,0) \right] - \frac{M}{2P^0}\,a_0^q.
\eeq
Possibilities for using this relation \emph{via} a Lattice calculation will be discussed in section \ref{secVII}.

\subsection{Canonical type angular momentum relations and sum rules involving polarized parton densities\label{secVIG}}

The Jaffe-Manohar decomposition of $\uvec J_\text{QCD}$ given in Eq.~\eqref{JMdec}, which corresponds to the canonical decomposition, and  which follows directly from Noether's theorem, contains spin and orbital terms for both quarks and gluons, but only the quark spin term is gauge invariant. Indeed it is the same as the Ji-type quark spin term $\uvec S^ q_\text{JM} = \uvec S^q_\text{Ji}$, so that from Eq.~\eqref{SJiexp}
\beq \label{SJMexp}
\la\la  P,\mathscr{S} | \uvec S^q_\text{JM}  | P, \mathscr{S}  \ra\ra = \frac{a_0^q}{ 2 P^0} \, \uvec{\msc{S}}.
\eeq

\subsubsection{Canonical type relation and a sum rule for longitudinally polarized nucleons\label{secVIG1}}

Consider the gauge-invariant expression for the polarized gluon density $\Delta G(x)$ given by Manohar \cite{Manohar:1990kr} and used by Jaffe\footnote{Note that there is a typographical error in the expression for $\Delta G(x)$  in these papers:  $\tilde{G}_{\alpha}^{\phantom{\alpha}+}(0)$ should be  $\tilde{G}^{+}_{\phantom{+}\alpha}(0)$.} \cite{Jaffe:1995an} \emph{i.e.}
\beq \label{DelG}
\Delta G (x)= \frac{i}{4\pi x P^+}\int \ud z^- e^{ixP^+z^-} \langle P,  \msc{S}_L | 2\uTr\!\left[G^{+\alpha}(0) \mathcal W(0,z^-)\tilde{G}^{+}_{\phantom{+}\alpha}(z^-)\mathcal W(z^-,0)\right]| P,  \msc{S}_L  \rangle + (x\mapsto-x),
\eeq
where the dual field-strength tensor is defined as
\beq
\tilde{G}^{\mu\nu} = \tfrac{1}{2}\,\epsilon^{\mu\nu\alpha\beta}G_{\alpha\beta}
\eeq
with $\epsilon_{0123}=1$, and $| P, \msc{S}_L  \rangle$ is, as before, a longitudinally polarized nucleon state. Because of its complicated non-local form, it has long been thought that $\Delta G(x)$ could not be computed on a lattice, but Ji, Zhang and Zhao recently proposed in Refs. \cite{Ji:2013fga,Ji:2013dva} some interesting new strategies for doing so.

Since the expression in Eq.~\eqref{DelG} is gauge invariant, we may evaluate it in the light-front gauge $A^+=0$ with antisymmetric boundary condition $A^j(+\infty^-)+A^j(-\infty^-)=0$. Then, following the argument in Ref. \cite{Jaffe:1995an} and integrating over $x$, one obtains\footnote{One integrates $x$ from $-1$ to $1$ because the definition \eqref{DelG} for $\Delta G(x)$ has been symmetrized in $x$ for convenience.}
\begin{align} \label{delG}
\Delta G &\equiv \int^1_{-1} \ud x\, \Delta G(x) \nn \\
&=\frac{1}{2P^+}\,\langle P,  \msc{S}_L |2\uTr\!\left[G^{1+}(0)A^2(0) -G^{2+}(0)A^1(0)\right]| P,  \msc{S}_L   \rangle \Big|_{A^+=0\,\text{+B.C.}}.
\end{align}
Consider now the possible tensorial structure for the matrix element $\langle P,\msc{S}_L |2\uTr\!\left[G^{1\mu}(0)A^2(0) -G^{2\mu}(0)A^1(0)\right]| P,  \msc{S}_L  \rangle$ with $\mu=0$ or $3$. It has mass dimension $[M]$, so the leading term can only come from $P^0$, $P^3$, $\msc{S}_L^0$ or $\msc{S}_L^3$. But to leading order, one has $P^0=P^3$ and $\msc{S}_L^0=\msc{S}^3_L$. Thus \emph{in leading twist}, \emph{i.e.} as $P_z\rightarrow \infty$
\begin{align} \label{delGLT}
\Delta G &=\lim_{P_z\rightarrow\infty} \,\frac{1}{2P^0}\,\langle P,  \msc{S}_L |2\uTr\!\left[G^{10}(0)A^2(0) -G^{20}(0)A^1(0)\right]| P, \msc{S}_L \rangle\Big|_{A^+=0\,\text{+B.C.}} \nn \\
&=\lim_{P_z\rightarrow\infty} \,\frac{1}{2P^0}\,\langle P, \msc{S}_L |\left[ \uvec E^a(0) \times \uvec A^a(0)\right]_z| P,  \msc{S}_L \rangle\Big|_{A^+=0\,\text{+B.C.}}.
\end{align}
Now since  the initial and final states have the same energy,  we may choose to work at time $t=0$. Then, for the gluon spin term, we have
\begin{align} \label{SGJM}
\langle P,  \msc{S}_L |\uvec S^G_\text{JM}   | P,  \msc{S}_L  \rangle& =  \int \ud^3 x \, \langle P,  \msc{S}_L |\uvec E^a(0,\uvec x)\times \uvec A^a(0,\uvec x)  | P,  \msc{S}_L  \rangle\nn \\
&= (2\pi)^3\, \delta^{(3)}(\uvec 0)\, \langle P, \msc{S}_L |\uvec E^a(0) \times \uvec A^a(0)  | P, \msc{S}_L \rangle.
\end{align}
Hence we have the important result for the expectation value of the longitudinal component of $\uvec S^G_\text{JM}$\footnote{Note that we are using the instant form of dynamics where dynamical quantities evolve with time $t$. In light-front dynamics, where dynamical quantities evolve with light-front time $x^+$, the RHS of Eq.~\eqref{delG} is already the gluon longitudinal spin operator.}
\beq \label{SGJML}
\lim_{P_z\rightarrow\infty}\,\la \la P, \msc{S}_L |S^G_{\text{JM}, z}   | P,  \msc{S}_L  \ra \ra\Big|_{A^+=0}= \Delta G.
\eeq
We have removed the mention of the boundary condition, since it has been shown by Bashinsky and Jaffe \cite{Bashinsky:1998if} that it does not affect the matrix element $\la \la P, \msc{S}_L |S^G_{\text{JM}, z}   | P,  \msc{S}_L  \ra \ra$ in the limit $P_z\to\infty$.

As stressed earlier, all the terms in the Jaffe-Manohar expression for $\uvec J_\text{QCD}$, with the exception of the quark spin term, are not gauge invariant. The above analysis tells us that we can identify $S^G_{\text{JM},z}$ with $\Delta G$ provided we choose the gauge $A^+=0$. This should not be considered a restriction, because the parton model is really a ``picture'' of QCD in the gauge $A^+=0$. So the correct way to state the Jaffe-Manohar relation for a longitudinally polarized nucleon moving along $OZ$, is:
\beq \label{CorJML}
\tfrac{1}{2} = \tfrac{1}{2}\, a_0 + \Delta G + \lim_{P_z\rightarrow \infty} \left[ \sum_q\la \la P,  \msc{S}_L |L^q_{\text{JM}, z}  | P,  \msc{S}_L \ra \ra\Big|_{A^+=0}+\la \la P, \msc{S}_L |L^G_{\text{JM}, z}   | P,  \msc{S}_L  \ra \ra\Big|_{A^+=0}\right],
\eeq
where the infinite-momentum frame limit $P_z\rightarrow \infty$ is a reflection of our use of the leading twist approximation in handling the expression for $\Delta G$\footnote{In the infinite-momentum frame, \emph{i.e.} for the leading-twist contribution, different gauge choices appear to give the same result. For example, since any four-vector $v^\mu$ transforms as $[v^+,v^-,\uvec v_\perp]\mapsto [Nv^+,\tfrac{1}{N}\,v^-,\uvec v_\perp]$ under a Lorentz boost along the $z$-direction, one can write $\uvec\nabla\cdot\uvec A\approx-\tfrac{N^2}{2}\,\partial^+A^+$ as $N\to \infty$, treating $A^\mu$ as a Lorentz four-vector. So, in the infinite-momentum frame, the light-front gauge $A^+=0$ implies automatically the Coulomb gauge $\uvec\nabla\cdot\uvec A=0$.}. Note that the Jaffe-Manohar orbital angular momentum terms are not gauge invariant and for consistency must be evaluated in the gauge $A^+=0$. Finally, it should not be forgotten that all terms in Eq.~\eqref{CorJML} are renormalization scale dependent.
\newline

Some people may feel uncomfortable with giving a physical meaning to a gauge non-invariant quantity like the gluon spin, and sometimes argue that a genuine ``physical'' interpretation should be gauge-invariant. So, since the gluon longitudinal spin coincides with the measurable quantity $\Delta G$ in the light-front gauge only, it is sometimes claimed that the the gluon longitudinal spin is not really a physical quantity.  This can be made more palatable using  the notion of \emph{gauge-invariant extension}. Since $\Delta G$ is gauge invariant and coincides with the Jaffe-Manohar gluon longitudinal spin in the light-front gauge, it can be considered as its light-front gauge-invariant extension. This means in particular that one should be able to rewrite Eq. \eqref{DelG} as the matrix element of some kind of longitudinal gluon spin operator \emph{without} fixing the gauge. This has been achieved explicitly by Hatta \cite{Hatta:2011zs} in a variant of the Chen \emph{et al.} approach,  where the physical gluon field  is defined by the following non-local expression\footnote{Hatta considered also the cases of advanced and retarded boundary conditions.}
\beq
A^\mu_\phys(x)\equiv-\int\ud z^-\,\tfrac{1}{2}\,\epsilon(z^--x^-)\,\mathcal W(x^-,z^-)G^{+\mu}(z^-)\mathcal W(z^-,x^-),
\eeq
with $\epsilon(z^--x^-)$ the sign function. This is simply the physical field associated with the gauge-invariant extension formalism defined by the light-front constraint with antisymmetric boundary conditions
\beq
A^+_\phys(x)=0,\qquad A^\mu_\phys(+\infty^-)+A^\mu_\phys(-\infty^-)=0.
\eeq
In other words, instead of fixing the gauge symmetry, one only fixes the Stueckelberg symmetry at the cost of dealing with implicitly non-local expressions. Now using the Cauchy principal value prescription for the $1/x$ factor in Eq.~\eqref{DelG}, one obtains
\beq
\Delta G=\frac{1}{2P^+}\,\langle P,  \msc{S}_L |2\uTr\!\left[G^{1+}(0)A^2_\phys(0) -G^{2+}(0)A^1_\phys(0)\right]| P,  \msc{S}_L \rangle,
\eeq
corresponding to the matrix element of the longitudinal gluon spin operator in the gauge-invariant form of the canonical decomposition. This means that the measurable quantity $\Delta G$ can be interpreted \emph{in any gauge} as a measure of the gluon longitudinal spin appearing in the Hatta decomposition, \emph{i.e.} in the light-front gauge-invariant canonical decomposition. Note that in the Chen \emph{et al.} decomposition \cite{Chen:2008ag}, the matrix element of the longitudinal gluon spin operator does not coincide with $\Delta G$, because they used a different definition for the physical field $A^\mu_\phys$, \emph{i.e.} a different Stueckelberg-fixing condition.

However, there is also a completely different way to approach this issue. It is always assumed, though not proved rigorously, that there exist ``in'' and ``out'' fields in both QED and QCD. On this basis, Leader \cite{Leader:2011za} showed that the expectation value of an operator taken between physical states, even if it is a non-conserved operator, can be evaluated using the ``in''  field form of the operator, and demonstrated that the \emph{expectation value} of the photon and gluon \emph{helicity} is actually gauge independent. This feature, that the physical matrix elements of a gauge non-invariant operator can be gauge-independent was discussed in section \ref{secIIE}. From this point of view, there is no reason to feel uncomfortable about the connection between $\Delta G$ and the expectation value of the gluon helicity.

\subsubsection{Canonical type relation for transversely polarized nucleons\label{secVIG2}}

The derivation of a transverse angular momentum sum rule, based on the canonical angular momentum operators, is rather complicated and does not follow the pattern used in the previous cases, where we took matrix elements of the various quark and gluon operator terms in the expression for $\uvec J$ and then succeeded in identifying some of those  matrix elements with measurable quantities. The approach for the transverse canonical case \cite{Bakker:2004ib} is based on writing down the general structure of the matrix elements of $\uvec J$ for a transversely polarized nucleon, and then inserting a Fock expansion for the nucleon state and identifying the various pieces in terms of number  densities of partons of a particular spin projection inside a transversely polarized nucleon. These densities can in principle be measured, but some of them do not correspond to matrix elements of any local gauge-invariant operator.\nl

Recall that in going from Eq.~\eqref{AM} to Eq.~\eqref{Jfinal}, we discarded a term which, in a wave-packet approach, represented the orbital angular momentum of the packet about the origin. In the following, we shall need to reinstate it in the form it takes in the limit of an infinitely narrow packet, and follow the discussion of Bakker, Leader and Trueman \cite{Bakker:2004ib} (BLT). For the nucleon in a state labelled by the $z$-component $m$ of the spin, one has
\beq
\la P',m'|\uvec J|P,m \ra =2P^0\, (2\pi)^3 \left[ \tfrac{1}{2}\, \uvec\sigma +(\uvec P\times i\uvec\nabla_{\uvec P})\right]_{m'm}\delta^{(3)}(\uvec P'-\uvec P). \label{NuclJ}
\eeq
In the following we shall loosely refer to the first term in Eq.~\eqref{NuclJ} as ``spin-like'' and the second as ``orbital-like'', but it should be noted that these do not correspond to the actual spin and orbital parts. Similarly, for quarks with $z$-component of spin $s_z$ and momentum $\uvec k$
\beq
\la k',s'_z|\uvec J|k,s_z \ra=2k^0\, (2\pi)^3 \left[ \tfrac{1}{2}\,\uvec\sigma +(\uvec k \times i \uvec\nabla_{\uvec k})\right]_{s'_zs_z}\delta^{(3)}(\uvec k'-\uvec k), \label{J11}
\eeq
and for gluons, in a state of definite Jacob-Wick helicity $\lambda$ \cite{Jacob:1959at}, one has
\begin{equation}
\langle k',\lambda' |\uvec J | k,\lambda \rangle = 2k^0\,(2\pi)^3\left[ \lambda\, \uvec\eta (\uvec k) + (\uvec k\times i\uvec\nabla_{\uvec k}) \right]\delta_{\lambda\lambda'}\,\delta^{(3)}(\uvec k' - \uvec k),
\label{eq.5.041}
\end{equation}
where
\beq
\eta_{x}= \cos(\phi)\tan(\tfrac{\theta}{2}),\qquad \eta_{y}= \sin(\phi)\tan(\tfrac{\theta}{2}), \qquad \eta_{z}=1 \label{helicity21}
\eeq
and $(\theta,\phi)$ are the polar angles of $\uvec k$. As already mentioned, the above classification into ``spin-like'' and ``orbital-like'' contributions should not be taken literally. Indeed, for a moving state of a spin-$1/2$ particle, one has in general
\begin{align}
  \la P',m'|\uvec S|P,m \ra &=2P^0\, (2\pi)^3 \left[\frac{M}{2P^0}\, \uvec\sigma+\frac{\uvec P\cdot\uvec\sigma}{2P^0(P^0+M)}\,\uvec P\right]_{m'm}\delta^{(3)}(\uvec P'-\uvec P),\label{Sgen}\\
\la P',m'|\uvec L|P,m \ra &=2P^0\, (2\pi)^3 \left[-\frac{\uvec P\times(\uvec P\times\uvec\sigma)}{2P^0(P^0+M)}+(\uvec P\times i\uvec\nabla_{\uvec P})\right]_{m'm}\delta^{(3)}(\uvec P'-\uvec P).\label{Lgen}
\end{align}

Returning to  Eq.~\eqref{NuclJ}, the nucleon state is expanded as a superposition of $n$-parton
Fock states\footnote{For simplicity we do not show color or flavor labels.},
\begin{align}
| \uvec P, m \rangle&  =  \sqrt{(2\pi)^3\,2P^0} \,\sum_n\sum_{\{\sigma_i\}} \int \frac{\ud^3k_1}{\sqrt{(2\pi)^3\, 2k^0_1}}\cdots \frac{\ud^3k_n}{\sqrt{(2\pi)^3 \,2k^0_n}} \nn\\
&\qquad\times \psi_{\uvec P, m} (\uvec k_1,\sigma_1, \cdots, \uvec k_n, \sigma_n)\,\delta^{(3)}(\uvec P - \uvec k_1 -  \cdots - \uvec k_n)\,|\uvec k_1,\sigma_1,  \cdots ,\uvec k_n, \sigma_n \rangle,\label{eq.5.08}
\end{align}
where $\sigma_i$ denotes either the spin projection on the
$z$-axis or the helicity, as appropriate. $\psi_{\uvec P,m}$ is the \emph{instant form}
partonic wave function of the nucleon normalized so that
\begin{equation}
\sum_{\{\sigma_i\}} \int \ud^3k_1 \dots \ud^3k_n\,| \psi_{\uvec P,m} (\uvec k_1,\sigma_1, \cdots, \uvec k_n, \sigma_n)|^2\,\delta^{(3)}(\uvec P - \uvec k_1 -  \cdots  - \uvec k_n) = {\cal P}_n \label{eq.5.09}
\end{equation}
with ${\cal P}_n$  denoting the probability of the $n$-parton
state and satisfying
\beq
\sum_n {\cal P}_n = 1.
\eeq
The $n$-parton contribution is then
\beq
\begin{split}
\la \la \uvec P', m' |\uvec J | \uvec P, m \ra \ra_{n\text{-parton}}  &= (2\pi)^3\,2\sqrt{P'^0P^0}  \sum_{\{\sigma'_i\},\{\sigma_i\}}\int [\ud^3 k'_1] \cdots [\ud^3 k'_n][\ud^3 k_1] \cdots [\ud^3 k_n]\\
&\qquad\times\psi^*_{\uvec P', m'} (\uvec k_1',\sigma_1', \cdots ,\uvec k_n', \sigma_n')\,\langle \uvec k'_1, \sigma'_1, \cdots, \uvec k'_n, \sigma'_n |\uvec J|\uvec k_1, \sigma_1, \cdots, \uvec k_n, \sigma_n  \rangle\\
&\qquad\times\psi_{\uvec P,m} (\uvec k_1,\sigma_1, \cdots ,\uvec k_n, \sigma_n)\,\delta^{(3)}(\uvec P' - \uvec k'_1- \cdots- \uvec k'_n)\,\delta^{(3)}(\uvec P - \uvec k_1- \cdots - \uvec k_n), \label{eq.5.10}
\end{split}
\eeq
where
\begin{equation}
[\ud^3 k] = \frac{\ud^3k}{\sqrt{(2\pi)^3\,2 k^0}}. \label{eq.5.11}
\end{equation}
Provided the operators are normal-ordered, the matrix elements are diagonal in parton number and the $n$-parton matrix element becomes a sum of two-particle matrix elements:
\beq
\langle \uvec k'_1, \sigma'_1, \cdots, \uvec k'_n, \sigma'_n |\uvec J|\uvec k_1, \sigma_1, \cdots, \uvec k_n, \sigma_n  \rangle= \sum_r \langle \uvec k'_r, \sigma'_r|\uvec J|\uvec k_r, \sigma_r  \rangle\prod_{l\neq r} (2\pi)^3 \,2 k^0_l\, \delta^{(3)}(\uvec k'_l -\uvec k_l)\,\delta_{\sigma'_l\sigma_l}.\label{eq.5.12}
\eeq
One therefore has
\beq
\begin{split}
\la \la \uvec P', m' |\uvec J | \uvec P, m \ra \ra&=(2\pi)^3\,2\sqrt{P'^0P^0}  \sum_a\sum_{\sigma, \sigma'}\int [\ud^3 k'] [\ud^3 k] \\
&\qquad\times\delta^{(3)}(\uvec P'-\uvec P+\uvec k-\uvec k')\,\rho^{m'm}_{\sigma'\sigma,a}(\uvec k',\uvec k)\,\langle \uvec k', \sigma'|\uvec J|\uvec k, \sigma \rangle,
\end{split}
\eeq
where we have introduced a density matrix
\beq
\begin{split}
\rho^{m'm}_{\sigma'\sigma,a}(\uvec k',\uvec k)=&\sum_n\sum_{r(a)}\sum_{\sigma'_r,\{\sigma_i\}}\delta_{\sigma'\sigma'_r}\,\delta_{\sigma\sigma_r}\int\ud^3k'_r\,\ud^3k_1\cdots\ud^3k_r\cdots\ud^3k_n\\
&\qquad\times\delta^{(3)}(\uvec k'-\uvec k'_r)\,\delta^{(3)}(\uvec k-\uvec k_r)\,\delta^{(3)}(\uvec P-\uvec k_1-\cdots-\uvec k_r-\cdots-\uvec k_n)\\
&\qquad\times\psi^*_{\uvec P', m'} (\uvec k_1,\sigma_1, \cdots ,\uvec k'_r,\sigma'_r,\cdots\uvec k_n, \sigma_n)\,\psi_{\uvec P,m} (\uvec k_1,\sigma_1, \cdots ,\uvec k_r,\sigma_r,\cdots,\uvec k_n, \sigma_n).
\end{split}
\eeq
The index $a=q,\overline q, G$ refers to quarks, antiquarks or gluons, and the sum over $r(a)$ means a sum over those $r$-values corresponding to partons of type $a$ in the multi-parton state.

The two terms in Eqs. \eqref{J11} and \eqref{eq.5.041} suggest a ``spin-like'' part and an ``orbital-like'' part for quarks and gluons.  The quark+antiquark ``spin-like'' term is defined as
\beq\label{qspinlike}
\la\la\uvec P,m'|\uvec J|\uvec P,m\ra\ra_{q+\overline q,\text{spin}}=\int\ud^3k\sum_{\sigma'\sigma}\tfrac{1}{2}\,\uvec\sigma_{\sigma'\sigma}\,\rho^{m'm}_{\sigma'\sigma,q+\overline q}(\uvec k,\uvec k),
\eeq
and the gluon ``spin-like'' term is defined analogously as
\beq\label{Gspinlike}
\la\la\uvec P,m'|\uvec J|\uvec P,m\ra\ra_{G,\text{spin}}=\int\ud^3k\sum_{\lambda}\uvec\eta(\uvec k)\,\lambda\,\rho^{m'm}_{\lambda\lambda,G}(\uvec k,\uvec k).
\eeq
The ``orbital-like'' terms $(\uvec k_r \times i\uvec \nabla_{\uvec k_r})\, \delta^{(3)}(\uvec k'_r - \uvec k_r)$ in $\langle \uvec k'_r, \sigma'_r|\uvec J|\uvec k_r, \sigma_r  \rangle$, once summed over all the $n$ partons, produce two terms, one of which exactly cancels the term $\left[(\uvec P \times i\uvec\nabla_{\uvec P})\right]_{m'm}\,\delta^{(3)}(\uvec P'-\uvec P)$ in Eq.~\eqref{NuclJ}. The other term produces an expression for the \emph{internal} ``orbital-like'' motion
\beq\label{totOAM}
\begin{split}
\la\la\uvec P,m'|\uvec J|\uvec P,m\ra\ra_\text{orb}&=\sum_n\sum_{\{\sigma_i\}}\int\ud^3k_1\cdots\ud^3k_n\,\delta^{(3)}(\uvec P-\uvec k_1-\cdots-\uvec k_n)\\
&\qquad\times\psi^*_{\uvec P,m'}(\uvec k_1,\sigma_1,\cdots,\uvec k_n,\sigma_n)\sum_r(\uvec k_r\times\tfrac{1}{i}\uvec\nabla_{\uvec k_r})\,\psi_{\uvec P,m}(\uvec k_1,\sigma_1,\cdots,\uvec k_n,\sigma_n).
\end{split}
\eeq
Bakker, Leader and Trueman further decomposed this ``orbital-like'' term into quark and gluon contributions by restricting the sum over partons to those of the corresponding type
\beq\label{partOAM}
\begin{split}
\la\la\uvec P,m'|\uvec J|\uvec P,m\ra\ra_{a,\text{orb}}&=\sum_n\sum_{\{\sigma_i\}}\int\ud^3k_1\cdots\ud^3k_n\,\delta^{(3)}(\uvec P-\uvec k_1-\cdots-\uvec k_n)\\
&\qquad\times\psi^*_{\uvec P,m'}(\uvec k_1,\sigma_1,\cdots,\uvec k_n,\sigma_n)\sum_{r(a)}(\uvec k_r\times\tfrac{1}{i}\uvec\nabla_{\uvec k_r})\,\psi_{\uvec P,m}(\uvec k_1,\sigma_1,\cdots,\uvec k_n,\sigma_n).
\end{split}
\eeq
The problem with this is that it does not represent the angular momentum about  the ``center'' of the nucleon. Indeed, the variables $\uvec k_i$ correspond to the extrinsic and not the intrinsic momenta of the partons. The definition of intrinsic variables, and therefore of intrinsic quark and gluon orbital angular momentum, in relativistic Quantum Mechanics is usually ambiguous, because of the difficulty of defining a unique consistent relativistic version of the center of mass, see \emph{e.g.} \cite{Alba:2006hs,Aguilar:2013} and references therein.
 Only for the \emph{total} orbital angular momentum \eqref{totOAM} does this ambiguity disappear \cite{Lorce:2011kn}.
Note however that  light-front quantization \cite{Brodsky:1997de} offers some simplifications thanks to the Galilean symmetry of the transverse plane and the kinematic nature of the light-front boosts. In this formulation, the nucleon state can be localized in the transverse plane, and one can identify the transverse center of longitudinal momentum with the transverse center of the nucleon \cite{Burkardt:2000za,Burkardt:2002hr,Burkardt:2005hp,Diehl:2005jf}, which is explained in section \ref{secVIIC}. The \emph{intrinsic} quark and gluon longitudinal orbital angular momentum is then naturally defined with respect to this transverse center. Brodsky, Hwang, Ma and Schmidt \cite{Brodsky:2000ii} do not use this transverse center of momentum and conclude that there are only $n-1$ \emph{relative} orbital angular momentum contributions in a $n$-parton Fock state, since there are only $n-1$ independent momenta owing to the delta function $\delta^{(3)}(\uvec P-\uvec k_1-\cdots-\uvec k_n)$. In this case, however, it is not so clear anymore how to attribute a particular contribution to quarks or gluons. Moreover, for $n>2$ vectors there exist infinitely many possible ways of defining $n-1$ relative vectors. The individual relative contributions are therefore totally ambiguous. For a discussion of these different types of orbital angular momentum, see Ref.~\cite{Lorce:2011kn}.

In summary, the decomposition proposed by Bakker, Leader and Trueman reads
\beq
\begin{split}
\tfrac{1}{2}\,\uvec\sigma_{m'm}&=\sum_q\la\la\uvec P,m'|\uvec J|\uvec P,m\ra\ra_{q+\overline q,\text{spin}}+\la\la\uvec P,m'|\uvec J|\uvec P,m\ra\ra_{G,\text{spin}}\\
&\qquad+\sum_q\la\la\uvec P,m'|\uvec J|\uvec P,m\ra\ra_{q+\overline q,\text{orb}}+\la\la\uvec P,m'|\uvec J|\uvec P,m\ra\ra_{G,\text{orb}}.
\end{split}
\eeq
We stress that, in general, the BLT decomposition differs from the various decompositions made at the operator level
\beq\label{comparison}
\begin{split}
\la\la\uvec P,m'|\uvec J|\uvec P,m\ra\ra_{q+\overline q,\text{spin}}&\neq \la\la\uvec P,m'|\uvec S_{q+\overline q}|\uvec P,m\ra\ra,\\
\la\la\uvec P,m'|\uvec J|\uvec P,m\ra\ra_{G,\text{spin}}&\neq \la\la\uvec P,m'|\uvec S_G|\uvec P,m\ra\ra,\\
\la\la\uvec P,m'|\uvec J|\uvec P,m\ra\ra_{q+\overline q,\text{orb}}&\neq \la\la\uvec P,m'|\uvec L_{q+\overline q}|\uvec P,m\ra\ra,\\
\la\la\uvec P,m'|\uvec J|\uvec P,m\ra\ra_{G,\text{orb}}&\neq \la\la\uvec P,m'|\uvec L_G|\uvec P,m\ra\ra.
\end{split}
\eeq
Note however that, in the infinite-momentum frame $P_z\to\infty$ and for the longitudinal component, the BLT decomposition coincides with the Jaffe-Manohar decomposition
\beq
\begin{split}
\lim_{P_z\to\infty}\,\la\la\uvec P,m'|J_z|\uvec P,m\ra\ra_{q+\overline q,\text{spin}}&= \lim_{P_z\to\infty}\,\la\la\uvec P,m'|S^{q+\overline q}_{\text{JM},z}|\uvec P,m\ra\ra,\\
\lim_{P_z\to\infty}\,\la\la\uvec P,m'|J_z|\uvec P,m\ra\ra_{G,\text{spin}}&= \lim_{P_z\to\infty}\,\la\la\uvec P,m'|S^G_{\text{JM},z}|\uvec P,m\ra\ra,\\
\lim_{P_z\to\infty}\,\la\la\uvec P,m'|J_z|\uvec P,m\ra\ra_{q+\overline q,\text{orb}}&= \lim_{P_z\to\infty}\,\la\la\uvec P,m'|L^{q+\overline q}_{\text{JM},z}|\uvec P,m\ra\ra,\\
\lim_{P_z\to\infty}\,\la\la\uvec P,m'|J_z|\uvec P,m\ra\ra_{G,\text{orb}}&= \lim_{P_z\to\infty}\,\la\la\uvec P,m'|L^G_{\text{JM},z}|\uvec P,m\ra\ra,
\end{split}
\eeq
as one can easily conclude from the collinear limit of Eqs. \eqref{Sgen} and \eqref{Lgen}.

For a nucleon moving along $OZ$ with infinite momentum and polarized in the $x$-direction
\begin{equation}\label{eq.S}
|\uvec P,\msc{S}_x\ra =
\tfrac{1}{\sqrt{2}} \left[ |\uvec P, m=+\tfrac{1}{2}\ra +|\uvec P, m=-\tfrac{1}{2}\ra \right],
\end{equation}
the $x$-component of the quark+antiquark ``spin-like'' piece does not coincide with the matrix elements of the spin operator
\beq
\lim_{P_z\to\infty}\,\la\la\uvec P,\msc{S}_x|J_x|\uvec P,\msc{S}_x\ra\ra_{q+\overline q,\text{spin}}\neq\lim_{P_z\to\infty}\,\la\la\uvec P,\msc{S}_x|S^{q+\overline q}_{\text{JM},x}|\uvec P,\msc{S}_x\ra\ra=0,
\eeq
since the transverse component of the spin operator is higher twist, and therefore does not survive in the infinite-momentum limit, as one can easily see from Eq. \eqref{Sgen}. At leading twist, the only transverse component that survives is the one associated with the \emph{transversity} operator, defined for quarks and antiquarks as (note the difference in signs)
\beq
\begin{split}
T^q_j&=\barpsi_q i\sigma^{j0}\gamma_5\psi_q,\\
T^{\overline q}_j&=-\barpsi_{\overline q}\,i\sigma^{j0}\gamma_5\psi_{\overline q},
\end{split}
\eeq
for which one has in general
\beq\label{Tgen}
\la P',m'|\uvec T^{q+\overline q}|P,m \ra =2P^0\, (2\pi)^3 \left[\frac{1}{2}\, \uvec\sigma-\frac{\uvec P\cdot\uvec\sigma}{2P^0(P^0+M)}\,\uvec P\right]_{m'm}\delta^{(3)}(\uvec P'-\uvec P).
\eeq
Considering the infinite-momentum limit, one sees that for the tranvserse component
\beq
\begin{split}
\lim_{P_z\to\infty}\,\la\la\uvec P,\msc{S}_x|J_x|\uvec P,\msc{S}_x\ra\ra_{q+\overline q,\text{spin}}&=\lim_{P_z\to\infty}\,\la\la\uvec P,\msc{S}_x|T^{q+\overline q}_x|\uvec P,\msc{S}_x\ra\ra\\
&=\delta q+\delta\overline q,
\end{split}
\eeq
where the quark and antiquark first moments of the transversity  distribution
\beq
\delta q=\int\ud x\,h^q_1(x),\qquad \delta \overline q=\int\ud x\,h^{\overline q}_1(x)
\eeq
represent the difference between the number density of quarks/antiquarks in a transversely polarized nucleon with polarization parallel or anti-parallel to the nucleon polarization. The gluon ``spin-like'' piece vanishes in the infinite-momentum frame
\beq
\lim_{P_z\to\infty}\,\la\la\uvec P,\msc S_x|J_x|\uvec P,\msc S_x\ra\ra_{G,\text{spin}}=0
\eeq
and comes from the fact that there cannot exist a transversity distribution associated with a massless parton of spin $s>1/2$ in a nucleon, by angular momentum conservation.

Since the same result holds when considering the component $J_y$ with the proton polarized along $OY$, we state the BLT transverse decomposition for a proton with polarization along the generic transverse direction $\uvec s_T$
\beq\label{Tcor}
\tfrac{1}{2}=\sum_q(\delta q+\delta\overline q)+\lim_{P_z\rightarrow \infty} \left[\sum_q\la\la\uvec P,\msc S_T|\uvec J\cdot\uvec s_T|\uvec P,\msc S_T\ra\ra_{q+\overline q,\text{orb}}+\la\la\uvec P,\msc S_T|\uvec J\cdot\uvec s_T|\uvec P,\msc S_T\ra\ra_{G,\text{orb}}\right].
\eeq
A couple of things should be noted in regard to this decomposition:
\begin{enumerate}
\item Based on a charge conjugation symmetry argument, it has been suggested that the BLT transverse decomposition in Eq.~\eqref{Tcor} cannot be correct. Indeed,
%because it involves the \emph{sum} of the transversities, whereas the tensor charge of the nucleon involves their difference. However,
the angular momentum operator is charge conjugation-even, \emph{i.e.} is of the form $J_q+J_{\overline q}$, whereas the tensor charge operator $\barpsi i\sigma^{j+}\gamma_5\psi$ is charge conjugation-odd, \emph{i.e.} gives the \emph{difference} $\delta q-\delta\overline q$. Note that it is actually the charge conjugation-even combination $\delta q+\delta\overline q$ that appears in the BLT transverse decomposition, and comes directly from the Fock expansion of the nucleon state. This means that the ``spin-like'' piece in the transverse case is \emph{not} directly given by the tensor charge operator. This criticism is therefore invalid.
\item The Bakker-Leader-Trueman relation has also been criticized based on the chirality argument. Indeed, the spin is associated with the axial-vector operator $\barpsi\gamma^\mu\gamma_5\psi$ which is chiral-even, \emph{i.e.} the corresponding Dirac matrix satisfies $\{\gamma^\mu\gamma_5,\gamma_5\}=0$. The quantities $\delta q$ and $\delta\overline q$ are associated with the tensor operator $\barpsi i\sigma^{\mu\nu}\gamma_5\psi$ which is chiral-odd, \emph{i.e.} the corresponding Dirac matrix satisfies $[i\sigma^{\mu\nu}\gamma_5,\gamma_5]=0$.  As already stressed in the paper \cite{Bakker:2004ib} and explicitly  stated in Eq. \eqref{comparison}, the definition of the ``orbital-like'' piece does not coincide in general with the matrix element of the orbital angular momentum operator. Similarly, the definition of the ``spin-like'' piece does not coincide with the matrix element of the spin operator. The BLT relation cannot therefore be considered as a direct relation amongst transverse spins.
\item Related to the previous comment, one may also be surprised that the longitudinal component of the ``spin-like'' piece coincides with the expectation value of the axial-vector operator, whereas the transverse component coincides with the expectation value of the tensor operator. This is a signal that the BLT decomposition is not Lorentz covariant. This can be understood as follows: under Lorentz boosts, the spin usually gets mixed with the orbital angular momentum, see Eq. \eqref{Sgen}, but in the definition of the ``spin-like'' piece, see Eqs. \eqref{qspinlike} and \eqref{Gspinlike}, this mixing is not taken into account. The ``spin-like'' piece therefore simply represents the \emph{non-covariant} spin contribution.
\end{enumerate}
Finally, since we know of no way to access the ``orbital-like'' terms experimentally, the BLT relation might only be useful as a testing ground for nucleon models based on instant form wave-functions.

\subsection{The Pauli-Lubanski vector transverse polarization sum rule\label{secVIH}}

We have argued several times in the above that for the case of transverse polarization, it is inevitable that the sum rules contain energy-dependent terms, \emph{i.e.} are not frame-independent. Nonetheless this has disturbed some people and attempts have been made to circumvent this problem. In particular Ji, Xiong and Yuan \cite{Ji:2012vj} (JXY) suggest that the energy dependence can be avoided by considering the transverse component of the Pauli-Lubanski vector
\beq \label{PLV}
W^\mu = \tfrac{1}{2}\,\epsilon^{\mu\nu\rho\sigma}M_{\nu\rho}P_\sigma
\eeq
rather than just the angular momentum itself. However,  Leader \cite{Leader:2012ar} and  Hatta, Tanaka and Yoshida \cite{Hatta:2012jm}
 pointed out some problems, and in particular a missing energy-dependent term in  the JXY results.

Here, and throughout this section, all momentum and angular momentum operators are of the Belinfante type, so we shall omit the labels ``Bel''. Now both the Belinfante momentum and angular momentum operators are \emph{additive} in the sense that
  \beq \label{add}
P^\mu =\sum_q P^\mu_q + P^\mu_G  \qquad \textrm{and} \qquad M^{\mu\nu} =\sum_q M^{\mu\nu} _q + M^{\mu\nu}_G.
\eeq
Clearly, then, the Pauli-Lubanski vector is \emph{not} additive
\begin{align}
W^\mu &= \tfrac{1}{2}\, \epsilon^{\mu\nu\rho\sigma}\left[\sum_qM_{\nu\rho}^q + M_{\nu\rho}^G \right]  \left[ \sum_q P_\sigma^q + P_\sigma^G \right] \nn \\
& \neq  \sum_q W_q^\mu  + W_G^\mu. \label{notadd}
\end{align}
In order to work with additive quantities, JXY \emph{define} operators which they call $W^\mu(\text{quark})$ and $W^\mu(\text{gluon})$, but which are not really the quark and gluon Pauli-Lubanski vectors since they contain the \emph{total} momentum operators:
 \beq \label{PLq}
W^\mu_\text{JXY}(\textrm{quark})=\tfrac{1}{2}\,\epsilon^{\mu\nu\rho\sigma}M_{\nu\rho}^qP_\sigma
\eeq
and
\beq \label{PLG}
W^\mu_\text{JXY}(\textrm{gluon})=\tfrac{1}{2}\, \epsilon^{\mu\nu\rho\sigma}M_{\nu\rho}^GP_\sigma,
\eeq
so that
\beq \label{Wconts}
W^\mu = \sum_qW^\mu_\text{JXY}(\text{quark}) + W^\mu_\text{JXY}(\text{gluon}).
\eeq
The question then is what is the physical interpretation of these operators? It is known on general grounds that the expectation value of the \emph{total} Pauli-Lubanski vector has a very simple and direct meaning in terms of the covariant spin vector $\msc{S}$, defined in Eq.~\eqref{S}, and used in specifying the state (see section $\bm{3.4}$ of \cite{Leader:2001gr}), namely, for a  spin-$1/2$ particle\footnote{An analogous relation holds for arbitrary spin: see Ref. \cite{Leader:2001gr}.}
\beq \label{ExPL}
\la\la P, \msc{S} | W^\mu| P,  \msc{S} \ra\ra = \tfrac{1}{2}\, \msc{S}^\mu.
\eeq
Clearly, then, for example,
\beq
\la\la P, \msc{S} |W^\mu_\text{JXY}(\text{quark})| P,  \msc{S} \ra\ra \neq \tfrac{1}{2}\, \msc{S}_q^\mu
\eeq
but
\begin{align}
\la\la P, \msc{S} |W^\mu _\text{JXY}(\text{quark}) | P,  \msc{S} \ra\ra& \equiv  \tfrac{1}{2}\, \msc{S}^\mu_q (\text{nucleon})\nn \\
& \equiv \text{contribution of $q$ to $\tfrac{1}{2}\, \msc{S}^\mu(\text{nucleon})$}.
\end{align}
Thus from Eq. \eqref{Wconts}
\beq
\msc{S}^\mu(\text{nucleon})= \sum_q\msc{S}^\mu_q (\text{nucleon})+ \msc{S}^\mu_G(\text{nucleon}).
\eeq

Let us now calculate the expectation value of the transverse component of the \emph{total} Pauli-Lubanski vector for a nucleon moving along $OZ$ with
$P^\mu=(P^0,0,0,P_z)$. One obtains
\beq \label{Wxdef}
W_x = W^1 = P^0\,M^{23} + P_z \, M^{02}.
\eeq
For $M^{23}$ we may use the expressions in Eqs.~\eqref{Jfinal} and \eqref{JFF} relating it to the matrix elements of the energy-momentum tensor. For $M^{02}$ we need to slightly modify the treatment that led to Eq.~\eqref{Jfinal}. From the first line of Eq.~\eqref{AM} we have
\begin{align}
\la P, \mathscr{S}  | M^{0k}  |  P, \mathscr{S}  \ra &= \lim_{\uvec\Delta\rightarrow \uvec 0}  \int \ud^3x \, e^{-i\uvec\Delta\cdot\uvec x} \,\langle P + \tfrac{\Delta}{2}, \mathscr{S}_f  |  x^0  T^{0k}(0)- x^k  T^{00}(0)| P - \tfrac{\Delta}{2}, \mathscr{S}_i \rangle \nn \\
&=\lim_{\uvec \Delta\rightarrow \uvec 0} \left[t   \int \ud^3x \, e^{-i\uvec\Delta\cdot\uvec x} \,\langle  P + \tfrac{\Delta}{2}, \mathscr{S}_f  |  T^{0k}(0) | P - \tfrac{\Delta}{2}, \mathscr{S}_i  \rangle \right. \nn \\
&\qquad\qquad \left. -  \int \ud^3x \, e^{-i\uvec\Delta\cdot\uvec x} \,\langle  P + \tfrac{\Delta}{2}, \mathscr{S}_f  |x^k T^{00}(0)| P - \tfrac{\Delta}{2}, \mathscr{S}_i \rangle \right].\label{Boost}
 \end{align}
The first term gives, \emph{via} Eq.~\eqref{Tmunu}
\beq
(2\pi)^3\, \delta^{(3)}(\uvec 0)\,  t \, \langle P    |  T^{0k}(0)| P   \rangle = (2\pi)^3\, \delta^{(3)}(\uvec 0)\, t \,2A\,P^0P^k =0 \qquad \text{for} \qquad k=1,2.
\eeq
Thus $ \la  P, \mathscr{S}  | M^{02}  |  P, \mathscr{S}  \ra $ is given entirely by the second term in Eq.~\eqref{Boost}, and using arguments analogous to those used in deriving Eq.~\eqref{JFF}, one obtains
\beq \label{BoostFF}
\la \la P, \mathscr{S} | M^{02}  |  P, \mathscr{S} \ra \ra=\frac{1}{2P^0} \left[i\frac{\partial}{\partial \Delta^2}\langle P + \tfrac{\Delta}{2}, \mathscr{S}_f  | T^{00}(0) | P - \tfrac{\Delta}{2}, \mathscr{S}_i \rangle \right]_{\uvec\Delta=\uvec 0}.
\eeq
Substituting the expression in Eq.~\eqref{Tmunu} for the matrix element of $T^{00}$ yields
\beq \label{PM02}
\la \la  P, \mathscr{S}  |P_z\,  M^{02} |  P, \mathscr{S}  \ra \ra =  P_z\, \la \la  P,  \mathscr{S}  | M^{02}  |  P, \mathscr{S}  \ra \ra  = \left[ \frac{M -P^0}{2M}\,A -\frac{P^2_z}{2M^2}\,B\right]\msc S_x.
\eeq
Note that there is no $\bar{C}$ term since we are dealing here with the total, conserved Pauli-Lubanski vector. Similarly, we find
\beq \label{P0M23}
\la \la  P, \mathscr{S}  |P^0\, M^{23}  |  P, \mathscr{S}  \ra \ra =  P^0\, \la \la  P, \mathscr{S} | M^{23}  |  P, \mathscr{S} \ra \ra = \left[\frac{P_0}{2M}\,A+ \frac{(P^0)^2}{2M^2}\,B\right]\msc S_x,
\eeq
so that finally, using Eq.~\eqref{Wxdef}
\beq \label{Wexptot}
\la \la  P, \mathscr{S}  |W_x |  P, \mathscr{S}  \ra \ra =\tfrac{1}{2}\left(A+B\right)\msc S_x.
\eeq
Now we showed in section \ref{secVID} that $A=1$ and $B=0$, so that, as expected on general grounds,
\beq \label{checkW}
\la \la  P, \mathscr{S}  |W_x |  P, \mathscr{S}  \ra \ra  = \tfrac{1}{2}\,\msc{S}_x
\eeq
in agreement with the general result in Eq.~\eqref{ExPL}.

Repeating the above calculation for the quark part alone, \emph{i.e.} for $\langle\langle P, \mathcal{S} | W^\text{JXY}_x(\text{quark})| P, \mathcal{S}  \rangle\rangle $, one gets
\beq \label{WJXYq}
\langle\langle P, \mathcal{S}|  W^\text{JXY}_x(\text{quark}) | P, \mathcal{S} \rangle\rangle = \left[\frac{1}{2}\left(A_q + B_q\right) + \frac{P^0 - M}{2P^0}\, \bar{C}_q \right] \msc{S}_x.
\eeq
Note that the term $\bar{C}^q$ appears here because $M^{\mu\nu}_q$ is not a conserved operator. A similar relation holds for the gluon contribution. It follows from Eqs.~\eqref{Hsum} and \eqref{Esum} and the analogous relations for gluons that\footnote{Note that the coefficients of $\bar{C}_q$ and $\bar{C}_G$ {obtained by Hatta, Tanaka and Yoshida} in Ref. \cite{Hatta:2012jm} are slightly different, because their treatment uses operators defined at light-front time $x^+$ rather than the instant form used here. In Ref. \cite{Harindranath:2013goa}, Harindranath, Kundu and Mukherjee obtained yet another result in the light-front form, by using light-front spinors instead of the standard instant-form ones. Note however that in the infinite-momentum frame $P_z\to\infty$, instant and light-front form coincide, and so do all the three results.}
\beq \label{SumSq}
\mathcal{S}^q_x(\text{nucleon})=  \frac{1}{2}\int_{-1}^{1} \ud x\, x\left[ H_q(x,0,0) +E_q(x,0,0)\right] + \frac{P^0 - M}{2P^0}\, \bar{C}_q
\eeq
and
\beq \label{SumSG}
\mathcal{S}^G_x(\text{nucleon})=  \frac{1}{2}\int_0^{1} \ud x\, x\left[ H_q(x,0,0) +E_q(x,0,0)\right]  + \frac{P^0 - M}{2P^0}\, \bar{C}_G,
\eeq
which disagrees with the relations given in Ref. \cite{Ji:2012vj}, where the energy-dependent terms involving $\bar{C}_q$ and $\bar{C}_G$ are missing.
Of course
\beq \label{additive}
M^{\mu\nu} = \sum_qM^{\mu\nu}_q + M^{\mu\nu}_G
\eeq
is a conserved operator so that
\beq \label{sumC}
\sum_q\bar{C}_q + \bar{C}_G =0,
\eeq
but that does not justify ignoring the $\bar{C}$ terms in Eqs.~\eqref{SumSq} and \eqref{SumSG} for the individual quark and gluon contributions. Since the individual quark and gluon angular momenta are definitely not conserved, it is certain that neither $\bar{C}_q$ nor $\bar{C}_G$ can be zero.

We shall now show that $\bar{C}_q $ can be expressed in terms of the higher-twist GPDs defined in Ref. \cite{Meissner:2009ww}. The analogue of Eq.~\eqref{GPD} with $\gamma^+ $ replaced by $\gamma^- $ on the left-hand side is
\begin{align}\label{GPD'}
\frac{1}{2}\int \frac{\ud z^-}{2\pi}\, e^{i x \bar{P}^+ z^-} \,\langle  P', \msc{S}' | \barpsi(-\tfrac{z^-}{2})&\gamma^- \mathcal W\psi (\tfrac{z^-}{2})  |  P,  \msc{S} \rangle \nn \\
&= \frac{M^2}{2 (\bar{P}^+)^3}\,\overline{u}(P',\mathscr S')\left[\gamma^+  H^q_3(x, \xi, t) +\tfrac{i\sigma^{+ \rho}\Delta_\rho}{2M}\, E^q_3(x,\xi, t) \right]u(P,\mathscr S),
\end{align}
and, multiplying by $x$ and integrating leads to the analogue of Eq.~\eqref{intGPD}
 \beq \label{intGPD'}
\frac{M^2}{2 (\bar{P}^+)^3}   \left[(\overline{u}'\gamma^+ u)\int \ud x \,x\, H^q_3(x, \xi, t)+\left( \overline{u}'\tfrac{i\sigma^{+ \rho}\Delta_\rho}{2M} u \right) \int \ud x\, x\, E^q_3(x,\xi, t) \right]= \frac{1}{4(\bar{P}^+)^2}\,\langle  P' ,\msc{S}' | \barpsi(0)  \gamma^- i\LRD^+ \psi (0)  |  P, \msc{S}\rangle,
\eeq
which can also be written in the form
 \beq \label{intGPDnew}
\bar P^+(\overline{u}' u)\int \ud x \,x\, H^q_3(x, \xi, t)+\left( \overline{u}'\tfrac{i\sigma^{+ \rho}\Delta_\rho}{2} u \right) \int \ud x\, x\left[ H^q_3(x, \xi, t)+ E^q_3(x,\xi, t)\right] = \frac{\bar P^+}{M}\,\langle  P' ,\msc{S}' |\tfrac{1}{2} \, \barpsi(0)  \gamma^- i\LRD^+ \psi (0)  |  P, \msc{S}\rangle.
\eeq
Consider now the structure of the matrix element on the RHS. Similarly to Eq.~\eqref{TMN}, we may write
\begin{multline} \label{minuplus}
\langle  P' ,\msc{S}' | \tfrac{1}{2}\,\barpsi(0)  \gamma^\mu  i\LRD^\nu \psi (0)  |  P , \msc{S }\rangle =2\left(A_q\, P^\mu P^\nu +  M^2\bar{C}_q\, g^{\mu\nu}\right)
+\frac{i\Delta_\rho}{M}\left[D_q\,\frac{P^\mu \epsilon^{\rho\nu\alpha\beta}- P^\nu \epsilon^{\rho\mu\alpha\beta}}{2}\right.\\
\left.+(A_q+ B_q)\,\frac{P^\mu \epsilon^{\rho\nu\alpha\beta} + P^\nu \epsilon^{\rho\mu\alpha\beta}}{2}+ \frac{A_q\,P^\mu P^\nu + M^2 \bar{C}_q \, g^{\mu\nu}}{P^0 + M}\,\epsilon^{0\rho\alpha\beta}\right] \frac{\mathscr{S}_\alpha}{M} \,P_\beta + \mathcal O(\uvec\Delta^2),
\end{multline}
where we have included a term $D_q$ antisymmetric under $\mu \leftrightarrow \nu$, not present in the symmetric $T^{\mu\nu}_{\text{Bel},q}$ \cite{Bakker:2004ib}. Note that the coefficients $A_q$, $B_q$ and $\bar{C}_q$ are the same as in the matrix element for $T^{\mu\nu}_{\text{Bel},q}$ because
\beq
T^{\mu\nu}_{\text{Bel},q} = \tfrac{1}{4}\,\barpsi(0)  \gamma^\mu  i\LRD^\nu \psi (0) + ( \mu \leftrightarrow \nu).
\eeq
After some algebra, one obtains
\beq
\langle P'  ,\msc{S}' | \tfrac{1}{2}\,\barpsi(0) \gamma^- i\LRD^+ \psi (0)  |  P , \msc{S }\rangle =  (\overline u'u) \left(\frac{M}{2}\,A_q + M\bar{C}_q\right) +\left( \overline{u}'\tfrac{i\sigma^{+ \rho}\Delta_\rho}{2} u \right) \frac{M}{P^+}\,D_q+ \mathcal O(\Delta^2) .
\eeq
Putting this into Eq.~\eqref{intGPDnew} and taking the limit $\Delta\rightarrow 0 $, we see that
\beq
\tfrac{1}{2}\,A_q + \bar{C}_q = \int \ud x\, x\,H_3^q(x, 0, 0).
\eeq
Inserting the expression for $A_q$ from Eq.~\eqref{Hsum}, we obtain\footnote{Note that the expression for $\bar{C}_q$ was given incorrectly in Ref. \cite{Leader:2012ar}.}
\beq \label{cbarfinal}
\bar{C}_q = \int \ud x\, x\left[H_3^q(x, 0, 0)- \tfrac{1}{2}\, H_q(x,0,0)\right].
\eeq
Finally, Eq.~\eqref{SumSq} becomes
\beq \label{SumSqfinal}
\mathcal{S}^q_x(\text{nucleon}) =  \frac{1}{2}\left[ \frac{P^0 + M}{2P^0}\int_{-1}^{1} \ud x\, x\, H_q(x,0,0) +  \int_{-1}^{1} \ud x\, x\, E_q(x,0,0) + \frac{P^0-M}{P^0}\int_{-1}^{1} \ud x\, x\, H_3^q(x,0,0)\right].
\eeq
The claim of Ji, Xiong and Yuan that they have produced a frame-independent relation is not valid, since they have discarded the annoying $\bar C$ term in their discussion. In a reply \cite{Ji:2013tva} to a critical comment \cite{PhysRevLett.111.039101} about this, they stress that the frame independence of their result remains true for the leading-twist part. This statement is a bit misleading, since it gives the impression that one can simply throw away the $\bar C$ term based on the twist expansion argument. This is not the case because the leading-twist result corresponds to the infinite-momentum limit of Eq. \eqref{WJXYq}, where the $\bar C$ term survives. In fact, using the light-front formalism instead of the instant form used here, and keeping all the terms, the result turns out to be frame independent, but still includes a $\bar C$ term \cite{Harindranath:2013goa}. In conclusion the JXY result, where the $\bar C$ term is missing, is correct only at $P^0=M$ in instant form, \emph{i.e.} \emph{in the rest frame} !

\section{Orbital angular momentum and the spin crisis in the parton model\label{secVII}}

Although Jaffe and Manohar only gave a formal derivation of their angular momentum sum rule in 1990 \cite{Jaffe:1989jz}, its structure is so intuitive that it had been used for many years in the form (for a proton with helicity $+1/2
$)
\beq \label{JMnaive}
\tfrac{1}{2}= \la \la S^q_z  \ra \ra + \la \la  S^G_z \ra \ra  + \la \la  L_z  \ra \ra .
\eeq
With $q_{\pm}(x) $ the number density of quarks with spin parallel/antiparallel to the spin of the proton, one has
\begin{align} \label{qspin}
\la \la  S^q_z  \ra \ra &=  \tfrac{1}{2}\sum_q\int^1_{-1} \ud x \left[q_+(x) -q_-(x)\right]\nn \\
&= \tfrac{1}{2}\sum_q\int^1_0 \ud x\left[\Delta q(x)+\Delta \overline{q}(x)\right]\nn \\
&\equiv \tfrac{1}{2} \,\int^1_0 \ud x\,\Delta \Sigma(x),
\end{align}
and analogously for the gluons
\beq \label{Gspin}
\la \la  S^G_z  \ra \ra =  \int^1_0 \ud x \left[ G_+(x) -G_-(x)\right]  =  \int^1_0 \ud x\,\Delta G(x).\eeq
Further, on the grounds that fairly successful models of the nucleon and its excited states utilised just three (constituent) quarks in an $s$-wave ground state and no gluons\footnote{Recall that one of the arguments for the need for color is based on the assumption that the wave function of the $N^*(1238)$, in the state $I_z=3/2$, $S_z=3/2$,  consists of 3 ``up'' quarks all with spin $S_z=1/2$, \emph{i.e.} has a symmetric wave function.}, it was naively assumed that gluons and orbital angular momentum were negligible so that Eq.~\eqref{JMnaive} was written
\beq \label{JMvnaive}
\tfrac{1}{2}= \tfrac{1}{2} \int^1_0 \ud x\,\Delta \Sigma(x).
\eeq
The famous European Muon Collaboration experiment on polarized deep-inelastic lepton-hadron scattering \cite{Ashman:1987hv}, in 1988,  which obtained $\int^1_0 \ud x\,\Delta \Sigma(x) \approx 0.06 $, showed that the equality  Eq.~\eqref{JMvnaive} was severely  broken and led to what was called a ``spin crisis in the parton model'' \cite{Leader:1988vd}. More recent analyses based on more statistics now indicate that $\int^1_0 \ud x\,\Delta \Sigma(x) \approx 0.3$ \cite{Ageev:2005gh,Alexakhin:2006vx,Airapetian:2007mh}, which is still much too low. Clearly the argument leading to Eq.~\eqref{JMvnaive} is, in retrospect, very misleading and the other terms in Eq.~\eqref{JMnaive} must be taken into consideration. $\Delta G(x)$ can be measured in various kinds of experiments, but its value has only been determined at a limited number of values of $x$. It is found to be very small, so that its first moment is likely to be small (for information about $\Delta G(x)$, see Refs. \cite{Kuhn:2008sy} and \cite{Leader:2010rb}). It  seems, therefore, that the parton orbital angular momentum (OAM) is important, and we shall here consider what is known about it. Just as there exist several versions of $\uvec J$, so, clearly, there will be a different expressions for the corresponding versions of the OAM $\uvec L$.

\subsection{Expressions for the Ji version of $\uvec L $ and $\uvec J$\label{secVIIA}}

We stress once again that the Eqs. \eqref{Jzsum}, \eqref{LsumL} and \eqref{LsumT} are not strictly speaking sum rules, since the LHS are not obtained independently but are \emph{expressed} by the RHS. In order to obtain a genuine sum rule, one  needs an alternative expression for the kinetic orbital angular momentum, which we shall discuss presently. However, Eqs. \eqref{Jzsum} and \eqref{LsumL} are extremely interesting since they provide a way to evaluate the Ji angular momentum and orbital angular momentum  carried by quarks of a given flavor, in terms of, in principle, measurable GPDs. To date, however, such an experimental evaluation has not been possible, but there have been several lattice calculations, directly, of the matrix elements of $J^q_{\text{Bel},z} $ and $J^G_{\text{Bel},z}$ to which we now turn.

\subsubsection{Lattice calculations of $J^q_{\text{Bel},z} $ and $L^q_{\text{Bel},z}$\label{secVIIA1}}

There are two steps. Firstly, the expectation value of $J^q_{\text{Bel},z}$ is written in terms of $A_q$ and $B_q$, as given in Eq.~\eqref{Jqexp}, which in lattice papers are written as $T_1(0)_q$ and $T_2(0)_q$, respectively. Secondly, $A_q$ and $B_q$ are evaluated on a lattice \emph{via} their relation \eqref{Tmunu} to the matrix elements of $T^{\mu\nu}_\text{Bel}$, and similarly for the gluon. The quark spin term is either evaluated separately on the lattice or taken from experiments on polarized deep inelastic scattering.

\begin{figure}[h!]
  \centering
  {\rotatebox{0}%
    {\includegraphics[width=.9\textwidth]{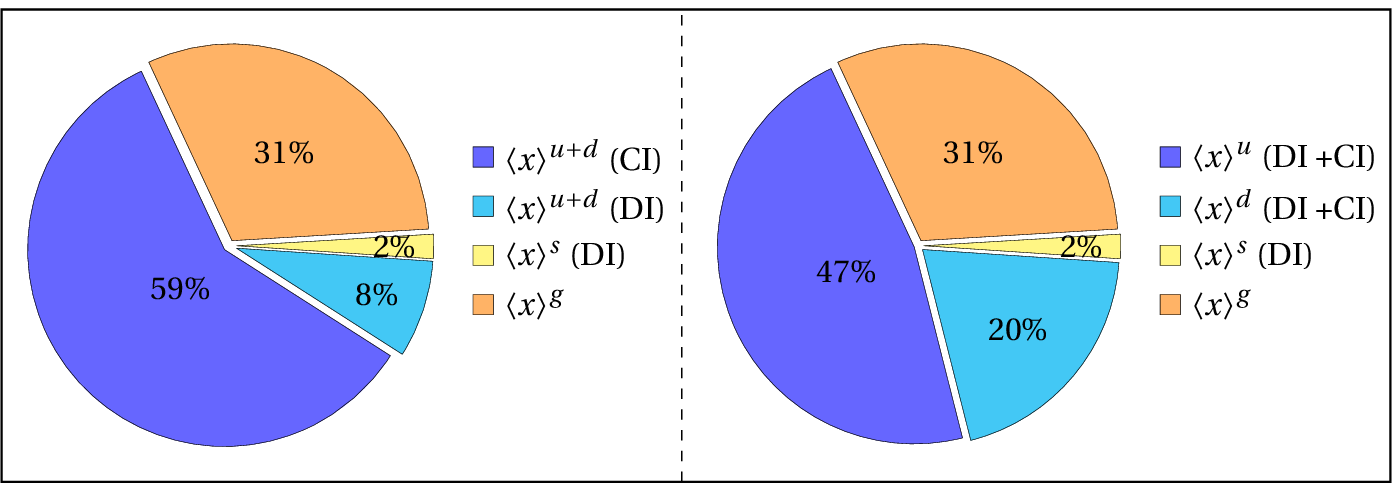}}
  }
  \caption{Preliminary results for the quark and gluon contributions to the proton momentum, presented at Lattice 2013. Courtesy of Keh-Fei Liu.}
  \label{fig:pie_diag_first_mom}
\end{figure}

\begin{figure}[h!]
  \centering
  {\rotatebox{0}%
    {\includegraphics[width=.9\textwidth]{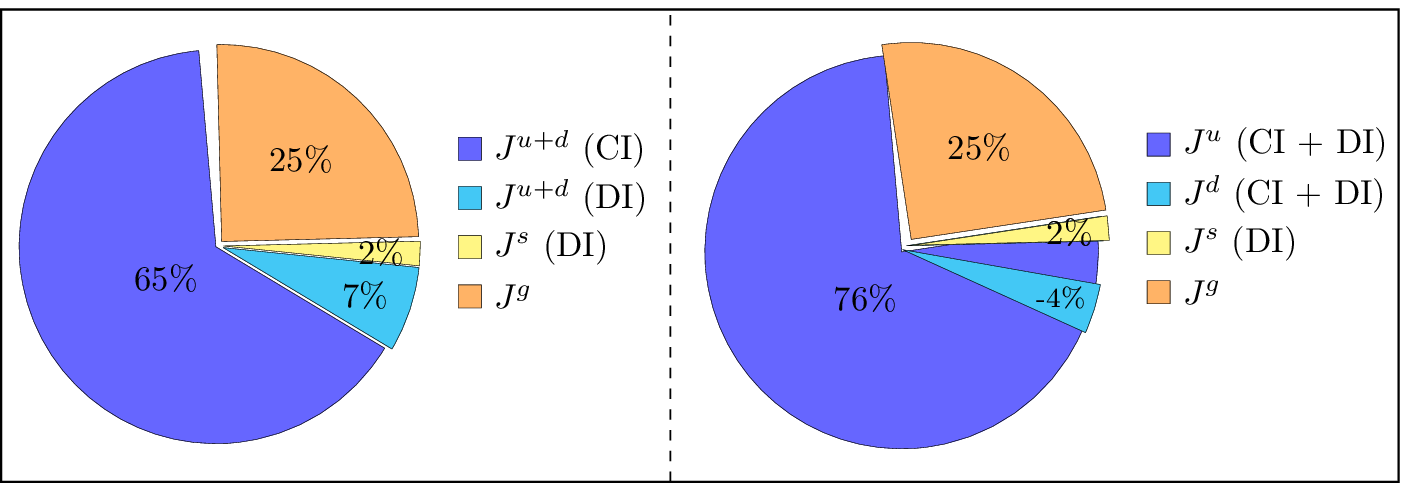}}
  }
  \caption{Preliminary results for the quark and gluon contributions to the angular momentum, presented at Lattice 2013. Courtesy of Keh-Fei Liu.}
  \label{fig:pie_diag_am}
\end{figure}

\begin{figure}[h!]
  \centering
  {\rotatebox{0}%
    {\includegraphics[width=.9\textwidth]{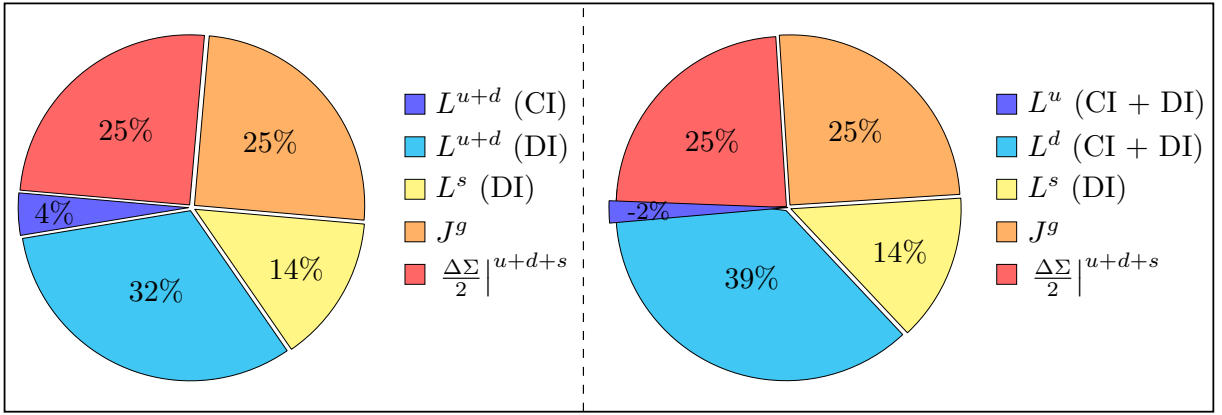}}
  }
  \caption{Preliminary results for the quark and gluon contributions to the orbital angular momentum, presented at Lattice 2013. Courtesy of Keh-Fei Liu.}
  \label{fig:pie_diag_orb_am}
\end{figure}

It is difficult, at present, to assess the reliability of the results. Different forms of the lattice action are used. There are two kinds of diagrams taken into account: so-called connected insertions (CI) corresponding to connected graphs, where the current is inserted on a valence quark line, and disconnected insertions (DI), where the current is inserted on a disconnected quark loop. Further, there is the quenched approximation, which ignores the fermion determinant which comes from vacuum fluctuation quark loops, and the dynamical fermion treatment which includes the latter. Dynamical fermions and DI contributions  are much more difficult to treat on a lattice, and were ignored in early calculations. Nonetheless, all calculations of the orbital angular momentum, based on subtracting the spin term from the total angular momentum, gave results for $\la\la   P,\mathscr{S}_L | L^q_{\text{Bel},z}  | P, \mathscr{S}_L  \ra \ra $, abbreviated as $L_q$ in the lattice papers, which had $L_u $ negative and $L_d$ positive. This was at first sight surprising since, to the best of our knowledge, relativistic constituent quark models, with the exception of the chiral quark-soliton model \cite{Lorce:2011kd}, yield the opposite result. An interesting attempt to reconcile these results was made by Thomas \cite{Thomas:2008ga} who argued that the constituent values evolved to the scales used in lattice calculations, were roughly compatible with the lattice results. This is however not supported by a similar analysis made by Wakamatsu \cite{Wakamatsu:2009gx} which led to some comments \cite{Thomas:2010zz,Wakamatsu:2010zza}. Studies using only CI contributions give values for $\la\la   P,\mathscr{S}_L | J^q_{\text{Bel},z}  | P, \mathscr{S}_L  \ra \ra$ (abbreviated as $J_q$) and $L_q$   like
 \beq \label{CILatt}
J_u \approx 0.4 - 0.5  \quad  | J_d | \leq 10\% \qquad L_{u+d} \approx 0 ,
\eeq
and thus require a relatively large gluon angular momentum contribution. It turns out that the DI contributions are large and significantly change the above results, as can be seen in Table~\ref{tab:chiral}.
\begin{table}[h]
\begin{center}
 \caption{Preliminary CI and DI values for the momentum fraction $\la x\ra$, $2J$, $a_0$ and $2L$ for $u$, $d$ and $s$ quarks, and $\la x\ra$ and $2J$ for gluons, in quenched approximation (from Liu \emph{et. al.} \cite{Liu:2012nz}). The conversion to $\overline{MS}$ renormalization has not yet been published, but is believed to alter the results by just a few percent.}
\begin{tabular}{@{\quad}c@{\quad}|@{\quad}c@{\quad}c@{\quad}|@{\quad}c@{\quad}c@{\quad}c@{\quad}c@{\quad}}\whline
    &CI($u$) & CI($d$)  & CI($u+d$) &  DI($u/d$) & DI($s$) & Glue \\\hline
$\langle x \rangle$ & 0.428(40)  &  0.156(20) & 0.586(45) & 0.038(7) & 0.024(6) & 0.313(56) \\
$2J$ &  0.726(128)  & -0.072(82) & 0.651(51) & 0.036(7) & 0.023(7) & 0.254(76)\\
$a_0$ &  0.91(11)  & -0.30(12)   & 0.61(8)  &  -0.12(1)  &  -0.12(1) & --\\
$2L$ &  -0.18(18)    &  0.23(14)   &  0.04(10)  &  0.16(2)  &  0.14(2) & --\\
   \whline
  \end{tabular}
  \label{tab:chiral}
\end{center}
\end{table}
The value of $L_u$ is still marginally negative and $L_d \approx 0.2$. Figures \ref{fig:pie_diag_first_mom}, \ref{fig:pie_diag_am} and \ref{fig:pie_diag_orb_am} give a  beautiful pictorial representation of how the various terms contribute to the momentum and angular momentum of the proton\footnote{ These diagrams were presented at Lattice 2013 by Syritsyn, albeit with an incorrect date attached.}
These figures give very interesting information as to how the proton spin is built up, in the Ji decomposition, from the contributions of its constituents. The values for the orbital angular momentum given here should be compared with the model calculations labeled $\ell^q_{\text{kin},z}$ in Table \ref{OAMtable} in section \ref{secVIIC}. However, it will be of far more fundamental interest when these numbers can be compared with the experimentally measured GPDs appearing in the relations in sections \ref{secVIF} and \ref{secVIIIC}.

\subsubsection{Lattice calculation of $J^q_{\text{Bel},T}$\label{secVIIA2}}

It should be noted that once $A_q$ and $B_q$ or, equivalently $T_1(0)_q$ and $T_2(0)_q$, have been evaluated from lattice studies of the longitudinal case, the Leader relation given by Eq.~\eqref{JqHET} for the transverse case can be used as a test of the consistency of the theory. Thus a direct evaluation on a lattice of
\beq
J_{q,T} \equiv \la\la   P,\mathscr{S}_T | J^q_{\text{Bel},T}  | P, \mathscr{S}_T  \ra \ra
\eeq
for a transversely polarized proton moving with a few different values of momentum, could be compared with the expected behavior
\beq
J_{q,T}  =\tfrac{1}{2}\left(A_q+\tfrac{P^0}{M}\,B_q\right).
\eeq

\subsubsection{Evaluation of $L^q_{\text{Ji},z}$ in a longitudinally polarized nucleon, from GPDs\label{secVIIA3}}

We shall now show how  the light-front version of $L^q_{\text{Ji},z}$  can be related to the    twist-3 GPDs defined according to the parametrization of Kiptily and Polyakov \cite{Kiptily:2002nx}, namely
\begin{multline} \label{Tw3GPD}
\frac{1}{2}\int \frac{\ud z^-}{2\pi}\,e^{i x \bar{P}^+ z^-} \langle P', \msc{S}' | \barpsi(-\tfrac{z^-}{2}) \gamma^j \mathcal W\psi (\tfrac{z^-}{2})  |  P,  \msc{S} \rangle\\
= \frac{1}{2 \bar{P}^+}\,\overline{u}(P',\mathscr S')\left[\frac{\Delta^j_\perp}{2M}\,G^q_1+\gamma^j\left(H_q+E_q+G^q_2\right)+\frac{\Delta^j_\perp\gamma^+}{\bar P^+}\,G^q_3+\frac{i\epsilon^{jk}_T\Delta^k_\perp\gamma^+\gamma_5}{\bar P^+}\,G^q_4 \right]u(P,\mathscr S),
\end{multline}
where the nucleon is moving in the $z$-direction, $\uvec\Delta_\perp$ is the transverse part of $\Delta^\mu$, and the indices $j,k=1,2$ are transverse indices. Analogously to Eq.~\eqref{intGPD}, if we multiply by $x$ and integrate, one obtains
\begin{align} \label{SecMom}
\langle  P' ,\msc{S}'| \tfrac{1}{2}\,\barpsi(0)  \gamma^j  i\LRD^+ \psi (0) | P , \msc{S } \rangle &=  \bar P^+ \,\overline{u}(P',\mathscr S')\left[\frac{\Delta^j_\perp}{2M}\int \ud x \, x\, G^q_1+\gamma^j \int\ud x \, x\left(H_q+E_q+G^q_2\right)\right.  \nn  \\
&\qquad\qquad\qquad\qquad+ \left.\frac{\Delta^j_\perp\gamma^+}{\bar P^+}\int \ud x \, x\,G^q_3+\frac{i\epsilon^{jk}_T\Delta^k_\perp\gamma^+\gamma_5}{\bar P^+}\int \ud x \, x\,G^q_4 \right]u(P,\mathscr S).
\end{align}
Now analogously to Eq.~\eqref{TMN}, one can write for $\mu \neq \nu$,
\begin{align}
\langle  P ',\msc{S}'|\tfrac{1}{2}\,\barpsi(0)  \gamma^\mu i\LRD^\nu \psi (0) | P , \msc{S } \rangle&=\overline{u}(P',\mathscr S')\left[\frac{\bar{P}^{\{\mu}\gamma^{\nu\}}}{2} \,A_q(\Delta^2) +\frac{\bar{P}^{\{\mu}i\sigma^{\nu\}\rho}\Delta_\rho}{4M}\,B_q(\Delta^2)\right. \nn \\
&\left.+\frac{\Delta^\mu \Delta^\nu - \Delta^2 g^{\mu\nu}}{M} \,C_q(\Delta^2)+Mg^{\mu\nu} \bar{C}_q(\Delta^2)+\frac{\bar{P}^{[\mu}\gamma^{\nu]}}{2} \,D_q(\Delta^2)\right]u(P,\mathscr S),
\end{align}
with the notations $a^{\{\mu}b^{\nu\}}=a^\mu b^\nu+a^\nu b^\mu$ and $a^{[\mu}b^{\nu]}=a^\mu b^\nu-a^\nu b^\mu$. Remembering that $\bar P^\mu=[\bar P^+,\bar P^-,\uvec 0_\perp]$, one obtains for $\mu = j$, $\nu = +$
\beq \label{anotherj}
\langle  P ',\msc{S}'| \tfrac{1}{2}\,\barpsi(0)  \gamma^j i\LRD^+ \psi (0) | P , \msc{S } \rangle = \overline{u}(P',\mathscr S') \left\{\frac{\gamma^j \bar{P}^+}{2}\left[A_q(\Delta^2) + B_q(\Delta^2) - D_q(\Delta^2)\right]+\frac{\Delta^+\Delta^j}{M} \, C_q(\Delta^2) \right\}u(P,\mathscr S),
\eeq
where we have used $\overline{u}'\frac{i\sigma^{j\rho}\Delta_\rho}{2M}\,u= \overline{u}'\gamma^ju$. Comparing with Eq.~\eqref{SecMom} and taking the limit $ \Delta^\mu \rightarrow 0$, we see that
\beq \label{SumSecMon}
\int \ud x \, x\,\left[H_q(x,0,0) +E_q(x,0,0)+G^q_2(x,0,0) \right] = \tfrac{1}{2}\left(A_q + B_q- D_q\right),
\eeq
so that
\beq \label{intG}
- \int \ud x \, x\,G^q_2(x,0,0)=  \la\la  P, \mathscr{S}_L  | J^q_{\text{Bel},z}  |  P, \mathscr{S}_L \ra \ra  +\tfrac{1}{2}\, D_q.
\eeq

Consider now the expression for the orbital angular momentum. From Eq.~\eqref{Bel4}
\beq
\langle  P ,\msc{S}_L | L^q_{\text{Ji},z} |  P , \msc{S }_L \rangle =   \langle  P ,\msc{S}_L | \int \ud x^- \ud^2x_\perp \left[\tfrac{1}{2}\,\barpsi(0)  \gamma^+  x^1i\LRD^2 \psi (0) - ( 1 \leftrightarrow 2) \right]| P , \msc{S }_L \rangle,
\eeq
so that analogously to Eq.~\eqref{JFF} we will have
\beq \label{LGPD}
\la\la  P, \mathscr{S}_L  | L^q_{\text{Bel},z}  |  P, \mathscr{S}_L \ra \ra=\frac{1}{2P^+}\left[-i\, \frac{\partial}{\partial \Delta^1}\langle  P + \tfrac{\Delta}{2}, \mathscr{S}_{Lf} |  \tfrac{1}{2}\,\barpsi(0)  \gamma^+  i\LRD^2 \psi (0)  | P - \tfrac{\Delta}{2}, \mathscr{S}_{Li}  \rangle    - (1\leftrightarrow 2)\right]_{\uvec\Delta= \uvec 0},
\eeq
where we have allowed for the fact that for light-front states are normalized so that
\beq
\la  P'  |  P  \ra = 2 P^+ \, (2\pi)^3 \, \delta(P'^+ - P^+)\, \delta ^{(2)} ( \uvec P'_\perp -\uvec P_\perp).
\eeq
Using the expression in Eq.~\eqref{minuplus} for the matrix elements on the RHS of Eq.~\eqref{LGPD} together with Eq.~\eqref{JqzAB}, yields
\begin{align} \label{LGPD1}
 \la\la  P, \mathscr{S}_L  | L^q_{\text{Ji},z}  |  P, \mathscr{S}_L \ra \ra&= \tfrac{1}{2} \left(A_q + B_q+ D_q\right)  \nn \\
&=    \la\la  P, \mathscr{S}_L  | J^q_{\text{Bel},z}  |  P, \mathscr{S}_L \ra \ra +\tfrac{1}{2}\, D_q,
\end{align}
so that
\beq \label{Dq}
-\tfrac{1}{2}\,D_q=\tfrac{1}{2}\,a^q_0=  \la\la  P, \mathscr{S}_L  | S^q_{z}  |  P, \mathscr{S}_L \ra \ra.
\eeq

Substituting in Eq.~\eqref{intG} yields an expression for the orbital angular momentum
 \beq\label{PPSS}
\la\la P,\mathscr{S}_L | L^q_{\text{Ji}, z}  | P, \mathscr{S}_L \ra \ra =- \int_{-1}^{1}  \ud x \,x\, G^q_2(x,0,0).
\eeq
Note that one gets exactly the same expression for the $z$ component of the instant form orbital angular momentum. This relation was first obtained by Penttinen, Polyakov, Shuvaev and Strikman in the parton model \cite{Penttinen:2000dg} and later confirmed in QCD by Hatta and Yoshida \cite{Hatta:2012cs}.

One has therefore the following genuine and non-trivial sum rule
\begin{equation}
\int_{-1}^{1}\ud x\,x\left[H_q(x,0,0)+E_q(x,0,0)+2G^q_2(x,0,0)\right]-a^q_0=0.
\end{equation}
The relation between twist-3 GPDs and the kinetic orbital angular momentum has also been discussed by Ji, Xiong and Yuan \cite{Ji:2012ba}, and similar expressions can be found in the gluon sector \cite{Hatta:2012cs,Ji:2012ba}. Unfortunately, it will be very difficult to extract the twist-3 GPD $G_2$ accurately from experiment.

\subsubsection{Evaluation of $L^q_{\text{Ji},z}$ in a longitudinally polarized nucleon, from GTMDs\label{secVIIA4}}

Another possibility to learn about the orbital angular momentum consists in considering twist-2 generalized transverse-momentum dependent distributions (GTMDs), also called unintegrated or $k_\perp$-dependent GPDs, defined as \cite{Meissner:2009ww}
\begin{multline} \label{GTMD}
\frac{1}{2}\int \frac{\ud z^-\,\ud^2z_\perp}{(2\pi)^3}\,e^{i x \bar{P}^+ z^--i\uvec k_\perp\cdot\uvec z_\perp} \langle P', \msc{S}' | \barpsi(-\tfrac{z}{2}) \gamma^+ \mathcal W\psi (\tfrac{z}{2})  |  P,  \msc{S} \rangle\big|_{z^+=0}\\
= \frac{1}{2M}\,\overline{u}(P',\mathscr S')\left[F^q_{1,1}+\frac{i\sigma^{j+}k^j_\perp}{\bar P^+} \,F^q_{1,2}+\frac{i\sigma^{j+}\Delta^j_\perp}{\bar P^+} \,F^q_{1,3}+\frac{i\sigma^{jk}k^j_\perp\Delta^k_\perp}{M^2} \,F^q_{1,4}\right]u(P,\mathscr S),
\end{multline}
where $j,k=1,2$ are transverse indices. The GTMDs are functions of the quark momentum $(x\bar P^+,\uvec k_\perp)$ and the momentum transfer $\Delta$, and depend upon the choice made for the Wilson line $\mathcal{W}$. For vanishing skewness $\xi=0$, their Fourier transform from $\uvec\Delta_\perp$-space to $\uvec b_\perp$-space can be interpreted as relativistic phase-space or Wigner distributions\footnote{The non-relativistic version of these phase-space or Wigner distributions was introduced in Refs. \cite{Ji:2003ak,Belitsky:2003nz}.} \cite{Lorce:2011kd}
\beq
\rho^q_{\msc S'\msc S}(x,\uvec k_\perp,\uvec b_\perp; \mathcal{W})=\int\frac{\ud^2\Delta_\perp}{(2\pi)^2}\,e^{-i\uvec\Delta_\perp\cdot\uvec b_\perp}\,\frac{1}{2}\int \frac{\ud z^-\,\ud^2z_\perp}{(2\pi)^3}\,e^{i x \bar{P}^+ z^--i\uvec k_\perp\cdot\uvec z_\perp} \,\langle P', \msc{S}' | \barpsi(-\tfrac{z}{2}) \gamma^+ \mathcal W\psi (\tfrac{z}{2})  |  P,  \msc{S} \rangle\big|_{z^+=0}.
\eeq
It can be shown that the longitudinal component of the Ji quark orbital angular momentum can be written as \cite{Lorce:2011kd,Lorce:2011ni}
\begin{align}
\la\la P,\mathscr{S}_L | L^q_{\text{Ji}, z}  | P, \mathscr{S}_L \ra \ra &=
\int\ud x\,\ud^2k_\perp\,\ud^2b_\perp\left(\uvec b_\perp\times\uvec k_\perp\right)_z\,\rho^q_{\msc S_L\msc S_L}(x,\uvec k_\perp,\uvec b_\perp; \mathcal W_\text{straight})\nn\\
&=-\int\ud x\,\ud^2k_\perp\,\frac{\uvec k^2_\perp}{M^2}\,F^q_{1,4}(x,\uvec k_\perp,\Delta=0; \mathcal W_\text{straight}),\label{LzGTMD}
\end{align}
where the Wilson line $\mathcal W_\text{straight}$ connects the points $-\tfrac{z}{2}$ and $\tfrac{z}{2}$ by a direct straight line \cite{Ji:2012sj,Lorce:2012ce}. One has therefore the following non-trivial sum rule
\begin{equation}
\int_{-1}^{1}\ud x\left[x\left[H_q(x,0,0)+E_q(x,0,0)\right]-2\int\ud^2k_\perp\,\frac{\uvec k^2_\perp}{M^2}\,F^q_{1,4}(x,\uvec k_\perp,0; \mathcal W_{\textrm{straight}})\right]-a^q_0=0.
\end{equation}
Note that, unfortunately, at present there is no clear way of extracting the twist-2 GTMDs from experimental data. This sum rule can therefore be only tested in lattice and model calculations.

\subsection{Expressions for the canonical version of $\uvec L$\label{secVIIB}}

It is extremely interesting that the gauge-invariant canonical orbital angular momentum  $L^q_{\text{gic},z}$, which equals the Jaffe-Manohar orbital angular momentum evaluated in the gauge $A^+=0$, \emph{i.e.}  $L^q_{\text{JM},z}|_{A^+=0}$, can be expressed in terms of the same twist-2 GTMD, but defined with a different choice of Wilson line \cite{Lorce:2011kd,Lorce:2011ni}. One obtains
\beq
\la\la P,\mathscr{S}_L | L^q_{\text{gic}, z}  | P, \mathscr{S}_L \ra \ra =-\int\ud x\,\ud^2k_\perp\,\frac{\uvec k^2_\perp}{M^2}\,F^q_{1,4}(x,\uvec k_\perp,\Delta=0; \mathcal{W}_{\textrm{LF}}),\label{LzcanGTMD}
\eeq
where the staple-like Wilson line $\mathcal W_{\textrm{LF}}$ connects the points $-\tfrac{z}{2}$ and $\tfrac{z}{2}$ \emph{via} the intermediary points $-\tfrac{z}{2}+\eta\infty^-$ and $\tfrac{z}{2}+\eta\infty^-$ by straight lines\footnote{More complicated Wilson lines can also be relevant depending on the process, see Refs. \cite{Buffing:2011mj,Buffing:2012sz,Buffing:2013kca}.}, with the parameter $\eta$ indicating whether the Wilson lines are future pointing ($\eta=+1$) or past pointing ($\eta=-1$) \cite{Meissner:2009ww}. In the light-front gauge, this Wilson line reduces  to a transverse straight link at light-front infinity $x^-=\eta\infty^-$. Because of parity and time-reversal symmetry, this residual gauge link does not contribute to the orbital angular momentum in the light-front gauge and can simply be ignored \cite{Hatta:2011ku,Hatta:2012cs,Ji:2012ba,Lorce:2012ce}.

So, by merely changing the shape of the Wilson line, one obtains either the kinetic or the canonical quark orbital angular momentum \cite{Ji:2012sj,Lorce:2012ce}. Hatta showed that a similar expression holds in the gluon sector \cite{Hatta:2011ku}. For a complete parametrization of the gluon GTMDs, see Ref. \cite{Lorce:2013pza}. The gauge-invariant canonical orbital angular momentum can also be accessed \emph{via} twist-3 GPDs by adding the potential angular momentum $L_{\text{pot},z}$ discussed in section \ref{secVB3} to the kinetic, \emph{i.e.} Ji, orbital angular momentum. Explicit, but cumbersome, expressions have been given in Refs.~\cite{Hatta:2012cs,Ji:2012ba}.

In lattice calculations, it is technically very difficult to fix a gauge. One is therefore forced to make calculations including explicitly the Wilson line. Note that very interesting pioneer calculations of naive T-odd TMDs on a lattice with staple-like Wilson lines have been carried out in Ref. \cite{Musch:2011er}. Moreover, Ji proposed a new strategy for computing parton distributions, in leading twist, on a lattice using space-like operators instead of light-like operators \cite{Ji:2013dva}, the former being more amenable to a lattice implementation than the latter. Because of these encouraging developments, we believe it is reasonable to expect the first lattice evaluations of the canonical orbital angular momentum \eqref{LzcanGTMD} within the coming years.

\subsection{The orbital angular momentum in quark models\label{secVIIC}}

Quark models are usually based on approximate or phenomenological expressions for the quark wave function inside a nucleon. Knowing this wave function, one has in principle all the information needed for computing the quark orbital angular momentum. Since gluons are absent in quark models, model results are commonly thought to be approximations of QCD at very low scale and in a fixed gauge, usually taken to be the light-front gauge $A^+=0$, to make contact with the partonic picture. Like in most studies, we shall consider only the longitudinal component of the quark orbital angular momentum
\beq
\ell^q_z\equiv\la\la P,\mathscr{S}_L | L^q_z | P, \mathscr{S}_L \ra \ra .
\eeq

The quark kinetic (or Ji) orbital angular momentum can be obtained in a model by calculating the GPDs involved in the Ji relation\footnote{Phenomenologically, the GPD $E$ is poorly known. Bacchetta and Radici \cite{Bacchetta:2011gx} tried to use the present information about the Sivers transverse momentum-dependent distribution (TMD) $f^{\perp}_{1T}$, to constrain better the GPD $E$ by assuming a phenomenological relation $\int\ud^2k_\perp\,f^{\perp q}_{1T}(x,\uvec k^2_\perp)\approx-L(x)E_q(x,0,0)$ at a fixed scale $Q^2_0$ and with a simple \emph{ansatz} for $L(x)$, inspired by the work of Burkardt \cite{Burkardt:2002ks,Burkardt:2003uw}. Even though such a phenomenological relation can hardly be put on firmer theoretical grounds, it seems to give reasonable results.}
\beq\label{OAM1}
\ell^q_{\text{kin},z}=\tfrac{1}{2}\int^1_{-1}\ud x\left\{x\left[H_q(x,0,0)+E_q(x,0,0)\right]-\tilde H_q(x,0,0)\right\},
\eeq
where $\tilde H$ is the helicity GPD satisfying
\beq
\int\ud x\,\tilde H_q(x,0,0)=a^q_0.
\eeq
In principle, one should get the same result from the twist-3 GPD relation \eqref{PPSS}
\beq\label{PPSS2}
\ell^q_{\text{kin},z}=- \int_{-1}^{1}  \ud x \,x\, G^q_2(x,0,0),
\eeq
and the GTMD relation \eqref{LzGTMD}
\beq
\ell^q_{\text{kin},z}=-\int\ud x\,\ud^2k_\perp\,\frac{\uvec k^2_\perp}{M^2}\,F^q_{1,4}(x,\uvec k_\perp,\Delta=0; \mathcal{W}_{straight})
\eeq
with a straight Wilson line. In practice, the last two approaches are often problematic and, to the best of our knowledge, the equivalence has never been checked explicitly in model calculations so far. The reason is that twist-3 GPDs contain interaction which is often missing or hard to compute realistically in quark models. Note also that the twist-3 GPD relation \eqref{PPSS2} is based on the QCD equations of motion which usually differ from the model equations of motion. Concerning the GTMD relation with straight Wilson lines, one cannot get rid of the straight Wilson line when $k_\perp \neq 0 $, by keeping the \emph{same} choice of gauge at each spatial point \cite{Lorce:2012ce}, so one cannot avoid the inclusion of gluons.

The quark canonical (or Jaffe-Manohar) orbital angular momentum can be calculated by means of the GTMD relation \eqref{LzcanGTMD}
\beq\label{OAM2}
\ell^q_{\text{can},z}=-\int\ud x\,\ud^2k_\perp\,\frac{\uvec k^2_\perp}{M^2}\,F^q_{1,4}(x,\uvec k_\perp,\Delta=0; \mathcal{W}_\text{LF}),
\eeq
where the staple-like Wilson line $\mathcal{W}_\text{LF}$ can basically be ignored in the light-front gauge, as explained earlier. This orbital angular momentum coincides with the partonic orbital angular momentum calculated in terms of light-front wave functions \cite{Lorce:2011ni,Lorce:2011kn}
\beq\label{partOAM}
\ell^q_{\text{can},z}=\sum_n\sum_{\{\lambda\}}\sum_{l,r(q)}(\delta_{rl}-x_l)\int[\ud x]_n\,[\ud^2k_\perp]_n\,\Psi^{*+}_n(\{x,\uvec k_{\perp},\lambda\})\left(\uvec k_{r\perp}\times\tfrac{1}{i}\uvec\nabla_{\uvec k_{l\perp}}\right)_z\Psi^+_n(\{x,\uvec k_{\perp},\lambda\}),
\eeq
defined with respect to the transverse center of momentum \cite{Burkardt:2000za,Burkardt:2002hr}
\beq
\uvec R_\perp=\sum^n_{l=1} x_l\,\uvec r_{l\perp},
\eeq
where in momentum space $\uvec r_{l\perp}=\tfrac{1}{i}\uvec\nabla_{\uvec k_{l\perp}}$. In Eq. \eqref{partOAM}, $\Psi^\Lambda_n$ is the $n$-parton light-front wave function for a nucleon with polarization $\Lambda$. The argument $\{x,\uvec k_{\perp},\lambda\}$ stands for the set of parton longitudinal momentum fractions, transverse momenta and polarizations, and the integration measures read
\beq\label{canOAM}
[\ud x]_n\equiv\left[\prod_{l=1}^n\ud x_l\right]\delta\!\left(1-\sum^n_{l=1} x_l\right),\qquad [\ud^2 k_\perp]_n\equiv\left[\prod_{l=1}^n\frac{\ud^2k_{l\perp}}{2(2\pi)^3}\right]2(2\pi)^3\,\delta^{(2)}\!\left(\sum^n_{l=1} \uvec k_{l\perp}\right).
\eeq
Similarly to section \ref{secVIG2}, $r(q)$ means that the sum is restricted to those values of $r$ corresponding to quarks of flavor $q$. \nl

In quark model calculations, like \emph{e.g.} \cite{Ma:1998ar}, one often defines the canonical orbital angular momentum with respect to the coordinate axes, so that
\beq\label{naiveOAM}
\uL^q_{\text{can},z}=\sum_n\sum_{\{\lambda\}}\sum_{r(q)}\int[\ud x]_n\,[\ud^2k_\perp]_n\,\Psi^{*+}_n(\{x,\uvec k_{\perp},\lambda\})\left(\uvec k_{r\perp}\times\tfrac{1}{i}\uvec\nabla_{\uvec k_{r\perp}}\right)_z\Psi^+_n(\{x,\uvec k_{\perp},\lambda\}).
\eeq
This expression is somewhat simpler than Eq. \eqref{canOAM} and comes from the fact that the transverse position of the nucleon is not properly taken into account. So, Eq. \eqref{naiveOAM} represents the \emph{naive} canonical orbital angular momentum, while Eq. \eqref{canOAM} represents the \emph{intrinsic} canonical orbital angular momentum \cite{Lorce:2011kn}. However, when summing over all the partons, the difference between naive and intrinsic orbital angular momentum disappears. In Ref. \cite{Brodsky:2000ii}, it is stressed that, owing to the momentum conservation constraint, there can only be $n-1$ \emph{relative} orbital angular momentum contributions in a $n$-parton Fock state. Note, however, that there is no unique way to define such $n-1$ relative constributions for $n>2$. Moreover, one cannot unambiguously attribute a given contribution to a particular parton. The transverse center of momentum provides the additional point with respect to which one can define unambiguously $n$ intrinsic contributions, each associated with a particular parton. Some quark-model calculations \cite{Ma:1998ar,She:2009jq,Avakian:2010br} suggested that the naive quark canonical orbital angular momentum defined in Eq. \eqref{naiveOAM} may be related to the so-called \emph{pretzelosity} transverse-momentum dependent distribution (TMD) $h_{1T}^\perp$ as follows
\beq\label{OAM3}
\uL^q_{\text{can},z}=-\int\ud x\,\ud^2k_\perp\,\frac{\uvec k_\perp^2}{2M^2}\,h_{1T}^{\perp q}(x,\uvec k^2_\perp).
\eeq
It appeared, unfortunately, that such a relation is only valid in a restricted class of models \cite{Lorce:2011dv,Lorce:2011kn}, where the instant-form wave function $\psi(\{\uvec k,\sigma\})$ is a pure $s$-wave and related to the light-front wave function $\Psi(\{x,\uvec k_\perp,\lambda\})$ by a mere Wigner rotation\footnote{In the case of a free quark, the Wigner rotation is also known as the Melosh rotation \cite{Melosh:1974cu}.}, which allows one to write canonical spin $\sigma$ in terms of light-front helicity $\lambda$
\beq\label{modelWF}
\Psi(\{x,\uvec k_\perp,\lambda\})=\psi(\{\uvec k,\sigma\})\prod_l D_{\lambda_l\sigma_l}(x_l,\uvec k_{l\perp}),\qquad D_{\lambda\sigma}(x,\uvec k_\perp)=\begin{pmatrix}\cos\tfrac{\theta}{2}&\hat k_L\sin\tfrac{\theta}{2}\\-\hat k_R\sin\tfrac{\theta}{2}&\cos\tfrac{\theta}{2}\end{pmatrix},
\eeq
with $\hat k_{R,L}=(k^1\pm ik^2)/|\uvec k_\perp|$ and $\theta\equiv \theta(x,\uvec k_\perp)$ a function of the quark momentum specific to the model. In this class of models, all the orbital angular momentum in the light-front wave function is simply generated by this Wigner rotation. Indeed, using the light-front wave function in Eq. \eqref{modelWF}, the naive canonical orbital angular momentum given by Eq. \eqref{naiveOAM} reads
\beq\label{OAMmod}
\uL^q_{\text{can},z}=\sum_n\sum_{\{\sigma\}}\sum_{r(q)}\int[\ud x]_n\,[\ud^2k_\perp]_n\left|\psi^+_n(\{\uvec k,\sigma\})\right|^2\left(1-\cos\theta_r\right)\sigma_r,
\eeq
where $\theta_r\equiv\theta(x_r,\uvec k_{r\perp})$ and $\sigma_r=\pm1/2$. The same Wigner rotation is responsible for the existence of a non-vanishing pretzelosity TMD $h_{1T}^\perp$. Using the generic results of Ref. \cite{Lorce:2011zta}, one can write in the restricted class of models
\beq\label{pretz}
\frac{\uvec k_\perp^2}{2M^2}\,h_{1T}^{\perp q}(x,\uvec k^2_\perp)=\sum_n\sum_{\{\sigma\}}\sum_{r(q)}\int[\ud x]_n\,[\ud^2k_\perp]_n\left|\psi^+_n(\{\uvec k,\sigma\})\right|^2\delta (x-x_r)\,\delta^{(2)}(\uvec k_\perp-\uvec k_{r\perp})\left(\cos\theta_r-1\right)\sigma_r
\eeq
where the pure $s$-wave nature of the instant-form wave function was crucial in the intermediate steps. Comparing Eqs. \eqref{OAMmod} and \eqref{pretz}, one obtains the model-dependent relation \eqref{OAM3} and understands it as originating from a pure Wigner rotation effect. The same effect is at the origin other relations among TMDs observed in the restricted class of models \cite{Lorce:2011zta}.

In Table \ref{OAMtable}, we reproduce the results for the three versions of orbital angular momentum given by Eqs. \eqref{OAM1}, \eqref{OAM2} and \eqref{OAM3} within the light-front constituent quark model (LFCQM) and the light-front chiral quark-soliton model (LF$\chi$QSM) restricted to the three-quark sector~\cite{Lorce:2011kd}. In these models, Eq. \eqref{OAM3} coincides with the naive definition of quark canonical orbital angular momentum \eqref{naiveOAM}.
\begin{table}[t]
\begin{center}
\caption{\footnotesize{Comparison between the kinetic ($\ell^q_{\text{kin},z}$), intrinsic canonical ($\ell^q_{\text{can},z}$) and naive canonical ($\mathcal L^q_{\text{can},z}$) orbital angular momentum in the light-front constituent quark model (LFCQM) and the light-front chiral quark-soliton model (LF$\chi$QSM) for $u$-, $d$- and total ($u+d$) quark contributions.}}\label{OAMtable}
\begin{tabular}{@{\quad}c@{\quad}|@{\quad}c@{\quad}c@{\quad}c@{\quad}|@{\quad}c@{\quad}c@{\quad}c@{\quad}}\whline
Model&\multicolumn{3}{c@{\quad}|@{\quad}}{LFCQM}&\multicolumn{3}{c@{\quad}}{LF$\chi$QSM}\\
$q$&$u$&$d$&Total&$u$&$d$&Total\\
\hline
$\ell^q_{\text{kin},z}$&$0.071$&$~~0.055$&$0.126$&$-0.008$&$~~0.077$&$0.069$\\
$\ell^q_{\text{can},z}$&$0.131$&$-0.005$&$0.126$&$~~0.073$&$-0.004$&$0.069$\\
$\mathcal L^q_{\text{can},z}$&$0.169$&$-0.042$&$0.126$&$~~0.093$&$-0.023$&$0.069$\\
\whline
\end{tabular}
\end{center}
\end{table}
 As expected in a pure quark model, all three versions of ``$\ell$''  in each model give the same value for the \emph{total} orbital angular momentum, but widely different results between the models. However, differences between the various definitions appear in the \emph{separate} quark-flavor contributions. The difference between the naive and intrinsic canonical orbital angular momentum naturally comes from the fact that the individual orbital angular momentum contributions are not defined with respect to the same point. What is more surprising is that $\ell^q_{\text{kin},z}\neq \ell^q_{\text{can},z}$, since it is generally believed that the kinetic and canonical orbital angular momentum should coincide in absence of gluons. Note that a similar observation has also been made in the instant-form version of the $\chi$QSM~\cite{Wakamatsu:2005vk}, where the Ji relation \eqref{OAM1} was shown not to hold for the isovector combination $u-d$ in the model. This should actually not be so surprising, since the Ji relation  is obtained using the QCD expression for the energy-momentum tensor, and not the quark model expression. Only in some cases, like \emph{e.g.} in the scalar diquark model \cite{Burkardt:2008ua}, one obtains $\ell^q_{\text{kin},z}= \ell^q_{\text{can},z}$. A comparison between the kinetic and canonical orbital angular momentum, where the gauge potential is included, can be found in Refs. \cite{Burkardt:2008ua,Burkardt:2010he}. For a physical interpretation of the difference $\ell^q_{\text{kin},z}- \ell^q_{\text{can},z}$, see Ref. \cite{Burkardt:2012sd}.

\subsection{The phase-space distribution of angular momentum\label{secVIID}}

We have essentially discussed the angular momentum integrated over all phase space. It is however possible to define densities of angular momentum as well. Many definitions have been proposed in the literature, creating a lot of confusion. One of the reasons is that \emph{different} densities can lead to the \emph{same} integrated quantity, a long as they differ by superpotential terms. The archetypical example is the total angular momentum of the system, which can be obtained either from the canonical density or the Belinfante-improved density. In the first case, the angular momentum density is naturally decomposed into an orbital angular momentum density and a spin density. In the second case, the spin density is converted into an orbital angular momentum density by adding a superpotential. It is therefore essential to keep track of these superpotential terms.
\nl

Since the $t$ dependence of twist-2 GPDs contains, \emph{via} two-dimensional Fourier transform, the information about the spatial distribution of partons \cite{Burkardt:2000za,Burkardt:2002hr}, it has been suggested by Goeke \emph{et al.} \cite{Goeke:2007fp}\footnote{Note that Goeke \emph{et al.} considered a three-dimensional Fourier transform, valid only in a non-relativistic interpretation.} that the Ji relation generalized to $t\neq 0$ contains the information about the spatial distribution of Belinfante angular momentum
\beq
J_q(t)=\tfrac{1}{2}\int\ud x\,x\left[H_q(x,0,t)+E_q(x,0,t)\right].
\eeq
It has also been suggested in Refs. \cite{Hoodbhoy:1998yb,Ji:2012sj,Ji:2012ba} that the integrand of the Ji relation
\beq
J_q(x)=\tfrac{x}{2}\left[H_q(x,0,0)+E_q(x,0,0)\right]
\eeq
can be interpreted as the density of angular momentum in $x$-space. In the scalar diquark model, where it is straightforward to maintain Lorentz invariance and where the absence of the gauge potential allows one to identify Ji and Jaffe-Manohar orbital angular momentum, Burkardt and Hikmat \cite{Burkardt:2008ua} checked that
\beq
\int\ud x\,J_q(x)=\int\ud x\,L_q(x)+\int\ud x\,S_q(x),
\eeq
but observed that
\beq\label{Jqx}
J_q(x)\neq L_q(x)+S_q(x),
\eeq
where $J_q(x)$, $L_q(x)$ and $S_q(x)$ are the angular momentum, orbital angular momentum and spin densities
\begin{align}
J_q(x)&=\tfrac{x}{2}\int\frac{\ud^2k_\perp}{2(2\pi)^3}\left[\left|\psi^+_+(x, \uvec k_\perp)\right|^2+\left|\psi^+_-(x, \uvec k_\perp)\right|^2-2M\,\psi^{+*}_+(x, \uvec k_\perp)\,\tfrac{\partial}{\partial \Delta^1}\psi^-_+(x, \uvec k_\perp-(1-x)\uvec\Delta_\perp)\big|_{\uvec\Delta_\perp=\uvec 0_\perp}\right],\\
L_q(x)&=\int\frac{\ud^2k_\perp}{2(2\pi)^3}\,(1-x)\left|\psi^+_-(x, \uvec k_\perp)\right|^2,\\
S_q(x)&=\tfrac{1}{2}\int\frac{\ud^2k_\perp}{2(2\pi)^3}\left[\left|\psi^+_+(x, \uvec k_\perp)\right|^2-\left|\psi^+_-(x, \uvec k_\perp)\right|^2\right],
\end{align}
expressed in terms of the light-front wave function $\psi^\Lambda_\lambda(x,\uvec k_\perp)$ of the scalar diquark model. The lack of equality in \eqref{Jqx} caused much puzzlement, since it seems difficult to understand. It cannot be explained by the extra contribution obtained by Hoodbhoy, Ji and Lu \cite{Hoodbhoy:1998yb}, since the latter is proportional to the gluon field strength $G^{\mu\nu}$, and therefore cannot contribute in the scalar diquark model. Upon examination, it turns out that belief in the equality of the LHS and RHS of \eqref{Jqx} is based on a paper of Hoodbhoy, Ji and Lu who actually showed that for the Mellin moments, defined as $f_n\equiv\int\ud x\,x^{n-1}f(x)$, one has
\beq
J_{qn}=L_{qn}+S_{qn},
\eeq
provided one drops systematically the surface terms for each $n$. In other words, they implicitly assumed that \emph{all} the Mellin moments of the superpotential term vanish, implying that the superpotential itself vanishes, which is impossible for a particle with spin. Similarly, Adhikari and Burkardt \cite{Adhikari:2013ima} checked in the same scalar diquark model that
\beq
\int\ud^2b_\perp\,J_q(\uvec b_\perp)=\int\ud^2b_\perp\,L_q(\uvec b_\perp)+\int\ud^2b_\perp\,S_q(\uvec b_\perp),
\eeq
but observed that
\beq\label{Jqb}
J_q(\uvec b_\perp)\neq L_q(\uvec b_\perp)+ S_q(\uvec b_\perp),
\eeq
where for $\xi=0$
\beq
f(\uvec b_\perp)\equiv\int\frac{\ud^2\Delta_\perp}{(2\pi)^2}\,e^{-i\uvec \Delta_\perp\cdot\uvec b_\perp}\,f(t=-\uvec\Delta^2_\perp).
\eeq
We note that the actual spatial density of angular momentum receives a contribution from the Fourier transform of \emph{both}\footnote{In the non-relativistic interpretation adopted by Goeke \emph{et al.} \cite{Goeke:2007fp} where $t=-\uvec \Delta^2$ and $\Delta^0=0$, the integrand of the three-dimensional Fourier transform is $J_q(t)+\tfrac{2t}{3}\,\tfrac{\partial}{\partial t}J_q(t)$. We note that Goeke \emph{et al.} implicitly discarded a quadrupole contribution by making the substitution $-\uvec\Delta^2_\perp\mapsto\tfrac{2t}{3}$. In the relativistic interpretation where $t=-\uvec \Delta^2_\perp$ and $\Delta^+=0$, the integrand of the two-dimensional Fourier transform would be $J_q(t)+t\,\tfrac{\partial}{\partial t}J_q(t)$.} $J_q(t)$ and $t\,\tfrac{\partial}{\partial t}J_q(t)$. The second term does not contribute to the total angular momentum since the integral over all space amounts to set $t=0$. Nevertheless, even with the correct spatial density associated with $J_q(t)$, we expect the non-equality Eq. \eqref{Jqb} to remain true. Indeed, the contribution from the superpotential term to the spatial density cannot identically vanish if the spin density does not identically vanish as well.

\begin{figure}[t]
\includegraphics[width=0.4\textwidth]{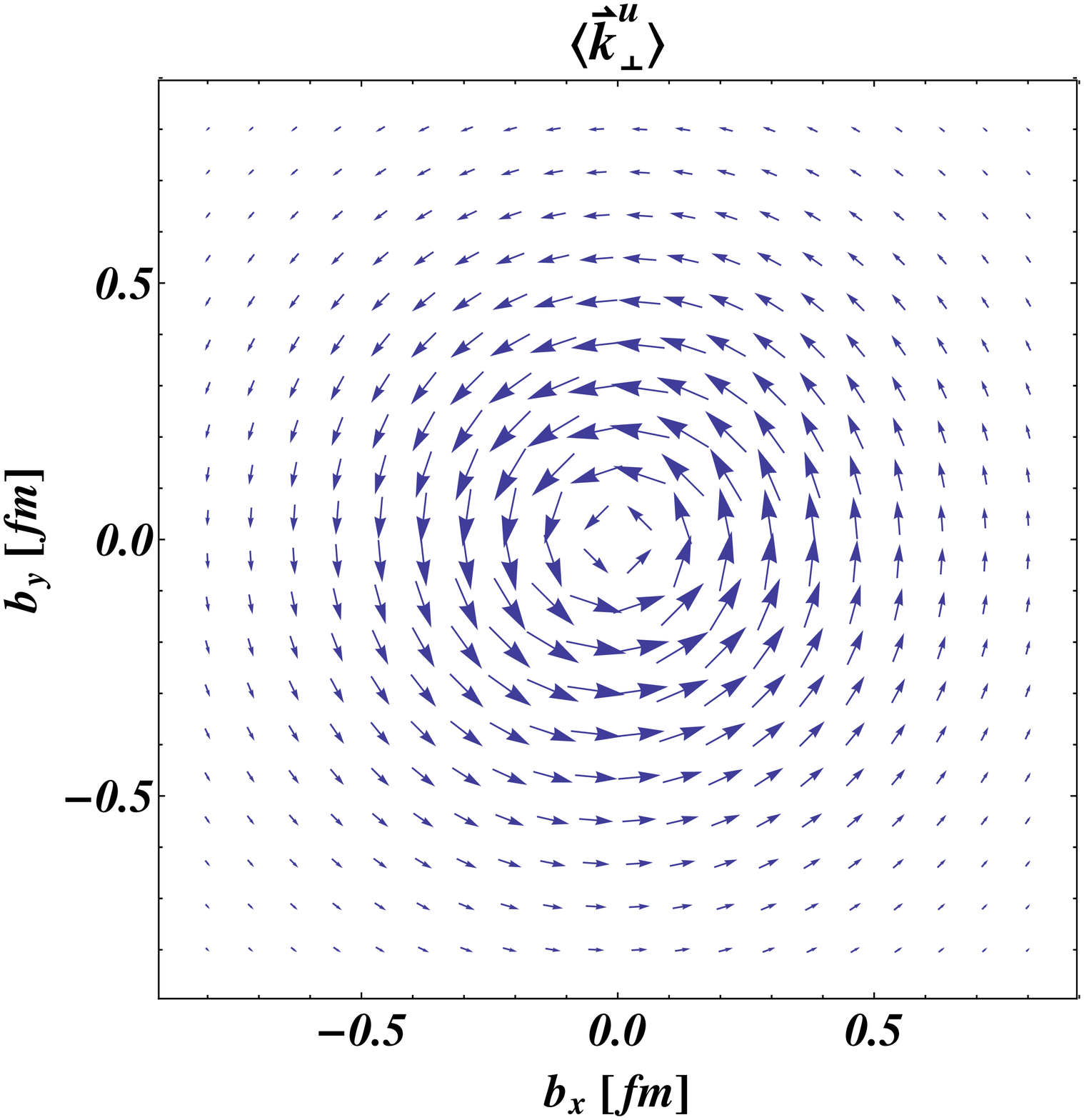}\hspace{1cm}
\includegraphics[width=0.4\textwidth]{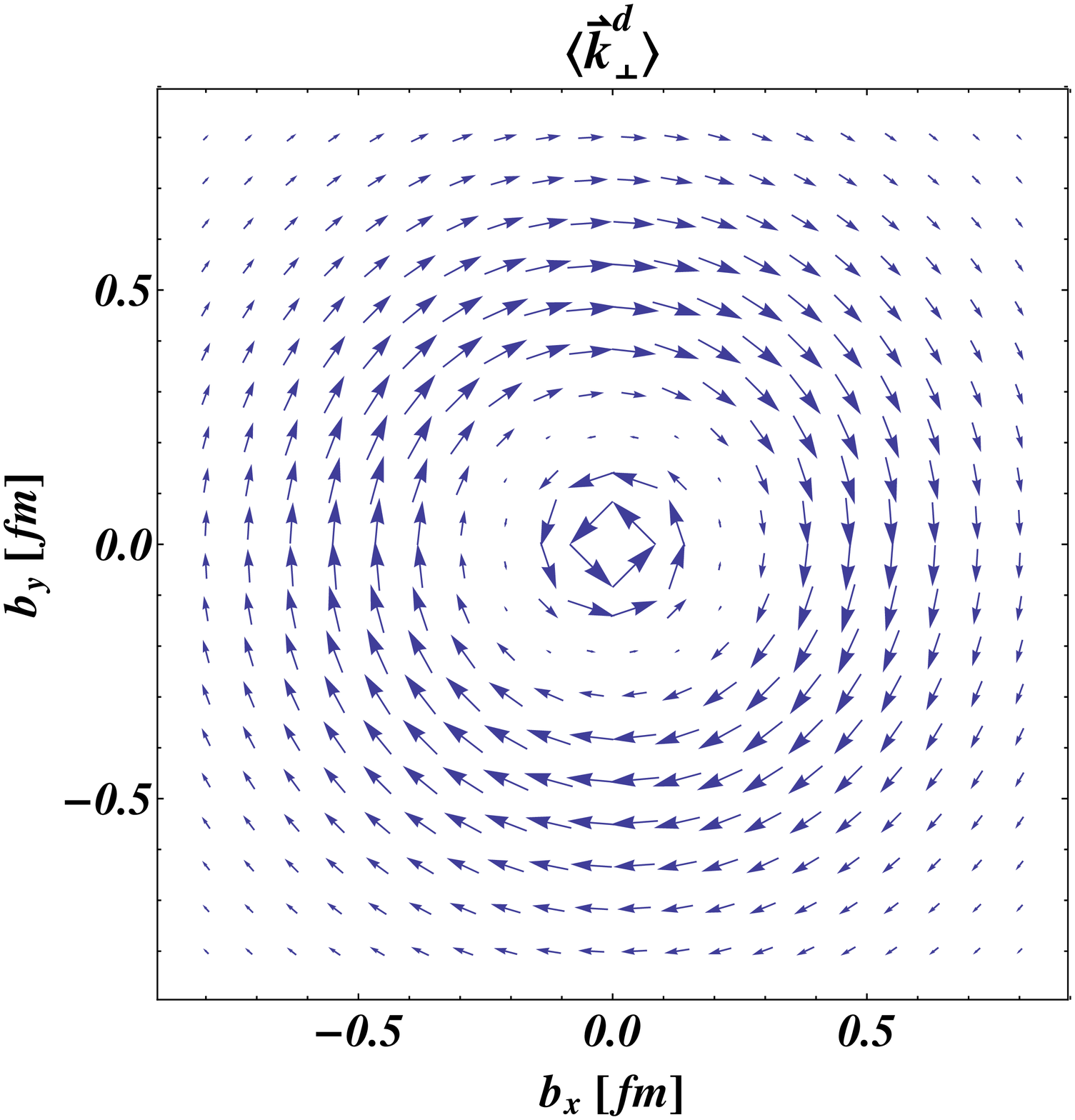}
\caption{Distributions in impact-parameter space of the average transverse momentum of up quarks (left panel) and down quarks (right panel) in a longitudinally polarized nucleon, computed within the light-front chiral quark-soliton model (LF$\chi$QSM). The nucleon polarization is pointing out of the plane, and the arrows show the size and direction of the average quark transverse momentum.}
\label{run1} \vspace{-0mm}
\end{figure}

The most natural definition of angular momentum density is based on phase-space distributions, also known as Wigner distributions. In the context of QCD, non-relativistic expressions have been introduced and discussed in Refs. \cite{Ji:2003ak,Belitsky:2003nz}. The corresponding relativistic expressions have been written down and studied by Lorc\'e and Pasquini in Ref. \cite{Lorce:2011kd}. They defined the relativistic quark phase-space density in a longitudinally polarized nucleon as
\beq
\rho^q(x,\uvec k_\perp,\uvec b_\perp;\mathcal W)=\int\frac{\ud^2\Delta_\perp}{(2\pi)^2}\,e^{-i\uvec\Delta_\perp\cdot\uvec b_\perp}\,\frac{1}{2}\int\frac{\ud z^-\ud^2z_\perp}{(2\pi)^3}\,e^{ik\cdot z}\,\la P^+,\tfrac{\uvec\Delta_\perp}{2},\mathscr{S}_L | \barpsi(-\tfrac{z}{2})\gamma^+\mathcal W\psi(\tfrac{z}{2}) | P^+,-\tfrac{\uvec\Delta_\perp}{2}, \mathscr{S}_L \ra\big|_{z^+=0},
\eeq
where the parametrization of the GTMD correlator is given by Eq. \eqref{GTMD}. Note that, using the light-front spinors, only the $F_{1,1}$ and $F_{1,4}$ GTMDs contribute in the case of longitudinal polarization. We can then write
\beq\label{Wignerdensity}
\rho^q(x,\uvec k_\perp,\uvec b_\perp;\mathcal W)=\mathcal F^q_{1,1}(x,\uvec k^2_\perp,\uvec k_\perp\cdot\uvec b_\perp,\uvec b^2_\perp;\mathcal W)-\tfrac{1}{M^2}\,\left(\uvec k_\perp\times\uvec\nabla_{\uvec b_\perp}\right)_z\mathcal F^q_{1,4}(x,\uvec k^2_\perp,\uvec k_\perp\cdot\uvec b_\perp,\uvec b^2_\perp;\mathcal W),
\eeq
where
\beq
\mathcal F^q_{1,i}(x,\uvec k^2_\perp,\uvec k_\perp\cdot\uvec b_\perp,\uvec b^2_\perp;\mathcal W)=\int\frac{\ud^2\Delta_\perp}{(2\pi)^2}\,e^{-i\uvec\Delta_\perp\cdot\uvec b_\perp}\,F^q_{1,i}(x,\xi=0,\uvec k^2_\perp,\uvec k_\perp\cdot\uvec\Delta_\perp,\uvec\Delta^2_\perp;\mathcal W)
\eeq
for $i=1,4$. The density of total longitudinal orbital angular momentum \cite{Lorce:2011kd, Lorce:2011ni} can then naturally be defined as
\beq
L^q_z(x,\uvec k_\perp,\uvec b_\perp;\mathcal W)
=\left(\uvec b_\perp\times\uvec k_\perp\right)_z\,\rho^q(x,\uvec k_\perp,\uvec b_\perp;\mathcal W).
\eeq
Like in the integrated case, for a staple-like Wilson line in the light-front direction, it coincides with the canonical or Jaffe-Manohar orbital angular momentum in the light-front gauge
\beq
L^q_{\text{can},z}(x,\uvec k_\perp,\uvec b_\perp)=L^q_z(x,\uvec k_\perp,\uvec b_\perp;\mathcal W_\text{LF}).
\eeq
For a straight Wilson line, as explained by Lorc\'e in Ref. \cite{Lorce:2012ce}, it is necessary to integrate over $\uvec k_\perp$ to interpret it as the kinetic (or Ji) orbital angular momentum. So we have
\beq
L^q_{\text{kin},z}(x,\uvec b_\perp)=\int\ud^2k_\perp\,L^q_z(x,\uvec k_\perp,\uvec b_\perp;\mathcal W_\text{straight}).
\eeq
From the quark phase-space distribution, one can also define in a natural way the distribution of average quark (canonical) transverse momentum in impact-parameter space as
\beq
\la\uvec k^q_\perp\ra(\uvec b_\perp)=\int\ud x\,\ud^2k_\perp\,\uvec k_\perp\,\rho^q(x,\uvec k_\perp,\uvec b_\perp;\mathcal W_\text{LF}).
\eeq
The longitudinal component of the quark canonical orbital angular momentum in impact-parameter space then reads
\begin{align}
L^q_{\text{can},z}&=\int\ud^2b_\perp\left[\uvec b_\perp\times\la\uvec k^q_\perp\ra(\uvec b_\perp)\right]_z\nn\\
&=-\int\ud x\,\ud^2k_\perp\,\ud^2b_\perp\,\frac{\uvec k^2_\perp}{M^2}\,\mathcal F^q_{1,4}(x,\uvec k^2_\perp,\uvec k_\perp\cdot\uvec b_\perp,\uvec b^2_\perp;\mathcal W_\text{LF})\nn\\
&=-\int\ud x\,\ud^2k_\perp\,\frac{\uvec k^2_\perp}{M^2}\,F^q_{1,4}(x,0,\uvec k^2_\perp,0,0;\mathcal W_\text{LF}).
\end{align}
The distribution of average quark transverse momentum has been computed within the LFCQM in Ref. \cite{Lorce:2011ni}. In Fig. \ref{run1}, we show the same distribution, obtained this time within the LF$\chi$QSM.
It clearly shows that in a longitudinally polarized nucleon, the quarks have on average nonzero orbital angular momentum. From Eq. \eqref{Wignerdensity}, it is clear that the polar component, and hence the average orbital angular momentum, originates solely from the $F_{1,4}$ GTMD. In principle, there can also be a radial component. From parity, time-reversal and hermiticity constraints, the latter can only receive contributions from the imaginary part of the GTMDs representing the effect of initial and final state interactions\footnote{A non-zero average radial momentum would indicate that the nucleon is not stable. It can therefore only come from initial and final state interactions which affect the distribution of the struck parton right before and after the hard interaction.} \cite{Meissner:2009ww}. Such interactions being absent in the model, no radial contribution was obtained.

\section{Qualitative summary and experimental implications\label{secVIII}}

In this section, we summarize the essential points concerning the issue of angular momentum decomposition and the experimental implications.

\subsection{Gauge invariance and measurability\label{secVIIIA}}

Gauge-invariant quantities are in principle measurable, whereas gauge non-invariant quantities are often claimed to be not measurable. It turns out, however, that gauge non-invariant quantities considered \emph{in some chosen gauge} can in principle also be measured. One way of understanding this is that there always exists a non-local gauge-invariant quantity (called gauge-invariant extension) that gives the same numerical answer as a local gauge non-invariant quantity in the chosen gauge. The archetypical example is the quantity called $\Delta G$. It is a measurable quantity whose gauge-invariant expression is clearly non-local. Only in the light-front gauge does this expression reduce to the local expression for the gluon spin contribution given by Jaffe and Manohar. So measuring $\Delta G$ is the same as obtaining the value of the gluon spin in the gauge $A^+=0$.

It is therefore essential to separate two things: the measured quantity and its physical interpretation. A measured quantity is necessarily gauge independent, but its interpretation need not be. This is nicely stated by Bashinsky and Jaffe: ``one should make clear what a quark or a gluon parton is in an interacting theory. The subtlety here is the issue of gauge invariance: a pure quark field in one gauge is a superposition of quarks and gluons in another. Different ways of gluon-field gauge fixing predetermine different decompositions of the coupled quark-gluon fields into quark and gluon degrees of freedom.'' Analogously, in particle physics one measures the four-momentum squared of a particle, which is a Lorentz-invariant quantity, but its physical interpretation as the mass squared of the particle is valid only in the rest frame.

The parton model is a picture of the nucleon in the infinite momentum frame, \emph{i.e.} in a frame where the nucleon moves with almost the speed of light. This picture is particularly useful to understand what is going on in scattering experiments involving high momentum transfer, like \emph{e.g.} deep-inelastic scattering, and predates QCD. The light-front gauge is particularly appealing in the infinite-momentum frame, since it relegates dynamical aspects to kinematically suppressed contributions. For this reason, the parton model picture is commonly identified with QCD in the light-front gauge and in the infinite-momentum frame (or equivalently at leading twist). Strictly speaking, the light-front gauge is not more ``physical'' than any other gauge, but turns out to be more \emph{convenient} for the interpretation of high-energy scattering experiments. One could of course choose to work in a different gauge, but the price to pay is high: far more complicated expressions and totally unclear physical interpretation.

The approach proposed a few years ago by Chen \emph{et al.} \cite{Chen:2008ag} reopened hordes of old controversies. But after many subsequent papers and discussions, one should recognize that nothing really new has appeared from the practical/experimental point of view. The new developments concern the physical interpretation point of view. Thus, the Chen \emph{et al.} approach provides a concrete non-local gauge-invariant alternative to the former local gauge non-invariant interpretation. In particular, it makes it clearer why a gauge non-invariant quantity in some chosen gauge can actually be ``measured''. It is therefore possible to interpret in any gauge $\Delta G$ as the gluon spin contribution, provided that the latter is re-defined in a non-local way.
\newline

In summary, our opinion is that there is no fundamental reason for insisting that the physical interpretation of a measurable quantity be gauge and/or frame independent. The important question is to determine whether a gauge non-invariant quantity with clear physical interpretation can be associated with a measurable quantity. Formally, this is always possible, since it suffices to write down the corresponding non-local gauge-invariant extension. The problematic but crucial point is actually to find a way of extracting the latter from experimental data. Since the nucleon internal structure is essentially probed in the infinite-momentum frame, the only experimentally relevant non-local gauge-invariant quantities are those defined with Wilson lines running essentially along the light-front direction.

\subsection{Two kinds of decompositions\label{secVIIIB}}

There are fundamentally two kinds of proton angular momentum decomposition:
\begin{itemize}
\item The \emph{canonical} decomposition is based on Noether's theorem and is written in terms of canonical generators of rotation. It is essentially the Jaffe-Manohar decomposition \cite{Jaffe:1989jz}, which has to be thought of in the light-front gauge in order to make contact with actual experiments.  This decomposition can be rewritten in a gauge-invariant but non-local way, as shown explicitly by Hatta \cite{Hatta:2011ku}.
\item The \emph{kinetic} decomposition is based on the gauge-invariant Belinfante-improved energy-momentum tensor. The form used by Ji \cite{Ji:1996ek} consists of three gauge-invariant local contributions (quark spin, quark orbital angular momentum and gluon total angular momentum) and is obtained using the QCD equations of motion and discarding surface terms. By either fixing a gauge or considering non-local expressions, the gluon total angular momentum can further be split into spin and orbital angular momentum contributions, as discussed by Wakamatsu \cite{Wakamatsu:2010qj}.
\end{itemize}
Opinions diverge as to which kind of decomposition should be considered as ``the" physical one. Both canonical and kinetic decompositions can in principle be accessed experimentally. It turns out that the canonical decomposition provides a clearer partonic interpretation, while the kinetic decomposition is easier to access experimentally. In our opinion, both are interesting.

By either fixing a gauge or working with non-local gauge-invariant expressions, the nucleon angular momentum can actually be decomposed into \emph{five} contributions, instead of the naively expected four. Beside quark spin, quark orbital angular momentum, gluon spin and gluon orbital angular momentum, there is a so-called potential angular momentum \cite{Wakamatsu:2010qj}. The canonical and kinetic decompositions simply differ in the attribution of this term to either quarks or gluons. The potential term is particularly interesting, as suggested by Burkardt \cite{Burkardt:2012sd}, since it is somehow related to the torque that acts on a quark in longitudinally polarized DIS. So, in some sense, the canonical and kinetic decompositions can be viewed as two different groupings of terms belonging to a single more general decomposition.

\subsection{Sum rules \emph{vs.} relations\label{secVIIIC}}

It is important to keep in mind the difference between a sum rule and a relation:
\begin{itemize}
\item A \emph{sum rule} is an identity between measurable quantities. An identity involves only objects that transform in the same manner under gauge and Lorentz transformations, and so is valid in any gauge and in any Lorentz frame. If experiment shows that such a sum rule does not hold, it implies a failure of the theory.
\item A \emph{relation} allows one to evaluate a theoretical quantity, itself not directly measurable from experiment, using the values of other measurable quantities. The interpretation is generally valid only in a fixed (or a restricted class of) gauge and/or  Lorentz frame. Using a \emph{relation} does not directly test a theory, though it might be used to test a model or lattice calculation of the theoretical quantity.
\end{itemize}
Prior to defining angular momentum, one has first to specify what form of dynamics one is working in, since instant form and light-front form operators usually differ, as explained in section \ref{secIIB3}. Similarly, one has also to specify what kind of definition for spin one is using in a moving frame (canonical spin, helicity, light-front helicity, \emph{etc.}). While a sum rule is insensitive to these choices, the particular quantity studied in a relation will in general depend on them. However, if one considers a relation in the infinite-momentum frame (or, equivalently, restricts it to its leading-twist part), these particular choices usually do not matter.

While there are few genuine sum rules, many relations have appeared recently and caused a lot of confusion. In the following, we list the most relevant ones in connection with the angular momentum decomposition. For convenience, we always choose the spatial axes such that the nucleon momentum lies along the $OZ$ axis, \emph{i.e.} we can write $P^\mu=(P^0,0,0,P_z)$ without loss of generality. We use the notation
\beq
\la\la P,\mathscr S|O|P,\mathscr S\ra\ra=\frac{\la P,\mathscr S|O|P,\mathscr S\ra}{\la P,\mathscr S|P,\mathscr S\ra}
\eeq
for the expectation value of an operator $O$, and work in the $\overline{MS}$ scheme.

\subsubsection{Sum rules\label{secVIIIC1}}

As shown in section \ref{secVI}, the nucleon matrix elements of the quark and gluon energy-momentum density are expressed in terms of form factors $A_i(t)$, $B_i(t)$, $C_i(t)$ and $\bar{C}_i(t)$, where  $t= \Delta^2$ is the invariant momentum transfer and the index $i$ labels specific quarks or gluons. Momentum conservation then implies the momentum sum rule
\beq\label{momSR}
\sum_{q,G} A_i(0)=1.
\eeq
From conservation of total angular momentum, one obtains the angular momentum sum rule
\beq
\sum_{q,G}\tfrac{1}{2}\left[A_i(0)+B_i(0)\right]=\tfrac{1}{2}.
\eeq
Combining momentum and angular momentum sum rules leads to the so-called anomalous gravitomagnetic moment sum rule
\beq\label{AGMSR}
\sum_{q,G} B_i(0)=0,\eeq
which can be shown to imply that the nucleon anomalous gravitomagnetic moment vanishes \cite{Brodsky:2000ii,Teryaev:1999su}. The $\bar C$ form factors appear only in the non-conserved individual quark and gluon energy-momentum tensors, so one has also the following sum rule
\beq\label{consSR}
\sum_{q,G} \bar C_i(t)=0,
\eeq
which is valid for any $t$.

The $A$ and $B$ form factors can be expressed in terms of the second $x$-moment of the twist-2 GPDs $H$ and $E$ \cite{Ji:1996ek,Diehl:2003ny}
\beq
A_{q,G}(t)=\int\ud x\,x\,H_{q,G}(x,0,t),\qquad B_{q,G}(t)=\int\ud x\,x\,E_{q,G}(x,0,t), \eeq
where $q$ refers to the sum of quark and antiquark of a given flavor. Integrals over quark GPDs run from $-1$ to $+1$, whereas integrals over gluon GPDs run from $0$ to $1$. For the $\bar C$ quark form factor, we find\footnote{Note that the expression was incorrectly given in Ref. \cite{Leader:2012ar}}
\beq
\bar C_q(t)=\int\ud x\,x\left[H^q_3(x,0,t)-\tfrac{1}{2}\,H_q(x,0,t)\right],
\eeq
where $H_3$ is a twist-four GPD. A similar expression is expected to hold in the gluon sector. Plugging these GPD expressions in the sum rules \eqref{momSR}, \eqref{AGMSR} and \eqref{consSR}, one arrives at
\begin{align}
\sum_{q,G}\int\ud x\,x\,H_i(x,0,0)&=1,\\
\sum_{q,G}\int\ud x\,x\,E_i(x,0,0)&=0,\label{ESR}\\
\sum_{q,G}\int\ud x\,x\,H^i_3(x,0,t)&=\tfrac{1}{2}\sum_{q,G}\int\ud x\,x\,H_i(x,0,t).\label{H3SR}
\end{align}

In addition, there is the Penttinen-Polyakov-Shuvaev-Strikman sum rule \cite{Penttinen:2000dg,Hatta:2012cs} derived in section \ref{secVIIA3}, which, in the parametrization of Kiptily and Polyakov for twist-3 GPDs \cite{Kiptily:2002nx}, reads
\beq
\int\ud x\,x\left[H_q(x,0,0)+E_q(x,0,0)+2G^q_2(x,0,0)\right]=\int\ud x\,\tilde H_q(x,0,0),
\eeq
where $G_2$ is a genuine twist-3 GPD and where one can write
\beq
\int\ud x\,\tilde H_q(x,0,0)=a^q_0,
\eeq
with $a^q_0$ the contribution of a quark plus antiquark of a given flavor to the nucleon flavor singlet axial charge $a_0$ (or $g_A^{(0)}$).

We stress that the above are genuine sum rules in that all the quantities they contain can, at least in principle, be determined from experiment. Of course, extracting them from experiment  may be extremely difficult in practice. Finally, one can also write the following sum rule
\beq
\int\ud x\,G^q_2(x,0,0)=\int\ud x\,\ud^2k_\perp\,\frac{\uvec k^2_\perp}{M^2}\,F^q_{1,4}(x,\uvec k_\perp,0;\mathcal W_\text{straight}),
\eeq
which is unfortunately not relevant from the experimental point of view, since there is no indication so far as to how one can extract this particular GTMD, especially since it has a Wilson line that does not run along the light-front direction. The best chance to check this sum rule is therefore with model and lattice calculations.

\subsubsection{Instant form relations\label{secVIIIC2}}

In instant form, as explained in section \ref{secIIB3}, the operator $Q$ associated with any density $j^\mu$ is defined as
\beq
Q=\int\ud^3x\,j^0(x),
\eeq
and the covariant spin four-vector is given by
\beq
\mathscr S^\mu(P,\uvec s)=(P_zs_z,M\uvec s_\perp,P^0s_z).
\eeq
The nucleon state normalization reads
\beq
\la P,\mathscr S|P,\mathscr S\ra=2P^0\,(2\pi)^3\,\delta^{(3)}(\uvec 0).
\eeq
Focusing on the \emph{longitudinal} component of the angular momentum in a \emph{longitudinally} polarized nucleon, one has the following explicit kinetic decompositions:
\begin{itemize}
\item the Belinfante decomposition
\beq
\tfrac{1}{2}=\sum_q\la\la P,\mathscr S_L|J^q_{\text{Bel},z}|P,\mathscr S_L\ra\ra+\la\la P,\mathscr S_L|J^G_{\text{Bel},z}|P,\mathscr S_L\ra\ra,
\eeq
\item the Ji decomposition
\beq
\tfrac{1}{2}=\sum_q\la\la P,\mathscr S_L|S^q_{z}|P,\mathscr S_L\ra\ra+\sum_q\la\la P,\mathscr S_L| L^q_{\text{Ji}, z}|P,\mathscr S_L\ra\ra+\la\la P,\mathscr S_L|J^G_{\text{Bel},z}|P,\mathscr S_L\ra\ra,
\eeq
\end{itemize}
where the various contributions are given explicitly in the following. Note that no explicit canonical decomposition is available in the instant form, unless one considers the infinite-momentum frame where instant and light-front forms simply coincide. For this reason, the explicit canonical decompositions will be considered only in the next subsection where the light-front expressions are discussed. In the following relations the first line of each equation involves  quantities,  form factors, that can be measured on a lattice, and the further lines involve experimentally measurable quantities.

The longitudinal component of the kinetic angular momentum carried by quarks and gluons is related to experimentally measurable quantities by the Ji relation \cite{Ji:1996ek}
\beq\label{JiIF}
\begin{split}
\la\la  P,\mathscr{S}_L |J^i_{\text{Bel},z} | P, \mathscr{S}_L  \ra \ra  \equiv  \la\la P,\mathscr S_L|M^{12}_{\text{Bel},i}|P,\mathscr S_L\ra\ra& =\tfrac{1}{2}\left[A_i(0)+B_i(0)\right]\\
&=\tfrac{1}{2}\int\ud x\,x\left[H_i(x,0,0)+E_i(x,0,0)\right],
\end{split}
\eeq
with $i=q,G$ and $\mathscr S_L\equiv \mathscr S(P,\uvec e_z)$.

The quark longitudinal spin is given by
\beq \label{Si}
\begin{split}
  \la\la  P, \mathscr{S}_L  | S^q_{z}  |  P, \mathscr{S}_L \ra \ra \equiv  \la\la P,\mathscr S_L|M^{12}_{q,\text{spin}}|P,\mathscr S_L \ra\ra&=\tfrac{1}{2}\,a^q_0\\
&=\tfrac{1}{2}\int\ud x\,\tilde H_q(x,0,0).
\end{split}
\eeq
This relation does not depend on the magnitude of the nucleon momentum, and remains therefore valid in the infinite-momentum frame $P_z\to\infty$. In this frame, the quarks move essentially collinearly with the nucleon, and so one can identify the quark longitudinal spin with the quark helicity.

Finally, there are several relations with experimental quantities for the longitudinal component of the quark kinetic orbital angular momentum \cite{Penttinen:2000dg,Ji:2012sj,Lorce:2012ce}
\beq \label{Li}
\begin{split}
\la\la P,\mathscr{S}_L | L^q_{\text{Ji}, z}  | P, \mathscr{S}_L \ra \ra \equiv  \la\la P,\mathscr S_L|M^{12}_{\text{Ji},q,\text{OAM}}|P,\mathscr S_L \ra\ra&=\tfrac{1}{2}\left[A_q(0)+B_q(0)-a^q_0\right]\\
&=\tfrac{1}{2}\int\ud x\,x\left[H_q(x,0,0)+E_q(x,0,0)\right]-\tfrac{1}{2}\int\ud x\,\tilde H_q(x,0,0)\\
&=-\int\ud x\,x\,G^q_2(x,0,0)\\
&=-\int\ud x\,\ud^2k_\perp\,\frac{\uvec k^2_\perp}{M^2}\,F^q_{1,4}(x,\uvec k_\perp,0;\mathcal W_\text{straight}).
\end{split}
\eeq

Turning now to the \emph{transverse} component of the kinetic angular momentum in a \emph{transversely} polarized nucleon, say in the $x$-direction, there is the Leader relation \cite{Leader:2011cr}
\beq\label{LeaderIF}
\begin{split}
\la\la P,\mathscr{S}_x | J^i_{\text{Bel},x}  | P, \mathscr{S}_x  \ra \ra  \equiv  \la\la P,\mathscr S_x|M^{23}_{\text{Bel},i}|P,\mathscr S_x\ra\ra&=\tfrac{1}{2}\left[A_i(0)+\tfrac{P^0}{M}\,B_i(0)\right]\\
&=\tfrac{1}{2}\int\ud x\,x\left[H_i(x,0,0)+\tfrac{P^0}{M}\,E_i(x,0,0)\right],
\end{split}
\eeq
with $i=q,G$ and $\mathscr S_x\equiv \mathscr S(P,\uvec e_x)$. Contrary to the longitudinal component, the transverse component of the kinetic angular momentum depends on the momentum of the nucleon. This should not be surprising, since longitudinal boosts commute with longitudinal rotations, but not with transverse rotations. In the rest frame $P^0=M$, the Ji and Leader relations naturally coincide by rotational invariance. Note also that the momentum-dependent term drops out when considering the total (\emph{i.e.} quark$+$gluon) angular momentum, thanks to the anomalous gravitomagnetic sum rule \eqref{AGMSR} or equivalently \eqref{ESR}.

To avoid momentum dependence in the case of transverse polarization, Ji, Xiong and Yuan \cite{Ji:2012vj} proposed to consider instead the quark and gluon contributions to the Pauli-Lubanski four-vector of the nucleon
\beq
W^\mu_{\text{Bel},i}\equiv\tfrac{1}{2}\,\epsilon^{\mu\nu\rho\sigma}M^{\text{Bel},i}_{\nu\rho}P_\sigma
\eeq
with $i=q,G$ and $\epsilon_{0123}=+1$, and $P_{\sigma}$ is the \emph{total} momentum operator. The Pauli-Lubanski four-vector has the welcome feature that
\beq
\la\la P,\mathscr S|\sum_{q,G}W^\mu_{\text{Bel},i}|P,\mathscr S\ra\ra=\tfrac{1}{2}\,\mathscr S^\mu.
\eeq
For the transverse component, say in the $x$-direction, Leader \cite{Leader:2012ar} obtained
\beq\label{PauliIF}
\begin{split}
\la\la P,\mathscr S_x|W^1_{\text{Bel},i}|P,\mathscr S_x\ra\ra&=\tfrac{1}{2}\left[A_i(0)+B_i(0)+\tfrac{P^0-M}{P^0}\,\bar C_i(0)\right]\mathscr S^1_x\\
&=\tfrac{1}{2}\int\ud x\,x\left[\tfrac{P^0+M}{2P^0}\,H_i(x,0,0)+E_i(x,0,0)+\tfrac{P^0-M}{P^0}\,H^i_3(x,0,0)\right]\mathscr S^1_x.
\end{split}
\eeq
So, even by using the Pauli-Lubanski four-vector for the transverse polarization, one cannot get rid of the momentum dependence for the separate quark and gluon contributions. Only for the total (\emph{i.e.} quark$+$gluon) Pauli-Lubanski four-vector does the momentum dependence drop out, owing to Eq. \eqref{consSR} or equivalently Eq. \eqref{H3SR}. In the rest frame $P^0=M$, the spatial components of the Pauli-Lubanski four-vector coincide with the angular momentum operators times the nucleon mass
\beq
\uvec W^{q,G}_\text{Bel}\big|_{P^0=M}=M\uvec J^{q,G}_\text{Bel},
\eeq
so it is quite natural that Eq. \eqref{PauliIF} agrees with the Ji \eqref{JiIF} and Leader \eqref{LeaderIF} relations in the rest frame.

\subsubsection{Light-front form relations\label{secVIIIC3}}

In light-front form, the operator $Q$ associated with a density $j^\mu$ is defined as
\beq
Q=\int\ud x^-\ud^2x_\perp\,j^+(x),
\eeq
and the covariant spin four-vector is given by
\beq
\mathscr S^\mu=[P^+s_z,-P^-s_z,M\uvec s_\perp].
\eeq
The nucleon state normalization reads
\beq
\la P,\mathscr S|P,\mathscr S\ra=2P^+\,(2\pi)^3\,\delta^{(3)}(\uvec 0).
\eeq
In this section, all operators and states are in light-front form, but to avoid a typographical mess, we shall not attach subscripts LF to the expressions.

Focusing on the \emph{longitudinal} component of the angular momentum in a \emph{longitudinally} polarized nucleon, one has the following explicit kinetic decompositions:
\begin{itemize}
\item the Belinfante decomposition
\beq
\tfrac{1}{2}=\sum_q\la\la P,\mathscr S_L|J^q_{\text{Bel},z}|P,\mathscr S_L\ra\ra+\la\la P,\mathscr S_L|J^G_{\text{Bel},z}|P,\mathscr S_L\ra\ra,
\eeq
\item the Ji decomposition
\beq
\tfrac{1}{2}=\sum_q\la\la P,\mathscr S_L|S^q_{z}|P,\mathscr S_L\ra\ra+\sum_q\la\la P,\mathscr S_L| L^q_{\text{Ji}, z}|P,\mathscr S_L\ra\ra+\la\la P,\mathscr S_L|J^G_{\text{Bel},z}|P,\mathscr S_L\ra\ra,
\eeq
\item the light-front gauge-invariant kinetic (gik) decomposition
\beq
\tfrac{1}{2}=\sum_q\la\la P,\mathscr S_L|S^q_{z}|P,\mathscr S_L\ra\ra+\sum_q\la\la P,\mathscr S_L|L^q_{\text{Ji}, z}|P,\mathscr S_L\ra\ra+\la\la P,\mathscr S_L|S^G_{z}|P,\mathscr S_L\ra\ra+\la\la P,\mathscr S_L|L^G_{\text{gik}, z}|P,\mathscr S_L\ra\ra,
\eeq
\end{itemize}
and the following explicit canonical decomposition:
\begin{itemize}
\item the light-front gauge-invariant canonical (gic) decomposition
\beq
\tfrac{1}{2}=\sum_q\la\la P,\mathscr S_L|S^q_{z}|P,\mathscr S_L\ra\ra+\sum_q\la\la P,\mathscr S_L|L^q_{\text{gic}, z}|P,\mathscr S_L\ra\ra+\la\la P,\mathscr S_L|S^G_{z}|P,\mathscr S_L\ra\ra+\la\la P,\mathscr S_L|L^G_{\text{gic}, z}|P,\mathscr S_L\ra\ra,
\eeq
\end{itemize}
where the various contributions are given explicitly in the following. Note that the light-front gauge-invariant canonical decomposition was written explicitly for the first time by Hatta \cite{Hatta:2011ku} and gives the same result as the Jaffe-Manohar decomposition \cite{Jaffe:1989jz} in the light-front gauge $A^+=0$.

The longitudinal component of the kinetic angular momentum is provided once again by the Ji relation \cite{Ji:1996ek}
\beq
\begin{split}
  \la\la  P,\mathscr{S}_L | J^i_{\text{Bel},z}  | P, \mathscr{S}_L  \ra \ra \equiv \la\la P,\mathscr S_L|M^{12}_{\text{Bel},i}|P,\mathscr S_L\ra\ra&=\tfrac{1}{2}\left[A_i(0)+B_i(0)\right]\\
&=\tfrac{1}{2}\int\ud x\,x\left[H_i(x,0,0)+E_i(x,0,0)\right],
\end{split}
\eeq
with $i=q,G$ and $\mathscr S_L\equiv \mathscr S(P,\uvec e_z)$.

The quark light-front helicity is given, as in Eq.~\eqref{Si},  by
\beq
\begin{split}
  \la\la  P, \mathscr{S}_L  | S^q_{z}  |  P, \mathscr{S}_L \ra \ra \equiv \la\la P,\mathscr S|M^{12}_{q,\text{spin}}|P,\mathscr S\ra\ra&=\tfrac{1}{2}\,a^q_0\\
&=\tfrac{1}{2}\int\ud x\,\tilde H_q(x,0,0).
\end{split}
\eeq
Similarly, the gluon light-front helicity is given by
\beq
  \la\la  P, \mathscr{S}_L  | S^G_{z}  |  P, \mathscr{S}_L \ra \ra \equiv \la\la P,\mathscr S|M^{12}_{G,\text{spin}}|P,\mathscr S\ra\ra=\int\ud x\,\tilde H_G(x,0,0),
\eeq
where $M^{12}_{G,\text{spin}}$ stands, equivalently, for the local Jaffe-Manohar operator in the light-front gauge \cite{Jaffe:1989jz} or the non-local gauge-invariant Hatta operator \cite{Hatta:2011zs}.

The longitudinal component of the kinetic orbital angular momentum  for quarks \cite{Penttinen:2000dg,Ji:2012sj,Lorce:2012ce} reads
\beq
\begin{split}
\la\la P,\mathscr{S}_L | L^q_{\text{Ji}, z}  | P, \mathscr{S}_L \ra \ra \equiv \la\la P,\mathscr S|M^{12}_{\text{Ji},q,\text{OAM}}|P,\mathscr S\ra\ra&=\tfrac{1}{2}\left[A_q(0)+B_q(0)-a^q_0\right]\\
&=\tfrac{1}{2}\int\ud x\,x\left[H_q(x,0,0)+E_q(x,0,0)\right]-\tfrac{1}{2}\int\ud x\,\tilde H_q(x,0,0)\\
&=-\int\ud x\,x\,G^q_2(x,0,0)\\
&=-\int\ud x\,\ud^2k_\perp\,\frac{\uvec k^2_\perp}{M^2}\,F^q_{1,4}(x,\uvec k_\perp,0;\mathcal W_\text{straight}),
\end{split}
\eeq
and similarly for gluons
\beq
\begin{split}
\la\la P,\mathscr{S}_L | L^G_{\text{gik}, z}  | P, \mathscr{S}_L \ra \ra \equiv \la\la P,\mathscr S|M^{12}_{\text{gik},G,\text{OAM}}|P,\mathscr S\ra\ra&=\tfrac{1}{2}\int\ud x\,x\left[H_G(x,0,0)+E_G(x,0,0)\right]-\int\ud x\tilde H_G(x,0,0)\\
&=-\int\ud x\,\ud^2k_\perp\,\frac{\uvec k^2_\perp}{M^2}\,F^G_{1,4}(x,\uvec k_\perp,0;\mathcal W_\text{straight}).
\end{split}
\eeq
Finally, the relation for the longitudinal component of the canonical orbital angular momentum has been obtained by Lorc\'e and Pasquini \cite{Lorce:2011kd} for quarks, and by Hatta \cite{Hatta:2011ku} for gluons
\beq
\la\la P,\mathscr{S}_L | L^i_{\text{gic}, z}  | P, \mathscr{S}_L \ra \ra \equiv\la\la P,\mathscr S|M^{12}_{\text{gic},i,\text{OAM}}|P,\mathscr S\ra\ra=-\int\ud x\,\ud^2k_\perp\,\frac{\uvec k^2_\perp}{M^2}\,F^i_{1,4}(x,\uvec k_\perp,0;\mathcal W_\text{staple}),
\eeq
where $M^{12}_{i,\text{OAM}}$ stands, equivalently, for the local Jaffe-Manohar operator in the light-front gauge \cite{Jaffe:1989jz} or the non-local gauge-invariant Hatta operator \cite{Hatta:2011zs}.

Turning to the \emph{transverse} component of the kinetic ``angular momentum''\footnote{This is, of course, not an angular momentum in the usual sense, since it involves both a rotation and a boost.} in a \emph{transversely} polarized nucleon, say in the $x$-direction, one finds
\beq
\begin{split}
\la\la P,\mathscr S_x|M^{-2}_{\text{Bel},i}|P,\mathscr S_x\ra\ra&=\tfrac{M}{2P^+}\left[A_i(0)+\tfrac{1}{2}\,B_i(0)+\bar C_i(0)\right]\\
&=\tfrac{M}{4P^+}\int\ud x\,x\left[H_i(x,0,0)+E_i(x,0,0)+2H^i_3(x,0,0)\right],
\end{split}
\eeq
with $i=q,G$ and $\mathscr S_x\equiv \mathscr S(P,\uvec e_x)$. Contrary to the longitudinal component, the transverse component of the kinetic angular momentum depends on the momentum of the nucleon, owing to the fact that longitudinal boosts commute with longitudinal rotations, but not with transverse rotations. Note however that, contrary to the instant form, the expressions for the longitudinal and transverse components do not coincide in the rest frame $P^+=M$. This comes from the fact that the generators of transverse rotations are dynamical (\emph{i.e.} contain interaction) in light-front form, whereas they are kinematic in instant form. The dynamical nature of the light-front generators of transverse rotations is manifested by the global $\tfrac{M}{P^+}$ factor indicating higher-twist, and by the presence of the $\bar C$ form factor. Note also that the momentum dependence does not drop out when considering the total (\emph{i.e.} quark$+$gluon) generators of transverse rotations: one gets $\tfrac{M}{2P^+}$ instead of $\tfrac{1}{2}$. There is actually nothing wrong with that. It simply signals that the light-front generators of transverse rotations cannot be interpreted as actual angular momentum operators. Indeed, these operators satisfy the algebra of two-dimensional Euclidean space
\beq
\left[M^{-1},M^{-2}\right]=0,\qquad\left[M^{12},M^{-i}\right]=i\epsilon^{ij}_\perp M^{-j},
\eeq
instead of the ordinary angular momentum commutation relations.

Using the Pauli-Lubanski four-vector to describe the transverse polarization, Harindranath, Kundu and Mukherjee \cite{Harindranath:2013goa} obtained\footnote{Hatta, Tanaka and Yoshida \cite{Hatta:2012jm} obtained a different result because they used the instant form polarization instead of the light-front polarization, in combination with the light-front operators.}
\beq\label{PauliLF}
\begin{split}
\la\la P,\mathscr S_x|W^1_{\text{Bel},i}|P,\mathscr S_x\ra\ra&=\tfrac{1}{2}\left[A_i(0)+B_i(0)+\bar C_i(0)\right]\mathscr S^1_x\\
&=\tfrac{1}{2}\int\ud x\,x\left[\tfrac{1}{2}\,H_i(x,0,0)+E_i(x,0,0)+H^i_3(x,0,0)\right]\mathscr S^1_x,
\end{split}
\eeq
where $\mathscr S^1_x=M$. This is the momentum-independent relation for the transverse component of the Pauli-Lubanski four-vector. Note that the $\bar C$ form factor was incorrectly discarded by Ji, Xiong and Yuan \cite{Ji:2012vj}. Only for the total (\emph{i.e.} quark$+$gluon) Pauli-Lubanski four-vector, it is justified to do so,  owing to Eq. \eqref{consSR} or equivalently Eq. \eqref{H3SR}.

In summary, in all the relations in this and the previous subsection, the LHS of the relations cannot be measured directly from experiment, and the RHSs simply provide information about the quantities on the LHS from experiment. The relations are thus not sum rules. However, the relations become particularly interesting when, as is sometimes possible, the LHSs can be calculated on a lattice or in models.

\section{Conclusions\label{secIX}}

The desire to find an expression for the photon or gluon total angular momentum, which contains separate, gauge-invariant spin and orbital pieces has led to an outpouring of  papers showing how this can be done, in, it turns out, an infinite number of different ways. None of these contradicts the age old textbook statement that this is impossible, since  all the new constructions involve non-local fields, which is outside the category of local fields considered in the textbooks. A somewhat surprising corollary to these constructions is that there are an infinite number of ways to \emph{define} what we mean by the quark or gluon momentum and angular momentum. It seems there is no absolutely compelling reason to prefer any one particular definition, though we feel that the fundamental versions are the canonical and the Belinfante ones, since they at least involve local fields, and it can be argued that the canonical version has certain properties that one would like to associate with an angular momentum. Moreover, they represent complementary information as to how the nucleon spin is built up from the angular momentum of its constituents, a key aspect in understanding the internal structure of the nucleon.

We have presented a pedagogical introduction to the subject and a detailed discussion of the new decompositions, and commented on their advantages and disadvantages. There have been many very interesting theoretical developments, but we have concluded that they contain no new important physical implications, and for that reason we have concentrated on experimental tests and measurements only with regard to the canonical and Belinfante versions of the angular momentum.

Because the subject is so technical we have presented a wholly pedagogical introduction and prepared a separate, largely non-technical section dealing with the physical implications. Therein we have placed particular stress on sum rules in which every term can be measured experimentally, so that a gross failure of such a sum rule would have severe implications for the validity of QCD. We have also highlighted important relations connecting some of the terms in the decompositions to experimentally measurable quantities.

\section*{Acknowledgements}
For this review, we greatly benefited from numerous discussions with many colleagues. E.L. is grateful to Mike Creutz and Keh-Fei Liu for discussions about the lattice studies, and to the latter for providing the beautiful diagrams displaying the latest lattice results. He has also benefited from helpful discussions with Ben Bakker, Vladimir Braun, Xiang-Song Chen, John Collins, Yoshitaka Hatta, Robert L. Jaffe, Dieter M\"{u}ller, Piet Mulders, Masashi Wakamatsu, and Feng Yuan. C.L. aknowledges particularly useful exchanges with Alessandro Baccheta, Matthias Burkardt, Xiang-Song Chen, Aurore Courtoy, Leonard Gamberg, Terry Goldman, Yoshitaka Hatta, Robert L. Jaffe, Xiangdong Ji, Andreas Metz, Asmita Mukherjee, Piet Mulders, Barbara Pasquini, Masashi Wakamatsu, Fan Wang, and Feng Yuan.

E.L. is also grateful to Jean-Philippe Lansberg and Hagop Sazdjian for hospitality at the Universit\'e Paris-Sud, to Jianwei Qiu, Raju Venugopalan, Werner Vogelsang  and Feng Yuan for hospitality at the Brookhaven National Laboratory and the Lawrence Berkeley National Laboratory, and to Jean-Ren\'e Cudell for hospitality at the Universit\'e de Li\`ege. C.L. would like to thank also Jordan Nash for hospitality at Imperial College London.

In this work, C.L. was supported by the P2I (``Physique des deux Infinis'') network and by the Belgian Fund F.R.S.-FNRS \emph{via} the contract of Charg\'e de recherches.

\bibliography{Elliot_General}

%Merlin.mbs v4.21 2009-07-09.
\begin{thebibliography}{100}%
\makeatletter
\providecommand \@ifxundefined [1]{%
 \ifx #1\undefined \expandafter \@firstoftwo
 \else \expandafter \@secondoftwo
\fi
}%
\providecommand \@ifnum [1]{%
 \ifnum #1\expandafter \@firstoftwo
 \else \expandafter \@secondoftwo
\fi
}%
\providecommand \enquote [1]{``#1''}%
\providecommand \bibnamefont  [1]{#1}%
\providecommand \bibfnamefont [1]{#1}%
\providecommand \citenamefont [1]{#1}%
\providecommand\href[0]{\@sanitize\@href}%
\providecommand\@href[1]{\endgroup\@@startlink{#1}\endgroup\@@href}%
\providecommand\@@href[1]{#1\@@endlink}%
\providecommand \@sanitize [0]{\begingroup\catcode`\&12\catcode`\#12\relax}%
\@ifxundefined \pdfoutput {\@firstoftwo}{%
 \@ifnum{\z@=\pdfoutput}{\@firstoftwo}{\@secondoftwo}%
}{%
 \providecommand\@@startlink[1]{\leavevmode\special{html:<a href="#1">}}%
 \providecommand\@@endlink[0]{\special{html:</a>}}%
}{%
 \providecommand\@@startlink[1]{%
  \leavevmode
  \pdfstartlink
   attr{/Border[0 0 1 ]/H/I/C[0 1 1]}%
   user{/Subtype/Link/A<</Type/Action/S/URI/URI(#1)>>}%
  \relax
 }%
 \providecommand\@@endlink[0]{\pdfendlink}%
}%
\providecommand \url  [0]{\begingroup\@sanitize \@url }%
\providecommand \@url [1]{\endgroup\@href {#1}{\urlprefix}}%
\providecommand \urlprefix [0]{URL }%
\providecommand \Eprint[0]{\href }%
\@ifxundefined \urlstyle {%
  \providecommand \doi [1]{doi:\discretionary{}{}{}#1}%
}{%
  \providecommand \doi [0]{doi:\discretionary{}{}{}\begingroup
  \urlstyle{rm}\Url }%
}%
\providecommand \doibase [0]{http://dx.doi.org/}%
\providecommand \Doi[1]{\href{\doibase#1}}%
\providecommand \bibAnnote [3]{%
  \BibitemShut{#1}%
  \begin{quotation}\noindent
    \textsc{Key:}\ #2\\\textsc{Annotation:}\ #3%
  \end{quotation}%
}%
\providecommand \bibAnnoteFile [2]{%
  \IfFileExists{#2}{\bibAnnote {#1} {#2} {\input{#2}}}{}%
}%
\providecommand \typeout [0]{\immediate \write \m@ne }%
\providecommand \selectlanguage [0]{\@gobble}%
\providecommand \bibinfo [0]{\@secondoftwo}%
\providecommand \bibfield [0]{\@secondoftwo}%
\providecommand \translation [1]{[#1]}%
\providecommand \BibitemOpen[0]{}%
\providecommand \bibitemStop [0]{}%
\providecommand \bibitemNoStop [0]{.\EOS\space}%
\providecommand \EOS [0]{\spacefactor3000\relax}%
\providecommand \BibitemShut [1]{\csname bibitem#1\endcsname}%
%</preamble>
\bibitem{Wentzel:1949}%
  \BibitemOpen
  \bibfield{author}{%
  \bibinfo {author} {\bibfnamefont{G.}~\bibnamefont{Wentzel}},\ }%
  \emph{\bibinfo {title} {{Quantum Theory of Fields}}}\ (\bibinfo {publisher}
  {Interscience},\ \bibinfo {address} {New York})%
  \bibAnnoteFile{NoStop}{Wentzel:1949}%
\bibitem{Gottfried:1966}%
  \BibitemOpen
  \bibfield{author}{%
  \bibinfo {author} {\bibfnamefont{K.}~\bibnamefont{Gottfried}},\ }%
  \emph{\bibinfo {title} {{Quantum Mechanics}}}\ (\bibinfo {publisher}
  {Benjamin},\ \bibinfo {address} {New York})%
  \bibAnnoteFile{NoStop}{Gottfried:1966}%
\bibitem{Merzbacher:1970}%
  \BibitemOpen
  \bibfield{author}{%
  \bibinfo {author} {\bibfnamefont{E.}~\bibnamefont{Merzbacher}},\ }%
  \emph{\bibinfo {title} {{Quantum Mechanics}}}\ (\bibinfo {publisher}
  {Wiley},\ \bibinfo {address} {New York})%
  \bibAnnoteFile{NoStop}{Merzbacher:1970}%
\bibitem{Lenstra:1982}%
  \BibitemOpen
  \bibfield{author}{%
  \bibinfo {author} {\bibfnamefont{D.}~\bibnamefont{Lenstra}}\ and\ \bibinfo
  {author} {\bibfnamefont{L.}~\bibnamefont{Mandel}},\ }%
  \bibfield{journal}{%
  \bibinfo {journal} {Phys.Rev.}\ }%
  \textbf{\bibinfo {volume} {A26}},\ \bibinfo {pages} {3428} (\bibinfo {year}
  {1982})%
  \bibAnnoteFile{NoStop}{Lenstra:1982}%
\bibitem{Allen:1992zz}%
  \BibitemOpen
  \bibfield{author}{%
  \bibinfo {author} {\bibfnamefont{L.}~\bibnamefont{Allen}}, \bibinfo {author}
  {\bibfnamefont{M.}~\bibnamefont{Beijersbergen}}, \bibinfo {author}
  {\bibfnamefont{R.}~\bibnamefont{Spreeuw}},\ and\ \bibinfo {author}
  {\bibfnamefont{J.}~\bibnamefont{Woerdman}},\ }%
  \bibfield{journal}{%
  \Doi{10.1103/PhysRevA.45.8185}{\bibinfo {journal} {Phys.Rev.}}\ }%
  \textbf{\bibinfo {volume} {A45}},\ \bibinfo {pages} {8185} (\bibinfo {year}
  {1992})%
  \bibAnnoteFile{NoStop}{Allen:1992zz}%
%%CITATION = PHRVA,A45,8185;%%
\bibitem{vanEnk:1992}%
  \BibitemOpen
  \bibfield{author}{%
  \bibinfo {author} {\bibfnamefont{S.}~\bibnamefont{van Enk}}\ and\ \bibinfo
  {author} {\bibfnamefont{G.}~\bibnamefont{Nienhuis}},\ }%
  \bibfield{journal}{%
  \bibinfo {journal} {Opt.Commun.}\ }%
  \textbf{\bibinfo {volume} {94}},\ \bibinfo {pages} {147} (\bibinfo {year}
  {1992})%
  \bibAnnoteFile{NoStop}{vanEnk:1992}%
\bibitem{Nienhuis:1993}%
  \BibitemOpen
  \bibfield{author}{%
  \bibinfo {author} {\bibfnamefont{G.}~\bibnamefont{Nienhuis}}\ and\ \bibinfo
  {author} {\bibfnamefont{L.}~\bibnamefont{Allen}},\ }%
  \bibfield{journal}{%
  \bibinfo {journal} {Phys.Rev.}\ }%
  \textbf{\bibinfo {volume} {A48}},\ \bibinfo {pages} {656} (\bibinfo {year}
  {1993})%
  \bibAnnoteFile{NoStop}{Nienhuis:1993}%
\bibitem{Barnett:1994}%
  \BibitemOpen
  \bibfield{author}{%
  \bibinfo {author} {\bibfnamefont{S.}~\bibnamefont{Barnett}}\ and\ \bibinfo
  {author} {\bibfnamefont{L.}~\bibnamefont{Allen}},\ }%
  \bibfield{journal}{%
  \bibinfo {journal} {Opt.Commun.}\ }%
  \textbf{\bibinfo {volume} {110}},\ \bibinfo {pages} {670} (\bibinfo {year}
  {1994})%
  \bibAnnoteFile{NoStop}{Barnett:1994}%
\bibitem{vanEnk:1994}%
  \BibitemOpen
  \bibfield{author}{%
  \bibinfo {author} {\bibfnamefont{S.}~\bibnamefont{van Enk}}\ and\ \bibinfo
  {author} {\bibfnamefont{G.}~\bibnamefont{Nienhuis}},\ }%
  \bibfield{journal}{%
  \bibinfo {journal} {Europhys.Lett.}\ }%
  \textbf{\bibinfo {volume} {25}},\ \bibinfo {pages} {497} (\bibinfo {year}
  {1994})%
  \bibAnnoteFile{NoStop}{vanEnk:1994}%
\bibitem{Barnett:2002}%
  \BibitemOpen
  \bibfield{author}{%
  \bibinfo {author} {\bibfnamefont{S.}~\bibnamefont{Barnett}},\ }%
  \bibfield{journal}{%
  \bibinfo {journal} {J.Opt.B:Quantum Semiclass. Opt.}\ }%
  \textbf{\bibinfo {volume} {4}},\ \bibinfo {pages} {S7} (\bibinfo {year}
  {2002})%
  \bibAnnoteFile{NoStop}{Barnett:2002}%
\bibitem{Jauregui:2005}%
  \BibitemOpen
  \bibfield{author}{%
  \bibinfo {author} {\bibfnamefont{R.}~\bibnamefont{J\'auregui}}\ and\ \bibinfo
  {author} {\bibfnamefont{S.}~\bibnamefont{Hacyan}},\ }%
  \bibfield{journal}{%
  \bibinfo {journal} {Phys.Rev.}\ }%
  \textbf{\bibinfo {volume} {A71}},\ \bibinfo {pages} {033411} (\bibinfo {year}
  {2005})%
  \bibAnnoteFile{NoStop}{Jauregui:2005}%
\bibitem{Calvo:2006}%
  \BibitemOpen
  \bibfield{author}{%
  \bibinfo {author} {\bibfnamefont{G.}~\bibnamefont{Calvo}}, \bibinfo {author}
  {\bibfnamefont{A.}~\bibnamefont{Pic\'on}},\ and\ \bibinfo {author}
  {\bibfnamefont{E.}~\bibnamefont{Bagan}},\ }%
  \bibfield{journal}{%
  \bibinfo {journal} {Phys.Rev.}\ }%
  \textbf{\bibinfo {volume} {A73}},\ \bibinfo {pages} {013805} (\bibinfo {year}
  {2006})%
  \bibAnnoteFile{NoStop}{Calvo:2006}%
\bibitem{Hacyan:2006}%
  \BibitemOpen
  \bibfield{author}{%
  \bibinfo {author} {\bibfnamefont{S.}~\bibnamefont{Hacyan}}\ and\ \bibinfo
  {author} {\bibfnamefont{R.}~\bibnamefont{J\'auregui}},\ }%
  \bibfield{journal}{%
  \bibinfo {journal} {J.Phys.B:At.Mol.Opt.Phys.}\ }%
  \textbf{\bibinfo {volume} {39}},\ \bibinfo {pages} {1669} (\bibinfo {year}
  {2006})%
  \bibAnnoteFile{NoStop}{Hacyan:2006}%
\bibitem{Chen:2008ag}%
  \BibitemOpen
  \bibfield{author}{%
  \bibinfo {author} {\bibfnamefont{X.-S.}\ \bibnamefont{Chen}}, \bibinfo
  {author} {\bibfnamefont{X.-F.}\ \bibnamefont{Lu}}, \bibinfo {author}
  {\bibfnamefont{W.-M.}\ \bibnamefont{Sun}}, \bibinfo {author}
  {\bibfnamefont{F.}~\bibnamefont{Wang}},\ and\ \bibinfo {author}
  {\bibfnamefont{T.}~\bibnamefont{Goldman}},\ }%
  \bibfield{journal}{%
  \Doi{10.1103/PhysRevLett.100.232002}{\bibinfo {journal} {Phys. Rev. Lett.}}\
  }%
  \textbf{\bibinfo {volume} {100}},\ \bibinfo {pages} {232002} (\bibinfo {year}
  {2008}),\ \Eprint{http://arxiv.org/abs/0806.3166}{arXiv:0806.3166 [hep-ph]}%
  \bibAnnoteFile{NoStop}{Chen:2008ag}%
%%CITATION = 0806.3166;%%
\bibitem{Nieminen:2008}%
  \BibitemOpen
  \bibfield{author}{%
  \bibinfo {author} {\bibfnamefont{T.}~\bibnamefont{Nieminen}}, \bibinfo
  {author} {\bibfnamefont{A.}~\bibnamefont{Stilgoe}}, \bibinfo {author}
  {\bibfnamefont{N.}~\bibnamefont{Heckenberg}},\ and\ \bibinfo {author}
  {\bibfnamefont{H.}~\bibnamefont{Rubinsztein-Dunlop}},\ }%
  \bibfield{journal}{%
  \bibinfo {journal} {J.Phys.A:Pure Appl.Opt.}\ }%
  \textbf{\bibinfo {volume} {10}},\ \bibinfo {pages} {115005} (\bibinfo {year}
  {2008})%
  \bibAnnoteFile{NoStop}{Nieminen:2008}%
\bibitem{Li:2009}%
  \BibitemOpen
  \bibfield{author}{%
  \bibinfo {author} {\bibfnamefont{C.}~\bibnamefont{Li}},\ }%
  \bibfield{journal}{%
  \bibinfo {journal} {Phys.Rev.}\ }%
  \textbf{\bibinfo {volume} {A80}},\ \bibinfo {pages} {063814} (\bibinfo {year}
  {2009})%
  \bibAnnoteFile{NoStop}{Li:2009}%
\bibitem{Berry:2009}%
  \BibitemOpen
  \bibfield{author}{%
  \bibinfo {author} {\bibfnamefont{M.}~\bibnamefont{Berry}},\ }%
  \bibfield{journal}{%
  \bibinfo {journal} {J.Phys.A:Pure Appl.Opt.}\ }%
  \textbf{\bibinfo {volume} {11}},\ \bibinfo {pages} {094001} (\bibinfo {year}
  {2009})%
  \bibAnnoteFile{NoStop}{Berry:2009}%
\bibitem{Mazilu:2009}%
  \BibitemOpen
  \bibfield{author}{%
  \bibinfo {author} {\bibfnamefont{M.}~\bibnamefont{Mazilu}},\ }%
  \bibfield{journal}{%
  \bibinfo {journal} {J.Phys.A:Pure Appl.Opt.}\ }%
  \textbf{\bibinfo {volume} {11}},\ \bibinfo {pages} {094005} (\bibinfo {year}
  {2009})%
  \bibAnnoteFile{NoStop}{Mazilu:2009}%
\bibitem{Aiello:2009a}%
  \BibitemOpen
  \bibfield{author}{%
  \bibinfo {author} {\bibfnamefont{A.}~\bibnamefont{Aiello}}, \bibinfo {author}
  {\bibfnamefont{N.}~\bibnamefont{Lindlein}}, \bibinfo {author}
  {\bibfnamefont{C.}~\bibnamefont{Marquardt}},\ and\ \bibinfo {author}
  {\bibfnamefont{G.}~\bibnamefont{Leuchs}},\ }%
  \bibfield{journal}{%
  \bibinfo {journal} {Phys.Rev.Lett.}\ }%
  \textbf{\bibinfo {volume} {103}},\ \bibinfo {pages} {100401} (\bibinfo {year}
  {2009})%
  \bibAnnoteFile{NoStop}{Aiello:2009a}%
\bibitem{Aiello:2009b}%
  \BibitemOpen
  \bibfield{author}{%
  \bibinfo {author} {\bibfnamefont{A.}~\bibnamefont{Aiello}}, \bibinfo {author}
  {\bibfnamefont{C.}~\bibnamefont{Marquardt}},\ and\ \bibinfo {author}
  {\bibfnamefont{G.}~\bibnamefont{Leuchs}},\ }%
  \bibfield{journal}{%
  \bibinfo {journal} {Phys.Rev.}\ }%
  \textbf{\bibinfo {volume} {A81}},\ \bibinfo {pages} {053838} (\bibinfo {year}
  {2009})%
  \bibAnnoteFile{NoStop}{Aiello:2009b}%
\bibitem{Barnett:2010}%
  \BibitemOpen
  \bibfield{author}{%
  \bibinfo {author} {\bibfnamefont{S.}~\bibnamefont{Barnett}},\ }%
  \bibfield{journal}{%
  \bibinfo {journal} {J.Mod.Opt.}\ }%
  \textbf{\bibinfo {volume} {57}},\ \bibinfo {pages} {1339} (\bibinfo {year}
  {2010})%
  \bibAnnoteFile{NoStop}{Barnett:2010}%
\bibitem{Stewart:2010ft}%
  \BibitemOpen
  \bibfield{author}{%
  \bibinfo {author} {\bibfnamefont{A.}~\bibnamefont{Stewart}},\ }%
  \bibfield{journal}{%
  \bibinfo {journal} {Int.J.Opt.}\ }%
  \textbf{\bibinfo {volume} {2011}},\ \bibinfo {pages} {728350} (\bibinfo
  {year} {2011}),\ \Eprint{http://arxiv.org/abs/1010.1056}{arXiv:1010.1056
  [physics.class-ph]}%
  \bibAnnoteFile{NoStop}{Stewart:2010ft}%
%%CITATION = ARXIV:1010.1056;%%
\bibitem{Bialynicki:2011}%
  \BibitemOpen
  \bibfield{author}{%
  \bibinfo {author} {\bibfnamefont{I.}~\bibnamefont{Bialynicki-Birula}}\ and\
  \bibinfo {author} {\bibfnamefont{Z.}~\bibnamefont{Bialynicka-Birula}},\ }%
  \bibfield{journal}{%
  \bibinfo {journal} {J.Opt.}\ }%
  \textbf{\bibinfo {volume} {13}},\ \bibinfo {pages} {064014} (\bibinfo {year}
  {2011})%
  \bibAnnoteFile{NoStop}{Bialynicki:2011}%
\bibitem{Tiwari:2008nz}%
  \BibitemOpen
  \bibfield{author}{%
  \bibinfo {author} {\bibfnamefont{S.}~\bibnamefont{Tiwari}}}%
   (\bibinfo {year} {2008}),\
  \Eprint{http://arxiv.org/abs/0807.0699}{arXiv:0807.0699 [physics.gen-ph]}%
  \bibAnnoteFile{NoStop}{Tiwari:2008nz}%
%%CITATION = ARXIV:0807.0699;%%
\bibitem{Ji:2009fu}%
  \BibitemOpen
  \bibfield{author}{%
  \bibinfo {author} {\bibfnamefont{X.}~\bibnamefont{Ji}}}%
   (\bibinfo {year} {2009}),\
  \Eprint{http://arxiv.org/abs/0910.5022}{arXiv:0910.5022 [hep-ph]}%
  \bibAnnoteFile{NoStop}{Ji:2009fu}%
%%CITATION = 0910.5022;%%
\bibitem{Ji:2010zza}%
  \BibitemOpen
  \bibfield{author}{%
  \bibinfo {author} {\bibfnamefont{X.}~\bibnamefont{Ji}},\ }%
  \bibfield{journal}{%
  \Doi{10.1103/PhysRevLett.104.039101}{\bibinfo {journal} {Phys.Rev.Lett.}}\ }%
  \textbf{\bibinfo {volume} {104}},\ \bibinfo {pages} {039101} (\bibinfo {year}
  {2010})%
  \bibAnnoteFile{NoStop}{Ji:2010zza}%
%%CITATION = PRLTA,104,039101;%%
\bibitem{Ji:2012gc}%
  \BibitemOpen
  \bibfield{author}{%
  \bibinfo {author} {\bibfnamefont{X.}~\bibnamefont{Ji}}, \bibinfo {author}
  {\bibfnamefont{Y.}~\bibnamefont{Xu}},\ and\ \bibinfo {author}
  {\bibfnamefont{Y.}~\bibnamefont{Zhao}},\ }%
  \bibfield{journal}{%
  \Doi{10.1007/JHEP08(2012)082}{\bibinfo {journal} {JHEP}}\ }%
  \textbf{\bibinfo {volume} {1208}},\ \bibinfo {pages} {082} (\bibinfo {year}
  {2012}),\ \Eprint{http://arxiv.org/abs/1205.0156}{arXiv:1205.0156 [hep-ph]}%
  \bibAnnoteFile{NoStop}{Ji:2012gc}%
%%CITATION = ARXIV:1205.0156;%%
\bibitem{Chen:2008gv}%
  \BibitemOpen
  \bibfield{author}{%
  \bibinfo {author} {\bibfnamefont{X.-S.}\ \bibnamefont{Chen}}, \bibinfo
  {author} {\bibfnamefont{X.-F.}\ \bibnamefont{Lu}}, \bibinfo {author}
  {\bibfnamefont{W.-M.}\ \bibnamefont{Sun}}, \bibinfo {author}
  {\bibfnamefont{F.}~\bibnamefont{Wang}},\ and\ \bibinfo {author}
  {\bibfnamefont{T.}~\bibnamefont{Goldman}}\ }%
  \textbf{\bibinfo {volume} {100}},\ \bibinfo {pages} {232002} (\bibinfo {year}
  {2008}),\ \Eprint{http://arxiv.org/abs/0807.3083}{arXiv:0807.3083 [hep-ph]}%
  \bibAnnoteFile{NoStop}{Chen:2008gv}%
%%CITATION = 0807.3083;%%
\bibitem{Chen:2008ja}%
  \BibitemOpen
  \bibfield{author}{%
  \bibinfo {author} {\bibfnamefont{X.-S.}\ \bibnamefont{Chen}}, \bibinfo
  {author} {\bibfnamefont{X.-F.}\ \bibnamefont{Lu}}, \bibinfo {author}
  {\bibfnamefont{W.-M.}\ \bibnamefont{Sun}}, \bibinfo {author}
  {\bibfnamefont{F.}~\bibnamefont{Wang}},\ and\ \bibinfo {author}
  {\bibfnamefont{T.}~\bibnamefont{Goldman}}}%
   (\bibinfo {year} {2008}),\
  \Eprint{http://arxiv.org/abs/0812.4336}{arXiv:0812.4336 [hep-ph]}%
  \bibAnnoteFile{NoStop}{Chen:2008ja}%
%%CITATION = 0812.4336;%%
\bibitem{Chen:2009dg}%
  \BibitemOpen
  \bibfield{author}{%
  \bibinfo {author} {\bibfnamefont{X.-S.}\ \bibnamefont{Chen}}, \bibinfo
  {author} {\bibfnamefont{W.-M.}\ \bibnamefont{Sun}}, \bibinfo {author}
  {\bibfnamefont{X.-F.}\ \bibnamefont{Lu}}, \bibinfo {author}
  {\bibfnamefont{F.}~\bibnamefont{Wang}},\ and\ \bibinfo {author}
  {\bibfnamefont{T.}~\bibnamefont{Goldman}}}%
   (\bibinfo {year} {2009}),\
  \Eprint{http://arxiv.org/abs/0911.0248}{arXiv:0911.0248 [hep-ph]}%
  \bibAnnoteFile{NoStop}{Chen:2009dg}%
%%CITATION = 0911.0248;%%
\bibitem{Chen:2011zzh}%
  \BibitemOpen
  \bibfield{author}{%
  \bibinfo {author} {\bibfnamefont{X.-S.}\ \bibnamefont{Chen}}, \bibinfo
  {author} {\bibfnamefont{W.-M.}\ \bibnamefont{Sun}}, \bibinfo {author}
  {\bibfnamefont{F.}~\bibnamefont{Wang}},\ and\ \bibinfo {author}
  {\bibfnamefont{T.}~\bibnamefont{Goldman}},\ }%
  \bibfield{journal}{%
  \Doi{10.1103/PhysRevD.83.071901}{\bibinfo {journal} {Phys.Rev.}}\ }%
  \textbf{\bibinfo {volume} {D83}},\ \bibinfo {pages} {071901} (\bibinfo {year}
  {2011}),\ \Eprint{http://arxiv.org/abs/1105.6304}{arXiv:1105.6304 [hep-ph]}%
  \bibAnnoteFile{NoStop}{Chen:2011zzh}%
%%CITATION = ARXIV:1105.6304;%%
\bibitem{Wong:2010rs}%
  \BibitemOpen
  \bibfield{author}{%
  \bibinfo {author} {\bibfnamefont{C.}~\bibnamefont{Wong}}, \bibinfo {author}
  {\bibfnamefont{F.}~\bibnamefont{Wang}}, \bibinfo {author}
  {\bibfnamefont{W.}~\bibnamefont{Sun}},\ and\ \bibinfo {author}
  {\bibfnamefont{X.}~\bibnamefont{Lu}}}%
   (\bibinfo {year} {2010}),\
  \Eprint{http://arxiv.org/abs/1010.4336}{arXiv:1010.4336 [hep-ph]}%
  \bibAnnoteFile{NoStop}{Wong:2010rs}%
%%CITATION = ARXIV:1010.4336;%%
\bibitem{Wang:2010ao}%
  \BibitemOpen
  \bibfield{author}{%
  \bibinfo {author} {\bibfnamefont{F.}~\bibnamefont{Wang}}, \bibinfo {author}
  {\bibfnamefont{X.-S.}\ \bibnamefont{Chen}}, \bibinfo {author}
  {\bibfnamefont{X.-F.}\ \bibnamefont{Lu}}, \bibinfo {author}
  {\bibfnamefont{W.-M.}\ \bibnamefont{Sun}},\ and\ \bibinfo {author}
  {\bibfnamefont{T.}~\bibnamefont{Goldman}},\ }%
  \bibfield{journal}{%
  \Doi{10.1016/j.nuclphysa.2010.05.019}{\bibinfo {journal} {Nucl.Phys.}}\ }%
  \textbf{\bibinfo {volume} {A844}},\ \bibinfo {pages} {85C} (\bibinfo {year}
  {2010})%
  \bibAnnoteFile{NoStop}{Wang:2010ao}%
%%CITATION = NUPHA,A844,85C;%%
\bibitem{Chen:2011gn}%
  \BibitemOpen
  \bibfield{author}{%
  \bibinfo {author} {\bibfnamefont{X.-S.}\ \bibnamefont{Chen}}, \bibinfo
  {author} {\bibfnamefont{W.-M.}\ \bibnamefont{Sun}}, \bibinfo {author}
  {\bibfnamefont{F.}~\bibnamefont{Wang}},\ and\ \bibinfo {author}
  {\bibfnamefont{T.}~\bibnamefont{Goldman}},\ }%
  \bibfield{journal}{%
  \Doi{10.1016/j.physletb.2011.04.045}{\bibinfo {journal} {Phys.Lett.}}\ }%
  \textbf{\bibinfo {volume} {B700}},\ \bibinfo {pages} {21} (\bibinfo {year}
  {2011}),\ \Eprint{http://arxiv.org/abs/1101.5358}{arXiv:1101.5358 [hep-ph]}%
  \bibAnnoteFile{NoStop}{Chen:2011gn}%
%%CITATION = ARXIV:1101.5358;%%
\bibitem{Chen:2012vg}%
  \BibitemOpen
  \bibfield{author}{%
  \bibinfo {author} {\bibfnamefont{X.-S.}\ \bibnamefont{Chen}}}%
   (\bibinfo {year} {2012}),\
  \Eprint{http://arxiv.org/abs/1203.1288}{arXiv:1203.1288 [hep-ph]}%
  \bibAnnoteFile{NoStop}{Chen:2012vg}%
%%CITATION = ARXIV:1203.1288;%%
\bibitem{Goldman:2011vs}%
  \BibitemOpen
  \bibfield{author}{%
  \bibinfo {author} {\bibfnamefont{T.}~\bibnamefont{Goldman}},\ }%
  \bibfield{journal}{%
  \Doi{10.1063/1.3667298}{\bibinfo {journal} {AIP Conf.Proc.}}\ }%
  \textbf{\bibinfo {volume} {1418}},\ \bibinfo {pages} {13} (\bibinfo {year}
  {2011}),\ \Eprint{http://arxiv.org/abs/1110.2533}{arXiv:1110.2533 [hep-ph]}%
  \bibAnnoteFile{NoStop}{Goldman:2011vs}%
%%CITATION = ARXIV:1110.2533;%%
\bibitem{Stoilov:2010pv}%
  \BibitemOpen
  \bibfield{author}{%
  \bibinfo {author} {\bibfnamefont{M.}~\bibnamefont{Stoilov}}}%
   (\bibinfo {year} {2010}),\
  \Eprint{http://arxiv.org/abs/1011.5617}{arXiv:1011.5617 [hep-th]}%
  \bibAnnoteFile{NoStop}{Stoilov:2010pv}%
%%CITATION = ARXIV:1011.5617;%%
\bibitem{Wakamatsu:2010qj}%
  \BibitemOpen
  \bibfield{author}{%
  \bibinfo {author} {\bibfnamefont{M.}~\bibnamefont{Wakamatsu}},\ }%
  \bibfield{journal}{%
  \Doi{10.1103/PhysRevD.81.114010}{\bibinfo {journal} {Phys. Rev.}}\ }%
  \textbf{\bibinfo {volume} {D81}},\ \bibinfo {pages} {114010} (\bibinfo {year}
  {2010}),\ \Eprint{http://arxiv.org/abs/1004.0268}{arXiv:1004.0268 [hep-ph]}%
  \bibAnnoteFile{NoStop}{Wakamatsu:2010qj}%
%%CITATION = 1004.0268;%%
\bibitem{Wakamatsu:2010cb}%
  \BibitemOpen
  \bibfield{author}{%
  \bibinfo {author} {\bibfnamefont{M.}~\bibnamefont{Wakamatsu}},\ }%
  \bibfield{journal}{%
  \Doi{10.1103/PhysRevD.83.014012}{\bibinfo {journal} {Phys.Rev.}}\ }%
  \textbf{\bibinfo {volume} {D83}},\ \bibinfo {pages} {014012} (\bibinfo {year}
  {2011}),\ \Eprint{http://arxiv.org/abs/1007.5355}{arXiv:1007.5355 [hep-ph]}%
  \bibAnnoteFile{NoStop}{Wakamatsu:2010cb}%
%%CITATION = ARXIV:1007.5355;%%
\bibitem{Wakamatsu:2011mb}%
  \BibitemOpen
  \bibfield{author}{%
  \bibinfo {author} {\bibfnamefont{M.}~\bibnamefont{Wakamatsu}},\ }%
  \bibfield{journal}{%
  \Doi{10.1103/PhysRevD.84.037501}{\bibinfo {journal} {Phys.Rev.}}\ }%
  \textbf{\bibinfo {volume} {D84}},\ \bibinfo {pages} {037501} (\bibinfo {year}
  {2011}),\ \Eprint{http://arxiv.org/abs/1104.1465}{arXiv:1104.1465 [hep-ph]}%
  \bibAnnoteFile{NoStop}{Wakamatsu:2011mb}%
%%CITATION = ARXIV:1104.1465;%%
\bibitem{Wakamatsu:2012ve}%
  \BibitemOpen
  \bibfield{author}{%
  \bibinfo {author} {\bibfnamefont{M.}~\bibnamefont{Wakamatsu}},\ }%
  \bibfield{journal}{%
  \Doi{10.1103/PhysRevD.85.114039}{\bibinfo {journal} {Phys.Rev.}}\ }%
  \textbf{\bibinfo {volume} {D85}},\ \bibinfo {pages} {114039} (\bibinfo {year}
  {2012}),\ \Eprint{http://arxiv.org/abs/1204.2860}{arXiv:1204.2860 [hep-ph]}%
  \bibAnnoteFile{NoStop}{Wakamatsu:2012ve}%
%%CITATION = ARXIV:1204.2860;%%
\bibitem{Wakamatsu:2013voa}%
  \BibitemOpen
  \bibfield{author}{%
  \bibinfo {author} {\bibfnamefont{M.}~\bibnamefont{Wakamatsu}}}%
   (\bibinfo {year} {2013}),\
  \Eprint{http://arxiv.org/abs/1302.5152}{arXiv:1302.5152 [hep-ph]}%
  \bibAnnoteFile{NoStop}{Wakamatsu:2013voa}%
%%CITATION = ARXIV:1302.5152;%%
\bibitem{Cho:2010cw}%
  \BibitemOpen
  \bibfield{author}{%
  \bibinfo {author} {\bibfnamefont{Y.}~\bibnamefont{Cho}}, \bibinfo {author}
  {\bibfnamefont{M.-L.}\ \bibnamefont{Ge}},\ and\ \bibinfo {author}
  {\bibfnamefont{P.}~\bibnamefont{Zhang}},\ }%
  \bibfield{journal}{%
  \Doi{10.1142/S0217732312300327}{\bibinfo {journal} {Mod.Phys.Lett.}}\ }%
  \textbf{\bibinfo {volume} {A27}},\ \bibinfo {pages} {1230032} (\bibinfo
  {year} {2012}),\ \Eprint{http://arxiv.org/abs/1010.1080}{arXiv:1010.1080
  [nucl-th]}%
  \bibAnnoteFile{NoStop}{Cho:2010cw}%
%%CITATION = ARXIV:1010.1080;%%
\bibitem{Cho:2011ee}%
  \BibitemOpen
  \bibfield{author}{%
  \bibinfo {author} {\bibfnamefont{Y.}~\bibnamefont{Cho}}, \bibinfo {author}
  {\bibfnamefont{M.-L.}\ \bibnamefont{Ge}}, \bibinfo {author}
  {\bibfnamefont{D.}~\bibnamefont{Pak}},\ and\ \bibinfo {author}
  {\bibfnamefont{P.}~\bibnamefont{Zhang}}}%
   (\bibinfo {year} {2011}),\
  \Eprint{http://arxiv.org/abs/1102.1130}{arXiv:1102.1130 [nucl-th]}%
  \bibAnnoteFile{NoStop}{Cho:2011ee}%
%%CITATION = ARXIV:1102.1130;%%
\bibitem{Hatta:2011zs}%
  \BibitemOpen
  \bibfield{author}{%
  \bibinfo {author} {\bibfnamefont{Y.}~\bibnamefont{Hatta}},\ }%
  \bibfield{journal}{%
  \Doi{10.1103/PhysRevD.84.041701}{\bibinfo {journal} {Phys.Rev.}}\ }%
  \textbf{\bibinfo {volume} {D84}},\ \bibinfo {pages} {041701} (\bibinfo {year}
  {2011}),\ \Eprint{http://arxiv.org/abs/1101.5989}{arXiv:1101.5989 [hep-ph]}%
  \bibAnnoteFile{NoStop}{Hatta:2011zs}%
%%CITATION = ARXIV:1101.5989;%%
\bibitem{Hatta:2011ku}%
  \BibitemOpen
  \bibfield{author}{%
  \bibinfo {author} {\bibfnamefont{Y.}~\bibnamefont{Hatta}},\ }%
  \bibfield{journal}{%
  \Doi{10.1016/j.physletb.2012.01.024}{\bibinfo {journal} {Phys.Lett.}}\ }%
  \textbf{\bibinfo {volume} {B708}},\ \bibinfo {pages} {186} (\bibinfo {year}
  {2012}),\ \Eprint{http://arxiv.org/abs/1111.3547}{arXiv:1111.3547 [hep-ph]}%
  \bibAnnoteFile{NoStop}{Hatta:2011ku}%
%%CITATION = ARXIV:1111.3547;%%
\bibitem{Hatta:2012cs}%
  \BibitemOpen
  \bibfield{author}{%
  \bibinfo {author} {\bibfnamefont{Y.}~\bibnamefont{Hatta}}\ and\ \bibinfo
  {author} {\bibfnamefont{S.}~\bibnamefont{Yoshida}},\ }%
  \bibfield{journal}{%
  \Doi{10.1007/JHEP10(2012)080}{\bibinfo {journal} {JHEP}}\ }%
  \textbf{\bibinfo {volume} {1210}},\ \bibinfo {pages} {080} (\bibinfo {year}
  {2012}),\ \Eprint{http://arxiv.org/abs/1207.5332}{arXiv:1207.5332 [hep-ph]}%
  \bibAnnoteFile{NoStop}{Hatta:2012cs}%
%%CITATION = ARXIV:1207.5332;%%
\bibitem{Hatta:2012jm}%
  \BibitemOpen
  \bibfield{author}{%
  \bibinfo {author} {\bibfnamefont{Y.}~\bibnamefont{Hatta}}, \bibinfo {author}
  {\bibfnamefont{K.}~\bibnamefont{Tanaka}},\ and\ \bibinfo {author}
  {\bibfnamefont{S.}~\bibnamefont{Yoshida}},\ }%
  \bibfield{journal}{%
  \Doi{10.1007/JHEP02(2013)003}{\bibinfo {journal} {JHEP}}\ }%
  \textbf{\bibinfo {volume} {1302}},\ \bibinfo {pages} {003} (\bibinfo {year}
  {2013}),\ \Eprint{http://arxiv.org/abs/1211.2918}{arXiv:1211.2918 [hep-ph]}%
  \bibAnnoteFile{NoStop}{Hatta:2012jm}%
%%CITATION = ARXIV:1211.2918;%%
\bibitem{Zhang:2011rn}%
  \BibitemOpen
  \bibfield{author}{%
  \bibinfo {author} {\bibfnamefont{P.}~\bibnamefont{Zhang}}\ and\ \bibinfo
  {author} {\bibfnamefont{D.}~\bibnamefont{Pak}},\ }%
  \bibfield{journal}{%
  \Doi{10.1140/epja/i2012-12091-8}{\bibinfo {journal} {Eur.Phys.J.}}\ }%
  \textbf{\bibinfo {volume} {A48}},\ \bibinfo {pages} {91} (\bibinfo {year}
  {2012}),\ \Eprint{http://arxiv.org/abs/1110.6516}{arXiv:1110.6516 [hep-ph]}%
  \bibAnnoteFile{NoStop}{Zhang:2011rn}%
%%CITATION = ARXIV:1110.6516;%%
\bibitem{Leader:2011za}%
  \BibitemOpen
  \bibfield{author}{%
  \bibinfo {author} {\bibfnamefont{E.}~\bibnamefont{Leader}},\ }%
  \bibfield{journal}{%
  \Doi{10.1103/PhysRevD.83.096012}{\bibinfo {journal} {Phys.Rev.}}\ }%
  \textbf{\bibinfo {volume} {D83}},\ \bibinfo {pages} {096012} (\bibinfo {year}
  {2011}),\ \Eprint{http://arxiv.org/abs/1101.5956}{arXiv:1101.5956 [hep-ph]}%
  \bibAnnoteFile{NoStop}{Leader:2011za}%
\bibitem{Lorce:2012ce}%
  \BibitemOpen
  \bibfield{author}{%
  \bibinfo {author} {\bibfnamefont{C.}~\bibnamefont{Lorc\'e}},\ }%
  \bibfield{journal}{%
  \Doi{10.1016/j.physletb.2013.01.007}{\bibinfo {journal} {Phys.Lett.}}\ }%
  \textbf{\bibinfo {volume} {B719}},\ \bibinfo {pages} {185} (\bibinfo {year}
  {2013}),\ \Eprint{http://arxiv.org/abs/1210.2581}{arXiv:1210.2581 [hep-ph]}%
  \bibAnnoteFile{NoStop}{Lorce:2012ce}%
%%CITATION = ARXIV:1210.2581;%%
\bibitem{Lorce:2012rr}%
  \BibitemOpen
  \bibfield{author}{%
  \bibinfo {author} {\bibfnamefont{C.}~\bibnamefont{Lorc\'e}},\ }%
  \bibfield{journal}{%
  \Doi{10.1103/PhysRevD.87.034031}{\bibinfo {journal} {Phys.Rev.}}\ }%
  \textbf{\bibinfo {volume} {D87}},\ \bibinfo {pages} {034031} (\bibinfo {year}
  {2013}),\ \Eprint{http://arxiv.org/abs/1205.6483}{arXiv:1205.6483 [hep-ph]}%
  \bibAnnoteFile{NoStop}{Lorce:2012rr}%
%%CITATION = ARXIV:1205.6483;%%
\bibitem{Lorce:2013gxa}%
  \BibitemOpen
  \bibfield{author}{%
  \bibinfo {author} {\bibfnamefont{C.}~\bibnamefont{Lorc\'e}}}%
   (\bibinfo {year} {2013}),\
  \Eprint{http://arxiv.org/abs/1302.5515}{arXiv:1302.5515 [hep-ph]}%
  \bibAnnoteFile{NoStop}{Lorce:2013gxa}%
%%CITATION = ARXIV:1302.5515;%%
\bibitem{Lorce:2013bja}%
  \BibitemOpen
  \bibfield{author}{%
  \bibinfo {author} {\bibfnamefont{C.}~\bibnamefont{Lorc\'e}}}%
   (\bibinfo {year} {2013}),\
  \Eprint{http://arxiv.org/abs/1306.0456}{arXiv:1306.0456 [hep-ph]}%
  \bibAnnoteFile{NoStop}{Lorce:2013bja}%
%%CITATION = ARXIV:1306.0456;%%
\bibitem{Noether:1918zz}%
  \BibitemOpen
  \bibfield{author}{%
  \bibinfo {author} {\bibfnamefont{E.}~\bibnamefont{Noether}},\ }%
  \bibfield{journal}{%
  \Doi{10.1080/00411457108231446}{\bibinfo {journal} {Gott.Nachr.}}\ }%
  \textbf{\bibinfo {volume} {1918}},\ \bibinfo {pages} {235} (\bibinfo {year}
  {1918}),\ \Eprint{http://arxiv.org/abs/physics/0503066}{arXiv:physics/0503066
  [physics]}%
  \bibAnnoteFile{NoStop}{Noether:1918zz}%
%%CITATION = PHYSICS/0503066;%%
\bibitem{Dirac:1928hu}%
  \BibitemOpen
  \bibfield{author}{%
  \bibinfo {author} {\bibfnamefont{P.~A.}\ \bibnamefont{Dirac}},\ }%
  \bibfield{journal}{%
  \bibinfo {journal} {Proc.Roy.Soc.Lond.}\ }%
  \textbf{\bibinfo {volume} {A117}},\ \bibinfo {pages} {610} (\bibinfo {year}
  {1928})%
  \bibAnnoteFile{NoStop}{Dirac:1928hu}%
%%CITATION = PRSLA,A117,610;%%
\bibitem{Leader:2001gr}%
  \BibitemOpen
  \bibfield{author}{%
  \bibinfo {author} {\bibfnamefont{E.}~\bibnamefont{Leader}},\ }%
  \emph{\bibinfo {title} {{Spin in particle physics}}}\ (\bibinfo {publisher}
  {Cambridge University Press},\ \bibinfo {address} {Cambridge, UK},\ \bibinfo
  {year} {2001, 2005})%
  \bibAnnoteFile{NoStop}{Leader:2001gr}%
%%CITATION = CMPCE,15,1;%%
\bibitem{Landau:1951}%
  \BibitemOpen
  \bibfield{author}{%
  \bibinfo {author} {\bibfnamefont{L.}~\bibnamefont{Landau}}\ and\ \bibinfo
  {author} {\bibfnamefont{E.}~\bibnamefont{Lifshitz}},\ }%
  \emph{\bibinfo {title} {{Classical Theory of Fields}}}\ (\bibinfo {publisher}
  {Addison-Wesley Press},\ \bibinfo {address} {Cambridge, Mass., USA},\
  \bibinfo {year} {1951})%
  \bibAnnoteFile{NoStop}{Landau:1951}%
\bibitem{Bjorken:1965zz}%
  \BibitemOpen
  \bibfield{author}{%
  \bibinfo {author} {\bibfnamefont{J.~D.}\ \bibnamefont{Bjorken}}\ and\
  \bibinfo {author} {\bibfnamefont{S.~D.}\ \bibnamefont{Drell}},\ }%
  \emph{\bibinfo {title} {{Relativistic quantum fields}}}\ (\bibinfo
  {publisher} {Mcgraw-Hill},\ \bibinfo {address} {New York, USA},\ \bibinfo
  {year} {1965})%
  \bibAnnoteFile{NoStop}{Bjorken:1965zz}%
\bibitem{Weinberg:1995mt}%
  \BibitemOpen
  \bibfield{author}{%
  \bibinfo {author} {\bibfnamefont{S.}~\bibnamefont{Weinberg}},\ }%
  \emph{\bibinfo {title} {{The Quantum theory of fields. Vol. 1:
  Foundations}}}\ (\bibinfo {publisher} {Cambridge University Press},\ \bibinfo
  {address} {Cambridge, U.K.},\ \bibinfo {year} {1995})%
  \bibAnnoteFile{NoStop}{Weinberg:1995mt}%
%%CITATION = INSPIRE-406190;%%
\bibitem{Lautrup}%
  \BibitemOpen
  \bibfield{author}{%
  \bibinfo {author} {\bibfnamefont{B.}~\bibnamefont{Lautrup}},\ }%
  \bibfield{journal}{%
  \bibinfo {journal} {Kgl. Danske Videnskap. Selskab, Mat.-fys. Medd.}\ }%
  \textbf{\bibinfo {volume} {35}},\ \bibinfo {pages} {1} (\bibinfo {year}
  {1967})%
  \bibAnnoteFile{NoStop}{Lautrup}%
\bibitem{Nakanishi:66}%
  \BibitemOpen
  \bibfield{author}{%
  \bibinfo {author} {\bibfnamefont{N.}~\bibnamefont{Nakanishi}},\ }%
  \bibfield{journal}{%
  \bibinfo {journal} {Prog. Theor. Phys.}\ }%
  \textbf{\bibinfo {volume} {35}},\ \bibinfo {pages} {1111} (\bibinfo {year}
  {1966})%
  \bibAnnoteFile{NoStop}{Nakanishi:66}%
\bibitem{Kugo:78}%
  \BibitemOpen
  \bibfield{author}{%
  \bibinfo {author} {\bibfnamefont{T.}~\bibnamefont{Kugo}}\ and\ \bibinfo
  {author} {\bibfnamefont{I.}~\bibnamefont{Ojima}},\ }%
  \bibfield{journal}{%
  \bibinfo {journal} {Prog. Theor. Phys.}\ }%
  \textbf{\bibinfo {volume} {60}},\ \bibinfo {pages} {1869} (\bibinfo {year}
  {1978})%
  \bibAnnoteFile{NoStop}{Kugo:78}%
\bibitem{Collins:1984xc}%
  \BibitemOpen
  \bibfield{author}{%
  \bibinfo {author} {\bibfnamefont{J.~C.}\ \bibnamefont{Collins}},\ }%
  \emph{\bibinfo {title} {{Renormalization}}}\ (\bibinfo {publisher} {Cambridge
  University Press},\ \bibinfo {address} {Cambridge, UK},\ \bibinfo {year}
  {1984})%
  \bibAnnoteFile{NoStop}{Collins:1984xc}%
%%CITATION = INSPIRE-209810;%%
\bibitem{Chen:1998iu}%
  \BibitemOpen
  \bibfield{author}{%
  \bibinfo {author} {\bibfnamefont{X.-S.}\ \bibnamefont{Chen}}\ and\ \bibinfo
  {author} {\bibfnamefont{F.}~\bibnamefont{Wang}}}%
   (\bibinfo {year} {1998}),\
  \Eprint{http://arxiv.org/abs/hep-ph/9802346}{arXiv:hep-ph/9802346 [hep-ph]}%
  \bibAnnoteFile{NoStop}{Chen:1998iu}%
%%CITATION = HEP-PH/9802346;%%
\bibitem{Hoodbhoy:1998bt}%
  \BibitemOpen
  \bibfield{author}{%
  \bibinfo {author} {\bibfnamefont{P.}~\bibnamefont{Hoodbhoy}}, \bibinfo
  {author} {\bibfnamefont{X.-D.}\ \bibnamefont{Ji}},\ and\ \bibinfo {author}
  {\bibfnamefont{W.}~\bibnamefont{Lu}},\ }%
  \bibfield{journal}{%
  \Doi{10.1103/PhysRevD.59.074010}{\bibinfo {journal} {Phys.Rev.}}\ }%
  \textbf{\bibinfo {volume} {D59}},\ \bibinfo {pages} {074010} (\bibinfo {year}
  {1999}),\ \Eprint{http://arxiv.org/abs/hep-ph/9808305}{arXiv:hep-ph/9808305
  [hep-ph]}%
  \bibAnnoteFile{NoStop}{Hoodbhoy:1998bt}%
%%CITATION = HEP-PH/9808305;%%
\bibitem{Chen:1999kz}%
  \BibitemOpen
  \bibfield{author}{%
  \bibinfo {author} {\bibfnamefont{X.-S.}\ \bibnamefont{Chen}}, \bibinfo
  {author} {\bibfnamefont{W.-M.}\ \bibnamefont{Sun}},\ and\ \bibinfo {author}
  {\bibfnamefont{F.}~\bibnamefont{Wang}},\ }%
  \bibfield{journal}{%
  \Doi{10.1088/0954-3899/25/10/302}{\bibinfo {journal} {J.Phys.}}\ }%
  \textbf{\bibinfo {volume} {G25}},\ \bibinfo {pages} {2021} (\bibinfo {year}
  {1999})%
  \bibAnnoteFile{NoStop}{Chen:1999kz}%
%%CITATION = INSPIRE-510896;%%
\bibitem{Sun:2000gc}%
  \BibitemOpen
  \bibfield{author}{%
  \bibinfo {author} {\bibfnamefont{W.-M.}\ \bibnamefont{Sun}}, \bibinfo
  {author} {\bibfnamefont{X.-S.}\ \bibnamefont{Chen}},\ and\ \bibinfo {author}
  {\bibfnamefont{F.}~\bibnamefont{Wang}},\ }%
  \bibfield{journal}{%
  \Doi{10.1016/S0370-2693(01)00245-3}{\bibinfo {journal} {Phys.Lett.}}\ }%
  \textbf{\bibinfo {volume} {B503}},\ \bibinfo {pages} {430} (\bibinfo {year}
  {2001}),\ \Eprint{http://arxiv.org/abs/hep-th/0012027}{arXiv:hep-th/0012027
  [hep-th]}%
  \bibAnnoteFile{NoStop}{Sun:2000gc}%
%%CITATION = HEP-TH/0012027;%%
\bibitem{Belinfante:1939}%
  \BibitemOpen
  \bibfield{author}{%
  \bibinfo {author} {\bibfnamefont{F.~J.}\ \bibnamefont{Belinfante}},\ }%
  \bibfield{journal}{%
  \bibinfo {journal} {Physica}\ }%
  \textbf{\bibinfo {volume} {6}},\ \bibinfo {pages} {887} (\bibinfo {year}
  {1939})%
  \bibAnnoteFile{NoStop}{Belinfante:1939}%
\bibitem{Rosenfeld:1940}%
  \BibitemOpen
  \bibfield{author}{%
  \bibinfo {author} {\bibfnamefont{L.}~\bibnamefont{Rosenfeld}},\ }%
  \bibfield{journal}{%
  \bibinfo {journal} {M\'em.\ Acad.\ Roy.\ Belg.}\ }%
  \textbf{\bibinfo {volume} {18}},\ \bibinfo {pages} {6} (\bibinfo {year}
  {1940})%
  \bibAnnoteFile{NoStop}{Rosenfeld:1940}%
\bibitem{Bakker:2004ib}%
  \BibitemOpen
  \bibfield{author}{%
  \bibinfo {author} {\bibfnamefont{B.~L.~G.}\ \bibnamefont{Bakker}}, \bibinfo
  {author} {\bibfnamefont{E.}~\bibnamefont{Leader}},\ and\ \bibinfo {author}
  {\bibfnamefont{T.~L.}\ \bibnamefont{Trueman}},\ }%
  \bibfield{journal}{%
  \Doi{10.1103/PhysRevD.70.114001}{\bibinfo {journal} {Phys. Rev.}}\ }%
  \textbf{\bibinfo {volume} {D70}},\ \bibinfo {pages} {114001} (\bibinfo {year}
  {2004}),\ \Eprint{http://arxiv.org/abs/hep-ph/0406139}{arXiv:hep-ph/0406139}%
  \bibAnnoteFile{NoStop}{Bakker:2004ib}%
%%CITATION = HEP-PH/0406139;%%
\bibitem{Shore:1999be}%
  \BibitemOpen
  \bibfield{author}{%
  \bibinfo {author} {\bibfnamefont{G.~M.}\ \bibnamefont{Shore}}\ and\ \bibinfo
  {author} {\bibfnamefont{B.~E.}\ \bibnamefont{White}},\ }%
  \bibfield{journal}{%
  \Doi{10.1016/S0550-3213(00)00288-1}{\bibinfo {journal} {Nucl. Phys.}}\ }%
  \textbf{\bibinfo {volume} {B581}},\ \bibinfo {pages} {409} (\bibinfo {year}
  {2000}),\ \Eprint{http://arxiv.org/abs/hep-ph/9912341}{arXiv:hep-ph/9912341}%
  \bibAnnoteFile{NoStop}{Shore:1999be}%
%%CITATION = HEP-PH/9912341;%%
\bibitem{Bessel:1921}%
  \BibitemOpen
  \bibfield{author}{%
  \bibinfo {author} {\bibfnamefont{E.}~\bibnamefont{Bessel-Hagen}},\ }%
  \bibfield{journal}{%
  \bibinfo {journal} {Math. Ann.}\ }%
  \textbf{\bibinfo {volume} {84}} (\bibinfo {year} {1921})%
  \bibAnnoteFile{NoStop}{Bessel:1921}%
\bibitem{Guo:2013jia}%
  \BibitemOpen
  \bibfield{author}{%
  \bibinfo {author} {\bibfnamefont{Z.-Q.}\ \bibnamefont{Guo}}\ and\ \bibinfo
  {author} {\bibfnamefont{I.}~\bibnamefont{Schmidt}},\ }%
  \bibfield{journal}{%
  \Doi{10.1103/PhysRevD.87.114017}{\bibinfo {journal} {Phys.Rev.}}\ }%
  \textbf{\bibinfo {volume} {D87}},\ \bibinfo {pages} {114017} (\bibinfo {year}
  {2013}),\ \Eprint{http://arxiv.org/abs/1303.7210}{arXiv:1303.7210 [hep-ph]}%
  \bibAnnoteFile{NoStop}{Guo:2013jia}%
%%CITATION = ARXIV:1303.7210;%%
\bibitem{Becchi:1975nq}%
  \BibitemOpen
  \bibfield{author}{%
  \bibinfo {author} {\bibfnamefont{C.}~\bibnamefont{Becchi}}, \bibinfo {author}
  {\bibfnamefont{A.}~\bibnamefont{Rouet}},\ and\ \bibinfo {author}
  {\bibfnamefont{R.}~\bibnamefont{Stora}},\ }%
  \bibfield{journal}{%
  \Doi{10.1016/0003-4916(76)90156-1}{\bibinfo {journal} {Annals Phys.}}\ }%
  \textbf{\bibinfo {volume} {98}},\ \bibinfo {pages} {287} (\bibinfo {year}
  {1976})%
  \bibAnnoteFile{NoStop}{Becchi:1975nq}%
%%CITATION = APNYA,98,287;%%
\bibitem{Tyutin}%
  \BibitemOpen
  \bibfield{author}{%
  \bibinfo {author} {\bibfnamefont{I.~V.}\ \bibnamefont{Tyutin}},\ }%
  \bibfield{journal}{%
  \bibinfo {journal} {Lebedev Institute preprint}\ }%
  \textbf{\bibinfo {volume} {39}} (\bibinfo {year} {1975})%
  \bibAnnoteFile{NoStop}{Tyutin}%
\bibitem{Kugo:1979gm}%
  \BibitemOpen
  \bibfield{author}{%
  \bibinfo {author} {\bibfnamefont{T.}~\bibnamefont{Kugo}}\ and\ \bibinfo
  {author} {\bibfnamefont{I.}~\bibnamefont{Ojima}},\ }%
  \bibfield{journal}{%
  \bibinfo {journal} {Prog. Theor. Phys. Suppl.}\ }%
  \textbf{\bibinfo {volume} {66}},\ \bibinfo {pages} {1} (\bibinfo {year}
  {1979})%
  \bibAnnoteFile{NoStop}{Kugo:1979gm}%
%%CITATION = PTPSA,66,1;%%
\bibitem{Jaffe:1989jz}%
  \BibitemOpen
  \bibfield{author}{%
  \bibinfo {author} {\bibfnamefont{R.~L.}\ \bibnamefont{Jaffe}}\ and\ \bibinfo
  {author} {\bibfnamefont{A.}~\bibnamefont{Manohar}},\ }%
  \bibfield{journal}{%
  \Doi{10.1016/0550-3213(90)90506-9}{\bibinfo {journal} {Nucl. Phys.}}\ }%
  \textbf{\bibinfo {volume} {B337}},\ \bibinfo {pages} {509} (\bibinfo {year}
  {1990})%
  \bibAnnoteFile{NoStop}{Jaffe:1989jz}%
%%CITATION = NUPHA,B337,509;%%
\bibitem{Chen:2009mr}%
  \BibitemOpen
  \bibfield{author}{%
  \bibinfo {author} {\bibfnamefont{X.-S.}\ \bibnamefont{Chen}}, \bibinfo
  {author} {\bibfnamefont{W.-M.}\ \bibnamefont{Sun}}, \bibinfo {author}
  {\bibfnamefont{X.-F.}\ \bibnamefont{Lu}}, \bibinfo {author}
  {\bibfnamefont{F.}~\bibnamefont{Wang}},\ and\ \bibinfo {author}
  {\bibfnamefont{T.}~\bibnamefont{Goldman}},\ }%
  \bibfield{journal}{%
  \Doi{10.1103/PhysRevLett.103.062001}{\bibinfo {journal} {Phys. Rev. Lett.}}\
  }%
  \textbf{\bibinfo {volume} {103}},\ \bibinfo {pages} {062001} (\bibinfo {year}
  {2009}),\ \Eprint{http://arxiv.org/abs/0904.0321}{arXiv:0904.0321 [hep-ph]}%
  \bibAnnoteFile{NoStop}{Chen:2009mr}%
%%CITATION = 0904.0321;%%
\bibitem{Lorce:2013fpa}%
  \BibitemOpen
  \bibfield{author}{%
  \bibinfo {author} {\bibfnamefont{C.}~\bibnamefont{Lorc\'e}}}%
   (\bibinfo {year} {2013}),\
  \Eprint{http://arxiv.org/abs/1307.4323}{arXiv:1307.4323 [hep-ph]}%
  \bibAnnoteFile{NoStop}{Lorce:2013fpa}%
%%CITATION = ARXIV:1307.4323;%%
\bibitem{Burkardt:2008ua}%
  \BibitemOpen
  \bibfield{author}{%
  \bibinfo {author} {\bibfnamefont{M.}~\bibnamefont{Burkardt}}\ and\ \bibinfo
  {author} {\bibfnamefont{B.~C.}\ \bibnamefont{Hikmat}},\ }%
  \bibfield{journal}{%
  \Doi{10.1103/PhysRevD.79.071501}{\bibinfo {journal} {Phys. Rev.}}\ }%
  \textbf{\bibinfo {volume} {D79}},\ \bibinfo {pages} {071501} (\bibinfo {year}
  {2009}),\ \Eprint{http://arxiv.org/abs/0812.1605}{arXiv:0812.1605 [hep-ph]}%
  \bibAnnoteFile{NoStop}{Burkardt:2008ua}%
%%CITATION = 0812.1605;%%
\bibitem{Schwinger:1962zz}%
  \BibitemOpen
  \bibfield{author}{%
  \bibinfo {author} {\bibfnamefont{J.}~\bibnamefont{Schwinger}},\ }%
  \bibfield{journal}{%
  \Doi{10.1103/PhysRev.125.1043}{\bibinfo {journal} {Phys.Rev.}}\ }%
  \textbf{\bibinfo {volume} {125}},\ \bibinfo {pages} {1043} (\bibinfo {year}
  {1962})%
  \bibAnnoteFile{NoStop}{Schwinger:1962zz}%
%%CITATION = PHRVA,125,1043;%%
\bibitem{Schwinger:1962fg}%
  \BibitemOpen
  \bibfield{author}{%
  \bibinfo {author} {\bibfnamefont{J.~S.}\ \bibnamefont{Schwinger}},\ }%
  \bibfield{journal}{%
  \Doi{10.1103/PhysRev.130.402}{\bibinfo {journal} {Phys.Rev.}}\ }%
  \textbf{\bibinfo {volume} {130}},\ \bibinfo {pages} {402} (\bibinfo {year}
  {1963})%
  \bibAnnoteFile{NoStop}{Schwinger:1962fg}%
%%CITATION = PHRVA,130,402;%%
\bibitem{Arnowitt:1962cv}%
  \BibitemOpen
  \bibfield{author}{%
  \bibinfo {author} {\bibfnamefont{R.~L.}\ \bibnamefont{Arnowitt}}\ and\
  \bibinfo {author} {\bibfnamefont{S.}~\bibnamefont{Fickler}},\ }%
  \bibfield{journal}{%
  \Doi{10.1103/PhysRev.127.1821}{\bibinfo {journal} {Phys.Rev.}}\ }%
  \textbf{\bibinfo {volume} {127}},\ \bibinfo {pages} {1821} (\bibinfo {year}
  {1962})%
  \bibAnnoteFile{NoStop}{Arnowitt:1962cv}%
%%CITATION = PHRVA,127,1821;%%
\bibitem{Goto:1966}%
  \BibitemOpen
  \bibfield{author}{%
  \bibinfo {author} {\bibfnamefont{T.}~\bibnamefont{Goto}},\ }%
  \bibfield{journal}{%
  \bibinfo {journal} {Prog.Theor.Phys.}\ }%
  \textbf{\bibinfo {volume} {36}},\ \bibinfo {pages} {1283} (\bibinfo {year}
  {1966})%
  \bibAnnoteFile{NoStop}{Goto:1966}%
\bibitem{Treat:1973yc}%
  \BibitemOpen
  \bibfield{author}{%
  \bibinfo {author} {\bibfnamefont{R.}~\bibnamefont{Treat}},\ }%
  \bibfield{journal}{%
  \Doi{10.1063/1.1665895}{\bibinfo {journal} {J.Math.Phys.}}\ }%
  \textbf{\bibinfo {volume} {13}},\ \bibinfo {pages} {1704} (\bibinfo {year}
  {1972})%
  \bibAnnoteFile{NoStop}{Treat:1973yc}%
%%CITATION = JMAPA,13,1704;%%
\bibitem{Treat:1975dz}%
  \BibitemOpen
  \bibfield{author}{%
  \bibinfo {author} {\bibfnamefont{R.~P.}\ \bibnamefont{Treat}},\ }%
  \bibfield{journal}{%
  \Doi{10.1103/PhysRevD.12.3145}{\bibinfo {journal} {Phys.Rev.}}\ }%
  \textbf{\bibinfo {volume} {D12}},\ \bibinfo {pages} {3145} (\bibinfo {year}
  {1975})%
  \bibAnnoteFile{NoStop}{Treat:1975dz}%
%%CITATION = PHRVA,D12,3145;%%
\bibitem{Duan:1979}%
  \BibitemOpen
  \bibfield{author}{%
  \bibinfo {author} {\bibfnamefont{Y.}~\bibnamefont{Duan}}\ and\ \bibinfo
  {author} {\bibfnamefont{M.}~\bibnamefont{Ge}},\ }%
  \bibfield{journal}{%
  \bibinfo {journal} {Sinica.Sci.}\ }%
  \textbf{\bibinfo {volume} {11}},\ \bibinfo {pages} {1072} (\bibinfo {year}
  {1979})%
  \bibAnnoteFile{NoStop}{Duan:1979}%
\bibitem{Duan:1984cb}%
  \BibitemOpen
  \bibfield{author}{%
  \bibinfo {author} {\bibfnamefont{Y.}~\bibnamefont{Duan}}\ and\ \bibinfo
  {author} {\bibfnamefont{S.-C.}\ \bibnamefont{Zhao}},\ }%
  \bibfield{journal}{%
  \bibinfo {journal} {Commun.Theor.Phys.}\ }%
  \textbf{\bibinfo {volume} {2}},\ \bibinfo {pages} {1553} (\bibinfo {year}
  {1983})%
  \bibAnnoteFile{NoStop}{Duan:1984cb}%
%%CITATION = CTPMD,2,1553;%%
\bibitem{Duan:1998um}%
  \BibitemOpen
  \bibfield{author}{%
  \bibinfo {author} {\bibfnamefont{Y.-S.}\ \bibnamefont{Duan}}\ and\ \bibinfo
  {author} {\bibfnamefont{X.-G.}\ \bibnamefont{Li}},\ }%
  \bibfield{journal}{%
  \bibinfo {journal} {Commun.Theor.Phys.}\ }%
  \textbf{\bibinfo {volume} {29}},\ \bibinfo {pages} {237} (\bibinfo {year}
  {1998})%
  \bibAnnoteFile{NoStop}{Duan:1998um}%
%%CITATION = CTPMD,29,237;%%
\bibitem{Duan:2002vh}%
  \BibitemOpen
  \bibfield{author}{%
  \bibinfo {author} {\bibfnamefont{Y.-s.}\ \bibnamefont{Duan}}\ and\ \bibinfo
  {author} {\bibfnamefont{P.-m.}\ \bibnamefont{Zhang}},\ }%
  \bibfield{journal}{%
  \Doi{10.1142/S0217732302008940}{\bibinfo {journal} {Mod.Phys.Lett.}}\ }%
  \textbf{\bibinfo {volume} {A17}},\ \bibinfo {pages} {2283} (\bibinfo {year}
  {2002}),\ \Eprint{http://arxiv.org/abs/hep-th/0212229}{arXiv:hep-th/0212229
  [hep-th]}%
  \bibAnnoteFile{NoStop}{Duan:2002vh}%
%%CITATION = HEP-TH/0212229;%%
\bibitem{Fulp:1983bt}%
  \BibitemOpen
  \bibfield{author}{%
  \bibinfo {author} {\bibfnamefont{R.}~\bibnamefont{Fulp}}\ and\ \bibinfo
  {author} {\bibfnamefont{L.}~\bibnamefont{Norris}},\ }%
  \bibfield{journal}{%
  \Doi{10.1063/1.525910}{\bibinfo {journal} {J.Math.Phys.}}\ }%
  \textbf{\bibinfo {volume} {24}},\ \bibinfo {pages} {1871} (\bibinfo {year}
  {1983})%
  \bibAnnoteFile{NoStop}{Fulp:1983bt}%
%%CITATION = JMAPA,24,1871;%%
\bibitem{Kashiwa:1996rs}%
  \BibitemOpen
  \bibfield{author}{%
  \bibinfo {author} {\bibfnamefont{T.}~\bibnamefont{Kashiwa}}\ and\ \bibinfo
  {author} {\bibfnamefont{N.}~\bibnamefont{Tanimura}}}%
   (\bibinfo {year} {1996}),\
  \Eprint{http://arxiv.org/abs/hep-th/9605207}{arXiv:hep-th/9605207 [hep-th]}%
  \bibAnnoteFile{NoStop}{Kashiwa:1996rs}%
%%CITATION = HEP-TH/9605207;%%
\bibitem{Kashiwa:1996hp}%
  \BibitemOpen
  \bibfield{author}{%
  \bibinfo {author} {\bibfnamefont{T.}~\bibnamefont{Kashiwa}}\ and\ \bibinfo
  {author} {\bibfnamefont{N.}~\bibnamefont{Tanimura}},\ }%
  \bibfield{journal}{%
  \Doi{10.1103/PhysRevD.56.2281}{\bibinfo {journal} {Phys.Rev.}}\ }%
  \textbf{\bibinfo {volume} {D56}},\ \bibinfo {pages} {2281} (\bibinfo {year}
  {1997}),\ \Eprint{http://arxiv.org/abs/hep-th/9612250}{arXiv:hep-th/9612250
  [hep-th]}%
  \bibAnnoteFile{NoStop}{Kashiwa:1996hp}%
%%CITATION = HEP-TH/9612250;%%
\bibitem{Guo:2012wv}%
  \BibitemOpen
  \bibfield{author}{%
  \bibinfo {author} {\bibfnamefont{Z.-Q.}\ \bibnamefont{Guo}}\ and\ \bibinfo
  {author} {\bibfnamefont{I.}~\bibnamefont{Schmidt}},\ }%
  \bibfield{journal}{%
  \Doi{10.1103/PhysRevD.87.114016}{\bibinfo {journal} {Phys.Rev.}}\ }%
  \textbf{\bibinfo {volume} {D87}},\ \bibinfo {pages} {114016} (\bibinfo {year}
  {2013}),\ \Eprint{http://arxiv.org/abs/1210.2263}{arXiv:1210.2263 [hep-ph]}%
  \bibAnnoteFile{NoStop}{Guo:2012wv}%
%%CITATION = ARXIV:1210.2263;%%
\bibitem{Konopinski}%
  \BibitemOpen
  \bibfield{author}{%
  \bibinfo {author} {\bibfnamefont{E.~J.}\ \bibnamefont{Konopinski}},\ }%
  \bibfield{journal}{%
  \bibinfo {journal} {Am.J. Phys.}\ }%
  \textbf{\bibinfo {volume} {46(5)}},\ \bibinfo {pages} {499} (\bibinfo {year}
  {1978})%
  \bibAnnoteFile{NoStop}{Konopinski}%
\bibitem{Stueckelberg:1900zz}%
  \BibitemOpen
  \bibfield{author}{%
  \bibinfo {author} {\bibfnamefont{E.}~\bibnamefont{Stueckelberg}},\ }%
  \bibfield{journal}{%
  \Doi{10.5169/seals-110852}{\bibinfo {journal} {Helv.Phys.Acta}}\ }%
  \textbf{\bibinfo {volume} {11}},\ \bibinfo {pages} {225} (\bibinfo {year}
  {1938})%
  \bibAnnoteFile{NoStop}{Stueckelberg:1900zz}%
%%CITATION = HPACA,11,225;%%
\bibitem{Stueckelberg:1938zz}%
  \BibitemOpen
  \bibfield{author}{%
  \bibinfo {author} {\bibfnamefont{E.}~\bibnamefont{Stueckelberg}},\ }%
  \bibfield{journal}{%
  \bibinfo {journal} {Helv.Phys.Acta}\ }%
  \textbf{\bibinfo {volume} {11}},\ \bibinfo {pages} {299} (\bibinfo {year}
  {1938})%
  \bibAnnoteFile{NoStop}{Stueckelberg:1938zz}%
%%CITATION = HPACA,11,299;%%
\bibitem{Stueckelberg:1957zz}%
  \BibitemOpen
  \bibfield{author}{%
  \bibinfo {author} {\bibfnamefont{E.}~\bibnamefont{Stueckelberg}},\ }%
  \bibfield{journal}{%
  \bibinfo {journal} {Helv.Phys.Acta}\ }%
  \textbf{\bibinfo {volume} {30}},\ \bibinfo {pages} {209} (\bibinfo {year}
  {1957})%
  \bibAnnoteFile{NoStop}{Stueckelberg:1957zz}%
%%CITATION = HPACA,30,209;%%
\bibitem{Pauli:1941zz}%
  \BibitemOpen
  \bibfield{author}{%
  \bibinfo {author} {\bibfnamefont{W.}~\bibnamefont{Pauli}},\ }%
  \bibfield{journal}{%
  \Doi{10.1103/RevModPhys.13.203}{\bibinfo {journal} {Rev.Mod.Phys.}}\ }%
  \textbf{\bibinfo {volume} {13}},\ \bibinfo {pages} {203} (\bibinfo {year}
  {1941})%
  \bibAnnoteFile{NoStop}{Pauli:1941zz}%
%%CITATION = RMPHA,13,203;%%
\bibitem{Ruegg:2003ps}%
  \BibitemOpen
  \bibfield{author}{%
  \bibinfo {author} {\bibfnamefont{H.}~\bibnamefont{Ruegg}}\ and\ \bibinfo
  {author} {\bibfnamefont{M.}~\bibnamefont{Ruiz-Altaba}},\ }%
  \bibfield{journal}{%
  \Doi{10.1142/S0217751X04019755}{\bibinfo {journal} {Int.J.Mod.Phys.}}\ }%
  \textbf{\bibinfo {volume} {A19}},\ \bibinfo {pages} {3265} (\bibinfo {year}
  {2004}),\ \Eprint{http://arxiv.org/abs/hep-th/0304245}{arXiv:hep-th/0304245
  [hep-th]}%
  \bibAnnoteFile{NoStop}{Ruegg:2003ps}%
%%CITATION = HEP-TH/0304245;%%
\bibitem{Jauch:1955jr}%
  \BibitemOpen
  \bibfield{author}{%
  \bibinfo {author} {\bibfnamefont{J.~M.}\ \bibnamefont{Jauch}}\ and\ \bibinfo
  {author} {\bibfnamefont{F.}~\bibnamefont{Rohrlich}},\ }%
  \emph{\bibinfo {title} {The Theory of Photons and Electrons}}\ (\bibinfo
  {publisher} {Addison-Wesley},\ \bibinfo {address} {Cambridge, USA},\ \bibinfo
  {year} {1955})%
  \bibAnnoteFile{NoStop}{Jauch:1955jr}%
\bibitem{Berestetskii}%
  \BibitemOpen
  \bibfield{author}{%
  \bibinfo {author} {\bibfnamefont{V.}~\bibnamefont{Berestetskii}}, \bibinfo
  {author} {\bibfnamefont{E.}~\bibnamefont{Lifshitz}},\ and\ \bibinfo {author}
  {\bibfnamefont{L.}~\bibnamefont{Pitaevskii}},\ }%
  \emph{\bibinfo {title} {{ Quantum Electrodynamics}}}\ (\bibinfo {publisher}
  {Pergamon},\ \bibinfo {address} {Oxford, U.K.},\ \bibinfo {year} {1982})%
  \bibAnnoteFile{NoStop}{Berestetskii}%
\bibitem{Cohen}%
  \BibitemOpen
  \bibfield{author}{%
  \bibinfo {author} {\bibfnamefont{C.}~\bibnamefont{Cohen-Tannoudji}}, \bibinfo
  {author} {\bibfnamefont{J.}~\bibnamefont{Dupont-Roc}},\ and\ \bibinfo
  {author} {\bibfnamefont{G.}~\bibnamefont{Grynberg}},\ }%
  \emph{\bibinfo {title} {{ Photons and Atoms}}}\ (\bibinfo {publisher} {John
  Wiley and Sons Inc.},\ \bibinfo {address} {New York, N.Y.},\ \bibinfo {year}
  {1989})%
  \bibAnnoteFile{NoStop}{Cohen}%
\bibitem{Bashinsky:1998if}%
  \BibitemOpen
  \bibfield{author}{%
  \bibinfo {author} {\bibfnamefont{S.}~\bibnamefont{Bashinsky}}\ and\ \bibinfo
  {author} {\bibfnamefont{R.}~\bibnamefont{Jaffe}},\ }%
  \bibfield{journal}{%
  \Doi{10.1016/S0550-3213(98)00559-8}{\bibinfo {journal} {Nucl.Phys.}}\ }%
  \textbf{\bibinfo {volume} {B536}},\ \bibinfo {pages} {303} (\bibinfo {year}
  {1998}),\ \Eprint{http://arxiv.org/abs/hep-ph/9804397}{arXiv:hep-ph/9804397
  [hep-ph]}%
  \bibAnnoteFile{NoStop}{Bashinsky:1998if}%
%%CITATION = HEP-PH/9804397;%%
\bibitem{Ivanov:1985np}%
  \BibitemOpen
  \bibfield{author}{%
  \bibinfo {author} {\bibfnamefont{S.}~\bibnamefont{Ivanov}}, \bibinfo {author}
  {\bibfnamefont{G.}~\bibnamefont{Korchemsky}},\ and\ \bibinfo {author}
  {\bibfnamefont{A.}~\bibnamefont{Radyushkin}},\ }%
  \bibfield{journal}{%
  \bibinfo {journal} {Yad.Fiz.}\ }%
  \textbf{\bibinfo {volume} {44}},\ \bibinfo {pages} {230} (\bibinfo {year}
  {1986})%
  \bibAnnoteFile{NoStop}{Ivanov:1985np}%
%%CITATION = YAFIA,44,230;%%
\bibitem{Belinfante:1962zz}%
  \BibitemOpen
  \bibfield{author}{%
  \bibinfo {author} {\bibfnamefont{F.~J.}\ \bibnamefont{Belinfante}},\ }%
  \bibfield{journal}{%
  \Doi{10.1103/PhysRev.128.2832}{\bibinfo {journal} {Phys.Rev.}}\ }%
  \textbf{\bibinfo {volume} {128}},\ \bibinfo {pages} {2832} (\bibinfo {year}
  {1962})%
  \bibAnnoteFile{NoStop}{Belinfante:1962zz}%
%%CITATION = PHRVA,128,2832;%%
\bibitem{Mandelstam:1962mi}%
  \BibitemOpen
  \bibfield{author}{%
  \bibinfo {author} {\bibfnamefont{S.}~\bibnamefont{Mandelstam}},\ }%
  \bibfield{journal}{%
  \Doi{10.1016/0003-4916(62)90232-4}{\bibinfo {journal} {Annals Phys.}}\ }%
  \textbf{\bibinfo {volume} {19}},\ \bibinfo {pages} {1} (\bibinfo {year}
  {1962})%
  \bibAnnoteFile{NoStop}{Mandelstam:1962mi}%
%%CITATION = APNYA,19,1;%%
\bibitem{Rohrlich:1965}%
  \BibitemOpen
  \bibfield{author}{%
  \bibinfo {author} {\bibfnamefont{F.}~\bibnamefont{Rohrlich}}\ and\ \bibinfo
  {author} {\bibfnamefont{F.}~\bibnamefont{Strocchi}},\ }%
  \bibfield{journal}{%
  \bibinfo {journal} {Phys.Rev.}\ }%
  \textbf{\bibinfo {volume} {139}},\ \bibinfo {pages} {B476} (\bibinfo {year}
  {1965})%
  \bibAnnoteFile{NoStop}{Rohrlich:1965}%
\bibitem{Yang:1985}%
  \BibitemOpen
  \bibfield{author}{%
  \bibinfo {author} {\bibfnamefont{K.}~\bibnamefont{Yang}},\ }%
  \bibfield{journal}{%
  \bibinfo {journal} {J.Phys.}\ }%
  \textbf{\bibinfo {volume} {A18}},\ \bibinfo {pages} {979} (\bibinfo {year}
  {1985})%
  \bibAnnoteFile{NoStop}{Yang:1985}%
%%CITATION = FPYKA,45,381;%%
\bibitem{Kashiwa:1994jn}%
  \BibitemOpen
  \bibfield{author}{%
  \bibinfo {author} {\bibfnamefont{T.}~\bibnamefont{Kashiwa}}\ and\ \bibinfo
  {author} {\bibfnamefont{Y.}~\bibnamefont{Takahashi}}}%
   (\bibinfo {year} {1994}),\
  \Eprint{http://arxiv.org/abs/hep-th/9401097}{arXiv:hep-th/9401097 [hep-th]}%
  \bibAnnoteFile{NoStop}{Kashiwa:1994jn}%
%%CITATION = HEP-TH/9401097;%%
\bibitem{Kashiwa:1997we}%
  \BibitemOpen
  \bibfield{author}{%
  \bibinfo {author} {\bibfnamefont{T.}~\bibnamefont{Kashiwa}}\ and\ \bibinfo
  {author} {\bibfnamefont{N.}~\bibnamefont{Tanimura}},\ }%
  \bibfield{journal}{%
  \bibinfo {journal} {Fortsch.Phys.}\ }%
  \textbf{\bibinfo {volume} {45}},\ \bibinfo {pages} {381} (\bibinfo {year}
  {1997})%
  \bibAnnoteFile{NoStop}{Kashiwa:1997we}%
%%CITATION = FPYKA,45,381;%%
\bibitem{Simmons}%
  \BibitemOpen
  \bibfield{author}{%
  \bibinfo {author} {\bibfnamefont{J.}~\bibnamefont{Simmons}}\ and\ \bibinfo
  {author} {\bibfnamefont{M.}~\bibnamefont{Guttmann}},\ }%
  \emph{\bibinfo {title} {{ States, Waves, and Photons}}}\ (\bibinfo
  {publisher} {Addison-Wesley},\ \bibinfo {address} {Reading, Mass.},\ \bibinfo
  {year} {1970})%
  \bibAnnoteFile{NoStop}{Simmons}%
\bibitem{DeWitt:1967ub}%
  \BibitemOpen
  \bibfield{author}{%
  \bibinfo {author} {\bibfnamefont{B.~S.}\ \bibnamefont{DeWitt}},\ }%
  \bibfield{journal}{%
  \Doi{10.1103/PhysRev.162.1195}{\bibinfo {journal} {Phys.Rev.}}\ }%
  \textbf{\bibinfo {volume} {162}},\ \bibinfo {pages} {1195} (\bibinfo {year}
  {1967})%
  \bibAnnoteFile{NoStop}{DeWitt:1967ub}%
%%CITATION = PHRVA,162,1195;%%
\bibitem{DeWitt:1967uc}%
  \BibitemOpen
  \bibfield{author}{%
  \bibinfo {author} {\bibfnamefont{B.~S.}\ \bibnamefont{DeWitt}},\ }%
  \bibfield{journal}{%
  \Doi{10.1103/PhysRev.162.1239}{\bibinfo {journal} {Phys.Rev.}}\ }%
  \textbf{\bibinfo {volume} {162}},\ \bibinfo {pages} {1239} (\bibinfo {year}
  {1967})%
  \bibAnnoteFile{NoStop}{DeWitt:1967uc}%
%%CITATION = PHRVA,162,1239;%%
\bibitem{DeWitt:1980jv}%
  \BibitemOpen
  \bibfield{author}{%
  \bibinfo {author} {\bibfnamefont{B.~S.}\ \bibnamefont{DeWitt}}}%
   (\bibinfo {year} {1980})%
  \bibAnnoteFile{NoStop}{DeWitt:1980jv}%
%%CITATION = NSF-ITP-80-31 ETC.;%%
\bibitem{'tHooft:1975vy}%
  \BibitemOpen
  \bibfield{author}{%
  \bibinfo {author} {\bibfnamefont{G.}~\bibnamefont{'t~Hooft}}}%
   (\bibinfo {year} {1975})%
  \bibAnnoteFile{NoStop}{'tHooft:1975vy}%
%%CITATION = INSPIRE-105214;%%
\bibitem{Grisaru:1975ei}%
  \BibitemOpen
  \bibfield{author}{%
  \bibinfo {author} {\bibfnamefont{M.~T.}\ \bibnamefont{Grisaru}}, \bibinfo
  {author} {\bibfnamefont{P.}~\bibnamefont{van Nieuwenhuizen}},\ and\ \bibinfo
  {author} {\bibfnamefont{C.}~\bibnamefont{Wu}},\ }%
  \bibfield{journal}{%
  \Doi{10.1103/PhysRevD.12.3203}{\bibinfo {journal} {Phys.Rev.}}\ }%
  \textbf{\bibinfo {volume} {D12}},\ \bibinfo {pages} {3203} (\bibinfo {year}
  {1975})%
  \bibAnnoteFile{NoStop}{Grisaru:1975ei}%
%%CITATION = PHRVA,D12,3203;%%
\bibitem{Boulware:1980av}%
  \BibitemOpen
  \bibfield{author}{%
  \bibinfo {author} {\bibfnamefont{D.~G.}\ \bibnamefont{Boulware}},\ }%
  \bibfield{journal}{%
  \Doi{10.1103/PhysRevD.23.389}{\bibinfo {journal} {Phys.Rev.}}\ }%
  \textbf{\bibinfo {volume} {D23}},\ \bibinfo {pages} {389} (\bibinfo {year}
  {1981})%
  \bibAnnoteFile{NoStop}{Boulware:1980av}%
%%CITATION = PHRVA,D23,389;%%
\bibitem{Abbott:1980hw}%
  \BibitemOpen
  \bibfield{author}{%
  \bibinfo {author} {\bibfnamefont{L.}~\bibnamefont{Abbott}},\ }%
  \bibfield{journal}{%
  \Doi{10.1016/0550-3213(81)90371-0}{\bibinfo {journal} {Nucl.Phys.}}\ }%
  \textbf{\bibinfo {volume} {B185}},\ \bibinfo {pages} {189} (\bibinfo {year}
  {1981})%
  \bibAnnoteFile{NoStop}{Abbott:1980hw}%
%%CITATION = NUPHA,B185,189;%%
\bibitem{'tHooft:1973us}%
  \BibitemOpen
  \bibfield{author}{%
  \bibinfo {author} {\bibfnamefont{G.}~\bibnamefont{'t~Hooft}},\ }%
  \bibfield{journal}{%
  \Doi{10.1016/0550-3213(73)90263-0}{\bibinfo {journal} {Nucl.Phys.}}\ }%
  \textbf{\bibinfo {volume} {B62}},\ \bibinfo {pages} {444} (\bibinfo {year}
  {1973})%
  \bibAnnoteFile{NoStop}{'tHooft:1973us}%
%%CITATION = NUPHA,B62,444;%%
\bibitem{'tHooft:1974bx}%
  \BibitemOpen
  \bibfield{author}{%
  \bibinfo {author} {\bibfnamefont{G.}~\bibnamefont{'t~Hooft}}\ and\ \bibinfo
  {author} {\bibfnamefont{M.}~\bibnamefont{Veltman}},\ }%
  \bibfield{journal}{%
  \bibinfo {journal} {Annales Poincare Phys.Theor.}\ }%
  \textbf{\bibinfo {volume} {A20}},\ \bibinfo {pages} {69} (\bibinfo {year}
  {1974})%
  \bibAnnoteFile{NoStop}{'tHooft:1974bx}%
%%CITATION = AHPAA,A20,69;%%
\bibitem{Deser:1977nt}%
  \BibitemOpen
  \bibfield{author}{%
  \bibinfo {author} {\bibfnamefont{S.}~\bibnamefont{Deser}}, \bibinfo {author}
  {\bibfnamefont{J.}~\bibnamefont{Kay}},\ and\ \bibinfo {author}
  {\bibfnamefont{K.}~\bibnamefont{Stelle}},\ }%
  \bibfield{journal}{%
  \Doi{10.1103/PhysRevLett.38.527}{\bibinfo {journal} {Phys.Rev.Lett.}}\ }%
  \textbf{\bibinfo {volume} {38}},\ \bibinfo {pages} {527} (\bibinfo {year}
  {1977})%
  \bibAnnoteFile{NoStop}{Deser:1977nt}%
%%CITATION = PRLTA,38,527;%%
\bibitem{Abbott:1981ff}%
  \BibitemOpen
  \bibfield{author}{%
  \bibinfo {author} {\bibfnamefont{L.}~\bibnamefont{Abbott}}\ and\ \bibinfo
  {author} {\bibfnamefont{S.}~\bibnamefont{Deser}},\ }%
  \bibfield{journal}{%
  \Doi{10.1016/0550-3213(82)90049-9}{\bibinfo {journal} {Nucl.Phys.}}\ }%
  \textbf{\bibinfo {volume} {B195}},\ \bibinfo {pages} {76} (\bibinfo {year}
  {1982})%
  \bibAnnoteFile{NoStop}{Abbott:1981ff}%
%%CITATION = NUPHA,B195,76;%%
\bibitem{Petrov:2007xva}%
  \BibitemOpen
  \bibfield{author}{%
  \bibinfo {author} {\bibfnamefont{A.}~\bibnamefont{Petrov}}}%
   (\bibinfo {year} {2007}),\
  \Eprint{http://arxiv.org/abs/0705.0019}{arXiv:0705.0019 [gr-qc]}%
  \bibAnnoteFile{NoStop}{Petrov:2007xva}%
%%CITATION = ARXIV:0705.0019;%%
\bibitem{Petrov:2012qn}%
  \BibitemOpen
  \bibfield{author}{%
  \bibinfo {author} {\bibfnamefont{A.~N.}\ \bibnamefont{Petrov}}\ and\ \bibinfo
  {author} {\bibfnamefont{R.~R.}\ \bibnamefont{Lompay}},\ }%
  \bibfield{journal}{%
  \Doi{10.1007/s10714-012-1487-4}{\bibinfo {journal} {Gen.Rel.Grav.}}\ }%
  \textbf{\bibinfo {volume} {45}},\ \bibinfo {pages} {545} (\bibinfo {year}
  {2013}),\ \Eprint{http://arxiv.org/abs/1211.3268}{arXiv:1211.3268 [gr-qc]}%
  \bibAnnoteFile{NoStop}{Petrov:2012qn}%
%%CITATION = ARXIV:1211.3268;%%
\bibitem{Dashen:1980vm}%
  \BibitemOpen
  \bibfield{author}{%
  \bibinfo {author} {\bibfnamefont{R.~F.}\ \bibnamefont{Dashen}}\ and\ \bibinfo
  {author} {\bibfnamefont{D.~J.}\ \bibnamefont{Gross}},\ }%
  \bibfield{journal}{%
  \Doi{10.1103/PhysRevD.23.2340}{\bibinfo {journal} {Phys.Rev.}}\ }%
  \textbf{\bibinfo {volume} {D23}},\ \bibinfo {pages} {2340} (\bibinfo {year}
  {1981})%
  \bibAnnoteFile{NoStop}{Dashen:1980vm}%
%%CITATION = PHRVA,D23,2340;%%
\bibitem{Abbott:1982jh}%
  \BibitemOpen
  \bibfield{author}{%
  \bibinfo {author} {\bibfnamefont{L.}~\bibnamefont{Abbott}}\ and\ \bibinfo
  {author} {\bibfnamefont{S.}~\bibnamefont{Deser}},\ }%
  \bibfield{journal}{%
  \Doi{10.1016/0370-2693(82)90338-0}{\bibinfo {journal} {Phys.Lett.}}\ }%
  \textbf{\bibinfo {volume} {B116}},\ \bibinfo {pages} {259} (\bibinfo {year}
  {1982})%
  \bibAnnoteFile{NoStop}{Abbott:1982jh}%
%%CITATION = PHLTA,B116,259;%%
\bibitem{Luscher:1995vs}%
  \BibitemOpen
  \bibfield{author}{%
  \bibinfo {author} {\bibfnamefont{M.}~\bibnamefont{Luscher}}\ and\ \bibinfo
  {author} {\bibfnamefont{P.}~\bibnamefont{Weisz}},\ }%
  \bibfield{journal}{%
  \Doi{10.1016/0550-3213(95)00346-T}{\bibinfo {journal} {Nucl.Phys.}}\ }%
  \textbf{\bibinfo {volume} {B452}},\ \bibinfo {pages} {213} (\bibinfo {year}
  {1995}),\ \Eprint{http://arxiv.org/abs/hep-lat/9504006}{arXiv:hep-lat/9504006
  [hep-lat]}%
  \bibAnnoteFile{NoStop}{Luscher:1995vs}%
%%CITATION = HEP-LAT/9504006;%%
\bibitem{Freire:2000bq}%
  \BibitemOpen
  \bibfield{author}{%
  \bibinfo {author} {\bibfnamefont{F.}~\bibnamefont{Freire}}, \bibinfo {author}
  {\bibfnamefont{D.~F.}\ \bibnamefont{Litim}},\ and\ \bibinfo {author}
  {\bibfnamefont{J.~M.}\ \bibnamefont{Pawlowski}},\ }%
  \bibfield{journal}{%
  \Doi{10.1016/S0370-2693(00)01231-4}{\bibinfo {journal} {Phys.Lett.}}\ }%
  \textbf{\bibinfo {volume} {B495}},\ \bibinfo {pages} {256} (\bibinfo {year}
  {2000}),\ \Eprint{http://arxiv.org/abs/hep-th/0009110}{arXiv:hep-th/0009110
  [hep-th]}%
  \bibAnnoteFile{NoStop}{Freire:2000bq}%
%%CITATION = HEP-TH/0009110;%%
\bibitem{Binosi:2009qm}%
  \BibitemOpen
  \bibfield{author}{%
  \bibinfo {author} {\bibfnamefont{D.}~\bibnamefont{Binosi}}\ and\ \bibinfo
  {author} {\bibfnamefont{J.}~\bibnamefont{Papavassiliou}},\ }%
  \bibfield{journal}{%
  \Doi{10.1016/j.physrep.2009.05.001}{\bibinfo {journal} {Phys.Rept.}}\ }%
  \textbf{\bibinfo {volume} {479}},\ \bibinfo {pages} {1} (\bibinfo {year}
  {2009}),\ \Eprint{http://arxiv.org/abs/0909.2536}{arXiv:0909.2536 [hep-ph]}%
  \bibAnnoteFile{NoStop}{Binosi:2009qm}%
%%CITATION = ARXIV:0909.2536;%%
\bibitem{Binosi:2012st}%
  \BibitemOpen
  \bibfield{author}{%
  \bibinfo {author} {\bibfnamefont{D.}~\bibnamefont{Binosi}}\ and\ \bibinfo
  {author} {\bibfnamefont{A.}~\bibnamefont{Quadri}},\ }%
  \bibfield{journal}{%
  \Doi{10.1103/PhysRevD.85.121702}{\bibinfo {journal} {Phys.Rev.}}\ }%
  \textbf{\bibinfo {volume} {D85}},\ \bibinfo {pages} {121702} (\bibinfo {year}
  {2012}),\ \Eprint{http://arxiv.org/abs/1203.6637}{arXiv:1203.6637 [hep-th]}%
  \bibAnnoteFile{NoStop}{Binosi:2012st}%
%%CITATION = ARXIV:1203.6637;%%
\bibitem{Abbott:1981ke}%
  \BibitemOpen
  \bibfield{author}{%
  \bibinfo {author} {\bibfnamefont{L.}~\bibnamefont{Abbott}},\ }%
  \bibfield{journal}{%
  \bibinfo {journal} {Acta Phys.Polon.}\ }%
  \textbf{\bibinfo {volume} {B13}},\ \bibinfo {pages} {33} (\bibinfo {year}
  {1982})%
  \bibAnnoteFile{NoStop}{Abbott:1981ke}%
%%CITATION = APPOA,B13,33;%%
\bibitem{Guay:2004zz}%
  \BibitemOpen
  \bibfield{author}{%
  \bibinfo {author} {\bibfnamefont{A.}~\bibnamefont{Guay}}}%
   (\bibinfo {year} {2004})%
  \bibAnnoteFile{NoStop}{Guay:2004zz}%
%%CITATION = INSPIRE-886454;%%
\bibitem{Smolin:2005mq}%
  \BibitemOpen
  \bibfield{author}{%
  \bibinfo {author} {\bibfnamefont{L.}~\bibnamefont{Smolin}}}%
   (\bibinfo {year} {2005}),\
  \Eprint{http://arxiv.org/abs/hep-th/0507235}{arXiv:hep-th/0507235 [hep-th]}%
  \bibAnnoteFile{NoStop}{Smolin:2005mq}%
%%CITATION = HEP-TH/0507235;%%
\bibitem{Rickles:2008}%
  \BibitemOpen
  \bibfield{author}{%
  \bibinfo {author} {\bibfnamefont{D.}~\bibnamefont{Rickles}},\ \bibinfo
  {pages} {133}}%
   (\bibinfo {year} {2008})%
  \bibAnnoteFile{NoStop}{Rickles:2008}%
\bibitem{Rozali:2008ex}%
  \BibitemOpen
  \bibfield{author}{%
  \bibinfo {author} {\bibfnamefont{M.}~\bibnamefont{Rozali}},\ }%
  \bibfield{journal}{%
  \Doi{10.1166/asl.2009.1031}{\bibinfo {journal} {Adv.Sci.Lett.}}\ }%
  \textbf{\bibinfo {volume} {2}},\ \bibinfo {pages} {244} (\bibinfo {year}
  {2009}),\ \Eprint{http://arxiv.org/abs/0809.3962}{arXiv:0809.3962 [gr-qc]}%
  \bibAnnoteFile{NoStop}{Rozali:2008ex}%
%%CITATION = ARXIV:0809.3962;%%
\bibitem{Barenz:2012av}%
  \BibitemOpen
  \bibfield{author}{%
  \bibinfo {author} {\bibfnamefont{M.}~\bibnamefont{Barenz}}}%
   (\bibinfo {year} {2012}),\
  \Eprint{http://arxiv.org/abs/1207.0340}{arXiv:1207.0340 [gr-qc]}%
  \bibAnnoteFile{NoStop}{Barenz:2012av}%
%%CITATION = ARXIV:1207.0340;%%
\bibitem{Collins}%
  \BibitemOpen
  \bibfield{author}{%
  \bibinfo {author} {\bibfnamefont{J.}~\bibnamefont{Collins}}, \bibinfo
  {author} {\bibfnamefont{D.}~\bibnamefont{Soper}},\ and\ \bibinfo {author}
  {\bibfnamefont{G.}~\bibnamefont{Sterman}},\ }%
  \emph{\bibinfo {title} {{Perturbative Quantum Chromodynamics}}}\ (\bibinfo
  {publisher} {World Scientific},\ \bibinfo {address} {Singapore})%
  \bibAnnoteFile{NoStop}{Collins}%
\bibitem{Ji:2013fga}%
  \BibitemOpen
  \bibfield{author}{%
  \bibinfo {author} {\bibfnamefont{X.}~\bibnamefont{Ji}}, \bibinfo {author}
  {\bibfnamefont{J.-H.}\ \bibnamefont{Zhang}},\ and\ \bibinfo {author}
  {\bibfnamefont{Y.}~\bibnamefont{Zhao}}}%
   (\bibinfo {year} {2013}),\
  \Eprint{http://arxiv.org/abs/1304.6708}{arXiv:1304.6708 [hep-ph]}%
  \bibAnnoteFile{NoStop}{Ji:2013fga}%
%%CITATION = ARXIV:1304.6708;%%
\bibitem{Gribov:1977wm}%
  \BibitemOpen
  \bibfield{author}{%
  \bibinfo {author} {\bibfnamefont{V.}~\bibnamefont{Gribov}},\ }%
  \bibfield{journal}{%
  \Doi{10.1016/0550-3213(78)90175-X}{\bibinfo {journal} {Nucl.Phys.}}\ }%
  \textbf{\bibinfo {volume} {B139}},\ \bibinfo {pages} {1} (\bibinfo {year}
  {1978})%
  \bibAnnoteFile{NoStop}{Gribov:1977wm}%
%%CITATION = NUPHA,B139,1;%%
\bibitem{Feynman:1969ej}%
  \BibitemOpen
  \bibfield{author}{%
  \bibinfo {author} {\bibfnamefont{R.~P.}\ \bibnamefont{Feynman}},\ }%
  \bibfield{journal}{%
  \Doi{10.1103/PhysRevLett.23.1415}{\bibinfo {journal} {Phys.Rev.Lett.}}\ }%
  \textbf{\bibinfo {volume} {23}},\ \bibinfo {pages} {1415} (\bibinfo {year}
  {1969})%
  \bibAnnoteFile{NoStop}{Feynman:1969ej}%
%%CITATION = PRLTA,23,1415;%%
\bibitem{Rovelli:2013fga}%
  \BibitemOpen
  \bibfield{author}{%
  \bibinfo {author} {\bibfnamefont{C.}~\bibnamefont{Rovelli}}}%
   (\bibinfo {year} {2013}),\
  \Eprint{http://arxiv.org/abs/1308.5599}{arXiv:1308.5599 [hep-th]}%
  \bibAnnoteFile{NoStop}{Rovelli:2013fga}%
%%CITATION = ARXIV:1308.5599;%%
\bibitem{Lepage:1980fj}%
  \BibitemOpen
  \bibfield{author}{%
  \bibinfo {author} {\bibfnamefont{G.~P.}\ \bibnamefont{Lepage}}\ and\ \bibinfo
  {author} {\bibfnamefont{S.~J.}\ \bibnamefont{Brodsky}},\ }%
  \bibfield{journal}{%
  \Doi{10.1103/PhysRevD.22.2157}{\bibinfo {journal} {Phys.Rev.}}\ }%
  \textbf{\bibinfo {volume} {D22}},\ \bibinfo {pages} {2157} (\bibinfo {year}
  {1980})%
  \bibAnnoteFile{NoStop}{Lepage:1980fj}%
%%CITATION = PHRVA,D22,2157;%%
\bibitem{Brodsky:1997de}%
  \BibitemOpen
  \bibfield{author}{%
  \bibinfo {author} {\bibfnamefont{S.~J.}\ \bibnamefont{Brodsky}}, \bibinfo
  {author} {\bibfnamefont{H.-C.}\ \bibnamefont{Pauli}},\ and\ \bibinfo {author}
  {\bibfnamefont{S.~S.}\ \bibnamefont{Pinsky}},\ }%
  \bibfield{journal}{%
  \Doi{10.1016/S0370-1573(97)00089-6}{\bibinfo {journal} {Phys.Rept.}}\ }%
  \textbf{\bibinfo {volume} {301}},\ \bibinfo {pages} {299} (\bibinfo {year}
  {1998}),\ \Eprint{http://arxiv.org/abs/hep-ph/9705477}{arXiv:hep-ph/9705477
  [hep-ph]}%
  \bibAnnoteFile{NoStop}{Brodsky:1997de}%
%%CITATION = HEP-PH/9705477;%%
\bibitem{Jackson}%
  \BibitemOpen
  \bibfield{author}{%
  \bibinfo {author} {\bibfnamefont{J.}~\bibnamefont{Jackson}},\ }%
  \emph{\bibinfo {title} {{ Classical Electrodynamics}}}\ (\bibinfo {publisher}
  {John Wiley \& Sons},\ \bibinfo {address} {New York, USA},\ \bibinfo {year}
  {1999})%
  \bibAnnoteFile{NoStop}{Jackson}%
\bibitem{Feynman}%
  \BibitemOpen
  \bibfield{author}{%
  \bibinfo {author} {\bibfnamefont{R.}~\bibnamefont{Feynman}}, \bibinfo
  {author} {\bibfnamefont{R.}~\bibnamefont{Leighton}},\ and\ \bibinfo {author}
  {\bibfnamefont{M.}~\bibnamefont{Sands}},\ }%
  \emph{\bibinfo {title} {{ The Feynman Lectures on Physics, Vol. III}}}\
  (\bibinfo {publisher} {Addison-Wesley},\ \bibinfo {address} {Reading, MA,
  USA},\ \bibinfo {year} {1965})%
  \bibAnnoteFile{NoStop}{Feynman}%
\bibitem{Manoukian:1987hy}%
  \BibitemOpen
  \bibfield{author}{%
  \bibinfo {author} {\bibfnamefont{E.~B.}\ \bibnamefont{Manoukian}},\ }%
  \bibfield{journal}{%
  \Doi{10.1088/0305-4616/13/8/008}{\bibinfo {journal} {J.Phys.}}\ }%
  \textbf{\bibinfo {volume} {G13}},\ \bibinfo {pages} {1013} (\bibinfo {year}
  {1987})%
  \bibAnnoteFile{NoStop}{Manoukian:1987hy}%
%%CITATION = INSPIRE-255146;%%
\bibitem{Moriyasu:1984mh}%
  \BibitemOpen
  \bibfield{author}{%
  \bibinfo {author} {\bibfnamefont{K.}~\bibnamefont{Moriyasu}}}%
   (\bibinfo {year} {1984})%
  \bibAnnoteFile{NoStop}{Moriyasu:1984mh}%
%%CITATION = INSPIRE-205399;%%
\bibitem{Coleman:1967ad}%
  \BibitemOpen
  \bibfield{author}{%
  \bibinfo {author} {\bibfnamefont{S.~R.}\ \bibnamefont{Coleman}}\ and\
  \bibinfo {author} {\bibfnamefont{J.}~\bibnamefont{Mandula}},\ }%
  \bibfield{journal}{%
  \Doi{10.1103/PhysRev.159.1251}{\bibinfo {journal} {Phys.Rev.}}\ }%
  \textbf{\bibinfo {volume} {159}},\ \bibinfo {pages} {1251} (\bibinfo {year}
  {1967})%
  \bibAnnoteFile{NoStop}{Coleman:1967ad}%
%%CITATION = PHRVA,159,1251;%%
\bibitem{Ji:1996ek}%
  \BibitemOpen
  \bibfield{author}{%
  \bibinfo {author} {\bibfnamefont{X.-D.}\ \bibnamefont{Ji}},\ }%
  \bibfield{journal}{%
  \Doi{10.1103/PhysRevLett.78.610}{\bibinfo {journal} {Phys. Rev. Lett.}}\ }%
  \textbf{\bibinfo {volume} {78}},\ \bibinfo {pages} {610} (\bibinfo {year}
  {1997}),\ \Eprint{http://arxiv.org/abs/hep-ph/9603249}{arXiv:hep-ph/9603249}%
  \bibAnnoteFile{NoStop}{Ji:1996ek}%
%%CITATION = HEP-PH/9603249;%%
\bibitem{Soper}%
  \BibitemOpen
  \bibfield{author}{%
  \bibinfo {author} {\bibfnamefont{D.}~\bibnamefont{Soper}},\ }%
  \emph{\bibinfo {title} {{ Classical Field Theory}}}\ (\bibinfo {publisher}
  {John Wiley and Sons},\ \bibinfo {address} {New York, USA},\ \bibinfo {year}
  {1976})%
  \bibAnnoteFile{NoStop}{Soper}%
\bibitem{Dirac:1955uv}%
  \BibitemOpen
  \bibfield{author}{%
  \bibinfo {author} {\bibfnamefont{P.~A.}\ \bibnamefont{Dirac}},\ }%
  \bibfield{journal}{%
  \bibinfo {journal} {Can.J.Phys.}\ }%
  \textbf{\bibinfo {volume} {33}},\ \bibinfo {pages} {650} (\bibinfo {year}
  {1955})%
  \bibAnnoteFile{NoStop}{Dirac:1955uv}%
%%CITATION = CJPHA,33,650;%%
\bibitem{DeWitt:1962mg}%
  \BibitemOpen
  \bibfield{author}{%
  \bibinfo {author} {\bibfnamefont{B.~S.}\ \bibnamefont{DeWitt}},\ }%
  \bibfield{journal}{%
  \Doi{10.1103/PhysRev.125.2189}{\bibinfo {journal} {Phys.Rev.}}\ }%
  \textbf{\bibinfo {volume} {125}},\ \bibinfo {pages} {2189} (\bibinfo {year}
  {1962})%
  \bibAnnoteFile{NoStop}{DeWitt:1962mg}%
%%CITATION = PHRVA,125,2189;%%
\bibitem{Mandelstam:1968hz}%
  \BibitemOpen
  \bibfield{author}{%
  \bibinfo {author} {\bibfnamefont{S.}~\bibnamefont{Mandelstam}},\ }%
  \bibfield{journal}{%
  \Doi{10.1103/PhysRev.175.1580}{\bibinfo {journal} {Phys.Rev.}}\ }%
  \textbf{\bibinfo {volume} {175}},\ \bibinfo {pages} {1580} (\bibinfo {year}
  {1968})%
  \bibAnnoteFile{NoStop}{Mandelstam:1968hz}%
%%CITATION = PHRVA,175,1580;%%
\bibitem{Mandelstam:1968ud}%
  \BibitemOpen
  \bibfield{author}{%
  \bibinfo {author} {\bibfnamefont{S.}~\bibnamefont{Mandelstam}},\ }%
  \bibfield{journal}{%
  \Doi{10.1103/PhysRev.175.1604}{\bibinfo {journal} {Phys.Rev.}}\ }%
  \textbf{\bibinfo {volume} {175}},\ \bibinfo {pages} {1604} (\bibinfo {year}
  {1968})%
  \bibAnnoteFile{NoStop}{Mandelstam:1968ud}%
%%CITATION = PHRVA,175,1604;%%
\bibitem{BialynickiBirula:1963}%
  \BibitemOpen
  \bibfield{author}{%
  \bibinfo {author} {\bibfnamefont{I.}~\bibnamefont{Bialynicki-Birula}},\ }%
  \bibfield{journal}{%
  \bibinfo {journal} {Bull.Acad.Pol.Sci.}\ }%
  \textbf{\bibinfo {volume} {11}},\ \bibinfo {pages} {135} (\bibinfo {year}
  {1963})%
  \bibAnnoteFile{NoStop}{BialynickiBirula:1963}%
\bibitem{Steinmann:1983ar}%
  \BibitemOpen
  \bibfield{author}{%
  \bibinfo {author} {\bibfnamefont{O.}~\bibnamefont{Steinmann}},\ }%
  \bibfield{journal}{%
  \Doi{10.1016/0003-4916(84)90053-8}{\bibinfo {journal} {Annals Phys.}}\ }%
  \textbf{\bibinfo {volume} {157}},\ \bibinfo {pages} {232} (\bibinfo {year}
  {1984})%
  \bibAnnoteFile{NoStop}{Steinmann:1983ar}%
%%CITATION = APNYA,157,232;%%
\bibitem{Steinmann:1985id}%
  \BibitemOpen
  \bibfield{author}{%
  \bibinfo {author} {\bibfnamefont{O.}~\bibnamefont{Steinmann}},\ }%
  \bibfield{journal}{%
  \bibinfo {journal} {Helv.Phys.Acta}\ }%
  \textbf{\bibinfo {volume} {58}},\ \bibinfo {pages} {995} (\bibinfo {year}
  {1985})%
  \bibAnnoteFile{NoStop}{Steinmann:1985id}%
%%CITATION = HPACA,58,995;%%
\bibitem{Skachkov:1985cz}%
  \BibitemOpen
  \bibfield{author}{%
  \bibinfo {author} {\bibfnamefont{N.}~\bibnamefont{Skachkov}}, \bibinfo
  {author} {\bibfnamefont{I.}~\bibnamefont{Solovtsov}},\ and\ \bibinfo {author}
  {\bibfnamefont{O.~Y.}\ \bibnamefont{Shevchenko}},\ }%
  \bibfield{journal}{%
  \Doi{10.1007/BF01560298}{\bibinfo {journal} {Z.Phys.}}\ }%
  \textbf{\bibinfo {volume} {C29}},\ \bibinfo {pages} {631} (\bibinfo {year}
  {1985})%
  \bibAnnoteFile{NoStop}{Skachkov:1985cz}%
%%CITATION = ZEPYA,C29,631;%%
\bibitem{Haagensen:1997pi}%
  \BibitemOpen
  \bibfield{author}{%
  \bibinfo {author} {\bibfnamefont{P.~E.}\ \bibnamefont{Haagensen}}\ and\
  \bibinfo {author} {\bibfnamefont{K.}~\bibnamefont{Johnson}}}%
   (\bibinfo {year} {1997}),\
  \Eprint{http://arxiv.org/abs/hep-th/9702204}{arXiv:hep-th/9702204 [hep-th]}%
  \bibAnnoteFile{NoStop}{Haagensen:1997pi}%
%%CITATION = HEP-TH/9702204;%%
\bibitem{Horan:1998im}%
  \BibitemOpen
  \bibfield{author}{%
  \bibinfo {author} {\bibfnamefont{R.}~\bibnamefont{Horan}}, \bibinfo {author}
  {\bibfnamefont{M.}~\bibnamefont{Lavelle}},\ and\ \bibinfo {author}
  {\bibfnamefont{D.}~\bibnamefont{McMullan}},\ }%
  \bibfield{journal}{%
  \Doi{10.1007/BF02828927}{\bibinfo {journal} {Pramana}}\ }%
  \textbf{\bibinfo {volume} {51}},\ \bibinfo {pages} {317} (\bibinfo {year}
  {1998}),\ \Eprint{http://arxiv.org/abs/hep-th/9810089}{arXiv:hep-th/9810089
  [hep-th]}%
  \bibAnnoteFile{NoStop}{Horan:1998im}%
%%CITATION = HEP-TH/9810089;%%
\bibitem{Masson:2010vx}%
  \BibitemOpen
  \bibfield{author}{%
  \bibinfo {author} {\bibfnamefont{T.}~\bibnamefont{Masson}}\ and\ \bibinfo
  {author} {\bibfnamefont{J.-C.}\ \bibnamefont{Wallet}}}%
   (\bibinfo {year} {2010}),\
  \Eprint{http://arxiv.org/abs/1001.1176}{arXiv:1001.1176 [hep-th]}%
  \bibAnnoteFile{NoStop}{Masson:2010vx}%
%%CITATION = ARXIV:1001.1176;%%
\bibitem{Fournel:2012cr}%
  \BibitemOpen
  \bibfield{author}{%
  \bibinfo {author} {\bibfnamefont{C.}~\bibnamefont{Fournel}}, \bibinfo
  {author} {\bibfnamefont{J.}~\bibnamefont{Francois}}, \bibinfo {author}
  {\bibfnamefont{S.}~\bibnamefont{Lazzarini}},\ and\ \bibinfo {author}
  {\bibfnamefont{T.}~\bibnamefont{Masson}}}%
   (\bibinfo {year} {2012}),\
  \Eprint{http://arxiv.org/abs/1212.6702}{arXiv:1212.6702 [math-ph]}%
  \bibAnnoteFile{NoStop}{Fournel:2012cr}%
%%CITATION = ARXIV:1212.6702;%%
\bibitem{Pervushin:2001kq}%
  \BibitemOpen
  \bibfield{author}{%
  \bibinfo {author} {\bibfnamefont{V.}~\bibnamefont{Pervushin}},\ }%
  \bibfield{journal}{%
  \bibinfo {journal} {Phys.Part.Nucl.}\ }%
  \textbf{\bibinfo {volume} {34}},\ \bibinfo {pages} {348} (\bibinfo {year}
  {2003}),\ \Eprint{http://arxiv.org/abs/hep-th/0109218}{arXiv:hep-th/0109218
  [hep-th]}%
  \bibAnnoteFile{NoStop}{Pervushin:2001kq}%
%%CITATION = HEP-TH/0109218;%%
\bibitem{Lantsman:2006ry}%
  \BibitemOpen
  \bibfield{author}{%
  \bibinfo {author} {\bibfnamefont{L.}~\bibnamefont{Lantsman}},\ }%
  \bibfield{journal}{%
  \bibinfo {journal} {Fizika}\ }%
  \textbf{\bibinfo {volume} {B18}},\ \bibinfo {pages} {99} (\bibinfo {year}
  {2009}),\ \Eprint{http://arxiv.org/abs/hep-th/0604004}{arXiv:hep-th/0604004
  [hep-th]}%
  \bibAnnoteFile{NoStop}{Lantsman:2006ry}%
%%CITATION = HEP-TH/0604004;%%
\bibitem{Taneja:2011sy}%
  \BibitemOpen
  \bibfield{author}{%
  \bibinfo {author} {\bibfnamefont{S.~K.}\ \bibnamefont{Taneja}}, \bibinfo
  {author} {\bibfnamefont{K.}~\bibnamefont{Kathuria}}, \bibinfo {author}
  {\bibfnamefont{S.}~\bibnamefont{Liuti}},\ and\ \bibinfo {author}
  {\bibfnamefont{G.~R.}\ \bibnamefont{Goldstein}},\ }%
  \bibfield{journal}{%
  \Doi{10.1103/PhysRevD.86.036008}{\bibinfo {journal} {Phys.Rev.}}\ }%
  \textbf{\bibinfo {volume} {D86}},\ \bibinfo {pages} {036008} (\bibinfo {year}
  {2012}),\ \Eprint{http://arxiv.org/abs/1101.0581}{arXiv:1101.0581 [hep-ph]}%
  \bibAnnoteFile{NoStop}{Taneja:2011sy}%
%%CITATION = ARXIV:1101.0581;%%
\bibitem{Ji:1997pf}%
  \BibitemOpen
  \bibfield{author}{%
  \bibinfo {author} {\bibfnamefont{X.-D.}\ \bibnamefont{Ji}},\ }%
  \bibfield{journal}{%
  \Doi{10.1103/PhysRevD.58.056003}{\bibinfo {journal} {Phys. Rev.}}\ }%
  \textbf{\bibinfo {volume} {D58}},\ \bibinfo {pages} {056003} (\bibinfo {year}
  {1998}),\ \Eprint{http://arxiv.org/abs/hep-ph/9710290}{arXiv:hep-ph/9710290}%
  \bibAnnoteFile{NoStop}{Ji:1997pf}%
%%CITATION = HEP-PH/9710290;%%
\bibitem{Ji:1996nm}%
  \BibitemOpen
  \bibfield{author}{%
  \bibinfo {author} {\bibfnamefont{X.-D.}\ \bibnamefont{Ji}},\ }%
  \bibfield{journal}{%
  \Doi{10.1103/PhysRevD.55.7114}{\bibinfo {journal} {Phys. Rev.}}\ }%
  \textbf{\bibinfo {volume} {D55}},\ \bibinfo {pages} {7114} (\bibinfo {year}
  {1997}),\ \Eprint{http://arxiv.org/abs/hep-ph/9609381}{arXiv:hep-ph/9609381}%
  \bibAnnoteFile{NoStop}{Ji:1996nm}%
%%CITATION = HEP-PH/9609381;%%
\bibitem{Leader:2011cr}%
  \BibitemOpen
  \bibfield{author}{%
  \bibinfo {author} {\bibfnamefont{E.}~\bibnamefont{Leader}},\ }%
  \bibfield{journal}{%
  \bibinfo {journal} {Phys.Rev.}\ }%
  \textbf{\bibinfo {volume} {D85}},\ \bibinfo {pages} {051501} (\bibinfo {year}
  {2012}),\ \Eprint{http://arxiv.org/abs/1109.1230}{arXiv:1109.1230 [hep-ph]}%
  \bibAnnoteFile{NoStop}{Leader:2011cr}%
%%CITATION = ARXIV:1109.1230;%%
\bibitem{Ji:2012vj}%
  \BibitemOpen
  \bibfield{author}{%
  \bibinfo {author} {\bibfnamefont{X.}~\bibnamefont{Ji}}, \bibinfo {author}
  {\bibfnamefont{X.}~\bibnamefont{Xiong}},\ and\ \bibinfo {author}
  {\bibfnamefont{F.}~\bibnamefont{Yuan}},\ }%
  \bibfield{journal}{%
  \Doi{10.1016/j.physletb.2012.09.027}{\bibinfo {journal} {Phys.Lett.}}\ }%
  \textbf{\bibinfo {volume} {B717}},\ \bibinfo {pages} {214} (\bibinfo {year}
  {2012}),\ \Eprint{http://arxiv.org/abs/1209.3246}{arXiv:1209.3246 [hep-ph]}%
  \bibAnnoteFile{NoStop}{Ji:2012vj}%
%%CITATION = ARXIV:1209.3246;%%
\bibitem{Harindranath:1998ve}%
  \BibitemOpen
  \bibfield{author}{%
  \bibinfo {author} {\bibfnamefont{A.}~\bibnamefont{Harindranath}}\ and\
  \bibinfo {author} {\bibfnamefont{R.}~\bibnamefont{Kundu}},\ }%
  \bibfield{journal}{%
  \Doi{10.1103/PhysRevD.59.116013}{\bibinfo {journal} {Phys.Rev.}}\ }%
  \textbf{\bibinfo {volume} {D59}},\ \bibinfo {pages} {116013} (\bibinfo {year}
  {1999}),\ \Eprint{http://arxiv.org/abs/hep-ph/9802406}{arXiv:hep-ph/9802406
  [hep-ph]}%
  \bibAnnoteFile{NoStop}{Harindranath:1998ve}%
%%CITATION = HEP-PH/9802406;%%
\bibitem{Harindranath:2001rc}%
  \BibitemOpen
  \bibfield{author}{%
  \bibinfo {author} {\bibfnamefont{A.}~\bibnamefont{Harindranath}}, \bibinfo
  {author} {\bibfnamefont{A.}~\bibnamefont{Mukherjee}},\ and\ \bibinfo {author}
  {\bibfnamefont{R.}~\bibnamefont{Ratabole}},\ }%
  \bibfield{journal}{%
  \Doi{10.1103/PhysRevD.63.045006}{\bibinfo {journal} {Phys.Rev.}}\ }%
  \textbf{\bibinfo {volume} {D63}},\ \bibinfo {pages} {045006} (\bibinfo {year}
  {2001})%
  \bibAnnoteFile{NoStop}{Harindranath:2001rc}%
%%CITATION = PHRVA,D63,045006;%%
\bibitem{Harindranath:1999ve}%
  \BibitemOpen
  \bibfield{author}{%
  \bibinfo {author} {\bibfnamefont{A.}~\bibnamefont{Harindranath}}, \bibinfo
  {author} {\bibfnamefont{A.}~\bibnamefont{Mukherjee}},\ and\ \bibinfo {author}
  {\bibfnamefont{R.}~\bibnamefont{Ratabole}},\ }%
  \bibfield{journal}{%
  \Doi{10.1016/S0370-2693(00)00148-9}{\bibinfo {journal} {Phys.Lett.}}\ }%
  \textbf{\bibinfo {volume} {B476}},\ \bibinfo {pages} {471} (\bibinfo {year}
  {2000}),\ \Eprint{http://arxiv.org/abs/hep-ph/9908424}{arXiv:hep-ph/9908424
  [hep-ph]}%
  \bibAnnoteFile{NoStop}{Harindranath:1999ve}%
%%CITATION = HEP-PH/9908424;%%
\bibitem{Harindranath:2013goa}%
  \BibitemOpen
  \bibfield{author}{%
  \bibinfo {author} {\bibfnamefont{A.}~\bibnamefont{Harindranath}}, \bibinfo
  {author} {\bibfnamefont{R.}~\bibnamefont{Kundu}},\ and\ \bibinfo {author}
  {\bibfnamefont{A.}~\bibnamefont{Mukherjee}}}%
   (\bibinfo {year} {2013}),\
  \Eprint{http://arxiv.org/abs/1308.1519}{arXiv:1308.1519 [hep-ph]}%
  \bibAnnoteFile{NoStop}{Harindranath:2013goa}%
%%CITATION = ARXIV:1308.1519;%%
\bibitem{Polyzou:2012ut}%
  \BibitemOpen
  \bibfield{author}{%
  \bibinfo {author} {\bibfnamefont{W.}~\bibnamefont{Polyzou}}, \bibinfo
  {author} {\bibfnamefont{W.}~\bibnamefont{Glockle}},\ and\ \bibinfo {author}
  {\bibfnamefont{H.}~\bibnamefont{Witala}}}%
   (\bibinfo {year} {2012}),\
  \Eprint{http://arxiv.org/abs/1208.5840}{arXiv:1208.5840 [nucl-th]}%
  \bibAnnoteFile{NoStop}{Polyzou:2012ut}%
%%CITATION = ARXIV:1208.5840;%%
\bibitem{Diehl:2003ny}%
  \BibitemOpen
  \bibfield{author}{%
  \bibinfo {author} {\bibfnamefont{M.}~\bibnamefont{Diehl}},\ }%
  \bibfield{journal}{%
  \Doi{10.1016/j.physrep.2003.08.002}{\bibinfo {journal} {Phys.Rept.}}\ }%
  \textbf{\bibinfo {volume} {388}},\ \bibinfo {pages} {41} (\bibinfo {year}
  {2003}),\ \bibinfo {note} {habilitation thesis},\
  \Eprint{http://arxiv.org/abs/hep-ph/0307382}{arXiv:hep-ph/0307382 [hep-ph]}%
  \bibAnnoteFile{NoStop}{Diehl:2003ny}%
\bibitem{Filippone:2001ux}%
  \BibitemOpen
  \bibfield{author}{%
  \bibinfo {author} {\bibfnamefont{B.}~\bibnamefont{Filippone}}\ and\ \bibinfo
  {author} {\bibfnamefont{X.-D.}\ \bibnamefont{Ji}},\ }%
  \bibfield{journal}{%
  \Doi{10.1007/0-306-47915-X_1}{\bibinfo {journal} {Adv.Nucl.Phys.}}\ }%
  \textbf{\bibinfo {volume} {26}},\ \bibinfo {pages} {1} (\bibinfo {year}
  {2001}),\ \Eprint{http://arxiv.org/abs/hep-ph/0101224}{arXiv:hep-ph/0101224
  [hep-ph]}%
  \bibAnnoteFile{NoStop}{Filippone:2001ux}%
%%CITATION = HEP-PH/0101224;%%
\bibitem{Brodsky:2000ii}%
  \BibitemOpen
  \bibfield{author}{%
  \bibinfo {author} {\bibfnamefont{S.~J.}\ \bibnamefont{Brodsky}}, \bibinfo
  {author} {\bibfnamefont{D.~S.}\ \bibnamefont{Hwang}}, \bibinfo {author}
  {\bibfnamefont{B.-Q.}\ \bibnamefont{Ma}},\ and\ \bibinfo {author}
  {\bibfnamefont{I.}~\bibnamefont{Schmidt}},\ }%
  \bibfield{journal}{%
  \Doi{10.1016/S0550-3213(00)00626-X}{\bibinfo {journal} {Nucl.Phys.}}\ }%
  \textbf{\bibinfo {volume} {B593}},\ \bibinfo {pages} {311} (\bibinfo {year}
  {2001}),\ \Eprint{http://arxiv.org/abs/hep-th/0003082}{arXiv:hep-th/0003082
  [hep-th]}%
  \bibAnnoteFile{NoStop}{Brodsky:2000ii}%
\bibitem{Teryaev:1999su}%
  \BibitemOpen
  \bibfield{author}{%
  \bibinfo {author} {\bibfnamefont{O.}~\bibnamefont{Teryaev}}}%
   (\bibinfo {year} {1999}),\
  \Eprint{http://arxiv.org/abs/hep-ph/9904376}{arXiv:hep-ph/9904376 [hep-ph]}%
  \bibAnnoteFile{NoStop}{Teryaev:1999su}%
\bibitem{Bass:1992ti}%
  \BibitemOpen
  \bibfield{author}{%
  \bibinfo {author} {\bibfnamefont{S.}~\bibnamefont{Bass}},\ }%
  \bibfield{journal}{%
  \Doi{10.1007/BF01561304}{\bibinfo {journal} {Z.Phys.}}\ }%
  \textbf{\bibinfo {volume} {C55}},\ \bibinfo {pages} {653} (\bibinfo {year}
  {1992})%
  \bibAnnoteFile{NoStop}{Bass:1992ti}%
%%CITATION = ZEPYA,C55,653;%%
\bibitem{Leader:1998nh}%
  \BibitemOpen
  \bibfield{author}{%
  \bibinfo {author} {\bibfnamefont{E.}~\bibnamefont{Leader}}, \bibinfo {author}
  {\bibfnamefont{A.~V.}\ \bibnamefont{Sidorov}},\ and\ \bibinfo {author}
  {\bibfnamefont{D.~B.}\ \bibnamefont{Stamenov}},\ }%
  \bibfield{journal}{%
  \Doi{10.1016/S0370-2693(98)01478-6}{\bibinfo {journal} {Phys.Lett.}}\ }%
  \textbf{\bibinfo {volume} {B445}},\ \bibinfo {pages} {232} (\bibinfo {year}
  {1998}),\ \Eprint{http://arxiv.org/abs/hep-ph/9808248}{arXiv:hep-ph/9808248
  [hep-ph]}%
  \bibAnnoteFile{NoStop}{Leader:1998nh}%
%%CITATION = HEP-PH/9808248;%%
\bibitem{Manohar:1990kr}%
  \BibitemOpen
  \bibfield{author}{%
  \bibinfo {author} {\bibfnamefont{A.~V.}\ \bibnamefont{Manohar}},\ }%
  \bibfield{journal}{%
  \Doi{10.1103/PhysRevLett.65.2511}{\bibinfo {journal} {Phys. Rev. Lett.}}\ }%
  \textbf{\bibinfo {volume} {65}},\ \bibinfo {pages} {2511} (\bibinfo {year}
  {1990})%
  \bibAnnoteFile{NoStop}{Manohar:1990kr}%
%%CITATION = PRLTA,65,2511;%%
\bibitem{Jaffe:1995an}%
  \BibitemOpen
  \bibfield{author}{%
  \bibinfo {author} {\bibfnamefont{R.}~\bibnamefont{Jaffe}},\ }%
  \bibfield{journal}{%
  \Doi{10.1016/0370-2693(95)01247-8}{\bibinfo {journal} {Phys.Lett.}}\ }%
  \textbf{\bibinfo {volume} {B365}},\ \bibinfo {pages} {359} (\bibinfo {year}
  {1996}),\ \Eprint{http://arxiv.org/abs/hep-ph/9509279}{arXiv:hep-ph/9509279
  [hep-ph]}%
  \bibAnnoteFile{NoStop}{Jaffe:1995an}%
\bibitem{Ji:2013dva}%
  \BibitemOpen
  \bibfield{author}{%
  \bibinfo {author} {\bibfnamefont{X.}~\bibnamefont{Ji}}}%
   (\bibinfo {year} {2013}),\
  \Eprint{http://arxiv.org/abs/1305.1539}{arXiv:1305.1539 [hep-ph]}%
  \bibAnnoteFile{NoStop}{Ji:2013dva}%
%%CITATION = ARXIV:1305.1539;%%
\bibitem{Jacob:1959at}%
  \BibitemOpen
  \bibfield{author}{%
  \bibinfo {author} {\bibfnamefont{M.}~\bibnamefont{Jacob}}\ and\ \bibinfo
  {author} {\bibfnamefont{G.}~\bibnamefont{Wick}},\ }%
  \bibfield{journal}{%
  \Doi{10.1016/0003-4916(59)90051-X}{\bibinfo {journal} {Annals Phys.}}\ }%
  \textbf{\bibinfo {volume} {7}},\ \bibinfo {pages} {404} (\bibinfo {year}
  {1959})%
  \bibAnnoteFile{NoStop}{Jacob:1959at}%
%%CITATION = APNYA,7,404;%%
\bibitem{Alba:2006hs}%
  \BibitemOpen
  \bibfield{author}{%
  \bibinfo {author} {\bibfnamefont{D.}~\bibnamefont{Alba}}, \bibinfo {author}
  {\bibfnamefont{H.~W.}\ \bibnamefont{Crater}},\ and\ \bibinfo {author}
  {\bibfnamefont{L.}~\bibnamefont{Lusanna}},\ }%
  \bibfield{journal}{%
  \Doi{10.1088/1751-8113/40/31/029}{\bibinfo {journal} {J.Phys.}}\ }%
  \textbf{\bibinfo {volume} {A40}},\ \bibinfo {pages} {9585} (\bibinfo {year}
  {2007}),\ \Eprint{http://arxiv.org/abs/hep-th/0610200}{arXiv:hep-th/0610200
  [hep-th]}%
  \bibAnnoteFile{NoStop}{Alba:2006hs}%
%%CITATION = HEP-TH/0610200;%%
\bibitem{Aguilar:2013}%
  \BibitemOpen
  \bibinfo {author} {\bibfnamefont{P.}~\bibnamefont{Aguiar}}, \bibinfo {author}
  {\bibfnamefont{C.}~\bibnamefont{Chryssomalakos}}, \bibinfo {author}
  {\bibfnamefont{H.}~\bibnamefont{Hernandez~Coronado}},\ and\ \bibinfo {author}
  {\bibfnamefont{E.}~\bibnamefont{Okon}}%
  \bibAnnoteFile{NoStop}{Aguilar:2013}%
\bibitem{Lorce:2011kn}%
  \BibitemOpen
\bibfield{author}{%
    }%
  \bibfield{author}{%
  \bibinfo {author} {\bibfnamefont{C.}~\bibnamefont{Lorc\'e}}\ and\ \bibinfo
  {author} {\bibfnamefont{B.}~\bibnamefont{Pasquini}},\ }%
  \bibfield{journal}{%
  \Doi{10.1016/j.physletb.2012.03.025}{\bibinfo {journal} {Phys.Lett.}}\ }%
  \textbf{\bibinfo {volume} {B710}},\ \bibinfo {pages} {486} (\bibinfo {year}
  {2012}),\ \Eprint{http://arxiv.org/abs/1111.6069}{arXiv:1111.6069 [hep-ph]}%
  \bibAnnoteFile{NoStop}{Lorce:2011kn}%
%%CITATION = ARXIV:1111.6069;%%
\bibitem{Burkardt:2000za}%
  \BibitemOpen
  \bibfield{author}{%
  \bibinfo {author} {\bibfnamefont{M.}~\bibnamefont{Burkardt}},\ }%
  \bibfield{journal}{%
  \Doi{10.1103/PhysRevD.62.071503, 10.1103/PhysRevD.66.119903}{\bibinfo
  {journal} {Phys.Rev.}}\ }%
  \textbf{\bibinfo {volume} {D62}},\ \bibinfo {pages} {071503} (\bibinfo {year}
  {2000}),\ \Eprint{http://arxiv.org/abs/hep-ph/0005108}{arXiv:hep-ph/0005108
  [hep-ph]}%
  \bibAnnoteFile{NoStop}{Burkardt:2000za}%
%%CITATION = HEP-PH/0005108;%%
\bibitem{Burkardt:2002hr}%
  \BibitemOpen
  \bibfield{author}{%
  \bibinfo {author} {\bibfnamefont{M.}~\bibnamefont{Burkardt}},\ }%
  \bibfield{journal}{%
  \Doi{10.1142/S0217751X03012370}{\bibinfo {journal} {Int.J.Mod.Phys.}}\ }%
  \textbf{\bibinfo {volume} {A18}},\ \bibinfo {pages} {173} (\bibinfo {year}
  {2003}),\ \Eprint{http://arxiv.org/abs/hep-ph/0207047}{arXiv:hep-ph/0207047
  [hep-ph]}%
  \bibAnnoteFile{NoStop}{Burkardt:2002hr}%
%%CITATION = HEP-PH/0207047;%%
\bibitem{Burkardt:2005hp}%
  \BibitemOpen
  \bibfield{author}{%
  \bibinfo {author} {\bibfnamefont{M.}~\bibnamefont{Burkardt}},\ }%
  \bibfield{journal}{%
  \Doi{10.1103/PhysRevD.72.094020}{\bibinfo {journal} {Phys.Rev.}}\ }%
  \textbf{\bibinfo {volume} {D72}},\ \bibinfo {pages} {094020} (\bibinfo {year}
  {2005}),\ \Eprint{http://arxiv.org/abs/hep-ph/0505189}{arXiv:hep-ph/0505189
  [hep-ph]}%
  \bibAnnoteFile{NoStop}{Burkardt:2005hp}%
%%CITATION = HEP-PH/0505189;%%
\bibitem{Diehl:2005jf}%
  \BibitemOpen
  \bibfield{author}{%
  \bibinfo {author} {\bibfnamefont{M.}~\bibnamefont{Diehl}}\ and\ \bibinfo
  {author} {\bibfnamefont{P.}~\bibnamefont{Hagler}},\ }%
  \bibfield{journal}{%
  \Doi{10.1140/epjc/s2005-02342-6}{\bibinfo {journal} {Eur.Phys.J.}}\ }%
  \textbf{\bibinfo {volume} {C44}},\ \bibinfo {pages} {87} (\bibinfo {year}
  {2005}),\ \Eprint{http://arxiv.org/abs/hep-ph/0504175}{arXiv:hep-ph/0504175
  [hep-ph]}%
  \bibAnnoteFile{NoStop}{Diehl:2005jf}%
%%CITATION = HEP-PH/0504175;%%
\bibitem{Leader:2012ar}%
  \BibitemOpen
  \bibfield{author}{%
  \bibinfo {author} {\bibfnamefont{E.}~\bibnamefont{Leader}},\ }%
  \bibfield{journal}{%
  \Doi{10.1016/j.physletb.2013.01.050}{\bibinfo {journal} {Phys.Lett.}}\ }%
  \textbf{\bibinfo {volume} {B720}},\ \bibinfo {pages} {120} (\bibinfo {year}
  {2013}),\ \Eprint{http://arxiv.org/abs/1211.3957}{arXiv:1211.3957 [hep-ph]}%
  \bibAnnoteFile{NoStop}{Leader:2012ar}%
%%CITATION = ARXIV:1211.3957;%%
\bibitem{Meissner:2009ww}%
  \BibitemOpen
  \bibfield{author}{%
  \bibinfo {author} {\bibfnamefont{S.}~\bibnamefont{Meissner}}, \bibinfo
  {author} {\bibfnamefont{A.}~\bibnamefont{Metz}},\ and\ \bibinfo {author}
  {\bibfnamefont{M.}~\bibnamefont{Schlegel}},\ }%
  \bibfield{journal}{%
  \Doi{10.1088/1126-6708/2009/08/056}{\bibinfo {journal} {JHEP}}\ }%
  \textbf{\bibinfo {volume} {0908}},\ \bibinfo {pages} {056} (\bibinfo {year}
  {2009}),\ \Eprint{http://arxiv.org/abs/0906.5323}{arXiv:0906.5323 [hep-ph]}%
  \bibAnnoteFile{NoStop}{Meissner:2009ww}%
%%CITATION = ARXIV:0906.5323;%%
\bibitem{Ji:2013tva}%
  \BibitemOpen
  \bibfield{author}{%
  \bibinfo {author} {\bibfnamefont{X.}~\bibnamefont{Ji}}, \bibinfo {author}
  {\bibfnamefont{X.}~\bibnamefont{Xiong}},\ and\ \bibinfo {author}
  {\bibfnamefont{F.}~\bibnamefont{Yuan}}}%
   (\bibinfo {year} {2013}),\
  \Eprint{http://arxiv.org/abs/1304.1009}{arXiv:1304.1009 [hep-ph]}%
  \bibAnnoteFile{NoStop}{Ji:2013tva}%
%%CITATION = ARXIV:1304.1009;%%
\bibitem{PhysRevLett.111.039101}%
  \BibitemOpen
  \bibfield{author}{%
  \bibinfo {author} {\bibfnamefont{E.}~\bibnamefont{Leader}}\ and\ \bibinfo
  {author} {\bibfnamefont{C.}~\bibnamefont{Lorc\'e}},\ }%
  \bibfield{journal}{%
  \bibinfo {journal} {Phys. Rev. Lett.}\ }%
  \textbf{\bibinfo {volume} {111}},\ \bibinfo {pages} {039101} (\bibinfo
  {month} {Jul}\ \bibinfo {year} {2013})%
  \bibAnnoteFile{NoStop}{PhysRevLett.111.039101}%
\bibitem{Ashman:1987hv}%
  \BibitemOpen
  \bibfield{author}{%
  \bibinfo {author} {\bibfnamefont{J.}~\bibnamefont{Ashman}} \emph{et~al.}
  (\bibinfo {collaboration} {European Muon Collaboration}),\ }%
  \bibfield{journal}{%
  \Doi{10.1016/0370-2693(88)91523-7}{\bibinfo {journal} {Phys.Lett.}}\ }%
  \textbf{\bibinfo {volume} {B206}},\ \bibinfo {pages} {364} (\bibinfo {year}
  {1988})%
  \bibAnnoteFile{NoStop}{Ashman:1987hv}%
%%CITATION = PHLTA,B206,364;%%
\bibitem{Leader:1988vd}%
  \BibitemOpen
  \bibfield{author}{%
  \bibinfo {author} {\bibfnamefont{E.}~\bibnamefont{Leader}}\ and\ \bibinfo
  {author} {\bibfnamefont{M.}~\bibnamefont{Anselmino}},\ }%
  \bibfield{journal}{%
  \Doi{10.1007/BF01566922}{\bibinfo {journal} {Z.Phys.}}\ }%
  \textbf{\bibinfo {volume} {C41}},\ \bibinfo {pages} {239} (\bibinfo {year}
  {1988})%
  \bibAnnoteFile{NoStop}{Leader:1988vd}%
%%CITATION = ZEPYA,C41,239;%%
\bibitem{Ageev:2005gh}%
  \BibitemOpen
  \bibfield{author}{%
  \bibinfo {author} {\bibfnamefont{E.}~\bibnamefont{Ageev}} \emph{et~al.}
  (\bibinfo {collaboration} {COMPASS Collaboration}),\ }%
  \bibfield{journal}{%
  \Doi{10.1016/j.physletb.2005.03.025}{\bibinfo {journal} {Phys.Lett.}}\ }%
  \textbf{\bibinfo {volume} {B612}},\ \bibinfo {pages} {154} (\bibinfo {year}
  {2005}),\ \Eprint{http://arxiv.org/abs/hep-ex/0501073}{arXiv:hep-ex/0501073
  [hep-ex]}%
  \bibAnnoteFile{NoStop}{Ageev:2005gh}%
%%CITATION = HEP-EX/0501073;%%
\bibitem{Alexakhin:2006vx}%
  \BibitemOpen
  \bibfield{author}{%
  \bibinfo {author} {\bibfnamefont{V.~Y.}\ \bibnamefont{Alexakhin}}
  \emph{et~al.} (\bibinfo {collaboration} {COMPASS Collaboration}),\ }%
  \bibfield{journal}{%
  \Doi{10.1016/j.physletb.2006.12.076}{\bibinfo {journal} {Phys.Lett.}}\ }%
  \textbf{\bibinfo {volume} {B647}},\ \bibinfo {pages} {8} (\bibinfo {year}
  {2007}),\ \Eprint{http://arxiv.org/abs/hep-ex/0609038}{arXiv:hep-ex/0609038
  [hep-ex]}%
  \bibAnnoteFile{NoStop}{Alexakhin:2006vx}%
%%CITATION = HEP-EX/0609038;%%
\bibitem{Airapetian:2007mh}%
  \BibitemOpen
  \bibfield{author}{%
  \bibinfo {author} {\bibfnamefont{A.}~\bibnamefont{Airapetian}} \emph{et~al.}
  (\bibinfo {collaboration} {HERMES Collaboration}),\ }%
  \bibfield{journal}{%
  \Doi{10.1103/PhysRevD.75.012007}{\bibinfo {journal} {Phys.Rev.}}\ }%
  \textbf{\bibinfo {volume} {D75}},\ \bibinfo {pages} {012007} (\bibinfo {year}
  {2007}),\ \Eprint{http://arxiv.org/abs/hep-ex/0609039}{arXiv:hep-ex/0609039
  [hep-ex]}%
  \bibAnnoteFile{NoStop}{Airapetian:2007mh}%
%%CITATION = HEP-EX/0609039;%%
\bibitem{Kuhn:2008sy}%
  \BibitemOpen
  \bibfield{author}{%
  \bibinfo {author} {\bibfnamefont{S.~E.}\ \bibnamefont{Kuhn}}, \bibinfo
  {author} {\bibfnamefont{J.~P.}\ \bibnamefont{Chen}},\ and\ \bibinfo {author}
  {\bibfnamefont{E.}~\bibnamefont{Leader}},\ }%
  \bibfield{journal}{%
  \Doi{10.1016/j.ppnp.2009.02.001}{\bibinfo {journal} {Prog. Part. Nucl.
  Phys.}}\ }%
  \textbf{\bibinfo {volume} {63}},\ \bibinfo {pages} {1} (\bibinfo {year}
  {2009}),\ \Eprint{http://arxiv.org/abs/0812.3535}{arXiv:0812.3535 [hep-ph]}%
  \bibAnnoteFile{NoStop}{Kuhn:2008sy}%
%%CITATION = 0812.3535;%%
\bibitem{Leader:2010rb}%
  \BibitemOpen
  \bibfield{author}{%
  \bibinfo {author} {\bibfnamefont{E.}~\bibnamefont{Leader}}, \bibinfo {author}
  {\bibfnamefont{A.~V.}\ \bibnamefont{Sidorov}},\ and\ \bibinfo {author}
  {\bibfnamefont{D.~B.}\ \bibnamefont{Stamenov}},\ }%
  \bibfield{journal}{%
  \Doi{10.1103/PhysRevD.82.114018}{\bibinfo {journal} {Phys.Rev.}}\ }%
  \textbf{\bibinfo {volume} {D82}},\ \bibinfo {pages} {114018} (\bibinfo {year}
  {2010}),\ \Eprint{http://arxiv.org/abs/1010.0574}{arXiv:1010.0574 [hep-ph]}%
  \bibAnnoteFile{NoStop}{Leader:2010rb}%
%%CITATION = ARXIV:1010.0574;%%
\bibitem{Lorce:2011kd}%
  \BibitemOpen
  \bibfield{author}{%
  \bibinfo {author} {\bibfnamefont{C.}~\bibnamefont{Lorc\'e}}\ and\ \bibinfo
  {author} {\bibfnamefont{B.}~\bibnamefont{Pasquini}},\ }%
  \bibfield{journal}{%
  \Doi{10.1103/PhysRevD.84.014015}{\bibinfo {journal} {Phys.Rev.}}\ }%
  \textbf{\bibinfo {volume} {D84}},\ \bibinfo {pages} {014015} (\bibinfo {year}
  {2011}),\ \Eprint{http://arxiv.org/abs/1106.0139}{arXiv:1106.0139 [hep-ph]}%
  \bibAnnoteFile{NoStop}{Lorce:2011kd}%
%%CITATION = ARXIV:1106.0139;%%
\bibitem{Thomas:2008ga}%
  \BibitemOpen
  \bibfield{author}{%
  \bibinfo {author} {\bibfnamefont{A.~W.}\ \bibnamefont{Thomas}},\ }%
  \bibfield{journal}{%
  \Doi{10.1103/PhysRevLett.101.102003}{\bibinfo {journal} {Phys.Rev.Lett.}}\ }%
  \textbf{\bibinfo {volume} {101}},\ \bibinfo {pages} {102003} (\bibinfo {year}
  {2008}),\ \Eprint{http://arxiv.org/abs/0803.2775}{arXiv:0803.2775 [hep-ph]}%
  \bibAnnoteFile{NoStop}{Thomas:2008ga}%
%%CITATION = ARXIV:0803.2775;%%
\bibitem{Wakamatsu:2009gx}%
  \BibitemOpen
  \bibfield{author}{%
  \bibinfo {author} {\bibfnamefont{M.}~\bibnamefont{Wakamatsu}},\ }%
  \bibfield{journal}{%
  \Doi{10.1140/epja/i2010-10954-6}{\bibinfo {journal} {Eur.Phys.J.}}\ }%
  \textbf{\bibinfo {volume} {A44}},\ \bibinfo {pages} {297} (\bibinfo {year}
  {2010}),\ \Eprint{http://arxiv.org/abs/0908.0972}{arXiv:0908.0972 [hep-ph]}%
  \bibAnnoteFile{NoStop}{Wakamatsu:2009gx}%
%%CITATION = ARXIV:0908.0972;%%
\bibitem{Thomas:2010zz}%
  \BibitemOpen
  \bibfield{author}{%
  \bibinfo {author} {\bibfnamefont{A.}~\bibnamefont{Thomas}}, \bibinfo {author}
  {\bibfnamefont{H.}~\bibnamefont{Matevosyan}},\ and\ \bibinfo {author}
  {\bibfnamefont{A.}~\bibnamefont{Casey}},\ }%
  \bibfield{journal}{%
  \Doi{10.1140/epja/i2010-11035-8}{\bibinfo {journal} {Eur.Phys.J.}}\ }%
  \textbf{\bibinfo {volume} {A46}},\ \bibinfo {pages} {325} (\bibinfo {year}
  {2010})%
  \bibAnnoteFile{NoStop}{Thomas:2010zz}%
%%CITATION = EPHJA,A46,325;%%
\bibitem{Wakamatsu:2010zza}%
  \BibitemOpen
  \bibfield{author}{%
  \bibinfo {author} {\bibfnamefont{M.}~\bibnamefont{Wakamatsu}},\ }%
  \bibfield{journal}{%
  \Doi{10.1140/epja/i2010-11036-7}{\bibinfo {journal} {Eur.Phys.J.}}\ }%
  \textbf{\bibinfo {volume} {A46}},\ \bibinfo {pages} {327} (\bibinfo {year}
  {2010})%
  \bibAnnoteFile{NoStop}{Wakamatsu:2010zza}%
%%CITATION = EPHJA,A46,327;%%
\bibitem{Liu:2012nz}%
  \BibitemOpen
  \bibfield{author}{%
  \bibinfo {author} {\bibfnamefont{K.}~\bibnamefont{Liu}}, \bibinfo {author}
  {\bibfnamefont{M.}~\bibnamefont{Deka}}, \bibinfo {author}
  {\bibfnamefont{T.}~\bibnamefont{Doi}}, \bibinfo {author}
  {\bibfnamefont{Y.}~\bibnamefont{Yang}}, \bibinfo {author}
  {\bibfnamefont{B.}~\bibnamefont{Chakraborty}}, \emph{et~al.},\ }%
  \bibfield{journal}{%
  \bibinfo {journal} {PoS}\ }%
  \textbf{\bibinfo {volume} {LATTICE2011}},\ \bibinfo {pages} {164} (\bibinfo
  {year} {2011}),\ \Eprint{http://arxiv.org/abs/1203.6388}{arXiv:1203.6388
  [hep-ph]}%
  \bibAnnoteFile{NoStop}{Liu:2012nz}%
%%CITATION = ARXIV:1203.6388;%%
\bibitem{Kiptily:2002nx}%
  \BibitemOpen
  \bibfield{author}{%
  \bibinfo {author} {\bibfnamefont{D.}~\bibnamefont{Kiptily}}\ and\ \bibinfo
  {author} {\bibfnamefont{M.}~\bibnamefont{Polyakov}},\ }%
  \bibfield{journal}{%
  \Doi{10.1140/epjc/s2004-01957-3}{\bibinfo {journal} {Eur.Phys.J.}}\ }%
  \textbf{\bibinfo {volume} {C37}},\ \bibinfo {pages} {105} (\bibinfo {year}
  {2004}),\ \Eprint{http://arxiv.org/abs/hep-ph/0212372}{arXiv:hep-ph/0212372
  [hep-ph]}%
  \bibAnnoteFile{NoStop}{Kiptily:2002nx}%
%%CITATION = HEP-PH/0212372;%%
\bibitem{Penttinen:2000dg}%
  \BibitemOpen
  \bibfield{author}{%
  \bibinfo {author} {\bibfnamefont{M.}~\bibnamefont{Penttinen}}, \bibinfo
  {author} {\bibfnamefont{M.~V.}\ \bibnamefont{Polyakov}}, \bibinfo {author}
  {\bibfnamefont{A.}~\bibnamefont{Shuvaev}},\ and\ \bibinfo {author}
  {\bibfnamefont{M.}~\bibnamefont{Strikman}},\ }%
  \bibfield{journal}{%
  \Doi{10.1016/S0370-2693(00)01035-2}{\bibinfo {journal} {Phys.Lett.}}\ }%
  \textbf{\bibinfo {volume} {B491}},\ \bibinfo {pages} {96} (\bibinfo {year}
  {2000}),\ \Eprint{http://arxiv.org/abs/hep-ph/0006321}{arXiv:hep-ph/0006321
  [hep-ph]}%
  \bibAnnoteFile{NoStop}{Penttinen:2000dg}%
%%CITATION = HEP-PH/0006321;%%
\bibitem{Ji:2012ba}%
  \BibitemOpen
  \bibfield{author}{%
  \bibinfo {author} {\bibfnamefont{X.}~\bibnamefont{Ji}}, \bibinfo {author}
  {\bibfnamefont{X.}~\bibnamefont{Xiong}},\ and\ \bibinfo {author}
  {\bibfnamefont{F.}~\bibnamefont{Yuan}},\ }%
  \bibfield{journal}{%
  \Doi{10.1103/PhysRevD.88.014041}{\bibinfo {journal} {Phys.Rev.}}\ }%
  \textbf{\bibinfo {volume} {D88}},\ \bibinfo {pages} {014041} (\bibinfo {year}
  {2013}),\ \Eprint{http://arxiv.org/abs/1207.5221}{arXiv:1207.5221 [hep-ph]}%
  \bibAnnoteFile{NoStop}{Ji:2012ba}%
%%CITATION = ARXIV:1207.5221;%%
\bibitem{Ji:2003ak}%
  \BibitemOpen
  \bibfield{author}{%
  \bibinfo {author} {\bibfnamefont{X.-d.}\ \bibnamefont{Ji}},\ }%
  \bibfield{journal}{%
  \Doi{10.1103/PhysRevLett.91.062001}{\bibinfo {journal} {Phys.Rev.Lett.}}\ }%
  \textbf{\bibinfo {volume} {91}},\ \bibinfo {pages} {062001} (\bibinfo {year}
  {2003}),\ \Eprint{http://arxiv.org/abs/hep-ph/0304037}{arXiv:hep-ph/0304037
  [hep-ph]}%
  \bibAnnoteFile{NoStop}{Ji:2003ak}%
%%CITATION = HEP-PH/0304037;%%
\bibitem{Belitsky:2003nz}%
  \BibitemOpen
  \bibfield{author}{%
  \bibinfo {author} {\bibfnamefont{A.~V.}\ \bibnamefont{Belitsky}}, \bibinfo
  {author} {\bibfnamefont{X.-d.}\ \bibnamefont{Ji}},\ and\ \bibinfo {author}
  {\bibfnamefont{F.}~\bibnamefont{Yuan}},\ }%
  \bibfield{journal}{%
  \Doi{10.1103/PhysRevD.69.074014}{\bibinfo {journal} {Phys.Rev.}}\ }%
  \textbf{\bibinfo {volume} {D69}},\ \bibinfo {pages} {074014} (\bibinfo {year}
  {2004}),\ \Eprint{http://arxiv.org/abs/hep-ph/0307383}{arXiv:hep-ph/0307383
  [hep-ph]}%
  \bibAnnoteFile{NoStop}{Belitsky:2003nz}%
%%CITATION = HEP-PH/0307383;%%
\bibitem{Lorce:2011ni}%
  \BibitemOpen
  \bibfield{author}{%
  \bibinfo {author} {\bibfnamefont{C.}~\bibnamefont{Lorc\'e}}, \bibinfo
  {author} {\bibfnamefont{B.}~\bibnamefont{Pasquini}}, \bibinfo {author}
  {\bibfnamefont{X.}~\bibnamefont{Xiong}},\ and\ \bibinfo {author}
  {\bibfnamefont{F.}~\bibnamefont{Yuan}},\ }%
  \bibfield{journal}{%
  \Doi{10.1103/PhysRevD.85.114006}{\bibinfo {journal} {Phys.Rev.}}\ }%
  \textbf{\bibinfo {volume} {D85}},\ \bibinfo {pages} {114006} (\bibinfo {year}
  {2012}),\ \Eprint{http://arxiv.org/abs/1111.4827}{arXiv:1111.4827 [hep-ph]}%
  \bibAnnoteFile{NoStop}{Lorce:2011ni}%
%%CITATION = ARXIV:1111.4827;%%
\bibitem{Ji:2012sj}%
  \BibitemOpen
  \bibfield{author}{%
  \bibinfo {author} {\bibfnamefont{X.}~\bibnamefont{Ji}}, \bibinfo {author}
  {\bibfnamefont{X.}~\bibnamefont{Xiong}},\ and\ \bibinfo {author}
  {\bibfnamefont{F.}~\bibnamefont{Yuan}},\ }%
  \bibfield{journal}{%
  \Doi{10.1103/PhysRevLett.109.152005}{\bibinfo {journal} {Phys.Rev.Lett.}}\ }%
  \textbf{\bibinfo {volume} {109}},\ \bibinfo {pages} {152005} (\bibinfo {year}
  {2012}),\ \Eprint{http://arxiv.org/abs/1202.2843}{arXiv:1202.2843 [hep-ph]}%
  \bibAnnoteFile{NoStop}{Ji:2012sj}%
%%CITATION = ARXIV:1202.2843;%%
\bibitem{Buffing:2011mj}%
  \BibitemOpen
  \bibfield{author}{%
  \bibinfo {author} {\bibfnamefont{M.}~\bibnamefont{Buffing}}\ and\ \bibinfo
  {author} {\bibfnamefont{P.}~\bibnamefont{Mulders}},\ }%
  \bibfield{journal}{%
  \Doi{10.1007/JHEP07(2011)065}{\bibinfo {journal} {JHEP}}\ }%
  \textbf{\bibinfo {volume} {1107}},\ \bibinfo {pages} {065} (\bibinfo {year}
  {2011}),\ \Eprint{http://arxiv.org/abs/1105.4804}{arXiv:1105.4804 [hep-ph]}%
  \bibAnnoteFile{NoStop}{Buffing:2011mj}%
%%CITATION = ARXIV:1105.4804;%%
\bibitem{Buffing:2012sz}%
  \BibitemOpen
  \bibfield{author}{%
  \bibinfo {author} {\bibfnamefont{M.}~\bibnamefont{Buffing}}, \bibinfo
  {author} {\bibfnamefont{A.}~\bibnamefont{Mukherjee}},\ and\ \bibinfo {author}
  {\bibfnamefont{P.}~\bibnamefont{Mulders}},\ }%
  \bibfield{journal}{%
  \Doi{10.1103/PhysRevD.86.074030}{\bibinfo {journal} {Phys.Rev.}}\ }%
  \textbf{\bibinfo {volume} {D86}},\ \bibinfo {pages} {074030} (\bibinfo {year}
  {2012}),\ \Eprint{http://arxiv.org/abs/1207.3221}{arXiv:1207.3221 [hep-ph]}%
  \bibAnnoteFile{NoStop}{Buffing:2012sz}%
%%CITATION = ARXIV:1207.3221;%%
\bibitem{Buffing:2013kca}%
  \BibitemOpen
  \bibfield{author}{%
  \bibinfo {author} {\bibfnamefont{M.}~\bibnamefont{Buffing}}, \bibinfo
  {author} {\bibfnamefont{A.}~\bibnamefont{Mukherjee}},\ and\ \bibinfo {author}
  {\bibfnamefont{P.}~\bibnamefont{Mulders}}}%
   (\bibinfo {year} {2013}),\
  \Eprint{http://arxiv.org/abs/1306.5897}{arXiv:1306.5897 [hep-ph]}%
  \bibAnnoteFile{NoStop}{Buffing:2013kca}%
%%CITATION = ARXIV:1306.5897;%%
\bibitem{Lorce:2013pza}%
  \BibitemOpen
  \bibfield{author}{%
  \bibinfo {author} {\bibfnamefont{C.}~\bibnamefont{Lorc\'e}}\ and\ \bibinfo
  {author} {\bibfnamefont{B.}~\bibnamefont{Pasquini}}}%
   (\bibinfo {year} {2013}),\
  \Eprint{http://arxiv.org/abs/1307.4497}{arXiv:1307.4497 [hep-ph]}%
  \bibAnnoteFile{NoStop}{Lorce:2013pza}%
%%CITATION = ARXIV:1307.4497;%%
\bibitem{Musch:2011er}%
  \BibitemOpen
  \bibfield{author}{%
  \bibinfo {author} {\bibfnamefont{B.}~\bibnamefont{Musch}}, \bibinfo {author}
  {\bibfnamefont{P.}~\bibnamefont{Hagler}}, \bibinfo {author}
  {\bibfnamefont{M.}~\bibnamefont{Engelhardt}}, \bibinfo {author}
  {\bibfnamefont{J.}~\bibnamefont{Negele}},\ and\ \bibinfo {author}
  {\bibfnamefont{A.}~\bibnamefont{Schafer}},\ }%
  \bibfield{journal}{%
  \Doi{10.1103/PhysRevD.85.094510}{\bibinfo {journal} {Phys.Rev.}}\ }%
  \textbf{\bibinfo {volume} {D85}},\ \bibinfo {pages} {094510} (\bibinfo {year}
  {2012}),\ \Eprint{http://arxiv.org/abs/1111.4249}{arXiv:1111.4249 [hep-lat]}%
  \bibAnnoteFile{NoStop}{Musch:2011er}%
%%CITATION = ARXIV:1111.4249;%%
\bibitem{Bacchetta:2011gx}%
  \BibitemOpen
  \bibfield{author}{%
  \bibinfo {author} {\bibfnamefont{A.}~\bibnamefont{Bacchetta}}\ and\ \bibinfo
  {author} {\bibfnamefont{M.}~\bibnamefont{Radici}},\ }%
  \bibfield{journal}{%
  \Doi{10.1103/PhysRevLett.107.212001}{\bibinfo {journal} {Phys.Rev.Lett.}}\ }%
  \textbf{\bibinfo {volume} {107}},\ \bibinfo {pages} {212001} (\bibinfo {year}
  {2011}),\ \Eprint{http://arxiv.org/abs/1107.5755}{arXiv:1107.5755 [hep-ph]}%
  \bibAnnoteFile{NoStop}{Bacchetta:2011gx}%
%%CITATION = ARXIV:1107.5755;%%
\bibitem{Burkardt:2002ks}%
  \BibitemOpen
  \bibfield{author}{%
  \bibinfo {author} {\bibfnamefont{M.}~\bibnamefont{Burkardt}},\ }%
  \bibfield{journal}{%
  \Doi{10.1103/PhysRevD.66.114005}{\bibinfo {journal} {Phys.Rev.}}\ }%
  \textbf{\bibinfo {volume} {D66}},\ \bibinfo {pages} {114005} (\bibinfo {year}
  {2002}),\ \Eprint{http://arxiv.org/abs/hep-ph/0209179}{arXiv:hep-ph/0209179
  [hep-ph]}%
  \bibAnnoteFile{NoStop}{Burkardt:2002ks}%
%%CITATION = HEP-PH/0209179;%%
\bibitem{Burkardt:2003uw}%
  \BibitemOpen
  \bibfield{author}{%
  \bibinfo {author} {\bibfnamefont{M.}~\bibnamefont{Burkardt}},\ }%
  \bibfield{journal}{%
  \Doi{10.1016/j.nuclphysa.2004.02.008}{\bibinfo {journal} {Nucl.Phys.}}\ }%
  \textbf{\bibinfo {volume} {A735}},\ \bibinfo {pages} {185} (\bibinfo {year}
  {2004}),\ \Eprint{http://arxiv.org/abs/hep-ph/0302144}{arXiv:hep-ph/0302144
  [hep-ph]}%
  \bibAnnoteFile{NoStop}{Burkardt:2003uw}%
%%CITATION = HEP-PH/0302144;%%
\bibitem{Ma:1998ar}%
  \BibitemOpen
  \bibfield{author}{%
  \bibinfo {author} {\bibfnamefont{B.-Q.}\ \bibnamefont{Ma}}\ and\ \bibinfo
  {author} {\bibfnamefont{I.}~\bibnamefont{Schmidt}},\ }%
  \bibfield{journal}{%
  \Doi{10.1103/PhysRevD.58.096008}{\bibinfo {journal} {Phys.Rev.}}\ }%
  \textbf{\bibinfo {volume} {D58}},\ \bibinfo {pages} {096008} (\bibinfo {year}
  {1998}),\ \Eprint{http://arxiv.org/abs/hep-ph/9808202}{arXiv:hep-ph/9808202
  [hep-ph]}%
  \bibAnnoteFile{NoStop}{Ma:1998ar}%
%%CITATION = HEP-PH/9808202;%%
\bibitem{She:2009jq}%
  \BibitemOpen
  \bibfield{author}{%
  \bibinfo {author} {\bibfnamefont{J.}~\bibnamefont{She}}, \bibinfo {author}
  {\bibfnamefont{J.}~\bibnamefont{Zhu}},\ and\ \bibinfo {author}
  {\bibfnamefont{B.-Q.}\ \bibnamefont{Ma}},\ }%
  \bibfield{journal}{%
  \Doi{10.1103/PhysRevD.79.054008}{\bibinfo {journal} {Phys.Rev.}}\ }%
  \textbf{\bibinfo {volume} {D79}},\ \bibinfo {pages} {054008} (\bibinfo {year}
  {2009}),\ \Eprint{http://arxiv.org/abs/0902.3718}{arXiv:0902.3718 [hep-ph]}%
  \bibAnnoteFile{NoStop}{She:2009jq}%
%%CITATION = ARXIV:0902.3718;%%
\bibitem{Avakian:2010br}%
  \BibitemOpen
  \bibfield{author}{%
  \bibinfo {author} {\bibfnamefont{H.}~\bibnamefont{Avakian}}, \bibinfo
  {author} {\bibfnamefont{A.}~\bibnamefont{Efremov}}, \bibinfo {author}
  {\bibfnamefont{P.}~\bibnamefont{Schweitzer}},\ and\ \bibinfo {author}
  {\bibfnamefont{F.}~\bibnamefont{Yuan}},\ }%
  \bibfield{journal}{%
  \Doi{10.1103/PhysRevD.81.074035}{\bibinfo {journal} {Phys.Rev.}}\ }%
  \textbf{\bibinfo {volume} {D81}},\ \bibinfo {pages} {074035} (\bibinfo {year}
  {2010}),\ \Eprint{http://arxiv.org/abs/1001.5467}{arXiv:1001.5467 [hep-ph]}%
  \bibAnnoteFile{NoStop}{Avakian:2010br}%
%%CITATION = ARXIV:1001.5467;%%
\bibitem{Lorce:2011dv}%
  \BibitemOpen
  \bibfield{author}{%
  \bibinfo {author} {\bibfnamefont{C.}~\bibnamefont{Lorc\'e}}, \bibinfo
  {author} {\bibfnamefont{B.}~\bibnamefont{Pasquini}},\ and\ \bibinfo {author}
  {\bibfnamefont{M.}~\bibnamefont{Vanderhaeghen}},\ }%
  \bibfield{journal}{%
  \Doi{10.1007/JHEP05(2011)041}{\bibinfo {journal} {JHEP}}\ }%
  \textbf{\bibinfo {volume} {1105}},\ \bibinfo {pages} {041} (\bibinfo {year}
  {2011}),\ \Eprint{http://arxiv.org/abs/1102.4704}{arXiv:1102.4704 [hep-ph]}%
  \bibAnnoteFile{NoStop}{Lorce:2011dv}%
%%CITATION = ARXIV:1102.4704;%%
\bibitem{Melosh:1974cu}%
  \BibitemOpen
  \bibfield{author}{%
  \bibinfo {author} {\bibfnamefont{H.}~\bibnamefont{Melosh}},\ }%
  \bibfield{journal}{%
  \Doi{10.1103/PhysRevD.9.1095}{\bibinfo {journal} {Phys.Rev.}}\ }%
  \textbf{\bibinfo {volume} {D9}},\ \bibinfo {pages} {1095} (\bibinfo {year}
  {1974})%
  \bibAnnoteFile{NoStop}{Melosh:1974cu}%
%%CITATION = PHRVA,D9,1095;%%
\bibitem{Lorce:2011zta}%
  \BibitemOpen
  \bibfield{author}{%
  \bibinfo {author} {\bibfnamefont{C.}~\bibnamefont{Lorc\'e}}\ and\ \bibinfo
  {author} {\bibfnamefont{B.}~\bibnamefont{Pasquini}},\ }%
  \bibfield{journal}{%
  \Doi{10.1103/PhysRevD.84.034039}{\bibinfo {journal} {Phys.Rev.}}\ }%
  \textbf{\bibinfo {volume} {D84}},\ \bibinfo {pages} {034039} (\bibinfo {year}
  {2011}),\ \Eprint{http://arxiv.org/abs/1104.5651}{arXiv:1104.5651 [hep-ph]}%
  \bibAnnoteFile{NoStop}{Lorce:2011zta}%
%%CITATION = ARXIV:1104.5651;%%
\bibitem{Wakamatsu:2005vk}%
  \BibitemOpen
  \bibfield{author}{%
  \bibinfo {author} {\bibfnamefont{M.}~\bibnamefont{Wakamatsu}}\ and\ \bibinfo
  {author} {\bibfnamefont{H.}~\bibnamefont{Tsujimoto}},\ }%
  \bibfield{journal}{%
  \Doi{10.1103/PhysRevD.71.074001}{\bibinfo {journal} {Phys.Rev.}}\ }%
  \textbf{\bibinfo {volume} {D71}},\ \bibinfo {pages} {074001} (\bibinfo {year}
  {2005}),\ \Eprint{http://arxiv.org/abs/hep-ph/0502030}{arXiv:hep-ph/0502030
  [hep-ph]}%
  \bibAnnoteFile{NoStop}{Wakamatsu:2005vk}%
%%CITATION = HEP-PH/0502030;%%
\bibitem{Burkardt:2010he}%
  \BibitemOpen
  \bibfield{author}{%
  \bibinfo {author} {\bibfnamefont{M.}~\bibnamefont{Burkardt}}\ and\ \bibinfo
  {author} {\bibfnamefont{A.}~\bibnamefont{Jarrah}}}%
   (\bibinfo {year} {2010}),\
  \Eprint{http://arxiv.org/abs/1011.1041}{arXiv:1011.1041 [hep-ph]}%
  \bibAnnoteFile{NoStop}{Burkardt:2010he}%
%%CITATION = ARXIV:1011.1041;%%
\bibitem{Burkardt:2012sd}%
  \BibitemOpen
  \bibfield{author}{%
  \bibinfo {author} {\bibfnamefont{M.}~\bibnamefont{Burkardt}}}%
   (\bibinfo {year} {2012}),\
  \Eprint{http://arxiv.org/abs/1205.2916}{arXiv:1205.2916 [hep-ph]}%
  \bibAnnoteFile{NoStop}{Burkardt:2012sd}%
%%CITATION = ARXIV:1205.2916;%%
\bibitem{Goeke:2007fp}%
  \BibitemOpen
  \bibfield{author}{%
  \bibinfo {author} {\bibfnamefont{K.}~\bibnamefont{Goeke}}, \bibinfo {author}
  {\bibfnamefont{J.}~\bibnamefont{Grabis}}, \bibinfo {author}
  {\bibfnamefont{J.}~\bibnamefont{Ossmann}}, \bibinfo {author}
  {\bibfnamefont{M.}~\bibnamefont{Polyakov}}, \bibinfo {author}
  {\bibfnamefont{P.}~\bibnamefont{Schweitzer}}, \emph{et~al.},\ }%
  \bibfield{journal}{%
  \Doi{10.1103/PhysRevD.75.094021}{\bibinfo {journal} {Phys.Rev.}}\ }%
  \textbf{\bibinfo {volume} {D75}},\ \bibinfo {pages} {094021} (\bibinfo {year}
  {2007}),\ \Eprint{http://arxiv.org/abs/hep-ph/0702030}{arXiv:hep-ph/0702030
  [hep-ph]}%
  \bibAnnoteFile{NoStop}{Goeke:2007fp}%
%%CITATION = HEP-PH/0702030;%%
\bibitem{Hoodbhoy:1998yb}%
  \BibitemOpen
  \bibfield{author}{%
  \bibinfo {author} {\bibfnamefont{P.}~\bibnamefont{Hoodbhoy}}, \bibinfo
  {author} {\bibfnamefont{X.-D.}\ \bibnamefont{Ji}},\ and\ \bibinfo {author}
  {\bibfnamefont{W.}~\bibnamefont{Lu}},\ }%
  \bibfield{journal}{%
  \Doi{10.1103/PhysRevD.59.014013}{\bibinfo {journal} {Phys.Rev.}}\ }%
  \textbf{\bibinfo {volume} {D59}},\ \bibinfo {pages} {014013} (\bibinfo {year}
  {1999}),\ \Eprint{http://arxiv.org/abs/hep-ph/9804337}{arXiv:hep-ph/9804337
  [hep-ph]}%
  \bibAnnoteFile{NoStop}{Hoodbhoy:1998yb}%
%%CITATION = HEP-PH/9804337;%%
\bibitem{Adhikari:2013ima}%
  \BibitemOpen
  \bibfield{author}{%
  \bibinfo {author} {\bibfnamefont{L.}~\bibnamefont{Adhikari}}\ and\ \bibinfo
  {author} {\bibfnamefont{M.}~\bibnamefont{Burkardt}}}%
   (\bibinfo {year} {2013}),\
  \Eprint{http://arxiv.org/abs/1307.2625}{arXiv:1307.2625 [hep-ph]}%
  \bibAnnoteFile{NoStop}{Adhikari:2013ima}%
%%CITATION = ARXIV:1307.2625;%%
\end{thebibliography}%

\end{document}